\documentclass{aa}  
\usepackage{graphicx}
\usepackage{txfonts}
\usepackage{float}
\usepackage{subfig}
\usepackage{xcolor}
\usepackage[normalem]{ulem}
\begin{document}

   \titlerunning{Extragalactic FXRT Candidates Discovered by \emph{Chandra} (2000--2014)}
   \authorrunning{Quirola-V\'asquez et al.}

   \title{Extragalactic fast \hbox{X-ray} transient candidates discovered by \emph{Chandra} (2000--2014)}

%   \subtitle{Extragalactic \hbox{X-ray} transients}

   \author{J. Quirola-V\'asquez
          \inst{1,2,3},
%          \and \fnmsep
          F. E. Bauer\inst{1,2,4},
          P.~G. Jonker\inst{3,5},
          W. N. Brandt\inst{6,7,8},
          G. Yang\inst{9,10},
          A. J. Levan\inst{3,11},
          Y. Q. Xue\inst{12,13},
          D. Eappachen\inst{5,3},
          X. C. Zheng\inst{14},
         \and
          B. Luo\inst{15,16}
          }
   \institute{Instituto de Astrof\'isica, Pontificia Universidad Cat\'olica de Chile, Casilla 306, Santiago 22, Chile\\
              \email{jquirola@astro.puc.cl}
         \and
             Millennium Institute of Astrophysics (MAS), Nuncio Monse$\tilde{\rm n}$or S\'otero Sanz 100, Providencia, Santiago, Chile
        \and
             Department of Astrophysics/IMAPP, Radboud University, P.O. Box 9010, 6500 GL, Nijmegen, The Netherlands
        \and
            Space Science Institute, 4750 Walnut Street, Suite 205, Boulder, Colorado 80301, USA
        \and
            SRON Netherlands Institute for Space Research, Niels Bohrweg 4, 2333 CA Leiden, The Netherlands
        \and
            Department of Astronomy \& Astrophysics, 525 Davey Laboratory, The Pennsylvania State University, University Park, PA 16802, USA
        \and
            Institute for Gravitation and the Cosmos, The Pennsylvania State University, University Park, PA 16802, USA
        \and
            Department of Physics, 104 Davey Laboratory, The Pennsylvania State University, University Park, PA 16802, USA
        \and
            Texas A\&M University, Physics and Astronomy, 4242 TAMU College Station, TX, 77843-4242, USA
        \and
            George P.\ and Cynthia Woods Mitchell Institute for Fundamental Physics and Astronomy, Texas A\&M University, College Station, TX, 77843-4242 USA
        \and
            Department of Physics, University of Warwick, Coventry, CV4 7AL, UK
        \and
            CAS Key Laboratory for Research in Galaxies and Cosmology, Department of Astronomy, University of Science and Technology of China, Hefei 230026, China
        \and
            School of Astronomy and Space Science, University of Science and Technology of China, Hefei 230026, China
        \and
            Leiden Observatory, Leiden University, PO Box 9513, NL-2300 RA, Leiden, the Netherlands
        \and
            School of Astronomy and Space Science, Nanjing University
        \and
            Key Laboratory of Modern Astronomy and Astrophysics (Nanjing University), Ministry of Education, Nanjing 210093, China
             }
   \date{Received January 05, 2022; accepted April 11, 2022}

% \abstract{}{}{}{}{} 
% 5 {} token are mandatory

  \abstract
   {Extragalactic fast X-ray transients (FXRTs) are short flashes of X-ray photons of unknown origin that last a few seconds to hours.}
    {Our ignorance about their physical mechanisms and progenitor systems is due in part to the lack of clear multiwavelength counterparts in most cases, because FXRTs have only been identified serendipitously.}
   {We develop a systematic search for FXRTs in the \emph{Chandra} Source Catalog (Data Release 2.0; 169.6\,Ms over 592.4~deg$^{2}$, using only observations with $|b|{>}10^{\circ}$ and before 2015), using a straightforward \hbox{X-ray} flare search algorithm  and incorporating various multiwavelength constraints to rule out Galactic contamination and characterize the candidates.}
    {We report the detection of 14 FXRT candidates from a parent sample of 214,701 sources. Candidates have peak 0.5--7 keV fluxes between 1$\times$10$^{-13}$ and 2$\times$10$^{-10}$~erg~cm$^{-2}$~s$^{-1}$ and $T_{90}$ values from 4 to 48~ks. 
    %We characterize their \hbox{X-ray} light curves and spectra. 
    The sample can be subdivided into two groups: six {"nearby"} FXRTs that occurred within $d{\lesssim}$100~Mpc and eight {"distant"} FXRTs with likely redshifts $\gtrsim$0.1. Three {distant} FXRT candidates exhibit light curves with a plateau (${\approx}$1--3~ks duration) followed by a power-law decay and X-ray spectral softening, similar to what was observed for the previously reported FXRT CDF-S~XT2, a proposed magnetar-powered binary neutron star merger event. After applying completeness corrections, we calculate event rates for the {nearby} and {distant} samples of 53.7$_{-15.1}^{+22.6}$ and 28.2$_{-6.9}^{+9.8}$~deg$^{-2}$~yr$^{-1}$, respectively.}
   {This novel sample of \emph{Chandra}-detected extragalactic FXRT candidates, although modest in size, breaks new ground in terms of characterizing the diverse properties, nature, and possible progenitors of these enigmatic events.}

    
   \maketitle
%
%-------------------------------------------------------------------

\section{Introduction}

The \emph{Chandra}, \emph{Swift}, and X-ray Multi-mirror Mission Newton (\emph{XMM-Newton}) observatories have accumulated sensitive 0.5--7~keV imaging observations over the past two decades that cover a sizeable fraction of the sky despite their relatively narrow fields of view. This has enabled the serendipitous discovery and characterization of several novel faint extragalactic transients \citep[e.g.,][]{Soderberg2008,Jonker2013,Glennie2015,Irwin2016,Bauer2017,Lin2018,Lin2019,Xue2019,Alp2020,Novara2020,Lin2020,Ide2020,Pastor2020,Lin2021,Sazonov2021,Lin2022}. The high angular resolution afforded by these space observatories has been critical for associating counterparts\footnote{We use the term ``counterpart'' throughout to denote the multiwavelength detection of emission from the transient.} (or lack thereof) and host galaxies with these transients, and hence elucidating their astrophysical nature.

In general, fast X-ray transients (FXRTs) produce short flashes of X-ray emission with durations from a few minutes to hours. Among the few extragalactic FXRTs that have been identified to date (mainly from systematic searches of serendipitous detections), in only one case, X-ray transient (XRT)\,080109/SN\,2008D \citep{Mazzali2008,Soderberg2008,Modjaz2009}\footnote{The most favored model for XRT\,080109/SN\,2008D is a breakout from a wind (regarding the breakout from the stellar surface), which changes the expected X-ray luminosity \citep[e.g.,][]{Balberg2011}.}, has it been possible to identify a multiwavelength counterpart after the initial detection. The most stringent limits come from deep optical Very Large Telescope imaging serendipitously acquired 80~minutes after the onset of XRT\,141001 \citep[$m_R{>}$25.7~AB~mag;][]{Bauer2017}. Moreover, only a few FXRTs have had clear host-galaxy associations, and even fewer have firm distance constraints \citep[e.g.,][]{Soderberg2008,Irwin2016,Bauer2017,Xue2019}. Hence, it is not trivial to discern their energetics and distance scale or, by extension, their physical origin.

Several scenarios could explain the X-ray flares of extragalactic FXRTs, including the following four. First, in nearby galaxies, X-ray binaries (XRBs) -- which includes ultra-luminous X-ray sources (ULXs) and quasi-periodic oscillations  -- soft gamma repeaters (SGRs), quasi-periodic eruptions, and anomalous X-ray pulsars (AXPs) are possible explanations of FXRTs with $L_{\rm X,peak}{\lesssim}$10$^{42}$~erg~s$^{-1}$ (\citealp[][]{Colbert1999,Kaaret2006,Woods2006,Miniutti2019}; and references therein).

A second scenario involves shock breakouts (SBOs; $L_{\rm X,peak}{\approx}$10$^{42}$--10$^{47}$~erg~s$^{-1}$) from a core-collapse supernova (CC-SN), whereby the X-ray emission is generated from the breakout of the supernova explosion shock once it crosses the surface of an evolved star \citep[e.g.,][]{Soderberg2008,Nakar2010,Waxman2017,Novara2020, Alp2020}. 
Third are tidal disruption events (TDEs; $L_{\rm X,peak}{\approx}$10$^{42}$--10$^{50}$~erg~s$^{-1}$ considering jetted emission) that involve a white dwarf (WD) and an intermediate-mass black hole (IMBH), whereby X-rays are produced by the tidal disruption and subsequent accretion of the compact WD in the gravitational field of the IMBH \citep[e.g.,][]{Jonker2013,Glennie2015}. 
The fourth is mergers of binary neutron stars \citep[BNSs; $L_{\rm X,peak}{\approx}$10$^{47}$--10$^{51}$~erg~s$^{-1}$ considering jetted emission; e.g.,][]{Dai2018,Jonker2013,Fong2015,Bauer2017,Xue2019}, whereby the X-rays are created by the accretion of fallback material onto the remnant magnetar or black hole (BH).

It has been argued that some of these FXRTs can be related to either long or short gamma-ray bursts (LGRBs or SGRBs, respectively) observed off-axis \cite[e.g.,][]{Jonker2013,Bauer2017,Xue2019,Alp2020}. \citet{Zhang2013} proposed a type of XRT associated with the merger product of a BNS, a rapidly spinning magnetar, where our line of sight is offset from the jet of an SGRB. Soon thereafter, \citet{Luo2014} and \citet{Zheng2017} identified two new unusual FXRTs in the 7~Ms Chandra Deep Field-South (CDF-S) data set, XRT~141001 and XRT~150321, denoted ``CDF-S~XT1'' and ``CDF-S~XT2.'' These two FXRTs were studied later in detail by \citet{Bauer2017} and \citet{Xue2019}, respectively. 
In the case of CDF-S~XT2, its multiwavelength constraints and host galaxy properties are consistent with the expected features of off-axis SGRBs \citep{Xue2019}, although other possibilities cannot be completely ruled out \citep[e.g., a TDE origin;][]{Peng2019}. CDF-S~XT2 is particularly intriguing because it exhibits a flat, extended \hbox{X-ray} light curve that suggests a magnetar wind origin \citep{Sun2019,Xiao2019,Lu2019}, similar to GRB~160821B \citep{Troja2019} and others and in line with the aforementioned predictions of \citet{Zhang2013}. The X-ray afterglows of gamma-ray bursts (GRBs) also show similar plateaus in their light curves \citep[e.g.,][]{Lyons2010,Rowlinson2013,Yi2014}, suggestive of a central engine related to a magnetar wind or an accreting BH \citep{Troja2007,Li2018}.

On the other hand, CDF-S~XT1 could be associated with a few possible scenarios: $(i)$ an ``orphan'' \hbox{X-ray} afterglow from an off-axis SGRB with weak optical emission \citep{Bauer2017,Sarin2021}, $(ii)$ a low-luminosity GRB at high redshift with no prompt gamma-ray emission below ${\sim}$20~keV rest frame \citep{Bauer2017}, or $(iii)$ a highly beamed IMBH--WD TDE \citep{Bauer2017,Peng2019}. More recently, \citet{Sun2019} proposed a possible origin as a magnetar remnant of a neutron star merger, viewed at a larger off-axis angle than CDF-S~XT2 and strongly obscured by ejecta material at early times. While none of these scenarios completely explain all observed properties, the large redshift uncertainty makes it difficult to discard them outright. Notably, the event rate of CDF-S~XT1-like events is comparable to those of orphan and low-luminosity GRBs, as well as TDEs, implying an untapped regime for a known transient class or a new type of variable phenomenon \citep{Bauer2017}.

In order to understand if, and if so how, FXRTs, GRBs, and gravitational wave (GW) events \citep[such as GW\,170817;][]{Abbott2017a,Nakar2020,Margutti2021,Hajela2021} are related, we need to enlarge the sample of FXRTs. To this end, \citet{Yang2019} conducted a systematic search for CDF-S~XT1- and CDF-S~XT2-like objects in ${\sim}$19~Ms of \emph{Chandra} blank-field survey data with good ancillary imaging. They constrained the event rate systematically but unfortunately found no new FXRTs. The discovery, confirmation, and characterization of more FXRTs and stricter limits on their number density can place valuable constraints on the unknown electromagnetic (EM) properties of several families of astronomical transients.

In this paper we extend the efforts of \citet{Yang2019} with a search of the entire \emph{Chandra Source Catalog 2.0} \citep[CSC2;][]{Evans2010}, identifying 14 extragalactic FXRTs, of which at least three share similar properties to CDF-S~XT2 and may be related with off-axis GRBs. We recover five events previously identified and classified as FXRTs by \citet{Jonker2013}, \citet{Glennie2015}, \citet{Bauer2017}, and \citet{Lin2019,Lin2022}. \\

This manuscript is organized as follows. We explain the methodology and selection criteria in Sect. \ref{sec:methodology}. We present the results of the search and the cross-match with other catalogs in Sect. \ref{sec:results}, a spectral and timing analysis of our final candidates in Sect. \ref{sec:time_spectra_prop}, and the properties of the identified potential host galaxies in Sect. \ref{sec:counterpart_SED}. In Sect. \ref{sec:flux} we discuss possible interpretations of some FXRTs and provide a comparison with other transients. We derive local and volumetric rates for the FXRTs in Sect. \ref{sec:rates} and the expected number in current and future X-ray missions. Finally, we present final comments and conclusions in Sect. \ref{sec:conclusion}.\\

Throughout the paper, a concordance cosmology with parameters $H_0{=}$70~km~s$^{-1}$~Mpc$^{-1}$, $\Omega_M{=}$0.30, and $\Omega_\Lambda{=}$0.70 is adopted. All magnitudes are quoted in the AB system.

\section{Methodology and sample selection}\label{sec:methodology}

We describe below our search algorithm for FXRT candidates in individual \emph{Chandra} exposures (Sect. \ref{sec:algorithm}), CSC2 data selection criteria (Sect. \ref{sec:data}), light curve extraction methodology (Sect. \ref{sec:LC}), initial candidate results (Sect. \ref{sec:initial_results}) and additional criteria to filter non-transient and Galactic-stellar events to clean our sample (Sect. \ref{sec:purity_criteria}), respectively. Finally, we explore tentative related EM sources using different catalogs (Sect. \ref{sec:results}).

\begin{figure*}
    \centering
    \includegraphics[scale=0.7]{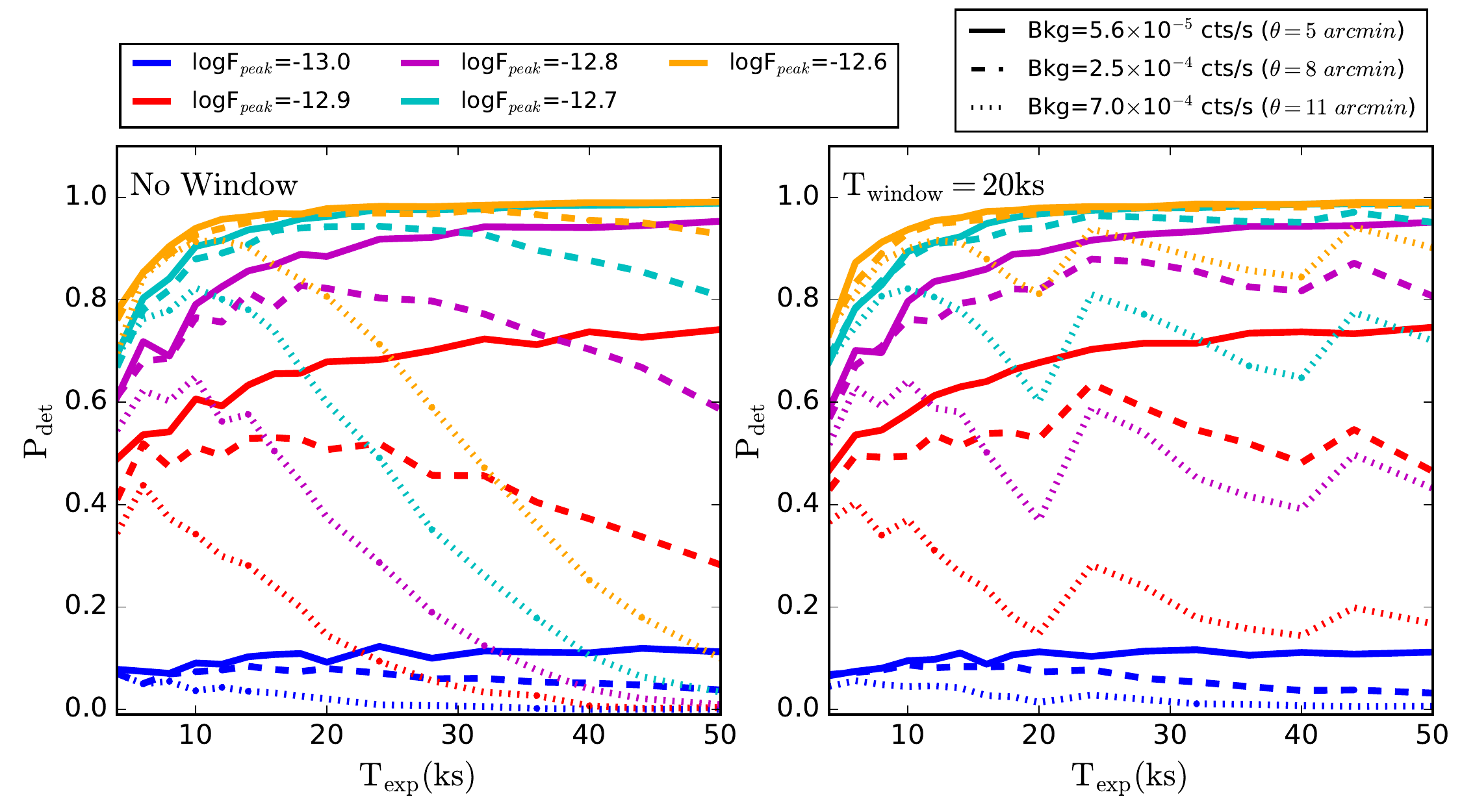}
    \vspace{-0.4cm}
    \caption{Detection probability ($P_{\rm det}$) as a function of the \emph{Chandra} exposure time ($T_{\rm exp}$) for typical instrumental off-axis angles of 5\farcm0 (\emph{solid lines}), 8\farcm0 (\emph{dashed lines}), and 11\farcm0 (\emph{dotted lines}). Different colors indicate different peak fluxes,  $\log[F_{\rm peak}~({\rm cgs})]$, as labeled (cgs units). The \emph{left} and \emph{right} panels show the probability assuming null and 20~ks windows, respectively (see Sect. \ref{sec:algorithm} for details).}
    \label{fig:LC_examples}
\end{figure*}

\subsection{Algorithm for transient-candidate selection} \label{sec:algorithm}

We adopt the algorithm presented in \citet{Yang2019} (see their Sect.~2.1 for more details), with some modifications to extend it to larger instrumental off-axis angles (as related to the position of the detector aimpoint) and/or higher background levels, which we discuss below. This method depends on the total ($N_{\rm tot}$) and background ($N_{\rm bkg}$) counts of the event, working on an unbinned \emph{Chandra} light curve (this is advantageous because it does not depend on how the light curve is built). Based on simulations, \citet{Yang2019} adopt an identification efficiency requirement [${\gtrsim}$~90\% for events with $\log(F_{\rm peak})~{>}-12.6$] located at ${<}$~8\farcm0. They enforce this instrumental off-axis angle limit because \emph{Chandra}'s detection sensitivity (as measured by, e.g., effective area and point-spread-function size) drops significantly beyond this limit \citep{Vito2016,Yang2016}.

The algorithm is split into two passes of the same light curve. Pass 1 calculates the total number of counts $N_1$ and $N_2$ in the two halves of the light curve at $t$=$(t_s,t_m)$ and $t$=$(t_m,t_e)$ respectively, where $t_s$ and $t_e$ are the {start} and {end} times of the \emph{Chandra} exposure, respectively, while $t_m$=$(t_s+t_e)/2$ is the {midpoint} of the observation. The method selects a source in an observation as a transient candidate if it satisfies all of the following criteria: $(i)$ $N_{\rm tot}$ is larger than the 5$\sigma$ Poisson upper limit of $N_{\rm bkg}$;
$(ii)$ $N_1$ and $N_2$ are statistically different at a ${>}$~4$\sigma$ significance level; and $(iii)$ $N_1~{>}$~5${\times}N_2$ or $N_2~{>}$~5${\times}N_1$.

Criterion $(i)$ rules out faint sources with low signal-to-noise (S/N) and helps to avoid false detections caused by rare background flares.
Criterion $(ii)$ selects sources that have significantly different counts between the first-half and second-half exposures. This comparison is made via an E-test \citep{Krishnamoorthy2004}, which assesses whether N$_1$ and N$_2$ are drawn from the same Poisson distribution, factoring in statistical fluctuations. Both criteria $(i)$ and $(ii)$ are based on statistical significance, and they chose high S/N sources with significant variability. On the other hand, criterion $(iii)$ permits events to be discarded, such as active galactic nuclei (AGNs) with a strong stochastic variability, requiring that the flux-variation amplitude be large.

The above sequence ({Pass 1}), however, will not efficiently select transients that occur around $t{\approx}t_m$, because $N_1$ and $N_2$ may have a similar number of counts. Thus, a second sequence ({Pass 2}) is used to account for transient events that occur near $t_m$, whereby the number of counts $N_1^\prime$ and $N_2^\prime$ within windows around the {edges} and {middle} of the light curve are computed, respectively. {Pass 2} identifies transient candidates in observations that satisfy all of the following criteria: $(i)$ $N_{\rm tot}$ is larger than the 5$\sigma$ Poisson upper limit of $N_{\rm bkg}$;
$(ii)$ $N_1^\prime$ and $N_2^\prime$ are statistically different at a ${>}$~4$\sigma$ significance level; and $(iii)$ $N_1^\prime~{>}$~5$\times$$N_2^\prime$ or $N_2^\prime~{>}$~5$\times$$N_1^\prime$.
%\end{enumerate}

This algorithm depends strongly on the background event rate and the degradation of the \emph{Chandra} point spread function (PSF) at high instrumental off-axis angles. To analyze the performance of the method, we simulate the {detection probability} ($P_{\rm det}$) of CDF-S~XT1 and CDF-S~XT2-like events at energies 0.5--7.0~keV as a function of the \emph{Chandra} exposure time ($T_{\rm exp}$). We consider the following conditions with instrumental  off-axis angles of 5\farcm0/8\farcm0/11\farcm0: a fiducial light-curve model similar to CDF-S~XT1 and CDF-S~XT2 \citep[identical to that used by][see their Sect.~2.2.1]{Yang2019}, taking into account their timing and spectral properties (power-law with photon index of $\Gamma{=}$~1.7), a conversion between $F_{\rm peak}$ and total net counts of $N_{\rm net}{\approx}$~1.6${\times}$10$^{14}F_{\rm peak}$~cts, aperture background count rates of 5.6${\times}$10$^{-5}$, 2.5${\times}$10$^{-4}$, and 7.0${\times}$10$^{-4}$~cts~s$^{-1}$ for 5\farcm0, 8\farcm0 and 11\farcm0, respectively, and $\log(F_{\rm peak})$ from $-13.0$ to $-12.6$. The ratio of aperture background count rates at 5\farcm0, 8\farcm0 and 11\farcm0 instrumental off-axis angles are ${\approx}$~9.5, 42, and 119 times larger than at 0\farcm5, respectively, highlighting the importance of defining the algorithm's effectiveness at different locations across \emph{Chandra's} field-of-view (FoV). For all simulations, we adopt as the background count rate the median value from the \emph{Chandra} Deep Field North/South surveys \citep{Xue2016,Luo2017,Yang2019}. 

Figure~\ref{fig:LC_examples}, \emph{left panel}, shows the detection probability $P_{\rm det}$ as a function of $T_{\rm exp}$, assuming instrumental off-axis angles of 5\farcm0 (\emph{solid lines}, representative of $\sim$20th--30th percentile), 8\farcm0 (\emph{dashed lines}, representative of $\sim$50th-70th percentile), or 11\farcm0 (\emph{dotted lines}, representative of worst case $\sim$100th percentile). It is clear that $P_{\rm det}$ decreases substantially for events at 8\farcm0 (by 20--50\%) and 11\farcm0 (by 50--100\%) at $\log(F_{\rm peak})~{\lesssim}-12.7$ (for reference $\log(F_{\rm peak})~{\lesssim}-12.7$ equates to ${\lesssim}$~32~counts for a CDF-S XT1-like event), especially at $T_{\rm exp}~{\gtrsim}$~30~ks. Thus, candidates with large instrumental  off-axis angles, which incur higher background levels, subsequently have worse flux sensitivity limits using this algorithm.

To mitigate this problem, we chop each light curve into segments of 20~ks ($T_{\rm window}{=}$~20~ks), and carry out {Passes 1} and {2} separately on each window. This reduces the integrated number of background counts and thus enables identification of fainter events at larger instrumental  off-axis angles. To maintain efficient selection of transients across the gaps between windows, we sequence through the entire light curve in three iterations: a forward division into 20~ks windows plus a remainder window, a backward division into 20~ks windows plus a remainder, and finally a forward division after a 10~ks shift into 20~ks windows plus a remainder window and the initial 10~ks window. As an example, for a 45~ks exposure, we divide it as follows: one iteration with windows of $T_{\rm exp}{=}$~20, 20, and 5~ks; another iteration with windows of $T_{\rm exp}{=}$~5, 20, and 20~ks, and a final iteration with windows of $T_{\rm exp}{=}$10, 20, 15~ks. Then for each separate window of 0--20 ks duration, we apply Passes 1 and 2. This {window time} is well matched to the expected durations for CDF-S~XT1 and CDF-S~XT2, which have $T_{90}$ of 5.0$_{-0.3}^{+4.2}$ and 11.1$_{-0.6}^{+0.4}$~ks, respectively; here, $T_{90}$ measures the time over which the event emits the central 90\% (i.e., from 5\% to 95\%) of the total measured number of counts \citep{Bauer2017, Xue2019}. We explored how $P_{\rm det}$ changes considering two other window sizes, $T_{\rm window}{=}$~10 and 25~ks. In the case of $T_{\rm exp}{=}$~10~ks, $P_{\rm det}$ decreases by ${\approx}$30\% at $T_{\rm exp}{=}$~10~ks, since the window size starts to become comparable to or smaller than the $T_{90}$ values of the simulated light curves. For $T_{\rm window}{=}$~25~ks, $P_{\rm det}$ does not change dramatically. 

This additional modification to the algorithm of \citet{Yang2019} (they only chopped observations with exposures longer than 50~ks) is crucial because it allows instrumental  off-axis FXRTs to be detected to fainter flux limits and across \emph{Chandra's} entire FoV. Indeed, FXRTs previously published by \citet{Jonker2013} and \citet{Glennie2015} were identified at large instrumental  off-axis angles (13\farcm0). Figure~\ref{fig:LC_examples}, \emph{right panel}, shows the detection probability $P_{\rm det}$ considering $T_{\rm window}{=}$~20~ks (but otherwise the same conditions as in the previous simulations). The $P_{\rm det}$ clearly improves by up to several tens of percent, especially for events fainter than $\log(F_{\rm peak}){\lesssim}~-12.7$ and $T_{\rm exp}{\gtrsim}$~20--30~ks. We note that \citet{Yang2019} adopted limits of $\log(F_{\rm peak})~{\gtrsim}~-12.6$, instrumental off-axis angles ${\lesssim}$8\farcm0, and $T_{\rm window}{\le}~50$~ks. With the above modification, we increase the chance to recover new FXRTs even at large instrumental off-axis (or high background levels), albeit at lower sensitivity and completeness thresholds.

We confirmed that our algorithm detects FXRTs with different light-curve shapes such as XRT~110103 \citep[where the flux-to-counts conversion factor for this transient is $N_{\rm net}{\approx}$~3.2$\times$10$^{12}F_{\rm peak}$~cts;][]{Yang2019}. For instance, those of CDF-S~XT1 and CDF-S~XT2, with main peak durations of ${\approx}$~5--11\,ks, are quite distinct from the events found by \citet{Jonker2013} and \citet{Glennie2015} with peak emission durations of only ${\approx}$~0.1-0.2~ks. Importantly, our algorithm successfully recovered all these events, and thus is flexible enough to recognize FXRTs with different light-curve shapes. We stress that this is a key advantage compared to matched filter techniques that assume an underlying model profile.

In this work, the false rate of spurious detections is inherited from the CSC2, which serves as our input catalog. The CSC2 includes real X-ray sources detected with flux estimates that are at least 3 times their estimated 1$\sigma$ uncertainties in at least one energy band (between 0.2--7.0~keV), while maintaining the number of spurious sources at a level of ${\lesssim}$1 false source per field for a 100~ks observation \citep{Evans2010,Evans2019,Evans2020}. Although this number seems small, spurious events could be an important source of contamination, especially for events without a clear optical or near-infrared (NIR) association. To avoid this problem, we adopt a more restrictive 5$\sigma$ cut, which should serve to remove all truly spurious sources (see above). Moreover, we make a final visual inspection to reject potential  spurious FXRTs that appear "constant" and associated with known diffuse/extended sources, or vary in the same way that the background varies with time (see Sect. \ref{sec:inst_effects}). To summarize, our strict cuts and visual review should produce a final sample that is largely free from spurious contamination.

\begin{figure}
    \centering
    \includegraphics[scale=0.65]{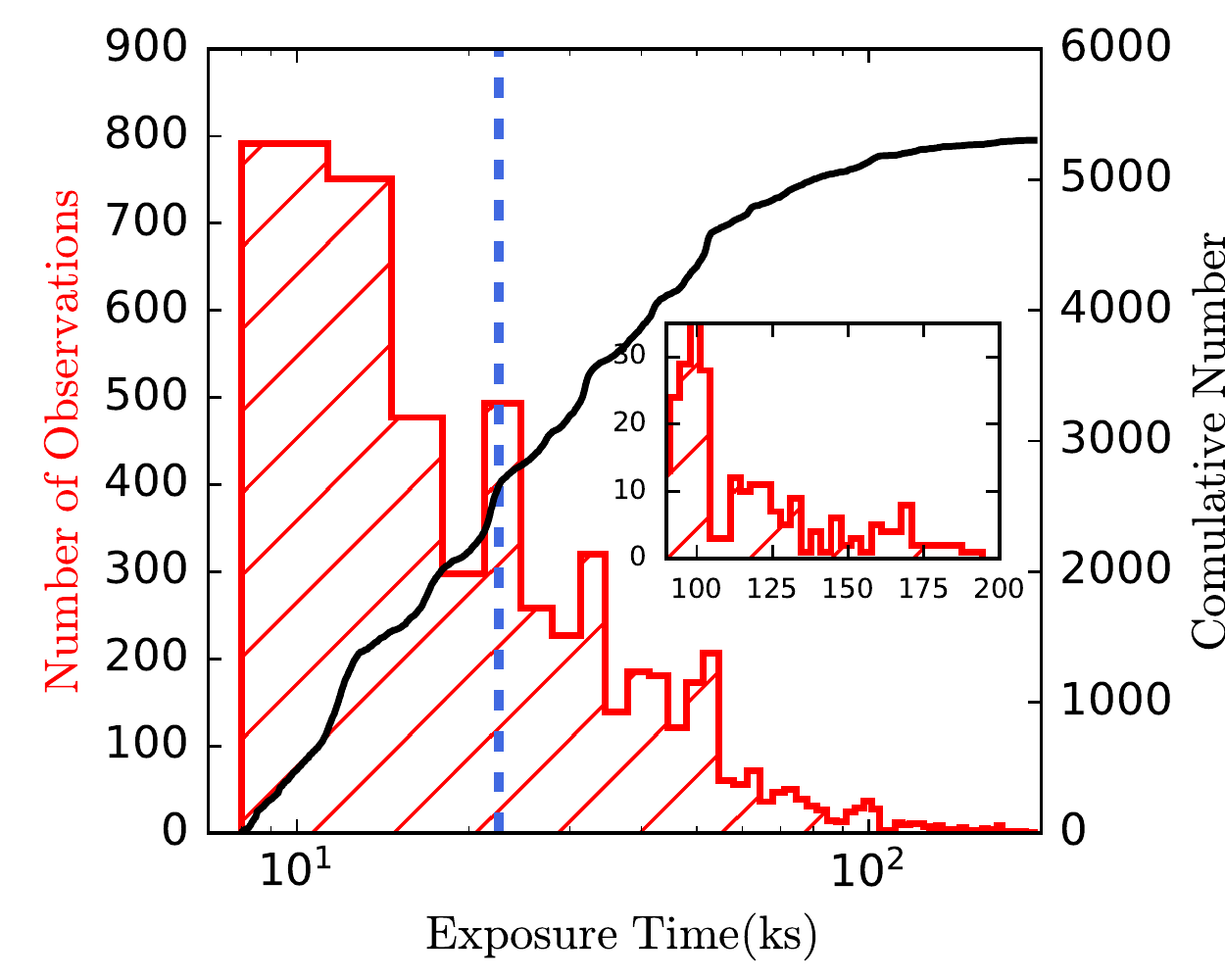}
    \vspace{-0.3cm}
    \caption{Histogram (\emph{red}; \emph{left} Y axis) and cumulative (\emph{black}; \emph{right} Y axis) distributions of the exposure time of the 5303 \emph{Chandra} observations used in this work. The inset provides a zoomed-in view to show the high-exposure-time tail of the distribution. The \emph{vertical dashed blue line} indicates the median exposure time (${=}~32$~ks) of the total sample. We adopt an exposure time of 8~ks as a lower bound due to the strongly decreasing probability of distinguishing FXRTs in short exposures.}
    \label{fig:OBS_features}
\end{figure}

\subsection{Data selection}\label{sec:data}

To extend previous efforts to search for FXRTs, we conducted a search through the CSC2,\footnote{https://cxc.harvard.edu/csc/} which provides uniformly extracted properties for 317,167 unique compact and extended \hbox{X-ray} sources (928,280 individual observation detections) identified in 10,382 \emph{Chandra} Advanced CCD Imaging Spectrometer (ACIS) and  High Resolution Camera (HRC-I) imaging observations released publicly through the end of 2014. The sensitivity limit for compact sources in CSC2 is ${\sim}$5 net counts (a factor of ${\geq}$2 better than the previous catalog release). For uniformity, we consider only ACIS observations in the energy range 0.5--7.0~keV, noting that HRC-I observations comprise only a few percent of the overall observations and have a poorer and softer response and limited energy resolution compared with the ACIS detectors.

The CSC2 database includes a wide variety of astrophysical objects, from galaxy clusters to stellar objects, although the CSC2 does not provide detailed source classifications. To this end, we apply the criteria explained in Sect. \ref{sec:algorithm} to select FXRT candidates, while the criteria explained below (Sect. \ref{sec:purity_criteria}) are chosen in order to discard objects that are considered contamination to our search. Given the extragalactic nature of the FXRTs CDF-S~XT1 and CDF-S~XT2 and the high contamination rate from flaring stars (e.g., \citealp{Yang2019} recovered CDF-S~XT1/XT2 but otherwise only found stellar flares in 19~Ms of data), we limit our initial light-curve search to CSC2 sources with Galactic latitudes $|b|{>}~10$~deg. A secondary benefit of considering objects with $|b|{>}~10$~deg is that it helps to minimize the effects of Galactic extinction in characterizing the spectral properties of our candidates. From the previous search developed by \citet{Yang2019}, the probability of detecting FXRTs such as CDF-S~XT1 or CDF-S~XT2 decreases dramatically in observations with exposure times ${<}$~8~ks (similar to our case, where $P_{\rm det}{\lesssim}$~0.9 for events $\log(F_{\rm peak}){\lesssim}-12.7$; see Fig.~\ref{fig:LC_examples}). Therefore, we exclude such short observations from further study in order to limit uncertainties associated with large completeness corrections when estimating the event rate (see Sect. \ref{sec:rates}). The above two criteria yield a sample of 214,701 \hbox{X-ray} sources detected within 5303 \emph{Chandra} observations, equating to ${\approx}$169.6~Ms of exposure over ${\approx}$592.4~deg$^{2}$; this is roughly nine times more than explored in \citet{Yang2019}.

To facilitate our search, we use the full-field per-observation \texttt{event files} available from the CSC2 data products\footnote{http://cxc.harvard.edu/csc2/data\_products} along with the detection properties provided in the CSC2 catalog \citep{Evans2010}. Figure~\ref{fig:OBS_features} shows the cumulative and histogram distributions of the \emph{Chandra} observations used in this work as a function of exposure time. 

\begin{figure*}
    \centering
    \includegraphics[scale=0.65]{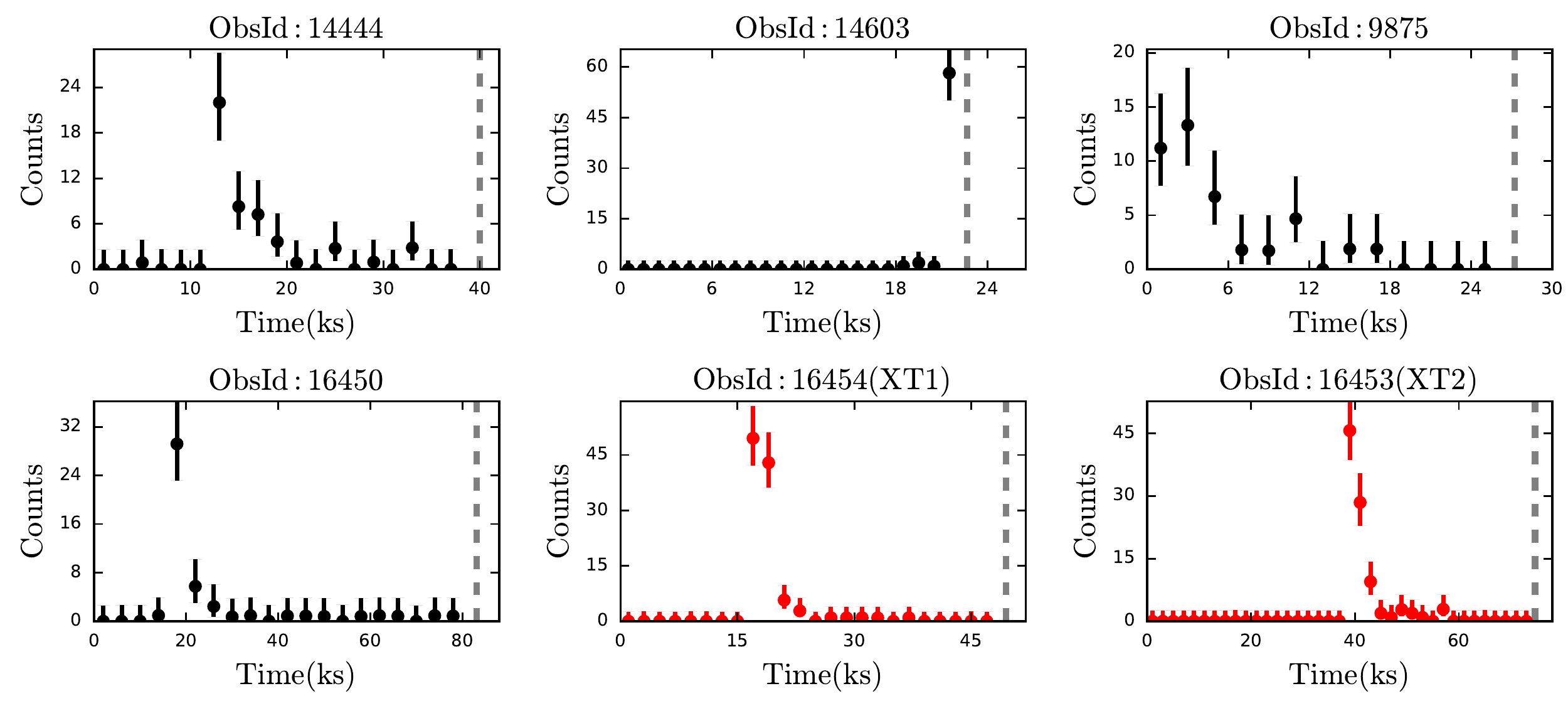}
    \vspace{-0.5cm}
    \caption{\hbox{X-ray} light curves extracted as described in Sect. \ref{sec:LC} and identified via our algorithm described in Sect. \ref{sec:algorithm}. The four light curves in black denote randomly selected sources from initial FXRTs found in the CSC2. For comparison, we show in red the FXRT sources CDF-S~XT1 and CDF-S~XT2. For visualization purposes, background-subtracted light curves are presented with either 1~ks or 2~ks bins with 1$\sigma$ errors. In all cases, the \emph{vertical dashed gray line} represents the end of the observation.}
    \label{fig:LC_examples_det}
\end{figure*}

\subsection{Generation of light curves} \label{sec:LC}

We began by downloading the \emph{Chandra} full-field per-observation data products from the CSC2 for all CSC2-detected sources with $|b|{>}10$~deg. These products are preprocessed following the standard methods developed by the CSC2 \citep{Evans2010,Evans2019,Evans2020}. We use the \texttt{astropy.io} \citep{Astropy2013,Astropy2018} package to extract the photon information.

The \texttt{event file} of full-field observations contains photon event data stored as a table, with information such as photon arrival time, energy, position on the detector, sky coordinates, and observing conditions. One advantage of using \emph{Chandra} over all other \hbox{X-ray} satellites currently in operation is the low average number of background counts, which enables a robust detection of transient candidates with as few as ${\gtrsim}$10 total counts \citep[at $\gtrsim$99\% confidence; e.g.,][]{Kraft1991}, allowing searches for faint FXRTs potentially in the CSC2 catalog. To construct light curves, we extract the photon arrival times in the 0.5--7.0~keV range from each \texttt{event file} using an aperture of 1.5$\times R_{90}$ \citep[following the same process developed by][]{Yang2019}, where $R_{90}$ is the radius encircling 90\% of the \hbox{X-ray} counts, which is a function of instrumental  off-axis \citep[and depends on the photon energy; for more details, see][]{Vito2016,Hickox2006}. We consider this aperture (1.5$\times R_{90}$) because, based on simulations by \citet{Yang2019}, it encircles ${\gtrsim}$98\% of \hbox{X-ray} counts regardless of instrumental off-axis angle. Meanwhile, we calculate $N_{\rm bkg}$ using an annulus with inner and outer aperture radii of 1.5${\times}R_{90}$ and 1.5${\times}R_{90}$+20 pixels, respectively. If the background region overlaps another nearby X-ray source, we mask the nearby source (with radius of 1.5${\times}R_{90}$), and do not include the masked area when estimating the background. To correct the source light curve for the effect that background photons would have, we weight $N_{\rm bkg}$ by the source-to-background area ratio. 

The typical counts of our candidates imply that we are in the Poissonian statistical regime, and therefore we adopt the distribution proposed by \citet{Kraft1991} to compute the confidence intervals of the background subtracted light curves \citep[we use the package \texttt{astropy.stats} from][]{Astropy2018}. Figure~\ref{fig:LC_examples_det} shows example light curves (\emph{black circles}) detected by our method, as well as light curves for CDF-S~XT1 and CDF-S~XT2 (\emph{red circles}) following our extraction methodology.

\subsection{Initial candidate results} \label{sec:initial_results}

To summarize, we apply the FXRT detection algorithm to the 0.5--7.0 keV light curves of 214,701 CSC2 sources outside of the Galactic plane ($|b|>$10~deg, 
%and with exposure times ${>}$8\,ks (
splitting up long exposures into sub-20\,ks segments), resulting in 728 FXRT candidates. This sample has total net counts, instrumental  off-axis angles and time-averaged fluxes spanning ${\approx}$6.5--42720 (mean value of 754), ${\approx}$0.3--20.5 (mean value of 4.4)~arcmin, and $F_X{\approx}$2.6$\times$10$^{-16}$--7.1$\times$10$^{-12}$ (mean value of 1.2$\times$10$^{-13}$)~erg~cm$^{-2}$~s$^{-1}$, respectively. As expected, our method selects FXRTs with a diverse range of light curve properties. 

\begin{table*}
    \centering
    \begin{tabular}{lllll}
    \hline\hline
        Criterion & \multicolumn{4}{c}{Candidates} \\
                  & \# constrained & \# total removed & \# uniquely removed & \# remaining \\ \hline
         (1) & (2) & (3) & (4) & (5) \\ \hline
        1) Archival \hbox{X-ray} data & 645$^*$ & 558 & 72 & 170 \\ 
        2) Cross-match with stars/\emph{Gaia} & 728 & 454 & 56 & 66 \\
        3) NED + SIMBAD + VizieR & 728 & 525 & 31 & 29 \\
        \hline
        4) Archival images$^{\dagger}$ & -- & 9 & 9 & 20 \\
        5) Instrumental effects$^{\dagger}$ & -- & 6 & 6 & 14 \\
    \hline
    \end{tabular}
    \caption{Breakdown of FXRT candidates as a function of the selection criteria proposed in Sect. \ref{sec:purity_criteria}. 
    \emph{Column 1:} Criterion. 
    \emph{Column 2:} Number of candidates constrained by this criterion.
    \emph{Column 3:} Number of candidates removed that would be cut at this stage if we disregard all previous stages.
    \emph{Column 4:} Number of candidates that are solely removed by this criterion, and not any other.
    \emph{Column 5:} Running total number of candidates that remain after applying this criterion. \\
    $^*$Candidates with additional \emph{Chandra}-ACIS, \emph{XMM-Newton}, or \emph{Swift}-XRT observations. \\
    $\dagger$Note that criteria 4 and 5 are only applied to the sources that remain after the first three criteria are applied.}
    \label{tab:criteria}
\end{table*}

\subsection{Initial purity criteria}\label{sec:purity_criteria}

\begin{figure*}
   \centering
   \includegraphics[scale=0.43]{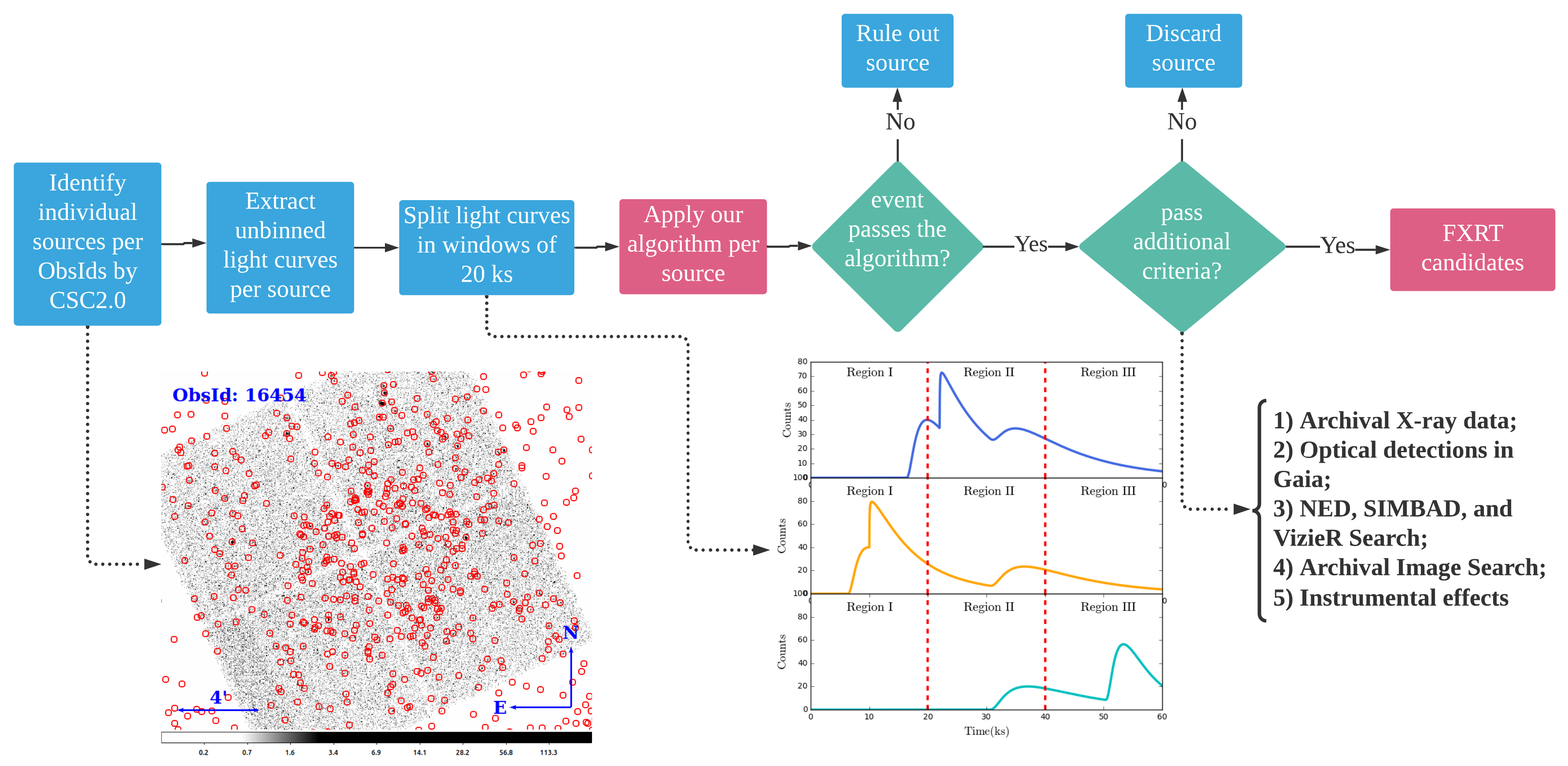}
    \caption{Methodology flowchart considered in this work to find FXRT candidates.}
    \label{fig:scheme}
\end{figure*}

It should be stressed that our search method does not guarantee a high-purity sample of real extragalactic FXRTs. Thus, we adopt some additional criteria based on archival \hbox{X-ray} data (prior and posterior \hbox{X-ray} detections of candidate FXRTs) and multiwavelength counterparts (e.g., bright stars) to help differentiate real extragalactic FXRTs from Galactic transients and variables among the 728 unique FXRT candidates. We explain and describe these additional criteria below. Table~\ref{tab:criteria} summarizes the number and percentage, relative to the total, of events that pass criteria (\emph{column 5}), as well as ignoring all previous steps (\emph{column 4}). Figure~\ref{fig:scheme} shows the steps to select/reject FXRTs taking into account our algorithm described in Sect. \ref{sec:algorithm} and the additional criteria that we explain below Sect. \ref{sec:previous_criteria}--\ref{sec:inst_effects}. We discuss the completeness of our search and selection criteria in Sect. \ref{sec:completness}.

\subsubsection{Criterion 1: Archival \hbox{X-ray} data} \label{sec:previous_criteria}

One important criterion to confirm the transient nature of the FXRT candidates is non-detection in prior and subsequent \hbox{X-ray} observations. We consider separately detections from: \emph{Chandra}, based on other observations in the CSC2; \emph{XMM-Newton}, based on individual observations of sources in the Serendipitous Source \citep[4XMM-DR9;][]{Rosen2016,Traulsen2019,Webb2020} and Slew Survey Source Catalogues \citep[XMMSL2;][]{Saxton2008}; and \emph{Swift}-XRT based on individual observations in the \emph{Swift}-XRT Point Source (2SXPS) catalog \citep{Evans2014}. In all cases, we require that 
the FXRT candidate remain undetected (consistent with zero counts) at 3$\sigma$ confidence in all observations outside of the one in which the FXRT candidate is found; we convert any detection or limit from the broadest original band to an equivalent 0.5--7.0~keV flux (using \texttt{PIMMS}) assuming a power-law (PL) with slope $\Gamma{=}2$. This requirement helps to exclude a large number of Galactic flaring sources, but may exclude FXRTs that occur in AGNs or strongly star-forming galaxies. For instance, CDF-S~XT1 has 105 additional \emph{Chandra} observations from the 7~Ms CDF-S survey, and its detection is ${>}$~5$\sigma$ higher than the limits from other observations and conforms with our adopted constraints.

The CSC2 provides uniform source extractions for all \emph{Chandra} observations associated with each candidate, at least up to 2014. For 33 candidates, more recent archival observations also exist. We downloaded and manually extracted photometry for these cases, adopting consistent source and background regions and aperture corrections compared to those used for the CSC2. In total, 580 FXRT candidates were observed in multiple \emph{Chandra} observation IDs, while 148 candidates have only a single \emph{Chandra} visit (available in CSC2).

To recover possible \emph{XMM-Newton} and \emph{Swift}-XRT detections, we match to the 4XMM-DR9, XMMSL2 and 2SXPS catalogs, adopting a search radius equivalent to the 3$\sigma$ combined positional errors of the \emph{Chandra} detection and tentative \emph{XMM-Newton} or \emph{Swift}-XRT match. 

We additionally search the \hbox{X-ray} upper limit servers
FLIX,\footnote{https://www.ledas.ac.uk/flix/flix.html} 2SXPS,\footnote{https://www.swift.ac.uk/2SXPS/ulserv.php}
and ULS.\footnote{http://xmmuls.esac.esa.int/upperlimitserver/} 
The latter provides upper limits for many \hbox{X-ray} observatory archives (including \emph{XMM-Newton} pointed observations and slew surveys; Swift pointed observations; R\"ontgen Satellite (\emph{ROSAT}) pointed observations and all-sky survey; \emph{Einstein} pointed observations), but does not necessarily use the same versions of the reduction pipeline as the first two and has somewhat different area coverage limits for the same observations.
Based on visual inspections, we found that the reported detections are not always reliable, and hence we require detections to be ${\geq}$5$\sigma$.
We found that: 
397 candidates are observed with \emph{XMM-Newton} 4XMM-DR9, with 206 candidates detected;
590 candidates are observed with \emph{XMM-Newton} XMMSL2, with 6 candidates detected;
351 candidates are observed with \emph{Swift}-XRT 2SXPS, with 31 candidates detected;
355 candidates are observed with \emph{ROSAT} pointed observations, with zero candidates detected;
443 candidates are observed with \emph{Einstein} pointed observations, with 1 candidate detected;
finally all candidates are observed with the \emph{ROSAT} All-Sky Survey, with 30 candidates detected.
The upper limits from \emph{Chandra} and \emph{XMM-Newton} pointed observations are all comparable to or lower than our FXRT candidate peak fluxes, such that further similar transient behavior would have been detectable in such observations if present. The \emph{Swift}-XRT, \emph{XMM-Newton}-Slew, \emph{ROSAT,} and \emph{Einstein} limits are not nearly as constraining.

In total, 645 candidates have multiple hard (meaning \emph{Chandra}, \emph{XMM-Newton}, or \emph{Swift}-XRT pointed observations) \hbox{X-ray} constraints, of which 580 candidates have been visited more than once by \emph{Chandra}. This implies re-detected fractions of at least ${\approx}$80\% among the candidate sample. On the other hand, 513 candidates have multiple soft (meaning \emph{ROSAT} or \emph{Einstein} pointed observations) \hbox{X-ray} constraints, of which 31 candidates have been detected more than once. The implied re-detection fractions are much lower, ${\approx}$4\%, among the candidate sample, presumably due to the much shallower sensitivities of these past observatories. The high \hbox{X-ray} re-detection fraction indicates that this is a very effective criterion if additional \emph{Chandra}, \emph{XMM-Newton} or \emph{Swift} observations are available.
For the remaining 215 candidates that show no additional \hbox{X-ray} detections, we note that, in general, their \hbox{X-ray} constraints are much shallower than the detected sources, and thus we might expect a significant fraction to be persistent/recurrent if observed again for similar exposure times with \emph{Chandra} or \emph{XMM-Newton}.

\begin{figure*}
    \centering
    \includegraphics[width=17cm,height=5.3cm]{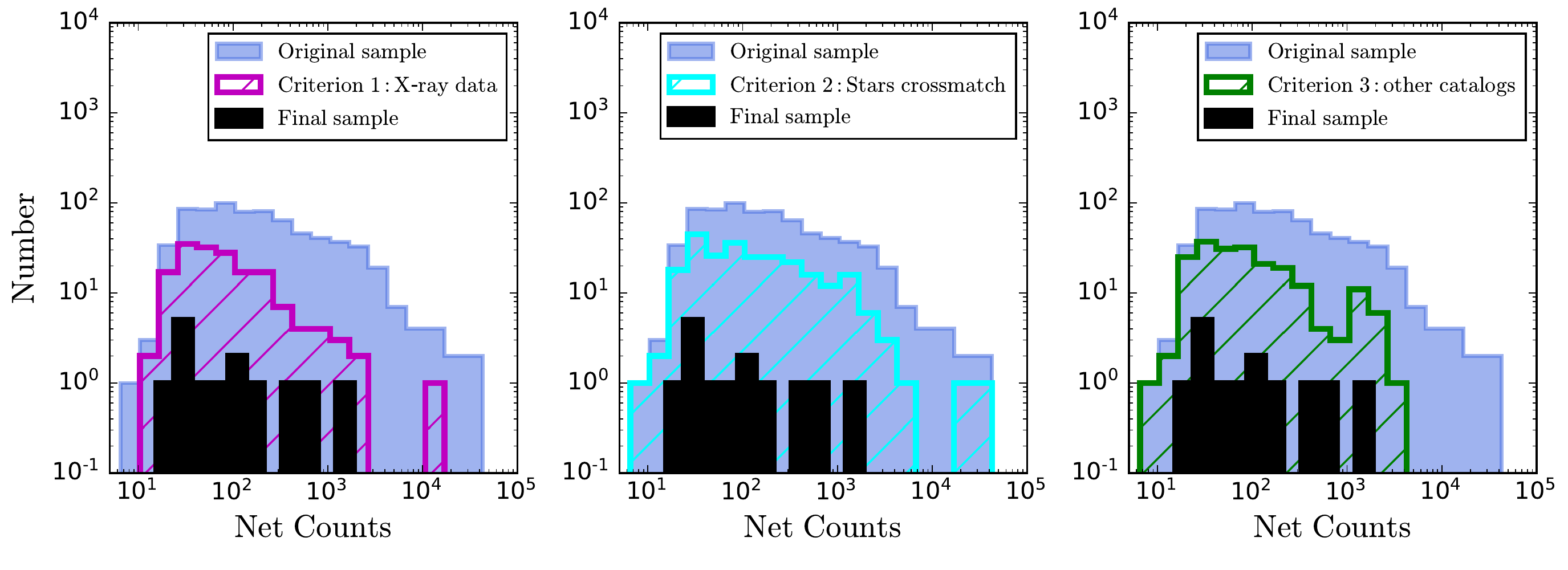}\\
    \includegraphics[width=17cm,height=5.3cm]{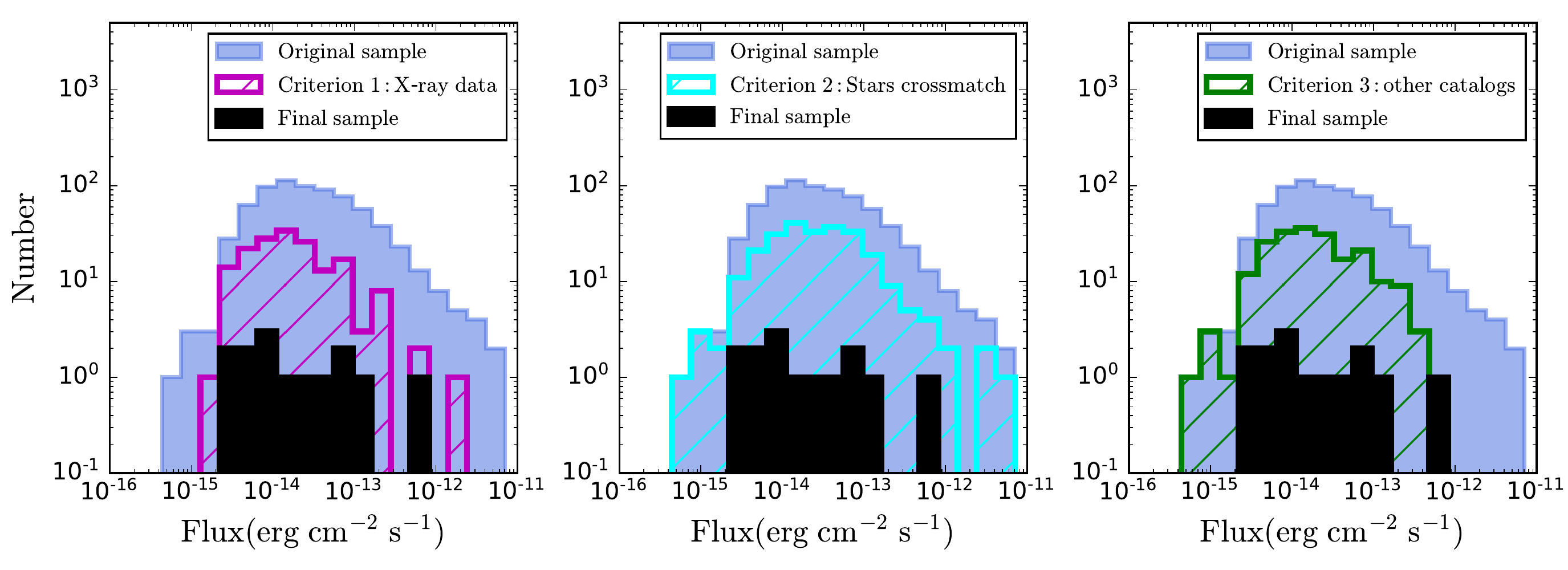}
    \caption{Comparison of 0.5--7.0~keV net-count (\emph{top panels}) and flux (\emph{bottom panels}; 0.5--7.0~keV) distributions for the initial (\emph{filled blue histograms}) and final (\emph{filled black histograms}) samples, as well as subsets covered by various purity criteria (\emph{colored, unfilled histograms}) for the sample. Net counts and fluxes are provided by the CSC2.}
    \label{fig:histo}
\end{figure*}

Finally, 170 candidates pass this criterion (see Table~\ref{tab:criteria}). Also, it is important to mention that 72 candidates are discarded by this criterion but not by the others. The \emph{left panels} of Fig.~\ref{fig:histo} show the net-count and flux distributions for the 170 events that pass this criterion. To conclude, this criterion appears to be an extremely effective means to identify persistent or repeat transients, when data are available.

\subsubsection{Criterion 2: Optical detections in \emph{Gaia}} \label{sec:gaia}

As discussed in \citet{Yang2019}, a large fraction of FXRT candidates are Galactic in origin, associated with relatively bright stellar sources. To identify these, we cross-match with 
the \emph{Gaia} Early Data Release 3 \citep[\emph{Gaia} EDR3;][]{Gaia_DR3_2020} catalog, which contains relatively uniform photometric and astrometric constraints for more than 1.8 billion sources in the magnitude range $G{=}3$--21~mag across the entire sky, based on observations collected during the first 34 months of its operational phase; these include parameters such as position, parallax, and proper motion in the Milky Way and throughout the Local Group \citep{Lindegren2018,Gaia2018}. 

We employ the \texttt{VizieR} package (EDR3 catalog), adopting the CSC2 3$\sigma$ positional uncertainty associated with each source as our search radius. In general, this search radius is sufficiently small to find a unique counterpart, given \emph{Chandra}'s high spatial resolution and demonstrated astrometric precision \citep[${\approx}$0\farcs5;][]{Rots2010}; 26 candidates show multiple \emph{Gaia} sources in their cone search area, for which we adopt the nearest \emph{Gaia} source. 

In total, 521 candidates have cross-matched sources in \emph{Gaia EDR3}. However, we only reject candidates matched to stellar \emph{Gaia} EDR3 optical detections (i.e., those with significant nonzero proper motion and/or parallax detected at $>$3$\sigma$ significance), which amounts to 454 candidates from the initial sample. These stellar counterparts span a wide range in magnitude $G{=}$10--20.8~mag ($\overline{G}{\approx}$16.9~mag) and proper motion $\mu{=}$[0.05--186]~mas~yr$^{-1}$ ($\overline{\mu}{\approx}$13.7~mas~yr$^{-1}$). To characterize better the X-ray sources classified as stars according to {Criterion 2}, we construct a color-magnitude diagram of their Panoramic Survey Telescope and Rapid Response System (Pan-STARRS) archive and Dark Energy Camera (DECam) counterparts (see Fig.~\ref{fig:color_color}) and compare to theoretical isochrones taken from the \emph{MESA Isochrones \& Stellar Tracks} (MIST) package (\citealp{Dotter2016}, \citealp{Choi2016}) with different metallicities (from $\rm{[Fe/H]}{=}-3.0$ to $+0.5$), ages ($\log(\rm{Age/yr}){=}$7.0, 9.0, 10.0, and 10.3) and attenuation ($A_V{=}$0.0 and 5.0). The sample of X-ray sources classified as stars covers a wide range in the parameter space (see Fig.~\ref{fig:color_color}), as expected for such an inhomogeneous sample of stars.

The central panels of Fig.~\ref{fig:histo} show the net-count and flux distributions of the 274 events that pass this criterion. Among the total sample, ${\approx}$65\% are associated with bright stars, highlighting the importance of this cross-match. Moreover, this criterion discards 56 sources that the other criteria do not. Nevertheless, due to the relatively bright magnitude limit and optical window of the \emph{Gaia} EDR3 objects with proper motion and parallax constraints, this criterion may not identify all persistent or recurring transient Galactic objects, as we discuss in the next subsection. As a running total, only 63 candidates successfully pass both this and the previous criterion (see Table~\ref{tab:criteria}). 

\begin{figure*}
    \centering
    \includegraphics[scale=0.68]{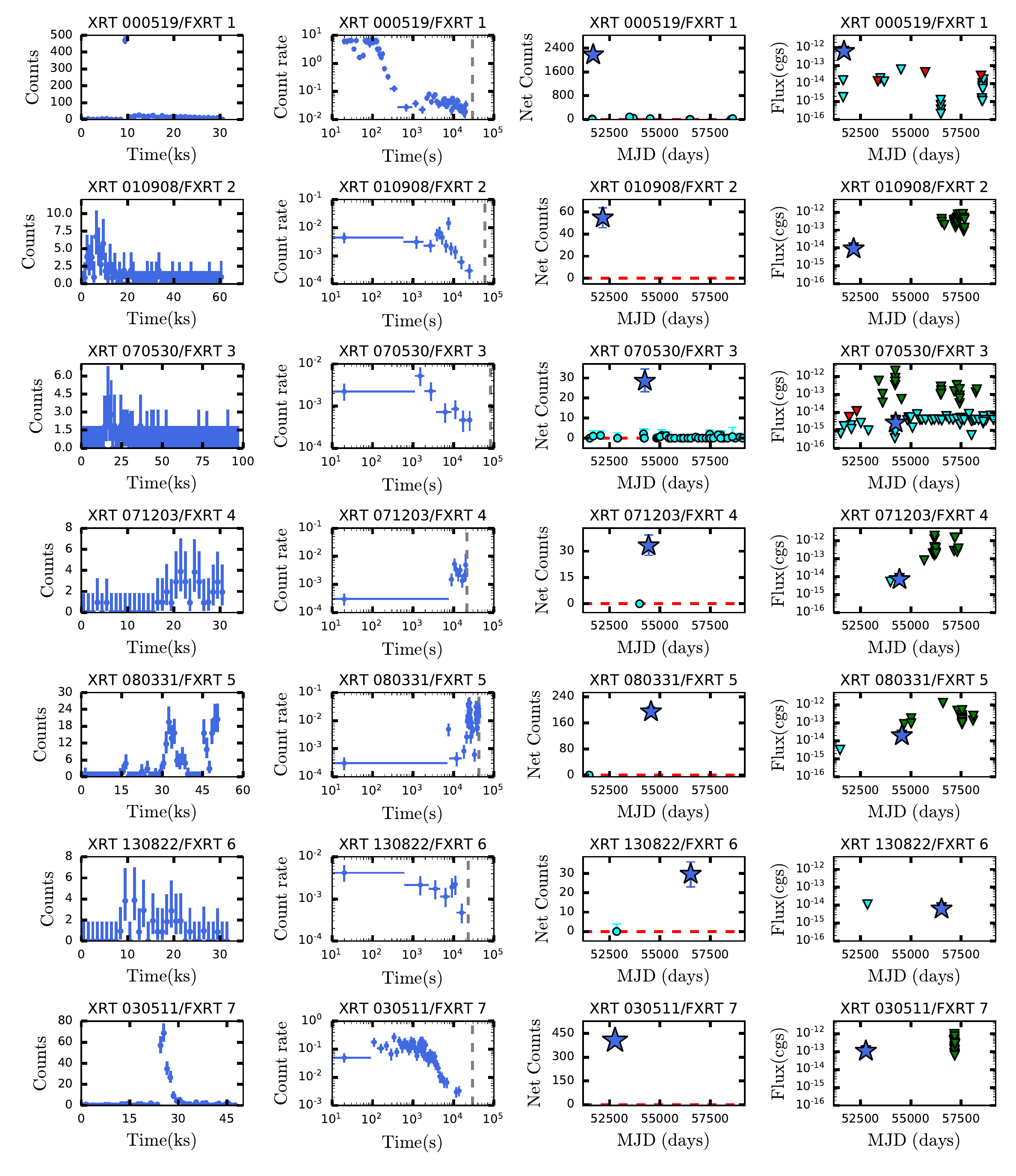}
    \vspace{-0.4cm}
    \caption{0.5--7\,keV light curves for each FXRT candidate: full exposure, in units of counts (\emph{first column 1}); zoomed-in view, from the detection of the first photon to the end of the exposure, in units of count rate (cts~s$^{-1}$), with log-log scaling and five counts per bin  (\emph{second column}); long-term light curve, with each point representing individual \emph{Chandra} exposures (cyan circles with 1$\sigma$ error bars) to highlight the significance of detections and non-detections, in units of counts (\emph{third column}); long-term light curve, with each point representing individual \emph{Chandra} (\emph{cyan}), {\it XMM-Newton} (\emph{red}), and \emph{Swift}-XRT (\emph{green}) exposures in units of flux (erg~s$^{-1}$~cm$^{-2}$) (fourth column). For the long-term light curves, the observation that includes the transient is denoted by a \emph{large blue star} (1$\sigma$ error bars), while triangles denote observations with (3$\sigma$) upper limits. All the fluxes are reported in the 0.5--7\,keV band in the observer's frame. In the case of FXRT~4 in Col. 4, additional data points are partially blocked by the \emph{blue} star.}
    \label{fig:light_curves_1}
\end{figure*}

\begin{figure*}
    \ContinuedFloat
    \centering
    \includegraphics[scale=0.68]{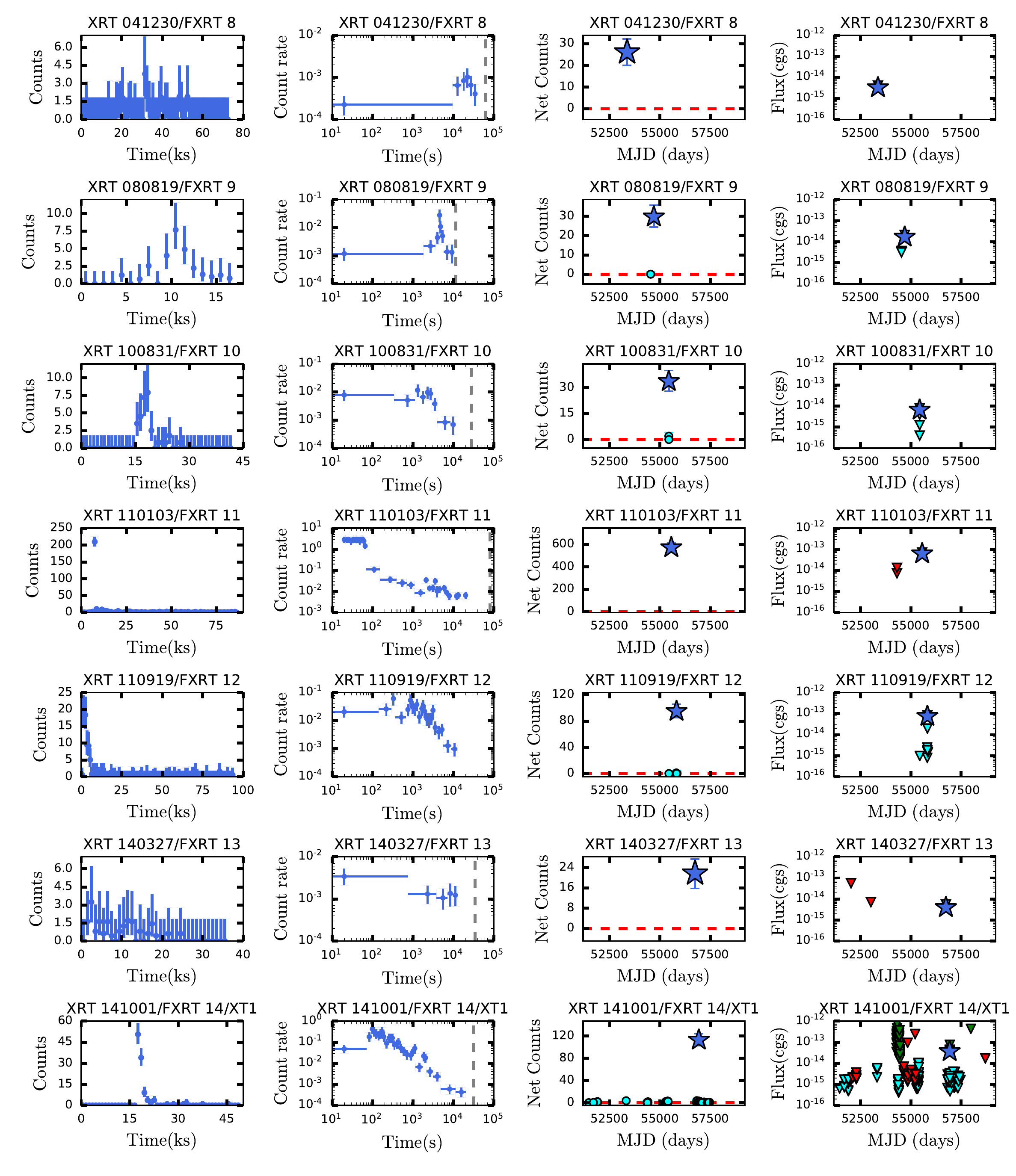}
    \caption[]{(continued)}
%    \label{fig:light_curves_1}
\end{figure*}

\begin{table*}
    \scriptsize
    \centering
    \advance\leftskip-0.5cm
    \begin{tabular}{llrclrrrrcccr}
    \hline\hline
     FXRT & Id & ObId & Exp. (ks) & Date & $T_{90}$ (ks) & RA (deg.) & DEC (deg.) & Off. Ang. & Flux & Pos. Unc. & HR & S/N\\ \hline
     (1) & (2) & (3) & (4) & (5) & (6) & (7) & (8) & (9) & (10) & (11) & (12) & (13) \\ \hline
    \multicolumn{13}{c}{Nearby extragalactic FXRT Candidates from CSC2} \\
    \hline
    % IdY9
    1   & XRT\,000519$\dagger$ & 803 & 31.0 & 2000-05-19 & 11.6$_{-0.9}^{+1.0}$ & 186.38125 & 13.06607 & 13\farcm3 & 6.4e-13 & 1\farcs8 & -0.59$\pm0.02$ & 35.1\\
    2   & XRT\,010908 & 2025 & 61.5 & 2001-09-08 & 25.7$_{-13.5}^{+27.2}$ & 167.86792 & 55.67253 & 2\farcm5 & 9.2e-15 & 1\farcs06 & -0.21$\pm$0.13 & 6.2 \\ 
    3   & XRT\,070530 & 8490 & 97.2 & 2007-05-30 & 29.8$_{-13.5}^{+47.5}$ & 201.24329 & -43.04060 & 4\farcm1 & 2.6e-15 & 1\farcs3 & -0.68$\pm$0.17 & 4.6 \\
    4   & XRT\,071203 & 9546 & 31.8 & 2007-12-03 & 25.3$_{-3.0}^{+14.3}$ & 211.25113 & 53.65706 & 0\farcm7 & 7.0e-15 & 1\farcs13 & -0.59$\pm$0.14 & 5.2 \\
    % IdY156
    5   & XRT\,080331 & 9548 & 51.7 & 2008-03-31/4-1 & 32.8$_{-0.9}^{+8.0}$ & 170.07296 & 12.97189 & 0\farcm9 & 2.0e-14 & 1\farcs0 & -0.73$\pm0.05$ & 12.0 \\
    % IdY209
    6   & XRT\,130822 & 14904 & 32.2 & 2013-08-22 & 12.1$_{-1.8}^{+8.5}$ & 345.49250 & 15.94871 & 1\farcm6 & 6.3e-15 & 0\farcs79 & -0.46$\pm$0.17 & 4.6 \\
    \hline
    \multicolumn{13}{c}{Distant extragalactic FXRT Candidates from CSC2} \\
    \hline
    % IdF43
    7   & XRT\,030511$\dagger$  & 4062 & 48.1 & 2003-05-10/11 & 6.5$_{-3.1}^{+3.0}$ & 76.77817 & -31.86980 & 10\farcm7 & 1.1e-13 & 1\farcs56 & -0.38$\pm$0.04 & 16.9 \\
    % IdY125
    8   & XRT\,041230 & 5885 & 73.4 & 2004-12-30/31 & 40.2$_{-6.7}^{+19.7}$ & 318.12646 & -63.49914 & 3\farcm4 & 3.3e-15 & 0\farcs93 & -0.46$\pm$0.20 & 4.3 \\
    % IdY158
    9   & XRT\,080819 & 9841 & 17.7 & 2008-08-19 & 8.3$_{-1.5}^{+4.9}$ & 175.00504 & -31.91743 & 5\farcm1 & 1.7e-14 & 1\farcs03 & -0.62$\pm$0.17 & 4.7 \\
    % IdF86
    10  & XRT\,100831  & 12264 & 43.0 & 2010-08-31 & 4.3$_{-1.0}^{+7.3}$ & 90.00450 & -52.71501 & 4\farcm8 & 3.9e-15 & 1\farcs16 & -0.66$\pm$0.14 & 5.0 \\
    % IdF96
    11  & XRT\,110103$\dagger$ & 12884 & 87.0 & 2011-01-03 & 40.1$_{-5.7}^{+6.8}$ & 212.12063 & -27.05784 & 13\farcm3 & 6.2e-14 & 2\farcs59 & -0.24$\pm0.04$ & 15.7 \\
    12  & XRT\,110919$\dagger$  & 13454 & 94.2 & 2011-09-19/20 & 17.2$_{-10.9}^{+50.4}$ & 15.93558 & -21.81272 & 7\farcm2 & 1.6e-14 & 1\farcs10 & -0.19$\pm0.11$ & 7.2 \\
    % IdY212
    13  & XRT\,140327 & 15113 & 36.4 & 2014-03-27 & 13.4$_{-3.4}^{+7.3}$ & 45.26725 & -77.88095 & 6\farcm6 & 3.9e-15 & 1\farcs76 & -0.63$\pm$0.19 & 3.9 \\
    14  & XRT\,141001/ & 16454 & 49.5 & 2014-10-01 & 5.1$_{-3.1}^{+15.0}$ & 53.16158 & -27.85940 & 4\farcm3 & 3.7e-14 & 0\farcs66 & -0.16$\pm0.09$ & 9.2 \\
      & CDF-S~XT1$\dagger$ &  &  &  &  &  &  &  &  &  &  \\ \hline
    \end{tabular}
    \caption{Properties of the extragalactic FXRT candidates detected and/or discussed in this work, ordered by subsample and date.
    \emph{Column 1:} Shorthand identifier (FXRT~\#) used throughout this work.
    \emph{Column 2:} \hbox{X-ray} transient identifier (XRT\,date), plus previous name when available. 
    \emph{Columns 3, 4, and 5:} \emph{Chandra} observation ID, exposure time in units of ks, and date. 
    \emph{Column 6:} $T_{90}$ duration, which measures the time over which the event emits the central 90\% (i.e., from 5\% to 95\%) of its total measured counts, in units of ks.
    \emph{Columns 7 and 8:} Right ascension and declination in J2000 equatorial coordinates. 
    \emph{Column 9:} Instrumental  off-axis angle of the FXRT candidates, with respect to the \emph{Chandra} aimpoint, in units of arcminutes. 
    \emph{Column 10:} Aperture-corrected, observation-averaged 0.5--7.0~keV flux inferred from the CSC2, in cgs units.
    \emph{Column 11:} Estimated 2$\sigma$ \hbox{X-ray} positional uncertainty from the CSC2, in units of arcseconds; as demonstrated in \citet{Bauer2017}, this can be improved by up to $\approx$40\% when sufficient optical/\hbox{X-ray} cross-matches are available. 
    \emph{Column 12:} Hardness ratio (HR) and 1$\sigma$ uncertainty, defined as HR=(H$-$S)/(H$+$S), where H=2--7~keV and S=0.5--2~keV energy bands, using the Bayesian estimation of \citet{Park2006}. 
    \emph{Column 13:} Approximate signal-to-noise ratio (S/N).\\
    $\dagger$ Previously reported as FXRTs by \citet{Jonker2013} in the case of FXRT~1 (or XRT\,000519), \citet{Glennie2015} for FXRT~11 (or XRT\,110103), \citet{Lin2019} for FXRT~7 (or XRT\,030511) and FXRT~12 (or XRT 110919), and \citet{Bauer2017} for FXRT~14 (or XRT\,141001/CDF-S~XT1).}
    \label{tab:my_detections}
\end{table*}

\subsubsection{Criterion 3: NED, SIMBAD, and VizieR Search} \label{sec:catal_cross}

To identify further known Galactic and Local Group objects, we search for associated objects (counterparts or host galaxies) in several large databases using the \texttt{astroquery} package: the NASA/IPAC Extragalactic Database \citep[NED;][]{Helou1991}, the Set of Identifications, Measurements, and Bibliography for Astronomical Data \citep[SIMBAD;][]{Wenger2000}, and VizieR \citep[which provides the most complete library of published astronomical catalogs;][]{Ochsenbein2000}. There is non-negligible redundancy here compared to the previous two searches, as these databases have ingested previous versions of \hbox{X-ray} serendipitous catalogs and \emph{Gaia} EDR3 in the case of VizieR. To begin, we performed a cone search per candidate considering a radius equivalent to the 3$\sigma$ positional error to find associated sources. These databases integrate many catalogs across the EM spectrum, helping rule out objects of our sample that were classified previously as stars, young stellar objects (YSOs), or objects associated with globular clusters, nebulae, or high-mass \hbox{X-ray} binaries (HMXBs) in either our Galaxy or the Local Group. However, we should stress that these catalogs are highly heterogeneous, and we must take care to not misinterpret candidate matches. Around 212 candidates have one or more entries in the various databases when cross-correlating to a region encompassing the 3$\sigma$ uncertainty of the FXRT positions. In all the cases, the multiple entries had the same source classification. We uniquely identify 31 objects in this way, either as YSOs embedded in nebulae or stars identified by other catalogs, for instance, the VISTA Hemisphere Survey (VHS), the United Kingdom InfraRed Telescope (UKIRT) Infrared Deep Sky Survey, the Sloan Digital Sky Survey (SDSS), or the catalog sources from combined the Wide-field Infrared Survey Explorer (WISE) and the near-Earth objects WISE (NEOWISE) all-sky survey data at 3.4 and 4.6 $\mu$m (CatWISE) \citep{McMahon2013,Dye2018,Marocco2021}. This step is also critical because ${\approx}$78\% of the initial sample show associated sources in these databases. The \emph{right panels} of Fig.~\ref{fig:histo} show the net-count and flux distribution for the 203 events that pass this criterion. Applying all criteria thus far, the sample is reduced to 29 candidates.

\subsubsection{Archival image search} \label{sec:archive_images}

In order to rule out fainter stellar counterparts, we carried out a search of ultraviolet (UV), optical, NIR, and mid-infrared (MIR) image archives; We perform a cone search within a radius equal to the 3$\sigma$ uncertainty on the \emph{Chandra} error position of the respective FXRTs (see Table~\ref{tab:my_detections}) in the following archives: 
the Hubble Legacy Archive;\footnote{https://hla.stsci.edu/hlaview.html}
the Pan-STARRS archive \citep{Flewelling2016};\footnote{http://ps1images.stsci.edu/cgi-bin/ps1cutouts}
the National Science Foundation's National Optical-Infrared Astronomy Research (NOIR) Astro Data Lab archive,\footnote{https://datalab.noao.edu/sia.php} which includes 
images from the Dark Energy Survey \citep[DES;][]{Abbott2016} and
the Legacy Survey (DR8);
the Gemini Observatory Archive;\footnote{https://archive.gemini.edu/searchform}
the National Optical Astronomy Observatory (NOAO) science archive;\footnote{http://archive1.dm.noao.edu/search/query/}
the ESO archive science portal;\footnote{http://archive.eso.org/scienceportal}
the VISTA Science Archive;\footnote{http://horus.roe.ac.uk/vsa/}
the Spitzer Enhanced Imaging Products archive \citep{Teplitz2010};\footnote{https://irsa.ipac.caltech.edu/data/SPITZER/Enhanced/SEIP/}
the UKIRT/Wide Field Camera (WFCAM) Science Archive;\footnote{http://wsa.roe.ac.uk/}
and the WISE archive \citep[][]{Wright2010}.

For images obtained under good seeing (${<}$\,1$"$) conditions, we visually search for counterparts or host galaxies in the 3$\sigma$ uncertainty on the \hbox{X-ray} location of the FXRT (ensuring that the optical images are co-aligned to \emph{Gaia} EDR3). We only undertake this step for the candidates that remain after the selection applied in Sect. \ref{sec:catal_cross}. If a source is found, we quantify its significance and assess its extent and radial profile visually. We identify sources as stellar if they are consistent with the spatial resolution of the imaging.
We reject nine candidates in this way: five sources are embedded in obvious Galactic nebulae with point-like NIR counterparts, and four candidates are identified as stars in \emph{Hubble Space Telescope} (HST) images. The latter have no clear nearby galaxy associations, suggesting that they are likely field stars, perhaps the fainter tail of the population probed by Gaia DR3. This reduces the number of candidates to 20.

\subsubsection{Instrumental effects} \label{sec:inst_effects}

As a final step, we perform additional manual and visual cross-checks to rule out false positive candidates that might arise from background flares, bad pixels or columns, or cosmic-ray afterglows. Again, we only undertake this step for the remaining candidates after Sect. \ref{sec:archive_images}.
To rule out events that occur during strong background flaring episodes (${\gtrsim}$3$\sigma$ mean value) in the energy range 0.5--7~keV, we employ the \texttt{dmextract} script (excluding counts associated with \hbox{X-ray} sources identified by CSC2 in the \emph{Chandra} FoV) to investigate the evolution of the background count rate during the observations. Using the \texttt{deflare} script, we identify and reject six candidate FXRTs found in a circular region with radius ${\approx}4\farcm0$ around the planetary nebula (PN) NGC~246 in the \emph{Chandra} observation ID 2565 that are affected by background flares, reducing the number of candidates to 14. We confirm that none of the remaining 14 sources is caused by detector artifacts (bad columns or hot pixels) or are associated with bad quality flags (confused source and background regions or saturation) in the CSC2 catalog entries. Furthermore, we confirm that the counts from all sources are detected in (many) dozens to hundreds of individual pixels tracing out portions of \emph{Chandra}'s Lissajous dither pattern (appearing as a sinusoidal-like evolution of $x$ and $y$ detector coordinates as a function of time; see Appendix Fig.~\ref{fig:lissajaus}) over their duration, which reinforces that they are real astrophysical sources.
Therefore, we have a final sample of 14 FXRTs.  

\subsubsection{Completeness}\label{sec:completness}

Below, we explore the probability that real FXRTs might have been discarded erroneously. To estimate this, we determine the likelihood that the position of a candidate FXRT overlaps, by chance, that of another X-ray source and/or star. The probability (assuming Poisson statistics; $P(k,\lambda)$) of one source ($k{=}1$) being found by chance inside the 3$\sigma$ localization uncertainty region of another is
\begin{equation}
    P(k=1,\lambda){=}\frac{e^{-\lambda}\lambda^k}{k!},
\end{equation}
where $\lambda$ is the source density of X-ray sources and/or stars on the sky multiplied by the 3$\sigma$ \emph{Chandra} localization uncertainty area. To measure the X-ray or optical source density, we consider X-ray detections from the CSC2, 4XMM-DR9 and 2SXPS catalogs \citep{Evans2010,Webb2020,Evans2014}, and the \emph{Gaia EDR3} catalog for stars \citep{Gaia_DR3_2020}, respectively. This probability is 0.0091 and 0.0071 for X-ray and optical sources, respectively. Taking the 72 and 56 X-ray sources that are discarded solely on the basis of {Criteria 1 or 2} (see Table~\ref{tab:criteria}), respectively, we expect ${\ll}$1 of these to be discarded erroneously. If we consider the 665 X-ray sources discarded by both {Criteria 1 and 2}, the combined probability is 6.5${\times}$10$^{-5}$, and thus the expected number of erroneously dismissed sources is also ${\ll}$1. The contribution of {Criterion 3} to the completeness is not easy to assess, given the highly distributed nature of the databases. Based on the high fraction of discarded sources that overlap with the other criteria, we assume that the databases used in {Criterion 3} are accurate and this criteria does not disproportionately discard real FXRTs (i.e., also ${\ll}$1). To summarize, our rejection of contaminating sources does not appear to impact the completeness of our FXRT candidate sample.

\subsubsection{Summary}\label{sec:filter_summary}

We discover 14 FXRT candidates in the CSC2, five of which had been discovered previously as FXRTs while an additional six had been detected in published works but not properly characterized (see Sect. \ref{sec:results} for more details).

Table~\ref{tab:my_detections} provides the coordinates, instrumental off-axis angle, flux, positional uncertainty, hardness ratio \citep[HR; computed following][]{Park2006}, and S/N ratio. Figure~\ref{fig:light_curves_1} shows the background-subtracted 0.5--7.0~keV light curves of our final sample of FXRT candidates: short-term, in units of counts (\emph{first column}) and count rates (\emph{second column}); long-term in units of counts for \emph{Chandra} only (\emph{third column}) and flux to compare uniformly \emph{Chandra}, {\it XMM-Newton} and \emph{Swift}-XRT data (\emph{fourth column}).
We highlight that the three criteria (\hbox{X-ray} archival data, \emph{Gaia} detection cross-match, and NED/SIMBAD/VizieR catalogs, respectively) contribute in complementary ways to clean the sample. We stress that the sample may still contain contamination from faint and/or extremely red Galactic objects, which we address below.

We designate each candidate by ``XRT'' followed by the date (the first two numbers correspond to the year, the second two numbers to the month, and the last two numbers to the day; see Table~\ref{tab:my_detections}, \emph{second column}). However, to identify each event quickly throughout this manuscript we also denominate them by ``FXRT''+\# (ordered by subsample and date; see Table~\ref{tab:my_detections}, \emph{first column}). Furthermore, from the final 14 events, 3 of them (FXRT~2, FXRT~4, and FXRT~5) were classified previously as HMXBs in galaxies at ${\gtrsim}$4~Mpc. Nevertheless, we keep them to be consistent with the selection criteria of this work (see Sect. \ref{sec:opt_NIR} for more details).

We note that FXRTs CDF-S~XT2 \citep[XRT~150321;][]{Xue2019}, XRT~170831 \citep{Lin2019,Lin2022},
and XRT~210423 \citep{Lin2021} are not part of this work because CSC2 only includes data released publicly up to the end of 2014.

\begin{figure*}
    \centering
    \includegraphics[width=18cm,height=2.5cm]{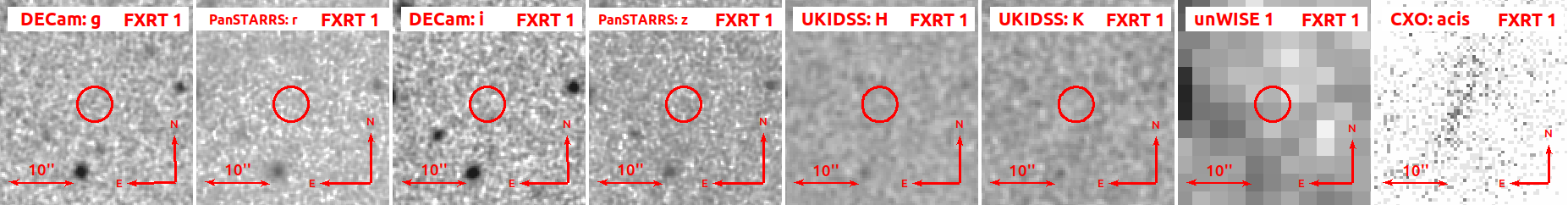}
    \includegraphics[width=18cm,height=2.5cm]{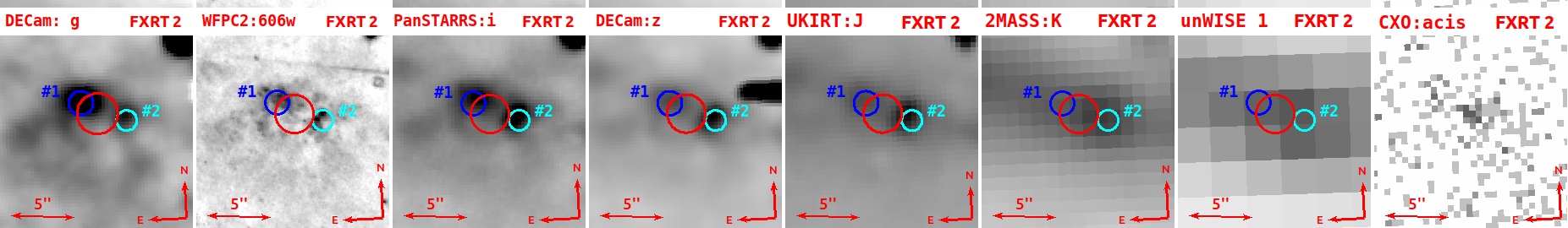}
    \includegraphics[width=18cm,height=2.5cm]{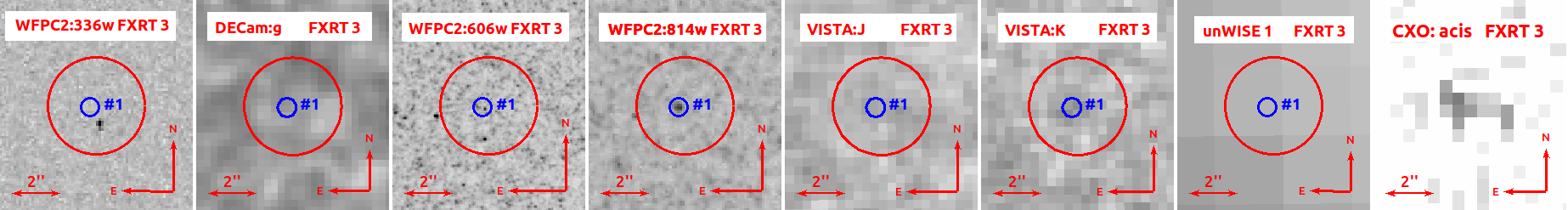}
    \includegraphics[width=18cm,height=2.5cm]{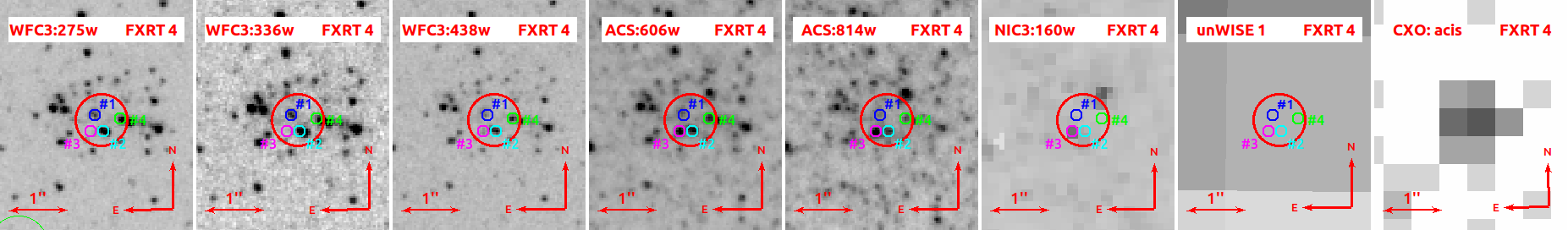}
    \includegraphics[width=18cm,height=2.5cm]{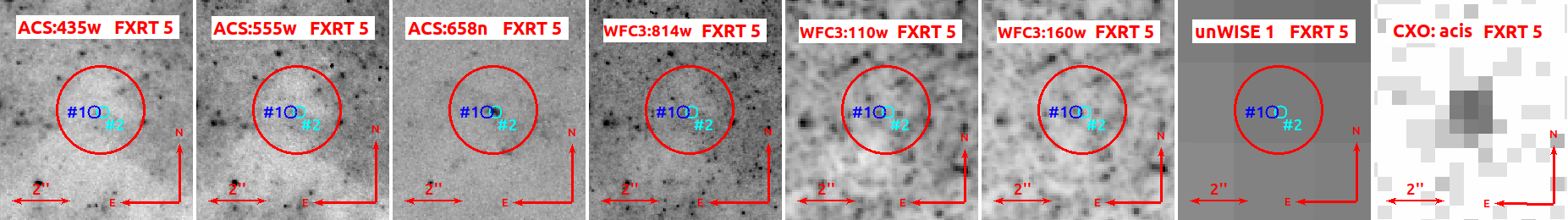}
    \includegraphics[width=18cm,height=2.5cm]{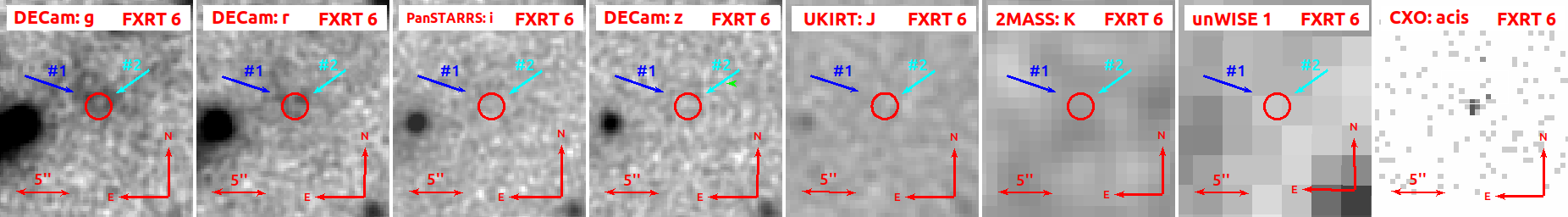}
    \includegraphics[width=18cm,height=2.5cm]{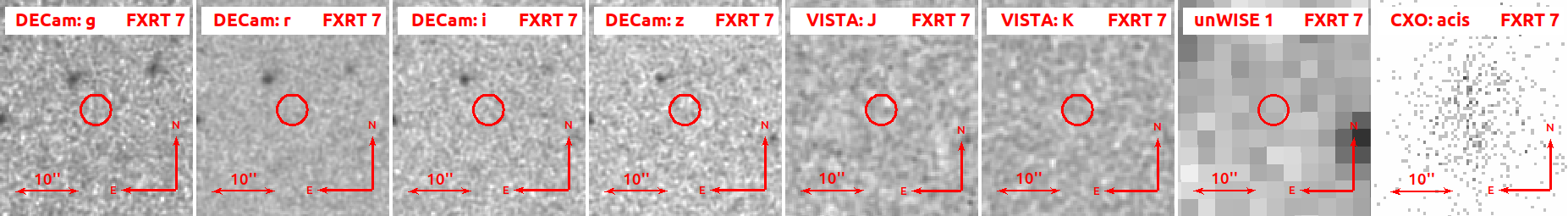}
    \caption{Archival optical, NIR, MIR, and \hbox{X-ray}  images of extragalactic FXRT candidates; the telescope or instrument plus filter and FXRT ID name are shown in the upper-left and upper-right corners, respectively. Each cutout is centered on the \hbox{X-ray} position, and red circles denote 3$\sigma$ \emph{Chandra} errors in the source localization. \emph{Columns 1, 2, 3, and 4: } Optical band (DECam, Pan-STARRS, and HST) images. \emph{Columns 5 and 6:} NIR $J$ or $H$ and $K$ (UKIRT or VISTA) images. \emph{Column 7:} 3.4$\mu$m (unWISE) images. \emph{Column 8:} \hbox{X-ray} \emph{Chandra} (ACIS) 0.5--7 keV images.}
    \label{fig:image_cutouts}
\end{figure*}
%We note that for the two FXRTs, the number of expected false matches is 0 [$P{<}0.004$ for offsets ${<}$0.54$^{\prime\prime}$ at $m_{\rm r}=23$\,mag].
\begin{figure*}
\ContinuedFloat
    \centering
    \includegraphics[width=18cm,height=2.5cm]{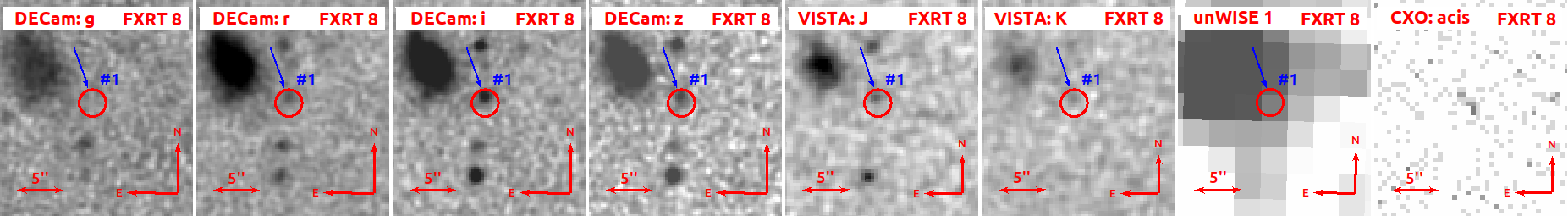}
    \includegraphics[width=18cm,height=2.5cm]{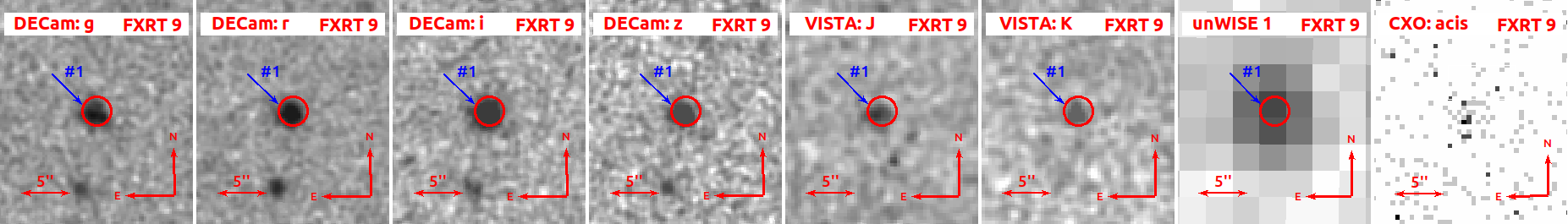}
    \includegraphics[width=18cm,height=2.5cm]{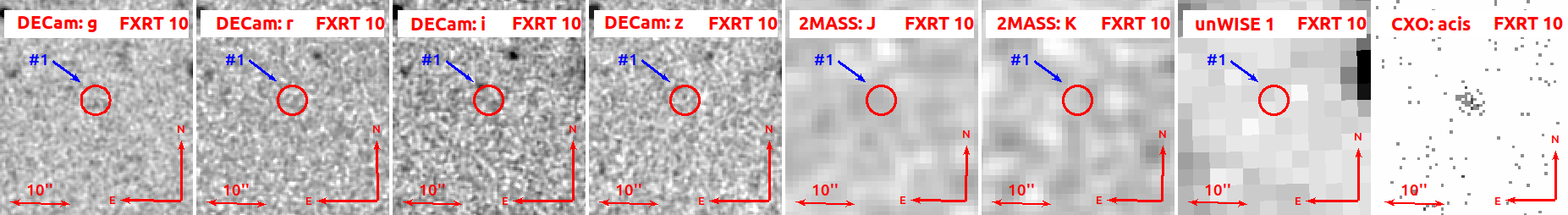}
    \includegraphics[width=18cm,height=2.5cm]{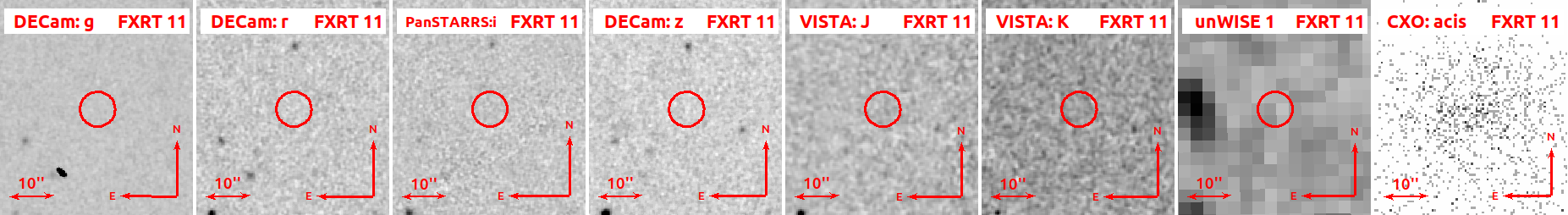}
    \includegraphics[width=18cm,height=2.5cm]{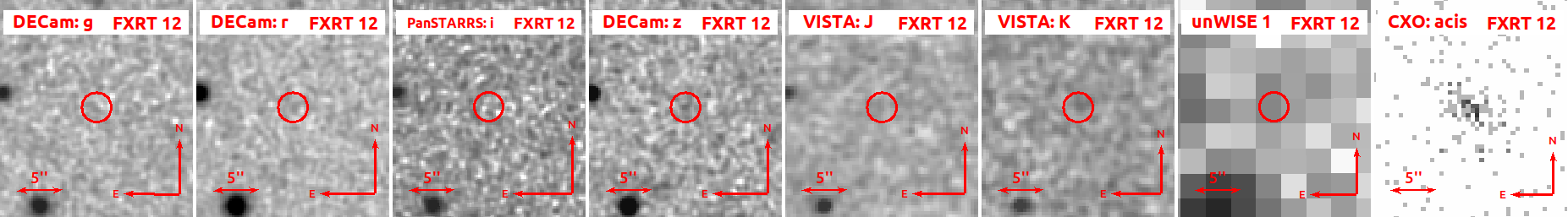}
    \includegraphics[width=18cm,height=2.5cm]{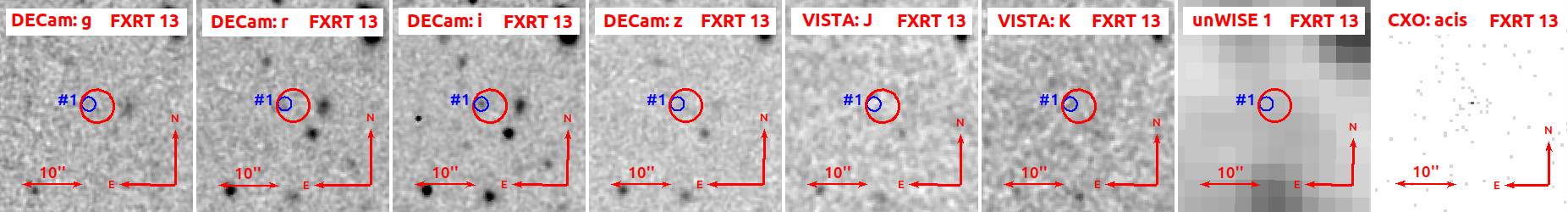}
    \includegraphics[width=18cm,height=2.5cm]{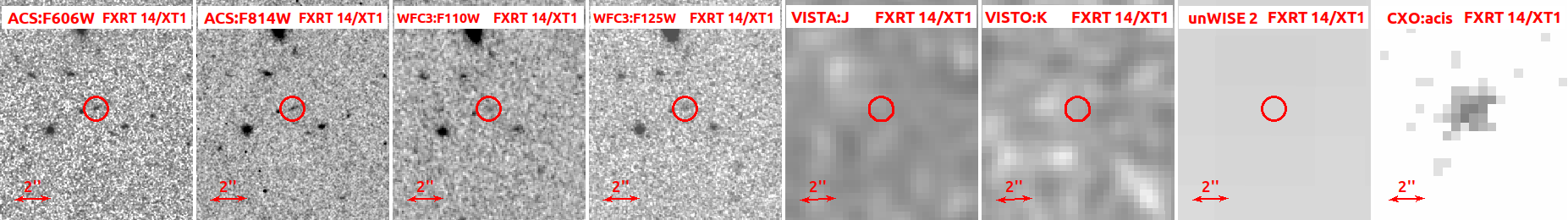}
    \caption[]{(continued)}
\end{figure*}

\subsection{Fainter electromagnetic detections}\label{sec:results}

Having ruled out obvious Galactic and spurious transients, we now focus on a detailed multiwavelength assessment of each remaining candidate using a variety of archival multiwavelength data, in order to try to understand their origin. In Sects. \ref{sec:opt_NIR} to \ref{sec:radio}, we describe a search  counterparts or host galaxies, from radio to gamma rays, of our final sample. To confirm that the final FXRT sample is consistent with real transient objects, in the next section we explain a cross-match with other catalogs.

\subsubsection{Ultraviolet, optical, and near-infrared sources} \label{sec:opt_NIR}

To search for possible UV, optical, NIR and MIR detections of a counterpart or host of each of the FXRTs, we perform a cone search within a radius equivalent to the 3$\sigma$ \emph{Chandra} error position (see Table~\ref{tab:my_detections}) in the following catalogs: 
GALEX Data Release 5 \citep[GR5;][]{Bianchi2011}, 
Pan-STARRS Data Release 2 \citep[Pan-STARRS--DR2;][]{Flewelling2018}, 
the DES Data Release 2 \citep[DES--DR2;][]{Abbott2021}, 
%the Guide Star Catalog \citep[GSC;][]{Lasker2008}, 
the SDSS Data Release 16 \citep[SDSS--DR16;][]{Ahumada2019},
the NOAO Source Catalog Data Release 2 \citep[NSC--DR2;][]{Nidever2020},
the \emph{Hubble} Source Catalog version 3 \citep[HSCv3;][]{Whitmore2016},
the UKIRT InfraRed Deep Sky Survey Data Release 11+\citep[UKIDSS--DR11+;][]{Warren2007},
the UKIRT Hemisphere Survey Data Release 1 \citep[UHS--DR1;][]{Dye2018},
the Two Micron All Sky Survey \citep[2MASS;][]{Skrutskie2006}, 
the VHS band-merged multi-waveband catalogs Data Release 5 \citep[DR5;][]{McMahon2013}, 
the Spitzer Enhanced Imaging Products Source List \citep{Teplitz2010}, 
and the unWide-field Infrared Survey Explorer catalog \citep[unWISE;][]{Schlafly2019}, 
as well as the 
ESO Catalogue Facility and 
the NED \citep{Helou1991}, 
SIMBAD \citep{Wenger2000}, and
VizieR \citep{Ochsenbein2000} databases. We supplement this with any large extended sources found during our archival image analysis in Sect. \ref{sec:archive_images}. We assume that uncertainties in the UV through MIR positions contribute negligibly to the overall error budget.
Figure~\ref{fig:image_cutouts} shows images of the FXRTs (one per row) from Pan-STARRS, DECam, or HST in the optical (\emph{1st--4th columns}, using $g$, $r$, $i$ and $z$ or the corresponding \emph{HST} filters), VISTA, UKIRT or 2MASS in the NIR (\emph{5th and 6th columns}, using $J$, $H$ or $K$ filters), unWISE in the MIR (\emph{7th column}, in the 3.6$\mu$m) band, and the \emph{Chandra}-ACIS image (\emph{8th column}, in the 0.5--7.0\,keV band).

We find clear optical/NIR/MIR extended sources in the above catalogs for two FXRT candidates: FXRT~8 and FXRT~9. In the case of FXRT~13 there is a faint point source inside the 2$\sigma$ localization uncertainty, but only in the $i$ band (see Fig.~\ref{fig:image_cutouts}).
A further six FXRT candidates lie in the immediate vicinity of large, nearby galaxies: FXRT~1, FXRT~2, FXRT~3, FXRT~4, FXRT~5, and FXRT~6. For FXRT~2, FXRT~3, FXRT~4, and FXRT~5, it was possible to identify potential counterparts.
This leaves four FXRT candidates (FXRT~7, FXRT~10, FXRT~11, and FXRT~12) where we could only derive upper limits to the presence of a host or counterpart in moderate-depth imaging; typical limits we derive are $m_{r}{>}$23.7 and $m_{z}{>}$22.4~AB~mag. We note that the fields of FXRT~1 and FXRT~14 have been observed by \citet{Jonker2013} and \citet{Bauer2017}, respectively. In Table~\ref{tab:photometric_data} we list the position, angular offset, and magnitudes of the candidate optical/NIR counterparts or host galaxies when available, and upper limits when not. We briefly describe the counterpart or host galaxy constraints for each FXRT below. 

FXRT~1/XRT\,000519 \citep[identified previously by][]{Jonker2013} is located in the outskirts of the galaxy M86 ($m_{R}{=}8.6$~AB~mag; ${\approx}$17\,Mpc) in the Virgo cluster, at an angular (projected) distance of 12\farcm2 (${\approx}$60\,kpc). This association is still under debate; the Poisson probability of a chance alignment is 3.6$\times$10$^{-4}$ based on its angular offset and the space density of $m_{R}{<}9$~mag galaxies \citep[using the GLADE catalog;][]{Dalya2018}, implying a possible association; however, the binomial probability that this FXRT is a background source is ${\approx}$0.3, indicating that the association with M86 is weak (see Sect. \ref{sec:population} for more details). The transient was previously reported by \citet{Jonker2013} to have two tentative counterparts with $m_{i}{=}24.3$~AB~mag (with an offset of 0\farcs8) and $m_{g}{=}26.8$~AB~mag (with an offset of 1\farcs2) in deeper images taken by the Isaac Newton Telescope (INT) and the Canada France Hawaii Telescope (CFHT), respectively \citep{Jonker2013}.

FXRT~2/XRT\,010908 (cataloged as an X-ray source by \citealp{Wang2016}, \citealp{Liu2011}, and \citealp{Mineo2012}, although never classified as an FXRT), a local FXRT, is located in the disk of the edge-on SB(s)cd galaxy M108 \citep[also known as NGC~3556; $m_{R}{\approx}9.2$~AB~mag and ${\approx}$9.0~Mpc;][]{Dalya2018,Tully2013}, at an angular (projected) distance of 0\farcm4 (${\approx}$1.1~kpc). The probability of a chance alignment is 3.2$\times$10$^{-6}$ based on its angular offset and the space density of $m_{R}{<}9.2$~AB~mag galaxies \citep[using the GLADE catalog;][]{Dalya2018}, thus implying a highly probable association; the binomial probability that this FXRT is a background source is ${\approx}$8.4$\times$10$^{-7}$, reinforcing an association with M108 (see Sect. \ref{sec:population} for more details). FXRT~2 appears to lie at the edge and intersection of two extended star-forming regions (see Fig.~\ref{fig:image_cutouts}, \emph{sources \#1 and \#2} in the northeast and southwest directions, respectively), with several potential, unresolved, optical/NIR candidate counterparts in the \emph{HST} F606W image inside the \emph{Chandra} 3$\sigma$ error circle. The estimated magnitudes of \emph{sources \#1} and \emph{\#2} are $m_{F606W}{=}18.4$ and 18.2~AB~mag (i.e., $M_{F606W}{\gtrsim}-11.4$ and $-11.6$~AB~mag), respectively \citep[taken from the HSCv3;][]{Whitmore2016}. As such, FXRT~2 is likely associated with a region of enhanced high-mass star formation.

FXRT~3/XRT\,070530 (cataloged as an X-ray source by \citealp{Liu2011} and \citealp{Wang2016}, although never classified as an FXRT) is located in the S0 peculiar galaxy NGC\,5128 (Cen\,A; $m_{R}{\approx}6.3$~AB~mag; ${\approx}3.1$\,Mpc), at an angular (projected) distance of 5\farcm5 (${\approx}$5.0~kpc). The probability of this association occurring by chance is 1.3$\times$10$^{-5}$ based on the FXRT--galaxy offset and the space density of $m_{R}{<}12$~AB~mag galaxies, thus implying a highly probable association; the binomial probability that this FXRT is a background source is ${\approx}$1.7$\times$10$^{-2}$, reinforcing an association with NGC\,5128 (see Sect. \ref{sec:population} for more details). There are several dozen possible faint counterpart candidates within the 3$\sigma$ \hbox{X-ray} error region in the \emph{HST} F606W and F814W images (typically $m_{F606W}$ and $m_{F814W}{\gtrsim}25$~AB~mag; see Fig.~\ref{fig:image_cutouts}), of which one very red object stands out near the center (\emph{source \#1} in Fig.~\ref{fig:image_cutouts}) with $m_{F606W}{=}25.4$ and $m_{F814W}{=}22.1$~AB~mag \citep[$M_{F606W}{=}-$2.1 and $M_{F814W}{=}-$5.4~AB~mag, respectively; taken from the HSCv3;][]{Whitmore2016} or from DECam $m_{z}{=}$22.3 and $m_{y}{=}$21.7~AB~mag ($M_{z}{=}-5.2$ and $M_{y}{=}-5.7$~AB~mag), which might be typical of either a small globular cluster or a red supergiant star. Based on the lack of young stars in the local host environment, we associate FXRT~3 with the former.

FXRT~4/XRT\,071203 (cataloged as an X-ray source by \citealp{Mineo2012} and \citealp{Wang2016}, although never classified as an FXRT) is located in the SA(s)cd peculiar dwarf galaxy NGC\,5474 ($m_{R}{=}10.8$~AB~mag; ${\approx}5.9$~Mpc), at an angular (projected) distance of 0\farcm4 (${\approx}$0.7\,kpc). NGC\,5474 is a highly asymmetric late-type peculiar dwarf galaxy in the M101 group, thought to be interacting with M101.
The probability of this occurring by chance is 1.9$\times$10$^{-6}$ based on its angular offset and the space density of $m_{R}{<}10.8$~AB~mag galaxies, thus implying a highly probable association; the binomial probability that this FXRT is a background source is ${\approx}$9.9$\times$10$^{-4}$, reinforcing an association with NGC\,5474 (see Sect. \ref{sec:population} for more details). The FXRT candidate appears to lie at the center of a resolved blue star cluster with a spatial extent of ${\approx}$40\,pc, with ${\approx}$10 candidate unresolved optical/NIR counterparts in \emph{HST} imaging inside the \emph{Chandra} 3$\sigma$ error circle (Fig.~\ref{fig:image_cutouts} shows the four most obvious optical and NIR counterparts). The majority of the candidate counterparts have blue colors, with brightness peaking in F275W and F606W with $m_{F275W}{\approx}21.6$--23.0 and $m_{F606W}{\approx}22.2$--22.9~AB~mag ($M_{F275W}{\approx}-5.9/-7.3$ and $M_{F606W}{\approx}-6.0/-6.7$~AB~mag, and hence consistent with O stars), while \emph{source~\#3} is redder, peaking between F814W and F160W, with $m_{F606W}{\approx}22.3$ and $m_{F814W}{\approx}22.1$~AB~mag ($M_{F606W}{\approx}-6.5$ and $M_{F814W}{\approx}-6.7$~AB~mag, respectively, typical of a massive red supergiant star). The photometric data are taken from the HSCv3 \citep{Whitmore2016}. As such, FXRT~4 is likely associated with a region of enhanced high-mass star formation. 

FXRT~5/XRT\,080331 (cataloged as an X-ray source by \citealp{Wang2016} and \citealp{Sazonov2017}, although never classified as an FXRT) is located in the disk of the SAB(s)b galaxy M66 ($m_{r}{=}9.6$~AB~mag, ${\approx}$11~Mpc), at an angular (projected) distance of 1\farcm3 (${\approx}$4.3~kpc). M66 is a barred spiral galaxy in the Leo group. The probability of this occurring by chance is 2.8$\times$10$^{-6}$ based on its angular offset and the space density of $m_{R}{<}9.6$~AB~mag galaxies, implying a highly probable association; the binomial probability that this FXRT is a background source is ${\approx}$3.9$\times$10$^{-3}$, reinforcing an association with M66 (see Sect. \ref{sec:population} for more details). The FXRT candidate error region is located in a high extinction region of the disk, at the edge of the bar, with very few optical counterpart candidates (${\lesssim}$10 sources). However, the \hbox{X-ray} centroid is notably well aligned with two knots of strong H$\alpha$ emission (\emph{sources 1 and 2} in the \emph{HST/ACS-F658N} image of Fig.~\ref{fig:image_cutouts}) with $m_{658N}{\approx}21.0$~AB~mag (or $M_{658N}{\approx}-9.2$~AB~mag). This suggests a link with a high-mass star formation region, while the 3$\sigma$ error circle encompasses at least ten fainter, unresolved candidate counterparts in the F110W and F160W images ($m_{160W}{\gtrsim}22.5$ or $M_{160W}{\gtrsim}-7.7$~AB~mag).

FXRT~6/XRT\,130822 (cataloged as an X-ray source by \citealp{Wang2016}, although never classified as an FXRT) is situated in the outskirts of the galaxy NGC~7465 ($m_{R}{=}12.0$~AB~mag; ${\approx}$27~Mpc), which is part of the merging NGC~7448 group, at an angular (projected) distance of 1\farcm2 (9.4~kpc). The probability of this occurring by chance is 1.5$\times$10$^{-4}$ based on its offset and the space density of $m_{R}{<}12$~AB~mag galaxies, thus implying a probable association; the binomial probability that this FXRT is a background source is ${\approx}$1.5$\times$10$^{-2}$, reinforcing an association with NGC~7465 (see Sect. \ref{sec:population} for more details). The FXRT position overlaps with a blue spiral arm and lies in between two diffuse blue candidate sources in DECam images (see Fig.~\ref{fig:image_cutouts}, sources \#1 and \#2 in $g$- and $r$-band images). These have offsets of ${\approx}$1\farcs3 to the northwest and 1\farcs5 to the northeast, respectively, which lie just slightly outside of the 3$\sigma$ \hbox{X-ray} error region, but their proximity suggests that FXRT~6 is likely associated with a region of high-mass star formation.

For FXRT~7/XRT\,030511 \citep[identified previously by][]{Lin2019,Lin2022}, no optical and NIR sources are detected within the \hbox{3$\sigma$} \hbox{X-ray} error region of this event in the DECam, VISTA, or unWISE images (see Fig.~\ref{fig:image_cutouts}). Upper limits are given in Table~\ref{tab:photometric_data}.

FXRT~8/XRT\,041230 lies close to a $m_{r}{\approx}23.1$~AB~mag source, at an angular distance of 0\farcs7, detected in DECam and VISTA images (see Fig.~\ref{fig:image_cutouts}, \emph{source \#1}). The probability of a false match \citep[adopting the formalism developed by][]{Bloom2002} is $P{<}0.003$ for such offsets from similar or brighter objects. We analyze the properties of this extended optical/NIR source in detail in Sect. \ref{sec:counterpart_SED}.

FXRT~9/XRT\,080819 lies close to a $m_{r}{\approx}21.1$~AB~mag source, at an angular distance of 0\farcs5, detected in DECam, VISTA, and unWISE images (see Fig.~\ref{fig:image_cutouts}, \emph{source \#1}). The probability of a false match is $P{<}0.0004$ for such offsets from similar or brighter objects. We analyze the properties of this extended optical/NIR source in detail in Sect. \ref{sec:counterpart_SED}.

Regarding FXRT~10/XRT\,100831, no optical/NIR sources are detected within the 3$\sigma$ \hbox{X-ray} error region of this event in the DECam or 2MASS images; upper limits are given in Table~\ref{tab:photometric_data}. There is a moderately bright, marginal DECam object, at an angular distance of 2\farcs6, just outside the 3$\sigma$ error region to the northeast (see Fig.~\ref{fig:image_cutouts}, \emph{source \#1} in the $i$-band DECam image).

Regarding FXRT~11/XRT\,110103 \citep[identified previously by][]{Glennie2015}, no optical/NIR sources are detected within the 3$\sigma$ \hbox{X-ray} error region of this event in the DECam, Pan-STARRS, or VISTA imaging (see Fig.~\ref{fig:image_cutouts}); upper limits are given in Table~\ref{tab:photometric_data}. This FXRT was discovered in an observation of the galaxy cluster Abell~3581 \citep[at a distance of ${\approx}$94.9~Mpc;][]{Johnstone2005,Glennie2015}, where the nearest known member of the cluster, LEDA\,760651 ($m_J{\approx}$16.7~AB~mag), is 2\farcm7 (${\approx}$71.4~kpc) from the \emph{Chandra} transient position \citep{Glennie2015}. The probability of this occurring by chance is 0.15 based on its offset and the space density of $m_{J}{<}16.7$~AB~mag galaxies, thus implying a low probability of association; the binomial probability that this FXRT is a background source is ${\approx}$7.8$\times$10$^{-2}$, reinforcing an unlikely association with LEDA\,760651 (see Sect. \ref{sec:population} for more details).

Regarding FXRT~12/XRT\,110919 \citep[identified previously by][]{Lin2019,Lin2022}, no significant optical and NIR sources are detected within the 3$\sigma$ \hbox{X-ray} error region of this event in the DECam, VISTA or unWISE imaging (see Fig.~\ref{fig:image_cutouts}) and catalogs, although we note that a marginal source (${\lesssim}$2$\sigma$) appears in red filters (DECam $z$-band and VISTA $K$-band); upper limits are given in Table~\ref{tab:photometric_data}.

FXRT~13/XRT\,140327 lies close to a faint, $m_{i}{\approx}24.7$~AB~mag source (see Fig.~\ref{fig:image_cutouts}, \emph{source \#1}), at an angular distance of 1\farcs5, detected in DECam $i$-band and marginally visible in $r$-band imaging. The probability of a false match is $P{<}0.004$ for such offsets from similar or brighter objects.

Finally, FXRT~14/XRT\,141001/CDF-S~XT1 \citep[identified previously by][]{Bauer2017} lies close to a faint ($m_{R}{=}27.2$ and $m_{J}{=}27.1$~AB~mag or $M_{R}{\approx}-19.0$ and $M_J{\approx}-19.1$~AB~mag, respectively, assuming $z_{\rm pho}{=}$2.23), extended ($r_{\rm Kron}{=}$0\farcs56) optical and NIR source in \emph{HST} imaging (see Fig.~\ref{fig:image_cutouts}), with an angular offset of 0\farcs13.

Overall, we find that six of the 14 FXRT candidates (FXRT~1--6) have high probabilities of being associated with nearby galaxies \citep[${<}$30~Mpc: FXRTs~2--5 show clear potential counterparts and FXRT~6 lies on top of faint optical emission, while FXRT~1 is still under consideration to be a distant event;][]{Eappachen2022}.\footnote{We caution that the probabilities calculated above could be overestimated, depending on the targeting biases among the \emph{Chandra} observations.}
Among the other eight candidates, three (FXRTs~8, 9, and 13) are coincident with moderately bright extended sources within the 3$\sigma$ position error, FXRT~14/CDF-S~XT1 is coincident with a faint extended source, and for three (FXRTs~7, 10, and 12) no optical or IR emission is detected to moderate-depth limits ($m_r{\lesssim}24.5$~AB~mag). In the case of FXRT~11, we do not discard its association with nearby galaxies completely (${\approx}$94.9~Mpc); however, a relation with a background source could be more likely.
Finally, based on arguments given in Sect. \ref{sec:galactic}, FXRTs~7, 10, 11, and 12 are highly likely to be extragalactic and have relatively distant and faint optical or NIR hosts similar to or fainter than CDF-S~XT1.

\begin{table*}
    \centering
    \advance\leftskip-0.1cm
    \scalebox{0.6}{
    \begin{tabular}{llllllllllllll}
    \hline\hline
    FXRT & Id & $m_u$ & $m_g$ & $m_r$ & $m_i$ & $m_z$ & $m_y$ & $m_Y$ & $m_J$ & $m_H$ & $m_K$ & W1 & W2 \\ \hline
    (1) & (2) & (3) & (4) & (5) & (6) & (7) & (8) & (9) & (10) & (11) & (12) & (13) & (14) \\ \hline
    \multicolumn{14}{c}{Nearby extragalactic FXRT Candidates from CSC2} \\ \hline
    1 & XRT\,000519 & ${>}$22.30$^{a}$ & 26.8${\pm}$0.1$^{i}$ & ${>}$21.60$^{a}$ & 24.3${\pm}$0.1$^{i}$ & ${>}$21.11$^{a}$ & ${>}$20.06$^{a}$ & ${>}$20.27$^{b}$ & ${>}$20.11$^{b}$ & ${>}$20.15$^{b}$ & ${>}$19.82$^{b}$ & ${>}$19.91$^{c}$ & ${>}$20.11$^{c}$ \\
    2 & XRT\,010908(S1)$^{\dagger}$ & ${>}$22.45$^{h}$  & 18.15$\pm$0.04$^{a}$ & 18.11$\pm$0.34$^{a}$ & 16.99$\pm$0.01$^{a}$ & 18.74$\pm$0.13$^{a}$ & 18.64$\pm$0.06$^{a}$ & -- & ${>}$17.32$^{f}$ & ${>}$16.61$^{f}$ & ${>}$16.68$^{f}$ & ${>}$19.91$^{c}$ & ${>}$20.82$^{c}$ \\ 
     & XRT\,010908(S2)$^{\dagger}$ & ${>}$22.45$^{h}$  & 18.77$\pm$0.01$^{a}$ & 17.80$\pm$0.06$^{a}$ & 17.12$\pm$0.14$^{a}$ & 17.50$\pm$0.25$^{a}$ & 16.22$\pm$0.10$^{a}$ & -- & ${>}$17.32$^{f}$ & ${>}$16.61$^{f}$ & 15.21$\pm$0.13$^{f}$ & 13.49$\pm$0.03$^{c}$ & 13.65$\pm$0.02$^{c}$ \\ 

    3 & XRT\,070530(S1)$^{\dagger}$ & ${>}$22.56$^{e}$ & ${>}$24.30$^{e}$ & ${>}$23.10$^{e}$ & ${>}$22.52$^{e}$ & 22.33$\pm$0.18$^{e}$ & 21.75$\pm$0.07$^{e}$ & -- & ${>}$19.82$^{g}$ & ${>}$17.05$^{f}$ & ${>}$19.45$^{g}$ & ${>}$18.68$^{c}$ & ${>}$20.41$^{c}$ \\ 
    6 & XRT\,130822 & ${>}$23.87$^{h}$ & ${>}$21.89$^{a}$ & ${>}$21.66$^{a}$ & ${>}$21.45$^{a}$ & ${>}$20.96$^{a}$ & ${>}$19.88$^{a}$ & -- & ${>}$17.46$^{f}$ & ${>}$17.21$^{f}$ & ${>}$17.24$^{f}$ & ${>}$19.75$^{c}$ & ${>}$20.17$^{c}$ \\ \hline
    \multicolumn{14}{c}{Distant extragalactic FXRT Candidates from CSC2} \\ \hline
    7 & XRT\,030511 & -- & ${>}$23.40$^{e}$ & ${>}$23.27$^{e}$ & ${>}$21.12$^{a}$ & ${>}$21.80$^{e}$ & ${>}$23.68$^{d}$ & -- & ${>}$21.59$^{g}$ & ${>}$17.23$^{f}$ & ${>}$19.74$^{g}$ & ${>}$20.36$^{c}$ & ${>}$20.48$^{c}$ \\
    8 & XRT\,041230 & -- & 25.74$\pm$0.84$^{d}$ & 23.01$\pm$0.10$^{d}$ & 21.96$\pm$0.07$^{d}$ & 21.65$\pm$0.10$^{d}$ & 22.96$\pm$1.17$^{d}$ & -- & 21.30$\pm$0.23$^{g}$ & ${>}$17.40$^{f}$ & ${>}$17.52$^{f}$ & ${>}$20.20$^{c}$ & ${>}$20.56$^{c}$ \\
    9 & XRT\,080819 & -- & 21.85$\pm$0.05$^{e}$ & 21.12$\pm$0.02$^{e}$ & 20.57$\pm$0.03$^{e}$ & 20.42$\pm$0.07$^{e}$ & 20.34$\pm$0.11$^{e}$ & -- & 20.21$\pm$0.20$^{g}$& ${>}$17.24$^{f}$ & 19.46$\pm$0.16$^{g}$ & 18.74$\pm$0.05$^{c}$ & 18.63$\pm$0.10$^{c}$ \\
    10 & XRT\,100831 & -- & ${>}$24.49$^{d}$ & ${>}$24.29$^{d}$ & ${>}$24.23$^{d}$ & ${>}$23.96$^{d}$ & ${>}$23.43$^{d}$ & -- & ${>}$17.35$^{f}$ & ${>}$17.16$^{f}$ & ${>}$17.28$^{f}$ & ${>}$20.59$^{c}$ & ${>}$20.87$^{c}$ \\
    11 & XRT\,110103 & -- & ${>}$23.19$^{e}$ & ${>}$22.95$^{e}$ & ${>}$21.06$^{a}$ & ${>}$21.84$^{e}$ & ${>}$19.67$^{a}$ & ${>}$20.61$^{g}$ & ${>}$20.07$^{g}$ & ${>}$17.30$^{f}$ & ${>}$19.22$^{g}$ & ${>}$20.01$^{c}$ & ${>}$20.27$^{c}$ \\
    12 & XRT\,110919 & -- & ${>}$24.78$^{d}$ & ${>}$24.40$^{d}$ & ${>}$24.36$^{d}$ & ${>}$24.12$^{d}$ & ${>}$23.56$^{d}$ & -- & ${>}$17.32$^{f}$ & ${>}$17.29$^{f}$ & ${>}$17.30$^{f}$ & ${>}$20.18$^{c}$ & ${>}$20.35$^{c}$ \\
    13 & XRT\,140327 & -- & ${>}$23.20$^{e}$ & ${>}$22.63$^{e}$ & 24.7$\pm$0.3$^{d\dagger\dagger}$ & ${>}$21.66$^{e}$ & -- & ${>}$20.88$^{g}$ & ${>}$20.45$^{g}$ & ${>}$17.01$^{f}$ & ${>}$19.86$^{g}$ & ${>}$20.24$^{c}$ & ${>}$20.65$^{c}$ \\
    14 & XRT\,141001 & 27.30$\pm$0.12$^{j}$ & 27.87$\pm$0.35$^{j}$ & 27.21$\pm$0.10$^{j}$ & 27.13$\pm$0.21$^{j}$ & 27.01$\pm$0.22$^{j}$ & ${>}$23.68$^{d}$ & 26.87$\pm$0.20$^{j}$ & 27.11$\pm$0.23$^{j}$ & 26.53$\pm$0.17$^{j}$ & 26.07$\pm$1.05$^{j}$ & 24.75$\pm$0.23$^{j}$ & 25.28$\pm$0.26$^{j}$ \\
     & /CDF-S~XT1 &  & & & & & & & & & & & \\ \hline
    \end{tabular}
    }
    \caption{Host and/or counterpart's photometric data or upper limits of FXRT candidates. All magnitudes are converted to the AB magnitude system using \citet{Gonzalez2018} for VHS and 2MASS data, \citet{Hewett2006} for UKIDSS data, and \citet{Wright2010} for unWISE data. If an optical/NIR counterpart candidate is detected, we list its magnitude and 1$\sigma$ error, otherwise we provide 3$\sigma$ limits from several catalogs: $^{a}$Pan-STARRS-DR2 \citep{Flewelling2018}, $^{b}$UKIDSS-DR11+ \citep{Warren2007}, $^{c}$unWISE \citep{Schlafly2019}, $^{d}$DES-DR2 \citep{Abbott2021}, $^{e}$NSC-DR2p \citep{Nidever2020}, $^{f}$2MASS \citep{Skrutskie2006}, $^{g}$VHS-DR5 \citep{McMahon2013}, $^{h}$SDSS-DR16 \citep{Ahumada2019}, $^{i}$ INT/CFHT \citep{Jonker2013}, $^{j}$ CANDELS \citep[nearest HST/Spitzer bands substituted: $g{=}F435W$, $r{=}F606W$, $i{=}F814W$, $z{=}F850LP$, $Y{=}F105W$, $J{=}F125W$, $H{=}F160W$, $W1{=}ch1$, $W2{=}ch2$;][]{Guo2013}. We omit entries for FXRTs~4 and 5, as both candidates have up to $\approx$~10 potential counterparts in \emph{HST} images. \\
    $\dagger$ Photometric data of FXRTs with counterpart(s) (S+\# means the \emph{source number}). \\
     $\dagger\dagger$ Obtained using a photometric aperture of 3.7 pixels.}
    \label{tab:photometric_data}
\end{table*}

\subsubsection{Higher energy counterparts}\label{sec:gamma}

To investigate if the sky locations of the FXRTs are covered by hard \hbox{X-ray} and $\gamma$-ray observations, we performed a cone search in the \emph{Swift}-Burst Alert Telescope \citep[\emph{Swift}-BAT;][]{Sakamoto2008}, INTErnational Gamma-Ray Astrophysics Laboratory \citep[\emph{INTEGRAL};][]{Rau2005}, High Energy Transient Explorer 2 \citep[\emph{HETE-2};][]{Hurley2011}, \emph{InterPlanetary Network} \citep{Ajello2019}, and \emph{Fermi} \citep{von_Kienlin2014,Narayana2016} archives. We adopt a 10\farcm0 search radius for the \emph{INTEGRAL}, \emph{Swift}-BAT, \emph{HETE-2} and \emph{Interplanetary Network} Gamma-Ray Bursts catalogs, while for the Gamma-ray Burst Monitor (\emph{GBM}) and the Large Area Telescope (LAT) \emph{Fermi} Burst catalogs we take a search radius of 4~deg \citep[which represents typical source positional uncertainties at the $\approx$68\% confidence level for those detectors;][]{Connaughton2015}. We find no hard \hbox{X-ray} or $\gamma$-ray counterparts associated with \emph{INTEGRAL}, \emph{Swift}-BAT, \emph{HETE-2,} and \emph{Interplanetary Network} catalogs. Some of the nearby (FXRTs~3, 4, and 6) and distant (FXRTs~7, 8, and 9) candidates have a potential gamma ray association in the \emph{GBM Fermi Burst} catalog; however, we rule out their association for FXRTs~3, 4, 6, 7, and 8 because of a large difference in time between the FXRT and gamma-ray detection (${\gtrsim}$4~years).

In the case of FXRT~9, it has a \emph{GBM Fermi} GRB detection (called GRB\,080812 at $\alpha$=11$^{\text{h}}$46$^{\text{m}}$48.\!\!$^{\text{s}}$00, $\delta$=--33$^\circ$12$^\prime$) seven days before the \emph{Chandra} trigger, with an offset of ${\approx}$1.9~deg, positional uncertainty of 4.1~deg, and $T_{90}{\approx}$15~sec \citep{Narayana2016}. In an on-axis scenario, the beamed X-ray emission should be detected effectively concurrently with the GRB; this is inconsistent with the observed light curve shown in Fig.~\ref{fig:light_curves_1}. For an off-axis scenario, a delay between the gamma-ray trigger and its peak \hbox{X-ray} afterglow depends on both intrinsic (e.g., the off-axis angle and the deceleration timescale of the outflow) and extrinsic (e.g., the low densities density of the BNS environment and the observer location) properties \citep[e.g.,][]{Granot2002,Granot2018b,Granot2018a,Troja2020,Lamb2021}, effectively spanning all timescales. Strong \hbox{X-ray} flares have been known to occur on top of \hbox{X-ray} afterglow emission, but these typically occur during the early phase of the afterglow \citep[${\lesssim}10^{3}$--$10^{4}$\,s; e.g.,][]{Yi2016}. As such, an association between GRB\,080812 and FXRT~9 seems unlikely.

In summary, none of our FXRT candidates has an associated detection at hard \hbox{X-ray} or gamma-ray wavelengths.

\subsubsection{Radio counterparts} \label{sec:radio}

To search for possible radio counterparts to our FXRT candidates, we utilize the \emph{RADIO--Master Radio Catalog}, which is a periodically revised master catalog that contains selected parameters from a number of the \texttt{HEASARC} database tables that hold information on radio sources from 34~MHz to 857~GHz. This catalog contains inputs from several telescopes and surveys such as the Australia Telescope Compact Array, the Very Large Array, the Very Long Baseline Array, and the Wilkinson Microwave Anisotropy Probe. Given the relatively poor angular resolution of some of these radio telescopes, 
we perform an initial cone search for radio sources within 60\arcsec. Only FXRTs~2, 4, and 5, all of which are associated with hosts at ${\lesssim}$10~Mpc, have radio sources within 60\arcsec. Following this initial 60$"$ cut, we refine our search using limiting radii consistent with the combined radio + \hbox{X-ray} 3$\sigma$ positional errors, which yields no matches. Due to their mutual association with nearby galaxies, we cannot rule out a chance association, as the radio emission could easily arise from other mechanisms within the host galaxies.
Therefore, we conclude that none of the FXRTs is unambiguously detected at radio wavelengths.

\begin{figure*}
    \centering
    \includegraphics[scale=0.7]{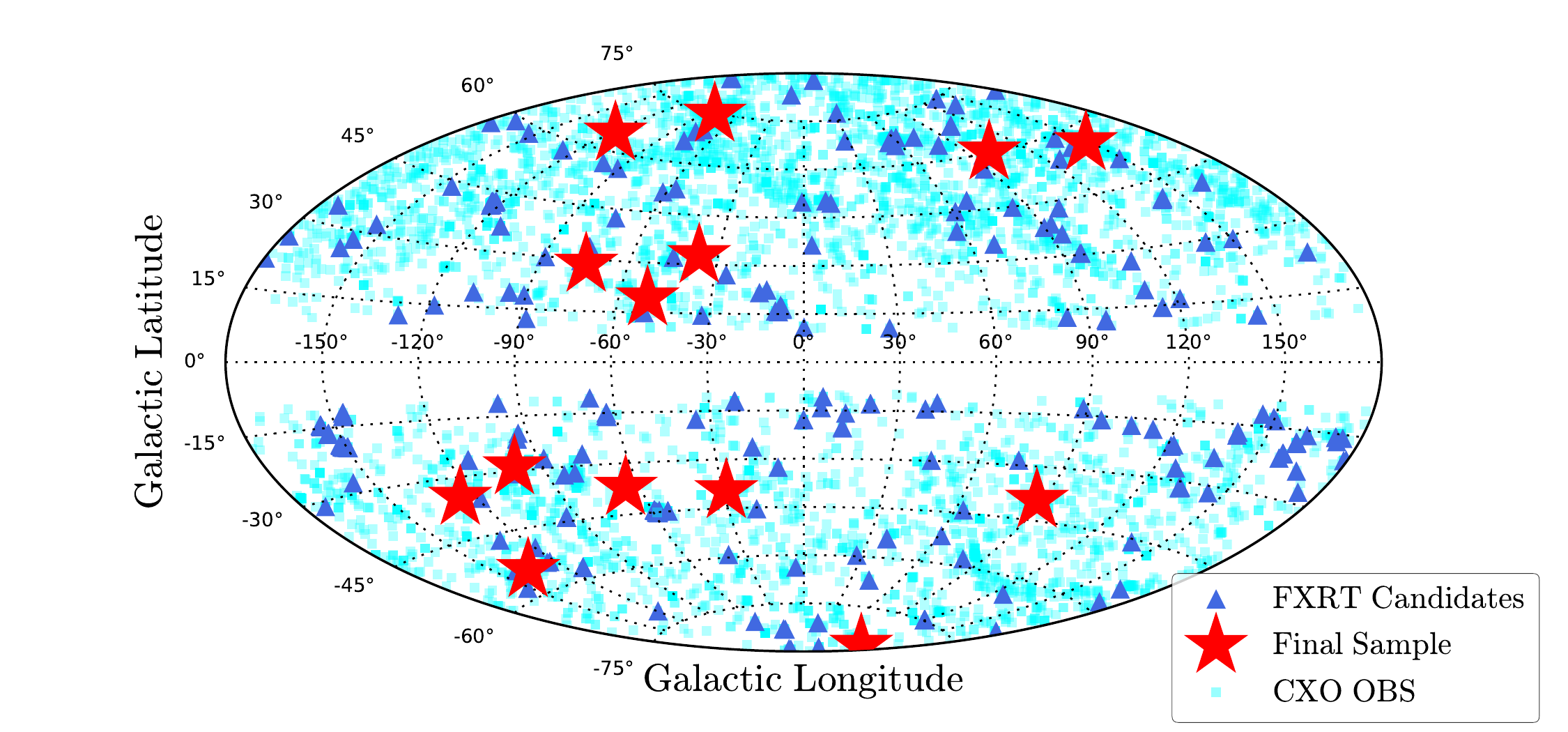}
    \vspace{-0.3cm}
    \caption{Positions on the sky, in Galactic coordinates, of FXRT candidates. The initial 728 candidates are represented by \emph{blue triangles}. The {final sample} of 14 extragalactic FXRT candidates from this work are denoted by \emph{large red stars}. The 5303 \emph{Chandra} observations used in this work are also shown (\emph{cyan squares}).}
    \label{fig:positions}
\end{figure*}

\begin{table*}
\centering
\scalebox{0.8}{
    \begin{tabular}{llllllllll}
    \hline\hline
    FXRT & ID & $T_0$(UTC) & Model & T$_{\text{break}}$(ks) & $\tau_1$ & $\tau_2$ & $F_0$ (erg~cm$^{-2}$~s$^{-1}$) & $\ln{\mathcal{L}}$(dof) & BIC \\ \hline
    (1) & (2) & (3) & (4) & (5) & (6) & (7) & (8) & (9) & (10)  \\ \hline
    \multicolumn{9}{c}{Nearby extragalactic FXRT Candidates from CSC2} \\ \hline
    1 & XRT\,000519 & 2000-05-19 10:39:36.50 & BPL & -- & -- & -- & -- & -- & -- \\
      & & & PL & -- & -- & -- & -- & -- & -- \\ \hline
    2 & XRT\,010908 & 2001-09-08 14:34:53.43 & BPL & 5.9${\pm}$0.1 & 0.04${\pm}$0.1 & 1.7${\pm}$0.3 & (3.6${\pm}$0.7)${\times}$10$^{-14}$ & 177.1(8) & $-344.2$ \\
     & & & PL & -- & 0.3${\pm}$0.1 & -- & (1.6${\pm}$1.2)${\times}$10$^{-13}$ & 168.9/10 & $-332.7$ \\ \hline
    3 & XRT\,070530 & 2007-05-30 06:15:13.58 & BPL & 1.5${\pm}$0.1 & $-0.1{\pm}0.1$ & 0.8${\pm}$0.2 & (1.9${\pm}$0.7)${\times}$10$^{-14}$ & 108.8(3) & $-212.2$ \\
     & & & PL & -- & -0.2${\pm}$0.1 & -- & (2.5${\pm}$1.5)${\times}$10$^{-14}$ & 108.0/5 & $-209.9$ \\ \hline
    4 & XRT\,071203 & 2007-12-03 08:49:55.59 & BPL & -- & -- & -- & -- & -- & -- \\
      & & & PL & -- & -- & -- & -- & -- & -- \\ \hline
    5 & XRT\,080331 & 2008-03-31 17:05:54.64 & BPL & -- & -- & -- & -- & -- & -- \\
      & & & PL & -- & -- & -- & -- & -- & -- \\ \hline
    6 & XRT\,130822 & 2013-08-22 16:27:24.82 & BPL & 12.3${\pm}$1.2 & 0.2${\pm}$0.1 & 4.1${\pm}$0.1 & (6.1${\pm}$0.8)${\times}$10$^{-15}$ & 109.3(3) & $-210.8$ \\ 
     & & & PL & -- & 0.3${\pm}$0.1 & -- & (4.8${\pm}$2.9)${\times}$10$^{-14}$ & 107.4/5 & $-210.8$ \\ \hline
    \multicolumn{9}{c}{Distant extragalactic FXRT Candidates from CSC2} \\ \hline
    7 & XRT\,030511 & 2003-05-11 04:39:39.66 & BPL & 1.1${\pm}$0.1 & $-0.2{\pm}0.1$ & 1.6${\pm}$0.1 & (1.4${\times}$0.1)${\times}$10$^{-12}$ & 719.3(53) & $-1422.5$ \\
    & & & PL & -- & 0.4${\pm}$0.1 & -- & (3.9${\pm}$2.1)${\times}$10$^{-12}$ & 611.1/51 & $-1214.2$ \\ \hline
    8 & XRT~041230 & 2004-12-30 15:40:07.36 & BPL & 23.8${\pm}$1.5 & $-0.2{\pm}0.1$ & 2.0${\pm}$0.1 & (5.6${\pm}$0.5)${\times}$10$^{-15}$ & 95.7(2) & $-184.1$ \\
    & & & PL & -- & $-0.1{\pm}$0.1 & -- & (1.1${\pm}$0.6)${\times}$10$^{-15}$ & 94.9/4 & $-186.3$ \\ \hline
    9 & XRT~080819 & 2008-08-19 03:22:21.83 & BPL & 5.3${\pm}$0.2 & $-0.2{\pm}0.2$ & 2.8${\pm}$1.9 & (1.8${\pm}$0.7)${\times}$10$^{-14}$ & 118.5(4) & $-229.7$ \\
    & & & PL & -- & $-0.1{\pm}0.1$ & -- & (5.3${\pm}$5.0)${\times}$10$^{-15}$ & 116.7/6 & $-228.1$ \\ \hline %arr
    10 & XRT\,100831 & 2010-08-31 12:03:28.53 & BPL & 2.7${\pm}$0.3 & $-0.0{\pm}0.1$ & 2.4${\pm}$0.4 & (2.0${\pm}$0.4)${\times}$10$^{-14}$ & 137.7(5) & $-266.7$ \\
    & & & PL & -- & 0.4${\pm}$0.1 & -- & (7.8${\pm}$6.6)${\times}$10$^{-14}$ & 131.4/7 & $-258.4$ \\ \hline
    11 & XRT\,110103 & 2011-01-03 21:13:02.14 & BPL & -- & -- & -- & -- & -- & -- \\
      & & & PL & -- & -- & -- & -- & -- & -- \\ \hline
    12 & XRT\,110919 & 2011-09-19 20:04:50.31 & BPL & 1.8${\pm}$0.2 & $-0.0{\pm}0.1$ & 1.9${\pm}$0.2 & (2.0${\pm}$0.3)${\times}$10$^{-13}$ & 307.5(18) & $-602.7$ \\
    & & & PL & -- & 0.5${\pm}$0.1 & -- & (1.2${\pm}$0.8)${\times}$10$^{-12}$ & 286.8/20 & $-567.5$ \\ \hline
    13 & XRT\,140327 & 2014-03-27 13:30:57.11 & BPL & 0.2${\pm}$0.2 & $-1.4{\pm}0.6$ & 0.2${\pm}$0.3 & (1.5${\pm}$1.6)${\times}$10$^{-14}$ & 77.6(1) & $-148.8$ \\
    & & & PL & -- & 0.2${\pm}$0.0 & -- & (3.3${\pm}$0.7)${\times}$10$^{-14}$ & 77.8/3 & $-152.4$ \\ \hline
    14 & XRT\,141001/ & 2014-10-01 07:04:26.20 & BPL & 0.2${\pm}$0.1 & $-0.4{\pm}0.1$ & 1.6${\pm}$0.1 & (8.2${\pm}$1.2)${\times}$10$^{-13}$ & 393.9(25) & $-774.3$ \\
    & CDF-XT1 & & PL & -- & 0.7${\pm}$0.1 & -- & (3.5${\pm}$2.5)${\times}$10$^{-12}$ & 357.5/27 & $-708.2$  \\ \hline 
    \end{tabular}
    }
    \caption{Best-fit parameters obtained using a broken power-law (BPL) and a power-law (PL) model fit to the \hbox{X-ray} light curves. 
    \emph{Columns 1 and 2:} FXRT\# and ID of the candidate, respectively. 
    \emph{Column 3:} Time when the count rate is 3$\sigma$ higher than the Poisson background level.
    \emph{Column 4:} Model used. 
    \emph{Column 5:} Break time for the  BPL model.
    \emph{Columns 6 and 7:} Slope(s) for the BPL or PL model. 
    \emph{Column 8:} Normalization for the BPL or PL model. 
    \emph{Columns 9 and 10:} Log-likelihood ($\ln{\mathcal{L}}$)/degrees-of-freedom (dof) and \emph{Bayesian information criterion} (BIC) of the fit, respectively. Errors are quoted at the 1$\sigma$ confidence level.}
    \label{tab:fitting_para}
\end{table*}

\begin{figure*}
    \centering
    \includegraphics[scale=0.65]{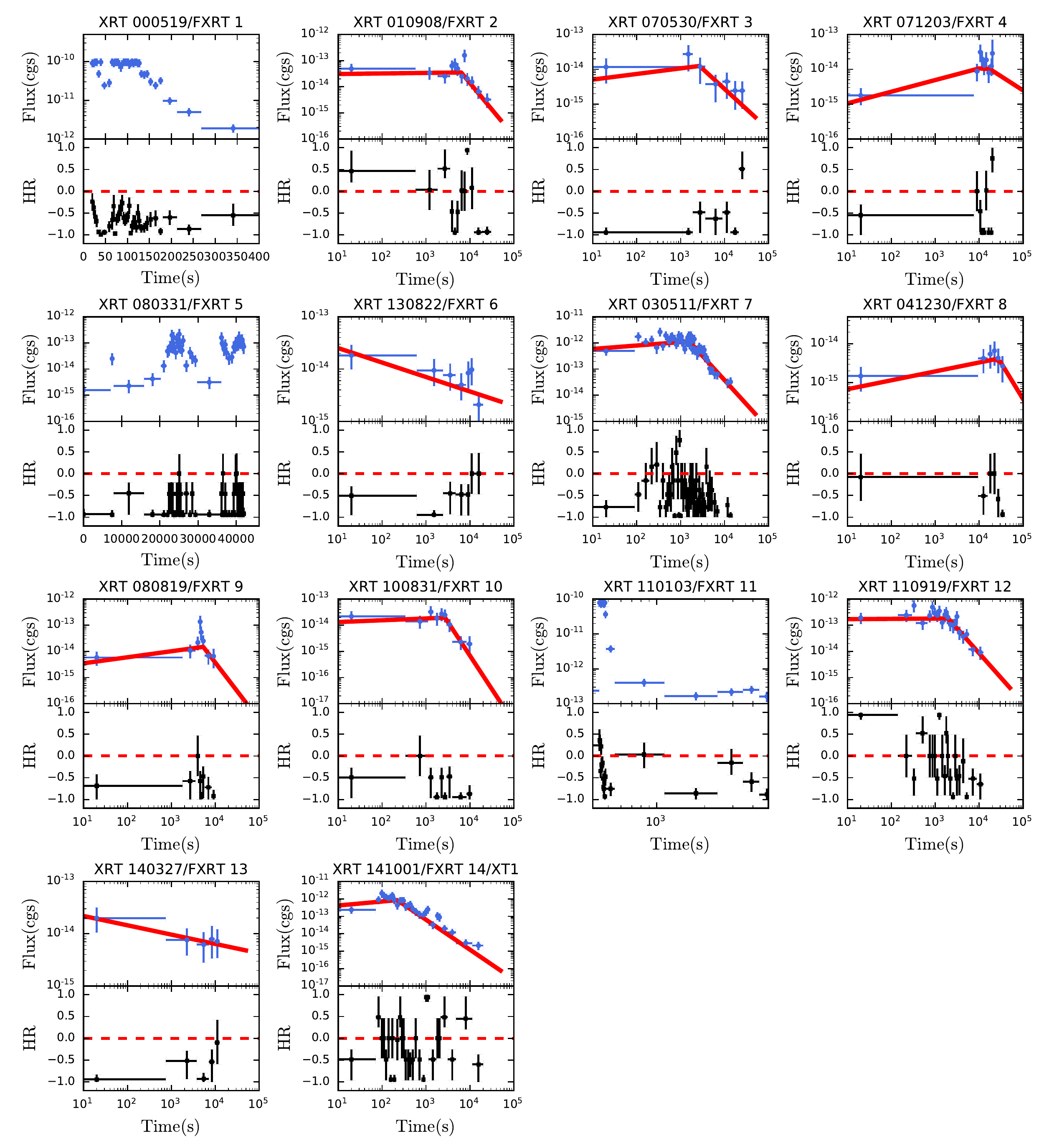}
    \vspace{-0.2cm} \caption{
    Light curves, the evolution of the HR over time, and the best fitting models of the FXRT sample. Top panels: Observed 0.5--7.0\,keV \hbox{X-ray} light curves in cgs units (\emph{blue points}), starting at $T{=}20$~seconds. For FXRTs~1 and 11, we only show the main event. For ten FXRT candidates, we also plot the best-fit BPL or simple PL model (\emph{red solid lines}), while for the remaining four FXRT candidates we do not because they are not well described by either model. The light curves contain five counts per bin (except that of FXRT~1, which has 20 counts per bin). \emph{Bottom panels:} HR evolution (the soft and hard energy bands are 0.5--2.0 keV and 2.0--7.0 keV, respectively), following the Bayesian method of \citet{Park2006}. The \emph{dashed red line} denotes an HR equal to zero. For XRT\,000519/FXRT~1 and XRT\,110103/FXRT~11, we show close-ups of the main flare to highlight in more detail their spectral behavior. Here, $T_0=0$~s is defined as the time when the count rate is 3$\sigma$ higher than the Poisson background level.}
    \label{fig:models_BPL}
\end{figure*}

\section{Spatial, temporal, and X-ray spectral properties}\label{sec:time_spectra_prop}

We investigate the spatial distribution of the final sample of FXRT candidates in Sect. \ref{sec:spatial}. Furthermore, the \hbox{X-ray} temporal and spectral properties can provide essential information about the origin and physical processes behind the FXRT candidates, and thus we describe these in Sects. \ref{sec:X-ray_LC} and  \ref{sec:X-ray_fitting}, respectively. With these in hand, we revisit whether any of the remaining FXRT candidates could be Galactic stellar flares in Sect. \ref{sec:galactic}. Finally, we explore the robustness of the existence of two populations of FXRTs in Sect. \ref{sec:population}.

\subsection{Spatial properties}\label{sec:spatial}

If the FXRT candidates are extragalactic, and given the isotropy of the universe on large scales, we expect the FXRT spatial distribution to be randomly distributed on the sky (see Fig~\ref{fig:positions}).
First, we investigate the sky distribution of all the \emph{Chandra} observations considered in this work using the nonparametric Kolmogorov--Smirnov (K-S) test \citep{Kolmogorov1933,Massey1951,Ishak2014}. We generate 5,303 points (equal to the total number of observations in the CSC2 at $|b|{>}$10~deg) randomly distributed (in Galactic coordinates), and we compare the generated random distributions and the real \emph{Chandra} observations using a 2D K-S test following \citet{Peacock1983} and \citet{Fasano1987}. We performed this process 10,000 times. As a result, we found that the null hypothesis $\mathcal{NH}$ that the random sample and the real data come from the same distribution
is rejected in ${\approx}$20\% of the draws (rejection of $\mathcal{NH}$ occurs when $P{<}$0.05). This is not surprising,  since the \emph{Chandra} pointings are not completely random and some sky regions are observed much more often than others (e.g., Magellanic clouds, Chandra Deep Field South/North; \citealp{Tananbaum2014}, \citealp{Wilkes2019}).

Next, we investigate whether the spatial distribution of the sample of FXRTs is random. Here we simulate 10,000 samples of 214,701 random sources (i.e., the number of \hbox{X-ray} sources analyzed in this work) distributed over the sky, taking as a prior distribution the CSC2 sky positions (which are functions of the pointings and exposures). Out of these 214,701 source we randomly select 14 sources, which we compare to the spatial distribution of the 14 FXRT candidates. We can reject the null hypothesis that these sources are drawn from the same (random) distribution only in ${\approx}$0.25\% of the draws. Therefore, we conclude that the sample of 14 FXRT candidates are randomly distributed over the \emph{Chandra} CSC2 observations of the sky.

\subsection{Temporal properties}\label{sec:X-ray_LC}

We characterize the \hbox{X-ray} light curves of the candidate FXRTs using single PL and broken power-law (BPL) models, and measure the break times and light-curve slopes. Both models describe the majority of the \hbox{X-ray} light curves well, although FXRTs~1, 4, 5, and 11 have more complex light curves and are not well described by these simple models.  Nevertheless, in what follows we describe the most important results of these fits. The PL model is given by
\begin{equation}
    F_{\rm X,PL}(t)=F_0 \times t^{-\tau_1},
\end{equation}
where $\tau_1$ and $F_0$ are the PL index and normalization, respectively. Moreover, the BPL model takes the form
\begin{equation}
    F_{\rm X,BPL}(t)=F_0\times\left\{
    \begin{matrix}
     \left(\frac{t}{T_{\text{break}}}\right)^{-\tau_1} & t{\leq}T_{\text{break}} \\
     \left(\frac{t}{T_{\text{break}}}\right)^{-\tau_2} & t{>}T_{\text{break}}
    \end{matrix}\right.,
\end{equation}
where T$_{\text{break}}$, $\tau_1$, $\tau_2$, and $F_0$ are the break time, the PL slope before and after the break, and normalization, respectively. 
The best-fit model parameters and statistics are given in Table~\ref{tab:fitting_para}, while the light curves (in flux units; light curves have five counts per bin, except FXRT~1, which has ten counts per bin) and best-fit models are shown in Fig.~\ref{fig:models_BPL}. We used the {Bayesian information criterion} (BIC)\footnote{BIC${=}-2\ln{\mathcal{L}}+k\ln{N}$, where $\mathcal{L}$ is the maximum value of the data likelihood, $k$ is the number of model parameters, and $N$ is the number of data points \citep{Ivezic2014}.} to determine which of the two models describes the data best.

For events where the adopted model does not provide a statistically good fit (because of the complex light curve shape), we only explain their main characteristics. We define the light curve zero point ($T{=}0$~sec) as the time when the count rate is 3$\sigma$ higher than the Poisson background level\footnote{It is important to note that the light curve parameters (slopes and break time) can change considering different zero points, especially for FXRTs with high background levels and/or high offset angles. For instance, in \citet{Bauer2017} and \citet{Xue2019}, the zero point is arbitrarily set to be 10 seconds before the arrival of the first photon. This is consistent with our method and does not change interpretations because of the low background level of both observations.}. The light curves and the fits (where applicable) are shown in Fig.~\ref{fig:models_BPL}, while the model fit results are given in Table~\ref{tab:fitting_para} for all the FXRTs. We briefly describe the timing properties for each candidate.

The light curve of FXRT~1/XRT\,000519 exhibits a strong flare at ${\approx}$9.6~ks into the observation. It has some faint precursor emission (not shown in Fig.~\ref{fig:models_BPL}) during the ${\sim}$4~ks prior to the flare at flux levels of ${\approx}$2--5${\times}$10$^{-13}$~erg~cm$^{-2}$~s$^{-1}$, followed by a sudden increase (in ${\approx}$20 seconds) reaching a peak flux of ${\approx}$1.0 ${\times}$10$^{-10}$~erg~cm$^{-2}$~s$^{-1}$. Using a bin-width of 10~s., the main flare is resolved into two peaks, as was also reported by \citet{Jonker2013}. From there, the flux decreases rapidly for ${\approx}$100~s., followed by a slow decline around ${\lesssim}$1--2 ${\times}$10$^{-12}$~erg~cm$^{-2}$~s$^{-1}$ for the next ${\approx}$15~ks \citep[with an index of $-$0.3$\pm$0.1;][]{Jonker2013} until the end of the observation. 

Based on the BIC, the light curve of FXRT~2/XRT\,010908 is described better by a BPL model than by a PL model. The plateau phase has a duration and flux of $T_{\rm break}{\approx}$6.0~ks and ${\approx}$5${\times}$10$^{-14}$~erg~cm$^{-2}$~s$^{-1}$, respectively, followed by a PL decay with an index of ${\approx}-1.7$.

The light curve of FXRT~3/XRT\,070530 is well described by a BPL model, although the $\Delta$BIC is only $-$2.3 with respect to the PL fit. Initially, the light curve increases slightly with an index of ${\approx}$0.1 until $T_{\rm break}{\approx}$1.5~ks, reaching a flux of ${\approx}$2${\times}$10$^{-14}$~erg~cm$^{-2}$~s$^{-1}$. After $T_{\rm break}$, the light curve decays slowly with a slope of ${\approx}0.8$.

The light curve of candidate FXRT~4/XRT\,071203 shows three counts during the first ${\approx}$9--10~ks (equivalent to a flux of ${\lesssim}$2${\times}$10$^{-15}$~erg~cm$^{-2}$~s$^{-1}$), before its flux increases to ${\approx}$4${\times}$10$^{-14}$~erg~cm$^{-2}$~s$^{-1}$ around ${\approx}$20--24~ks, for a duration of ${\approx}$12--14~ks.

    The light curve of FXRT~5/XRT\,080331 shows multiple peaks. In the first ${\approx}$20~ks prior to the bright flares, the flux is around ${\gtrsim}$2$\times$10$^{-15}$~erg~cm$^{-2}$~s$^{-1}$. However, the main flares appear at ${\approx}$20 and 40~ks after the start of the \emph{Chandra} observation, reaching fluxes of $\approx$(1--2)$\times$10$^{-13}$~erg~cm$^{-2}$~s$^{-1}$. Between both flares, there is a quiescent epoch where the flux diminishes by a factor of ${\approx}$7, with large errors, with respect to the main flares.

The light curve of FXRT~6/XRT\,130822 is well described by a PL model with an index of ${\approx}0.3$, although at ${\approx}$10~ks into the event, a slight enhancement in flux beyond that expected for a PL decay occurs.

The light curve of FXRT~7/XRT\,030511
    is described well by a BPL model ($\Delta$BIC${=}-208.3$). The flux duration of the plateau phase until the break is $T_{\rm break}{\approx}$1.1~ks with a rough flux of ${\approx}$1${\times}$10$^{-12}$~erg~cm$^{-2}$~s$^{-1}$, followed by a PL decay with an index of ${\approx}$1.6.

The light curve  of FXRT~8/XRT\,041230
    is described slightly better by a BPL than by a PL model (although $\Delta$BIC${=}2.2$). The source flux is consistent with being constant at a value of ${\approx}$2${\times}$10$^{-15}$~erg~cm$^{-2}$~s$^{-1}$ for about ${\approx}$10~ks, then it rises.

   The light curve of FXRT~9/XRT\,080819
 is relatively symmetric in time, and hence not perfectly described by a BPL model ($\Delta$BIC${=}-1.6$), with a flux rising from ${\lesssim}$5${\times}$10$^{-14}$ to ${\approx}$1${\times}$10$^{-13}$~erg~cm$^{-2}$~s$^{-1}$. After 10~ks into the observation, the flux decreases to ${\approx}$1${\times}$10$^{-14}$~erg~cm$^{-2}$~s$^{-1}$ with a PL index of ${\approx}2.8$ for ${\approx}$5~ks.

The light curve of FXRT~10/XRT\,100831
     is well fitted by a BPL model ($\Delta$BIC${=}-8.3$), with a clear plateau and a subsequent PL decay. The plateau duration is $T_{\rm break}{\approx}$2.7~ks, with a flux of ${\approx}$2$\times$10$^{-14}$~erg~cm$^{-2}$~s$^{-1}$. The decay has an index of ${\approx}1.9$.

The light curve of FXRT~11/XRT\,110103 is similar to that of FXRT~1/XRT~000519. The flux is ${\lesssim}$1${\times}$10$^{-13}$~erg~cm$^{-2}$~s$^{-1}$ until a sudden increase to a flux of ${\approx}$1--2${\times}$10$^{-10}$~erg~cm$^{-2}$~s$^{-1}$. The main burst lasts just a few hundred seconds (but without a double-peak structure as in FXRT~1/XRT~000519) followed by a slow PL decay over the remainder of the observation \citep{Glennie2015}. 

The light curve of FXRT~12/XRT\,110919 is well fitted by a BPL model, with a plateau phase duration of $T_{\rm break}{\approx}$1.8~ks and flux of ${\approx}$2${\times}$10$^{-13}$~erg~cm$^{-2}$~s$^{-1}$. The decays follows a PL index of ${\approx}1.9$.

The light curve of FXRT~13/XRT\,140327 is similar to that of FXRT~6/XRT\,130822 (i.e., a PL describes the data well). The decay index is ${\approx}0.2$.

The light curve of FXRT~14/XRT\,141001/CDF-S~XT1 is well described by a BPL model, although there is no plateau phase. The flux rises rapidly until $T_{\rm break}{\approx}$100--200~s., reaching a flux of $\approx$3${\times}$10$^{-12}$~erg~cm$^{-2}$~s$^{-1}$. The flux subsequently decreases following a PL slope of index ${\approx}1.6$ until $T{\approx}$20~ks, after which no counts are detected. These values agree at the 1$\sigma$ confidence level with the values reported by \citet{Bauer2017}.

In summary, the lights curves of two nearby (FXRTs~2 and 3) and four distant (FXRTs~7, 10, 12, 14/CDF-S~XT1) extragalactic FXRTs are well described by BPL models, with mean PL indexes of $\overline{\tau_1}{\approx}$0.1 and $\overline{\tau_2}{\approx}$1.7 before and after the break, respectively. Among these, all except FXRT~14/CDF-S~XT1 show a few ks plateau phase. On the other hand, FXRTs~8 and 9 are not well described by BPL (see Table~\ref{tab:fitting_para}). Meanwhile, the light curves of FXRTs~6 and 13 follow pure PL decays, with mean PL indexes of $\overline{\tau_1}{\approx}0.3$. The slow decay until the end of the \emph{Chandra} observation after the main flare for FXRTs~1 and 11 (previously reported by \citealp{Jonker2013} and \citealp{Glennie2015}, respectively) is not seen in any of the other candidate FXRTs. Finally, the light curve of FXRT~5 shows clear multiple flares, while weaker events like FXRTs~4 and 6 show marginal hints of multiple-flare structure.

\begin{table*}
    \centering
\scalebox{0.8}{
    \begin{tabular}{lllllllllll}
    \hline\hline
    FXRT & ID & Model & $N_{\rm H,Gal}$ & $N_H$ ($z{=}$0.0) & $\Gamma$ & $kT$ & $\log{\rm Norm}$ & Flux & C-stat(dof) & $\ln \mathcal{Z}$ \\ \hline
    (1) & (2) & (3) & (4) & (5) & (6) & (7) & (8) & (9) & (10) & (11) \\ \hline
    \multicolumn{11}{c}{Nearby extragalactic FXRT Candidates from CSC2} \\ \hline
    1 & XRT\,000519 & \texttt{phabs*zphabs*po} & 1.0 & 0.3${\pm}$0.1 & 2.2${\pm}$0.1 & -- & $-3.5{\pm}$0.03 & 90.5${\pm}$1.8 & 95.3(122) & $-63.4{\pm}0.02$ \\
     &  & \texttt{phabs*zphabs*bremss} & 1.0 & 0.1${\pm}$0.0 & -- & 2.6${\pm}$0.17 & $-3.5{\pm}$0.02 & 84.2${\pm}$2.2 & 117.0(122) & $-77.9{\pm}0.02$ \\
     &  & \texttt{phabs*zphabs*bb} & 1.0 & 0.1${\pm}$0.0 & -- & 0.4${\pm}$0.02 & $-5.1{\pm}$0.01 & 60.9${\pm}$1.5 & 402.1(122) & $-223.2{\pm}0.02$ \\ \hline

    2 & XRT\,010908 & \texttt{phabs*zphabs*po} & 3.0 & 5.8${\pm}$3.6 & 2.1${\pm}$0.6 & -- & $-5.3{\pm}$0.3 & 1.0${\pm}$0.1 & 23.6(15) & $-20.4{\pm}0.02$ \\
     &  & \texttt{phabs*zphabs*bremss} & 3.0 & 2.1${\pm}$1.0 & -- & 36.3${\pm}$26.1 & $-5.4{\pm}$0.1 & 1.2${\pm}$0.2 & 24.3(15) & $-20.4{\pm}0.03$ \\
     &  & \texttt{phabs*zphabs*bb} & 3.0 & 1.2${\pm}$1.0 & -- & 0.8${\pm}$0.1 & $-6.8{\pm}$0.1 & 1.0${\pm}$0.1 & 29.9(15) & $-28.8{\pm}0.3$ \\ \hline

    3 & XRT\,070530 & \texttt{phabs*zphabs*po} & 5.0 & 2.7$_{-2.0}^{+3.8}$ & 5.9${\pm}$1.6 & -- & $-5.2{\pm}0.4$ & 0.2${\pm}$0.1 & 17.7(6) & $-17.4{\pm}0.02$ \\
     &  & \texttt{phabs*zphabs*bremss} & 5.0 & 0.9$_{-0.5}^{+1.3}$ & -- & 3.0$_{-2.4}^{+39.5}$ & $-5.7_{-0.2}^{+0.7}$ & 0.2${\pm}$0.1 & 20.6(6) & $-23.4{\pm}0.02$ \\
     &  & \texttt{phabs*zphabs*bb} & 5.0 & 1.7$_{-1.2}^{+2.5}$ & -- & 0.2${\pm}$0.1 & $-6.9{\pm}0.4$ & 0.1${\pm}$0.1 & 22.0(6) & $-25.4{\pm}0.1$ \\ \hline

    4 & XRT\,071203 & \texttt{phabs*zphabs*po} & 0.6 & 1.0${\pm}$1.0 & 2.6${\pm}$0.5 & -- & $-5.6{\pm}0.2$ & 0.7${\pm}$0.1 & 11.8(7) & $-15.9{\pm}0.02$ \\
     &  & \texttt{phabs*zphabs*bremss} & 0.6 & 0.4${\pm}$0.3 & -- & 11.8$_{-8.9}^{+38.2}$ & $-5.6{\pm}0.1$ & 1.0${\pm}$0.2 & 13.6(7) & $-18.4{\pm}0.02$ \\
     &  & \texttt{phabs*zphabs*bb} & 0.6 & 0.4${\pm}$0.3 & -- & 0.4${\pm}$0.1 & $-7.1{\pm}0.1$ & 0.6${\pm}$0.1 & 24.2(7) & $-28.1{\pm}0.1$ \\ \hline

    5 & XRT\,080331 & \texttt{phabs*zphabs*po} & 0.6 & 6.8${\pm}$1.0 & 3.9${\pm}$0.4 & -- & $-4.4{\pm}0.2$ & 2.0${\pm}$0.1 & 18.9(30) & $-20.1{\pm}0.02$ \\
     &  & \texttt{phabs*zphabs*bremss} & 0.6 & 3.5${\pm}$1.0 & -- & 1.1${\pm}$0.2 & $-4.5{\pm}0.2$ & 2.0${\pm}$0.1 & 19.8(30) & $-24.1{\pm}0.03$ \\
     &  & \texttt{phabs*zphabs*bb} & 0.6 & 0.8${\pm}$0.6 & -- & 0.4${\pm}$0.03 & $-6.5{\pm}$0.1 & 1.9${\pm}$0.1 & 23.1(30) & $-28.1{\pm}0.02$ \\ \hline

    6 & XRT\,130822 & \texttt{phabs*zphabs*po} & 0.4 & 8.1${\pm}$5.3 & 3.6${\pm}$0.9 & -- & $-5.0{\pm}0.4$ & 0.6${\pm}$0.1 & 6.3(5) & $-10.4{\pm}0.02$ \\
     &  & \texttt{phabs*zphabs*bremss} & 0.4 & 2.1${\pm}$2.0 & -- & 3.1$_{-1.8}^{+22.5}$ & $-5.5{\pm}0.2$ & 0.7${\pm}$0.1 & 5.9(5) & $-12.8{\pm}0.03$ \\
     &  & \texttt{phabs*zphabs*bb} & 0.4 & 2.7${\pm}$2.6 & -- & 0.5${\pm}$0.1 & $-7.0{\pm}0.1$ & 0.6${\pm}$0.1 & 6.0(5) & $-15.7{\pm}0.2$ \\ \hline
    \multicolumn{11}{c}{Distant extragalactic FXRT Candidates from CSC2} \\ \hline

    7 & XRT\,030511 & \texttt{phabs*zphabs*po} & 0.1 & 1.4${\pm}$0.1 & 2.1${\pm}$0.2 & -- & $-4.6{\pm}0.1$ & 8.9${\pm}$0.5 & 59.9(61) & $-42.1{\pm}0.02$  \\
     &  & \texttt{phabs*zphabs*bremss} & 0.1 & 0.5${\pm}$0.4 & -- & 4.1${\pm}$1.1 & $-4.6{\pm}0.1$ & 8.9${\pm}$0.6 & 60.8(61) & $-44.3{\pm}0.02$ \\
     &  & \texttt{phabs*zphabs*bb} & 0.1 & 0.2${\pm}0.2$ & -- & 0.5${\pm}$0.04 & $-6.1{\pm}0.03$ & 6.9${\pm}$0.3 & 95.4(61) & $-66.4{\pm}0.02$ \\ \hline

    8 & XRT\,041230 & \texttt{phabs*zphabs*po} & 0.4 & 9.4${\pm}$6.4 & 2.7${\pm}$1.3 & -- & $-5.6{\pm}0.5$ & 0.3${\pm}$0.1 & 1.1(3) & $-7.1{\pm}0.02$ \\
     &  & \texttt{phabs*zphabs*bremss} & 0.4 & 3.9${\pm}$3.0 & -- & 39.0${\pm}$27.7 & $-5.9{\pm}0.1$ & 0.4${\pm}$0.1 & 1.1(3) & $-6.8{\pm}0.1$ \\
     &  & \texttt{phabs*zphabs*bb} & 0.4 & 1.8${\pm}$1.8 & -- & 8.0$_{-7.4}^{+49.0}$ & $-4.9{\pm}2.5$ & 0.5${\pm}$0.2 & 0.9(3) & $-11.1{\pm}0.02$ \\ \hline

    9 & XRT\,080819 & \texttt{phabs*zphabs*po} & 1.4 & 6.0${\pm}$5.0 & 3.0${\pm}$1.1 & -- & $-4.8{\pm}0.4$ & 1.2${\pm}$0.2 & 9.1(5) & $-12.0{\pm}0.02$ \\
     &  & \texttt{phabs*zphabs*bremss} & 1.4 & 1.7${\pm}$1.6 & -- & 35.2${\pm}$28.0 & $-5.2{\pm}0.1$ & 2.0${\pm}$0.3 & 9.2(5) & $-12.8{\pm}0.02$ \\
     &  & \texttt{phabs*zphabs*bb} & 1.4 & 1.5$_{-1.7}^{+3.9}$ & -- & 0.5${\pm}$0.2 & $-6.7{\pm}0.1$ & 1.1${\pm}$0.2 & 9.7(5) & $-17.8{\pm}0.02$ \\ \hline

    10 & XRT\,100831 & \texttt{phabs*zphabs*po} & 1.0 & 5.0${\pm}$3.3 & 3.4${\pm}$0.9 & -- & $-5.0{\pm}0.3$ & 0.8${\pm}$0.1 & 4.9(7) & $-10.3{\pm}0.02$ \\
     &  & \texttt{phabs*zphabs*bremss} & 1.0 & 1.1${\pm}$1.0 & -- & 15.6$_{-13.5}^{+40.4}$ & $-5.4{\pm}0.1$ & 1.3${\pm}$0.2 & 5.1(7) & $-12.4{\pm}0.02$ \\ 
     &  & \texttt{phabs*zphabs*bb} & 1.0 & 0.5${\pm}$0.4 & -- & 53.3${\pm}$26.3 & $-2.1_{-1.5}^{+0.4}$ & 2.0${\pm}$0.5 & 43.7(7) & $-31.3{\pm}0.1$ \\ \hline

    11 & XRT\,110103 & \texttt{phabs*zphabs*po} & 0.9 & 2.7${\pm}$1.0 & 2.2${\pm}$0.2 & -- & $-4.4{\pm}$0.1 & 10.4${\pm}$0.5 & 27.7(40) & $-25.8{\pm}0.02$  \\
     & & \texttt{phabs*zphabs*bremss} & 0.9 & 1.1${\pm}$0.6 & -- & 4.5${\pm}$1.3 & $-4.4{\pm}0.1$ & 10.6${\pm}$0.5 & 29.9(40) & $-28.0{\pm}0.12$ \\
     &  & \texttt{phabs*zphabs*bb} & 0.9 & 0.3${\pm}$0.2 & -- & 0.6${\pm}$0.03 & $-5.9{\pm}$0.03 & 8.2${\pm}$0.4 & 56.1(40) & $-46.5{\pm}0.02$  \\ \hline

    12 & XRT\,110919 & \texttt{phabs*zphabs*po} & 0.8 & 1.3$_{-0.9}^{+1.9}$ & 2.4${\pm}$0.5 & -- & $-5.4{\pm}0.1$ & 1.0${\pm}$0.1 & 19.1(14) & $-19.7{\pm}0.1$  \\
     & & \texttt{phabs*zphabs*bremss} & 0.8 & 0.7${\pm}$0.6 & -- & 8.5$_{-4.9}^{+29.2}$ & $-5.5{\pm}0.1$ & 1.2${\pm}$0.2 & 20.4(14) & $-21.4{\pm}0.03$  \\
     &  & \texttt{phabs*zphabs*bb} & 0.8 & 1.0${\pm}$1.0 & -- & 0.6${\pm}$0.1 & $-6.9{\pm}0.1$ & 0.9${\pm}$0.2 & 28.5(14) & $-28.5{\pm}0.2$  \\ \hline

    13 & XRT\,140327 & \texttt{phabs*zphabs*po} & 0.7 & 6.7${\pm}$5.0 & 3.7${\pm}$0.9 & -- & $-5.0{\pm}0.4$ & 0.5${\pm}$0.1 & 20.4(17) & $-17.1{\pm}0.03$  \\
     &  & \texttt{phabs*zphabs*bremss} & 0.7 & 1.8${\pm}$1.5 & -- & 14.4$_{-12.7}^{+41.2}$ & $-5.5{\pm}0.2$ & 0.8${\pm}$0.2 & 20.0(17) & $-19.1{\pm}0.03$  \\
     &  & \texttt{phabs*zphabs*bb} & 0.7 & 3.7${\pm}$3.0 & -- & 0.4$\pm$0.1 & $-6.9{\pm}0.2$ & 0.5${\pm}$0.1 & 19.5(17) & $-22.2{\pm}$0.105 \\ \hline

    14 & XRT\,141001/ & \texttt{phabs*zphabs*po} & 0.2 & 1.1${\pm}$0.1 & 1.9${\pm}$0.3 & -- & $-5.1{\pm}0.1$ & 2.8${\pm}$0.2 & 26.8(17) & $-24.6{\pm}0.02$  \\
     & CDF-S~XT1 & \texttt{phabs*zphabs*bremss} & 0.2 & 0.5${\pm}$0.4 & -- & 10.4$_{-5.3}^{+24.5}$ & $-5.1{\pm}0.1$ & 3.1${\pm}$0.3 & 27.6(17) & $-25.1{\pm}0.1$  \\
     & & \texttt{phabs*zphabs*bb} & 0.2 & 0.4${\pm}$0.3 & -- & 0.7${\pm}$0.1 & $-6.5{\pm}0.1$ & 2.6${\pm}$0.2 & 41.9(17) & $-37.7{\pm}0.02$  \\ \hline
    \end{tabular}
    }
    \caption{Results of the 0.5--7 keV \hbox{X-ray} spectral fits for the CSC2 FXRT candidates. 
    \emph{Column 1:} Number of the candidate. 
    \emph{Column 2:} Transient ID used in this work. 
    \emph{Column 3:} Spectral model considered. 
    \emph{Columns 4 and 5:} Galactic and intrinsic column density absorption ($\times$10$^{21}$), respectively, in units of cm$^{-2}$. The former is kept fixed during the fit. 
    \emph{Column 6:} Photon index from the PL model. 
    \emph{Column 7:} Temperature in units of keV from the BR or BB models. 
    \emph{Column 8:} Normalization parameter (in units of photons~keV$^{-1}$~cm$^{-2}$~s$^{-1}$). 
    \emph{Column 9:} Absorbed fluxes ($\times$10$^{-14}$) in units of erg~cm$^{-2}$~s$^{-1}$ (0.5--7.0~keV). 
    \emph{Column 10:} C-stat value and the number of degrees of freedom. 
    \emph{Column 11:} Log-evidence ($\ln \mathcal{Z}$) values for each model. Errors are quoted at the 1$\sigma$ confidence level.}
    \label{tab:spectral_para}
\end{table*}

\begin{figure*}
    \centering
    \includegraphics[scale=0.65]{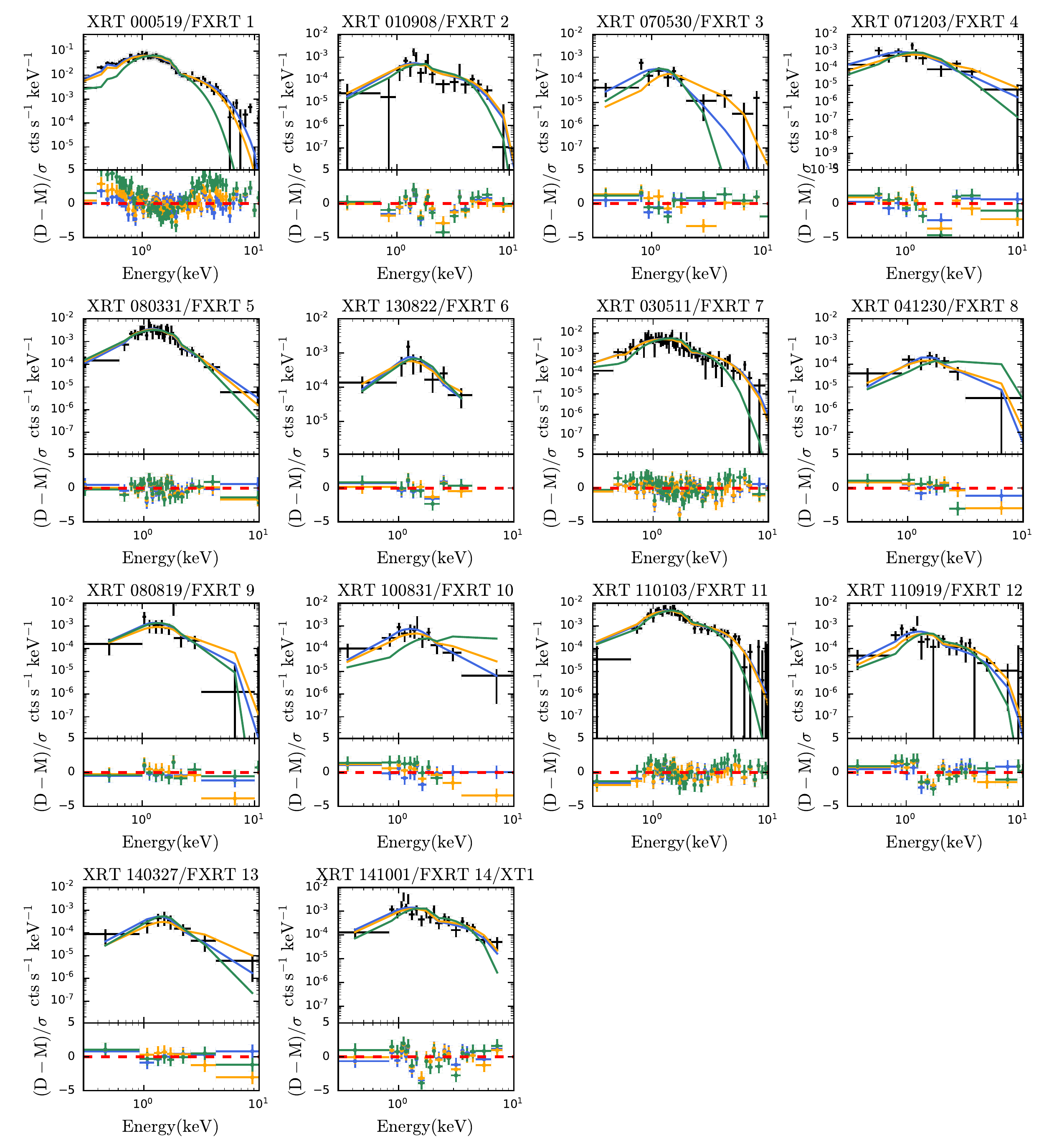}
    \vspace{-0.2cm} \caption{Observed time-integrated spectra fitted with different spectral models.
    \emph{Top panels:} X-ray spectra (\emph{black dots}; the data were grouped to at least one count per bin), in units of counts (cts~s$^{-1}$~keV$^{-1}$). We also plot the best-fit PO (\emph{blue lines}), BR (\emph{orange lines}), and BB (\emph{green lines}) spectral models; see Table~\ref{tab:spectral_para} for the corresponding best fitting parameters. \emph{Bottom panels:} Residuals (defined as data-model normalized by the uncertainty; $(D-M)/\sigma$) of each spectral model.}
    \label{fig:X_ray_spectra}
\end{figure*}

\subsection{Spectral properties}\label{sec:X-ray_fitting}

In this section we describe the spectral properties of the sample of FXRT candidates using some basic models, as well as their HR and photon index evolution with time.

\subsubsection{Spectral parameters}

We generate \hbox{X-ray} spectra and response matrices following standard procedures for point-like sources using \texttt{CIAO} with the \texttt{specextract} script. The source and background regions are the same as those for generating the light curves (see Sect. \ref{sec:LC}). Due to the low number of counts per bin, we adopt maximum likelihood statistics for a Poisson distribution, the so-called Cash-statistics (C-stat, with $C{=}-2\ln{L_{\rm Poisson}}{+}{\rm const}$; \citealt{Cash1979}) to find the best-fit model. Although C-stat is not distributed like $\chi^2$, meaning that the standard goodness-of-fit is not applicable \citep{Buchner2014,Kaastra2017a}. Thus, to evaluate if there are differences in the goodness-of-fit between models, we use {the Bayesian \hbox{X-ray} Astronomy (BXA) package} \citep{Buchner2014}, which joins the Monte Carlo nested sampling algorithm \texttt{MultiNest} \citep{Feroz2009} with the fitting environment of \texttt{XSPEC} \citep{Arnaud1996}. BXA computes the integrals over parameter space, called the evidence ($\mathcal{Z}$), which is maximized for the best-fit model. For BXA, we assume uniform model priors.

We consider three simple continuum models: 
$i)$ an absorbed PL model  (\texttt{phabs*zphabs*po}, hereafter the PO model); 
$ii)$ an absorbed thermal Bremsstrahlung model (\texttt{phabs*zphabs*bremss}, hereafter the BR model); and 
$iii)$ an absorbed black-body model (\texttt{phabs*zphabs*bb}, hereafter the BB model). 
The PO model is typically thought to be produced by a nonthermal electron distribution, while the other two models have a thermal origin. We chose these models because we do not know the origin and the processes behind the spectral properties of FXRTs, while the limited numbers of counts do not warrant more complex models. The spectral components \texttt{phabs} and \texttt{zphabs} represent the Galactic and intrinsic contribution to the total absorption, respectively. The Galactic absorption ($N_{\rm H,Gal}$) was fixed at the values of \citet{Kalberla2005} and \citet{Kalberla2015} during the fit, while for the intrinsic redshifted absorption, we adopt $z{=}0$, which provides a strict lower bound.

\begin{figure}
    \centering
    \includegraphics[scale=0.6]{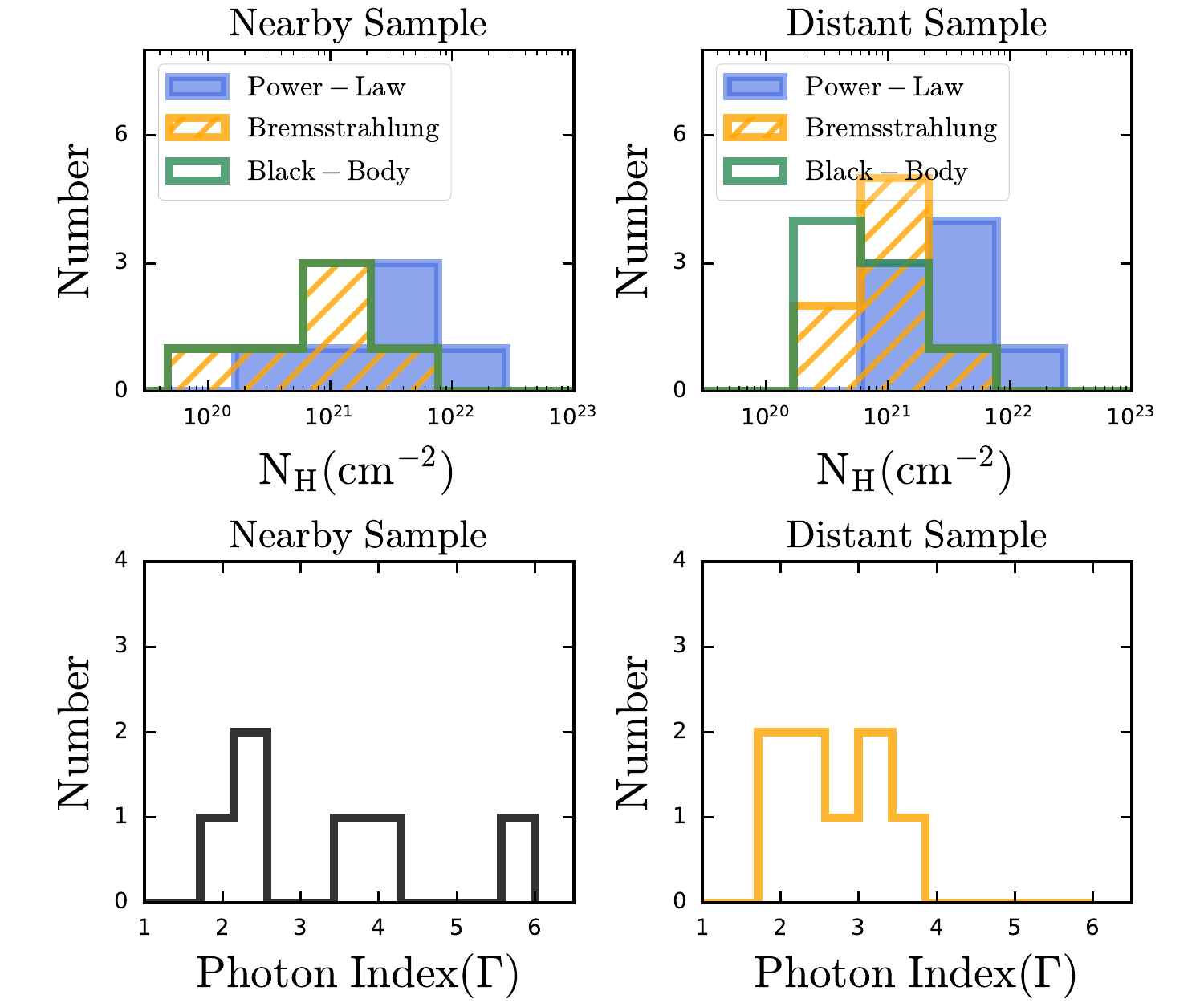}
    \vspace{-0.1cm}
    \caption{Distribution of best-fit \hbox{X-ray} parameters for nearby (\emph{left panels}) and distant (\emph{right panels}) FXRT candidates. \emph{Top panels:} Histogram of neutral hydrogen column densities, in units of cm$^{-2}$, obtained using the {PL}, {BR}, and {BB} models. \emph{Bottom panels:} Histogram of the photon indices obtained using a {PL} model.
    } 
    \label{fig:X-ray-params}
\end{figure}

The best fitting spectral models (and residuals) and their parameters are provided in Fig.~\ref{fig:X_ray_spectra} and Table~\ref{tab:spectral_para}, respectively, while Fig.~\ref{fig:X-ray-params} shows the histograms of the best-fit intrinsic neutral hydrogen column densities in addition to the Galactic value ($N_H$; \emph{top panels}) and photon index ($\Gamma$; \emph{bottom panels}) for nearby (\emph{left panels}) and distant (\emph{right panels}) extragalactic FXRTs candidates. The $N_H$ covers ranges for {nearby} ({distant}) candidates of $N_{\rm H,PO}{=}$0.3--8.1(1.1--9.4), $N_{\rm H,BR}{=}$0.1--3.5(0.5--3.9), and $N_{\rm H,BB}{=}$0.1--2.7(0.2--3.7)$\times$10$^{21}$~cm$^{-2}$, and mean values of $\overline{N}_{\rm H,PO}{=}$4.1(4.2), $\overline{N}_{\rm H,BR}{=}$1.5(1.4), and $\overline{N}_{\rm H,BB}{=}$1.2(1.2)$\times$10$^{21}$~cm$^{-2}$, respectively. Furthermore, we compare the best-fit $N_{\rm H,PO}$ with the HI constraints from
\citet{Kalberla2005} and \citet{Kalberla2015} and note that in all cases aside from FXRT~1 and FXRT~3, the bulk of the measured $N_{\rm H,PO}$ are higher than $N_{\rm H,Gal}$ (a factor of $\approx$2--15 higher).

The best-fit PL photon index ranges between $\Gamma{=}$2.1--5.9 (1.9--3.7) for the {nearby} ({distant}) candidate FXRTs, with mean values of $\overline{\Gamma}{=}$3.4 (2.7). According to \citet{Lin2012} (which classified sources detected by \emph{XMM-Newton}), the photon index covers a wide range for different types of sources such as stars, AGNs or compact objects; however, only stars and compact objects have photon indices as high as $\Gamma{\sim}$6.
For BR models, the best-fit temperatures range from $kT_{\rm BR}{=}$1.1--36.2(4.1--39.0)~keV for nearby (distant) candidates, while BB temperatures span $kT_{\rm BB}{=}$0.2--0.8 (0.4--53.2)~keV for {nearby} ({distant}) candidate FXRTs. The events with BR temperatures $kT_{\rm BR}{\gtrsim}$10~keV are FXRTs~2, 4, 8, 9, 10, 13, and 14, while with BB temperatures $kT_{\rm BB}{\gtrsim}$5~keV are FXRTs~8, and 10. Both temperatures (especially $kT_{\rm BB}$) are important to eventually analyze a possible association with SBOs ($kT_{\rm{SBOs}}{\approx}$0.03--3.0~keV, based on the progenitor star; \citealp{Matzner1999}, \citealp{Nakar2010}, \citealp{Sapir2013}).

\begin{figure}
    \centering
    \includegraphics[scale=0.63]{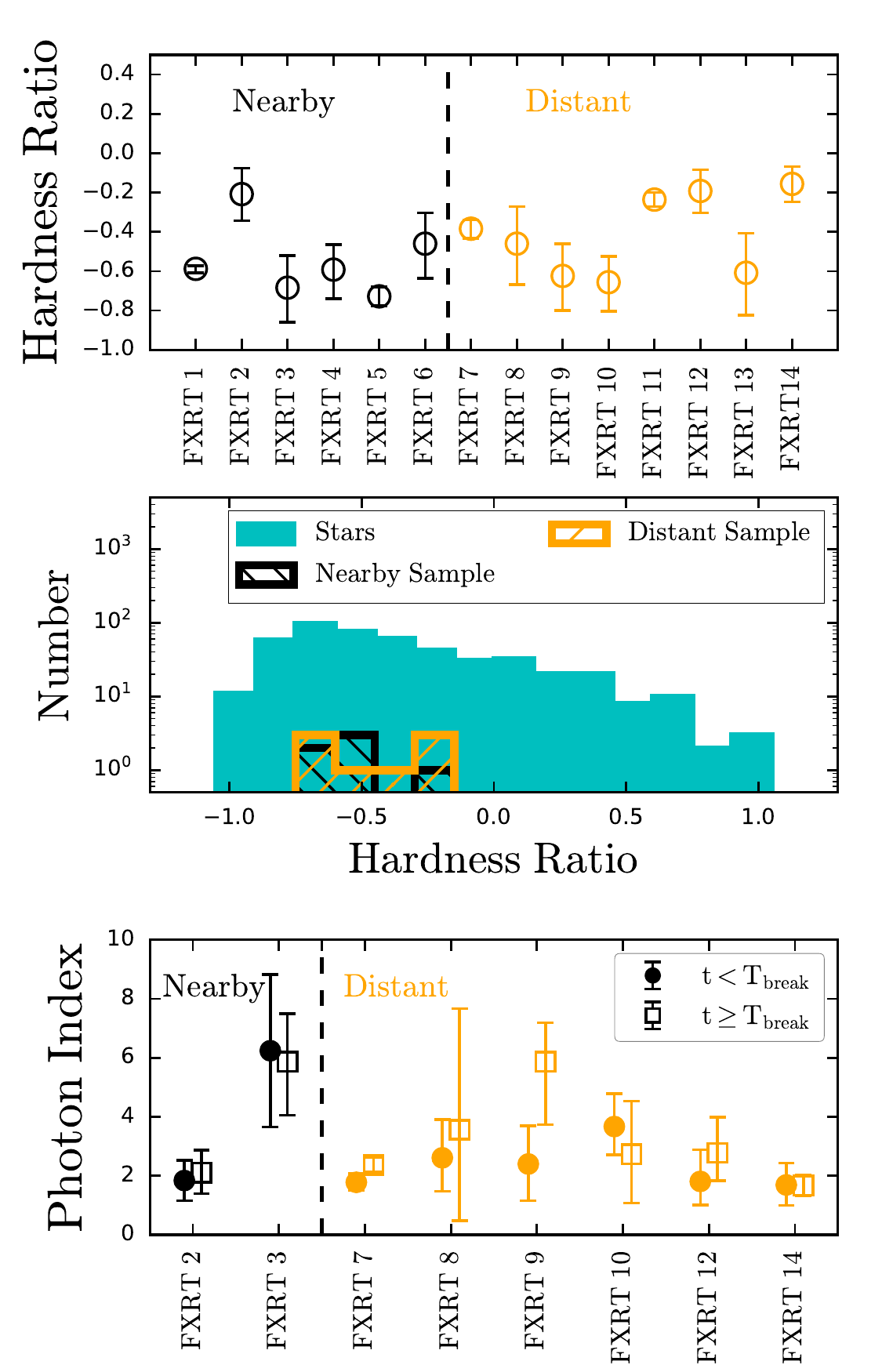}
    \vspace{-0.3cm}
    \caption{Photon index and HR distribution of our selected transient candidates.
    \emph{Top panel:} HR of each FXRT candidate \citep[using the Bayesian \texttt{BEHR} code;][]{Park2006}. \emph{Middle panel:} HR distributions of our final samples of {nearby} (hashed \emph{black} histogram) and {distant} (hashed \emph{orange} histogram) FXRTs, compared to the XRTs classified as ``stars'' according to {Criterion 2} using \emph{Gaia} (filled \emph{cyan} histogram). \emph{Bottom panel}: Photon indices for FXRT candidates where the light curve was best fit by a BPL model, {before} (\emph{filled circles}) and {after} (\emph{open squares}) the break time ($T_{\rm break}$) taken from Table~\ref{tab:fitting_para}. In all cases, errors bars are at the 90\% confidence level.}
    \label{fig:hardness}
\end{figure}

\subsubsection{Hardness ratio and photon index evolution}\label{sec:hardness_ratio}

The HR can be used to classify \hbox{X-ray} sources and study their spectral evolution, particularly when low number statistics prevail \citep[e.g.,][]{Lin2012,Peretz2018}. Below, we investigate the HR for the population of FXRTs, compare these to candidates previously classified as ``stars,'' and look at the evolution of the HR and photon indices over the duration of the flare. The HR is defined as
\begin{equation}
    HR=\frac{H-S}{H+S},
\end{equation}
where $H$ and $S$ are the number of \hbox{X-ray} photons in the soft and hard energy bands, defined as the 0.5--2.0 and 2.0--7.0~keV bands, respectively. 
For each candidate, we calculate the HR using the Bayesian code \texttt{BEHR} \citep[][]{Park2006}, which we list in Table~\ref{tab:my_detections}, \emph{column 13}, and plot in Fig.~\ref{fig:hardness} (\emph{top panel}).

Notably, \citet{Yang2019} found differences between stellar objects and the FXRT CDF-S~XT1/XT2 (see their Fig.~5), where the latter has an average HR${\gtrsim}-0.16$. Taking into account the 472 objects identified as stars according to {Criterion 2} in Sect. \ref{sec:gaia}, we compare the HRs of these objects (see Fig.~\ref{fig:hardness}, \emph{middle panel, cyan histogram}) to the final sample of nearby and distant FXRTs (\emph{black and orange histograms}). Stars typically have very soft \hbox{X-ray} spectra \citep{Gudel2009}, with some notable exceptions like Be stars \citep[e.g., Be star HD 110432 has an HR${\gtrsim}$0.0;][]{Lopes2007}.
We find that the HR distribution of ``star'' candidates also strongly skews toward softer HRs, but demonstrates that stars associated with \hbox{X-ray} flares cover essentially all HRs, ranging from $-0.99$ to $+0.97$ (see Fig.~\ref{fig:hardness}, \emph{middle panel}). Importantly, there is a smooth, non-negligible tail to harder values, with $\approx$20\% of stars having HR${\gtrsim}$0.0 (possibly related to magnetic cataclysmic variables). Given this, we conclude that the \hbox{X-ray} HR is not a useful discriminator on its own.

Next, we analyze if, and if so how, the HR and PL index of the \hbox{X-ray} spectrum evolve with time. To start, we compute the HR for each bin of the light curves using the \texttt{BEHR} code of \citet{Park2006}, which we show in the  \emph{lower panels} of Fig.~\ref{fig:models_BPL}. For light curves that are well fit by a BPL model, we additionally split the \texttt{event files} at $T_{\rm break}$ and extract "before" and "after" spectra to compute the spectral slopes ($\Gamma_{\rm before}$ and $\Gamma_{\rm after}$, respectively; see Table~\ref{tab:fitting_para_Tbreak}) using the best-fit PO model (see Table~\ref{tab:spectral_para}). We fit both intervals together assuming fixed constant $N_{\rm H,Gal}$ and $N_H$ (taken from Table~\ref{tab:spectral_para}).

The resulting evolution of the HRs and the PL spectral indices are shown in Figs.~\ref{fig:models_BPL} and \ref{fig:hardness} (\emph{bottom panel}), respectively. FXRTs~1 and 11 show significant early softening in their HR evolution (Figs.~\ref{fig:models_BPL}) during the $\sim$50\,s following their main peaks \citep[consistent with][]{Glennie2015}, while FXRTs~2, 7, and 12 show marginal ($\sim$90\% confidence) spectral softening after the plateau stage, like CDF-S~XT2 trends, and FXRT~8 appears to soften marginally ($\sim$90\% confidence) throughout its light curve. None of the other FXRTs show any evidence of spectral evolution.

\begin{table}
    \centering
    \begin{tabular}{llll}
    \hline\hline
    FXRT & ID & $\Gamma_{\rm before}$($T{<}T_{\rm break}$) & $\Gamma_{\rm after}$($T{\geq}T_{\rm break}$)\\ \hline
    (1) & (2) & (3) & (4) \\ \hline
    \multicolumn{4}{c}{Nearby extragalactic FXRT Candidates from CSC2} \\ \hline
    2 & XRT\,010908 & 1.8$\pm$0.7 & 2.1$\pm$0.7 \\
    3 & XRT\,070530 & 6.2$\pm$2.6 & 5.9$\pm$1.6 \\ \hline
    \multicolumn{4}{c}{Distant extragalactic FXRT Candidates from CSC2} \\ \hline
    7 & XRT\,030511 & 1.8$\pm$0.3 & 2.4$\pm$0.2 \\
    8 & XRT\,041230 & 2.6$\pm$1.2 & 3.6$\pm$3.5 \\
    9 & XRT\,080819 & 2.4$\pm$1.2 & 5.9$\pm$2.2 \\
    10 & XRT\,100831 & 3.7$\pm$1.0 & 2.7$\pm$1.7 \\
    12 & XRT\,110919 & 1.8$\pm$0.9 & 2.8$\pm$1.1 \\ 
    14 & XRT\,141001 & 1.7$\pm$0.8 & 1.7$\pm$0.3 \\ \hline 
    \end{tabular}
    \caption{Spectral slope computed "before" and "after" the $T_{\rm break}$. \emph{Columns 1 and 2:} FXRT\# and ID of the candidate, respectively. \emph{Columns 3 and 4:} Spectrum photon index computed before and after the $T_{\rm break}$ for light curves that are well fit with a BPL. Errors are quoted at the 90\% confidence level.} 
    \label{tab:fitting_para_Tbreak}
\end{table}

\subsection{Galactic origin}\label{sec:galactic}

FXRTs~2, 3, 4, 5, 6, 8, 9, and 14 can be associated with extended host galaxies, proving and/or strengthening their extragalactic origin (see Sect. \ref{sec:results} for more details). FXRTs~1 and 11 are located near the outskirts of M86 and Abell~3581 \citep{Jonker2013,Glennie2015}, respectively, which, while suggestive of an extragalactic nature, is not definitive. Below, we investigate whether some FXRTs could still be associated with Galactic M- or brown-dwarf flares.

Magnetically active dwarfs (which comprise around 30\% of M dwarfs and 5\% of brown dwarfs) are known to exhibit flares on timescales of minutes to hours, with flux increases (not only in \hbox{X-ray}) by one or two orders of  magnitude \citep{Schmitt2004,Mitra-Kraev2005,Berger2006,Welsh2007}. The coldest object observed to flare in \hbox{X-ray}s is an L1 dwarf \citep{DeLuca2020}. Flares can be classified in two groups according to a time--luminosity relation \citep[following previous efforts by][]{Bauer2017}: $i)$ short ``compact'' flares ($L{\lesssim}$10$^{30}$~erg~s$^{-1}$ and $\Delta t{\lesssim}$1~h), and $ii)$ ``long'' flares ($L{\lesssim}$10$^{32}$~erg~s$^{-1}$ and $\Delta t{\gtrsim}$1~h). The flaring episodes often occur recurrently on timescales from hours to years. The flares typically have thermal spectra with temperatures of $kT{=}$0.5--1~keV. M-dwarf stars have optical and NIR absolute magnitudes in the range of $M_z{\sim}$8--13~mag \citep{Hawley2002} and $M_{K_s}{\sim}$3--10~mag \citep{Avenhaus2012}, respectively, while brown dwarfs have $M_z{\sim}$13--18~mag \citep{Hawley2002} and $M_J{\sim}$15--25~mag \citep{Tinney2014}, respectively. In the case of \hbox{X-ray} emission, M dwarfs show flares in the range of $L_X^{\rm M-dwarf}{\approx}$10$^{28}$--10$^{32}$~erg~s$^{-1}$ \citep{Pallavicini1990,Pandey2008,Pye2015}, while brown dwarf flares span $L_X^{\rm B-dwarf}{\approx}$10$^{27}$--10$^{30}$~erg~s$^{-1}$ \citep{Berger2006,Robrade2010}. Furthermore, cold M dwarfs and L dwarfs typically exhibit ratios no larger than $\log(L_X/L_{\rm bol}){\lesssim}$0.0 and ${\lesssim}-3.0$ (the dwarf star flare saturation limit), respectively, where $L_X$ and $L_{\rm bol}$ are the \hbox{X-ray} flare and average (non-flare) bolometric luminosities, respectively \citep[e.g.,][]{Garcia2008,DeLuca2020}.

Thus, it is possible to discard a stellar flare explanation for FXRTs using their optical and NIR detections and/or upper limits compared to the expected absolute magnitudes in these bands (see above), as well as the ratio $\log(L_X/L_{\rm bol}){=}\log(F_X/F_{\rm bol}){\lesssim}-3.0$ \citep{Garcia2008}\footnote{To compute the ratio $\log(L_X/L_{\rm bol})$, we normalize stellar synthetic models of dwarf stars \citep[taken from][$1000{\lesssim}T_{\rm eff}{\lesssim}3000$~K and $2.5{\lesssim}\log{g}{\lesssim}5.5$]{Phillips2020} to the deepest photometric upper limits and/or detections (as listed in Table~\ref{tab:photometric_data}), and compute bolometric fluxes by integrating the normalized models at optical/NIR wavelengths.}.

We derive a lower limit to the distance for each source using the expected $z$-band absolute magnitude ranges for M-dwarf and brown dwarf stellar flares listed above. We subsequently convert the \hbox{X-ray} flux to a lower limit on the luminosity using these distance limits. If this lower limit is above the maximum luminosity observed for M-dwarf and brown dwarf  stellar flares, we rule out this explanation for the FXRT. In this way, we explore the possible Galactic origin of each FXRT without a clear extragalactic host. 

For FXRT~1/XRT\,000519, the deep detections $m_g{=}$26.8 and $m_i{=}$24.3 imply limits to the distance of putative M- and brown dwarfs responsible for the \hbox{X-ray} flares of 0.6--6.5~kpc and 0.06--0.7~kpc, respectively. The corresponding \hbox{X-ray} flare luminosities are $L_X^{\rm M-dwarf}{\sim}$(8.7--880)$\times$10$^{33}$ and $L_X^{\rm B-dwarf}{\sim}$(8.8--876)$\times$10$^{31}$~erg~s$^{-1}$, respectively, at least 1.5\,dex higher than the known range. Furthermore, the ratio $\log(F_X/F_{\rm bol}){\approx}$2.7--3.2 is well above the known range. Thus, FXRT~1 is unlikely to be a stellar flare, consistent with the conclusions drawn in \citet{Jonker2013}.

For FXRT~7/XRT\,030511, the limit of $m_y{>}23.7$ implies distance lower limits of ${>}$1.7--17~kpc and ${>}$0.2--1.7~kpc for M- and brown-dwarfs, respectively, and corresponding \hbox{X-ray} flare luminosities are $L_X^{\rm M-dwarf}{\gtrsim}$(8.3--800)$\times$10$^{32}$ and $L_X^{\rm B-dwarf}{\gtrsim}$(8.3--831)$\times$10$^{30}$~erg~s$^{-1}$, respectively, at least 0.9\,dex higher than the known range. The ratio $\log(F_X/F_{\rm bol}){\gtrsim}$1.6--2.1 is also well above the known range, ruling out a stellar flare origin.

For FXRT~10/XRT\,100831, the limit of $m_g{>}$24.5 yields distance lower limits of ${>}$0.2--2300 and ${>}$0.02--0.2~kpc for M- and brown-dwarfs, respectively, and corresponding \hbox{X-ray} flare luminosities are $L_X^{\rm M-dwarf}{\gtrsim}$(2.9--300)$\times$10$^{30}$ and $L_X^{\rm B-dwarf}{\gtrsim}$(2.9--290)$\times$10$^{28}$~erg~s$^{-1}$, respectively. The lower bound of the $g$-band estimate remains consistent with the known range. The ratio $\log(F_X/F_{\rm bol}){\gtrsim}-0.8$ to $-0.3$ also remains mildly consistent with for example, the extreme spectral type L1 J0331-27 star \citep{DeLuca2020}, implying that we cannot completely rule out an extreme stellar flare origin for FXRT~10. 

For FXRT~11/XRT\,110103, the limit of $m_g{>}$23.2 implies distance lower limits of ${>}$0.1--1.3 and ${>}$0.01--0.1~kpc for M- and brown dwarfs, respectively, and corresponding \hbox{X-ray} flare luminosities are $L_X^{\rm M-dwarf}{\gtrsim}$(4.8--480)$\times$10$^{32}$ and $L_X^{\rm B-dwarf}{\gtrsim}$(4.8--500)$\times$10$^{30}$~erg~s$^{-1}$, respectively, at least 0.7\,dex higher than the known range. The ratio $\log(F_X/F_{\rm bol}){\gtrsim}$1.4--1.9 also implies that FXRT~11 is not caused by a stellar flare, consistent with the conclusions drawn in \citet{Glennie2015}.

For FXRT~12/XRT\,110919, the limit of $m_z{>}$24.8 leads to distance lower limits of ${>}$0.3--2.6 and ${>}$0.02--0.3~kpc, respectively, and corresponding \hbox{X-ray} flare luminosities are $L_X^{\rm M-dwarf}{\gtrsim}$(2.4--240)$\times$10$^{30}$ and $L_X^{\rm B-dwarf}{\gtrsim}$(2.4--242)$\times$10$^{28}$~erg~s$^{-1}$, respectively. The lower bound of the $g$-band estimate is marginally overlaps at the known range of luminosities. The ratio $\log(F_X/F_{\rm bol}){\gtrsim}-0.9$ to $-0.4$ also remains mildly consistent with for instance, the extreme type L1 J0331-27 star \citep{DeLuca2020}, implying that we cannot completely rule out an extreme stellar flare origin for FXRT~12.

For FXRT~13/XRT\,140327, the detection at $m_i{=}$24.7 implies a distance range of ${\approx}$1.2--12000 and ${\sim}$0.1--1.2~kpc, respectively, and corresponding \hbox{X-ray} flare luminosities of $L_X^{\rm M-dwarf}{\sim}$(2.1--200)$\times$10$^{31}$ and $L_X^{\rm B-dwarf}{\sim}$(2.1--208)$\times$10$^{29}$~erg~s$^{-1}$, respectively. This is not enough to discard a Galactic stellar flare nature. However, we find a ratio $\log(F_X/F_{\rm bol}){\approx}$0.0--0.6, implying that FXRT~13 is not caused by a stellar flare.

As a summary, the multiwavelength photometry for four FXRTs appears inconsistent with expectations for flares from Galactic M dwarfs and brown dwarfs, while deeper limits are still required to completely rule out this out for FXRTs 10 and 12.

\begin{table}
    \centering
    \scalebox{0.75}{
    \begin{tabular}{llllll}
    \hline\hline
    FXRT & $d_{\rm offset}$ & $P_{\rm ch}$ & $P_{\rm dist.FXRT}^\dagger$ & $N$ & $P(X=1|M<M_{\rm host}$  \\ 
     & (arcmin) & & & $(M<M_{\rm host})$ & $\wedge d\leq d_{\rm offset})^{\dagger\dagger}$ \\ \hline
    (1) & (2) & (3) & (4) & (5) & (6) \\ \hline
    1 & 12.2 & 3.5e$\times$10$^{-4}$ & 3.0$\times$10$^{-3}$ & 134 & 0.27  \\
    2 & 0.4 & 8.4$\times$10$^{-7}$ & 3.3$\times$10$^{-6}$ & 216 & 7.0$\times$10$^{-4}$ \\ 
    3 & 5.5 & 1.3$\times$10$^{-5}$ & 6.2$\times$10$^{-4}$ & 29 & 1.7$\times$10$^{-2}$ \\ 
    4 & 0.4 & 1.9$\times$10$^{-6}$ & 3.3$\times$10$^{-6}$ & 303 & 9.9$\times$10$^{-4}$ \\
    5 & 1.3 & 2.8$\times$10$^{-6}$ & 3.5$\times$10$^{-5}$ & 114 & 3.9$\times$10$^{-3}$ \\
    6 & 1.2 & 1.5$\times$10$^{-4}$ & 2.9$\times$10$^{-5}$ & 508 & 1.5$\times$10$^{-2}$ \\ 
    11 & 2.7 & 0.15 & 1.5$\times$10$^{-4}$ & 571 & 7.8$\times$10$^{-2}$ \\ \hline 
    \end{tabular}
    }
    \caption{Probabilities related to the {local} sample. 
    \emph{Column 1:} FXRT candidate number.
    \emph{Column 2:} Angular offset of the {local} FXRT, in arcmin, from its associated host galaxy center. 
    \emph{Column 3:} Probability of a random chance alignment between an FXRT and a nearby galaxy at least as bright as the associated host galaxy.
    \emph{Column 4:} Probability of finding a \emph{distant FXRT} at an angular offset $d_{\rm offset}$. 
    \emph{Column 5:} Number of \emph{Chandra} observations that include a nearby galaxy at least as bright as the associated host galaxy.
    \emph{Column 6:} Probability of finding one ($X=1$) \emph{distant FXRT} given a cut distant offset ($d<d_{\rm offset}$) and host galaxy magnitude ($M<M_{\rm host}$).\\
    $^\dagger$ Using Poisson statistic, $P(k;\lambda)$, for one source, $k=1$, and $\lambda$ is equal to the density of {distant} FXRTs,  ${\approx}$6.5${\times}$10$^{-6}$~arcmin$^{-2}$, multiplied by the circular area defined with a radius equal to the angular offset.\\
    $^{\dagger\dagger}$ Using Binomial statistic, $P(k;N;P_{\rm dis.FXRT})$, for number of trials, $N(M<M_{\rm host})$, number of success events (in this case $X{=}1$), and success probability of $P_{\rm dist.FXRT}$ (see Sect. \ref{sec:population}).}
    \label{tab:prob}
\end{table}

\subsection{One or two populations of FXRTs?}\label{sec:population}

A key question is how robust individual FXRT associations with local or distant populations are. In particular, there remains some probability that FXRTs associated with the {local} sample are in fact background {distant} FXRTs that simply lie in projection with nearby large-scale structures. A first consideration here is to identify and isolate the fraction of \emph{Chandra} observations that actively target nearby galaxies. While {distant} FXRTs can be detected in any \emph{Chandra} observation (i.e., in the background of nearby galaxy observations), {nearby} FXRTs can only be detected if nearby galaxies lie within the \emph{Chandra} FoV. To this end, the fraction of useful\footnote{That is, ignoring all observations with $|b|{<}$10~deg.} \emph{Chandra} observations that target nearby galaxies at ${\leq}$100~Mpc is ${\approx}$20\% of the total sample or ${\approx}$36.7~Ms (based on a match with the GLADE catalog; \citealt{Dalya2018}), while 80\% is spent observing distant extragalactic sky, respectively. In these fractions, we find 6 and 8 FXRTs, respectively.

Extrapolating from 8 FXRTs in 80\% of the observations, we can expect ${\approx}$2 {distant} FXRTs should occur in the 20\% fraction dedicated to nearby objects, and thus the true number of {nearby} FXRTs would be $6{-}2{\approx}$4. However, given that we are in the Poisson regime, we need to know whether this excess is significant. If we assume a null hypothesis, $\mathcal{NH}$, whereby the sample consists of just one population of FXRTs, such that there are 14 distant FXRTs detected in 100\% of our data, then we expect 2.8 sources in the 20\% of the extragalactic fields that overlap with local galaxies, and yet we observe 6. This corresponds to an excess at 90\% confidence level (i.e., detecting 6 is inconsistent with 2.8 at 90\% confidence), which is  
% 0.1--7.9 using \citet{Kraft1991} statistics) 
likely related with an additional {nearby} population of FXRTs. So there is tentative (at 90\% confidence level) evidence for two different population of FXRTs.

More systematically, we explore the likelihood of whether each individual {local} FXRT could in fact be a {distant} FXRT in projection with a nearby galaxy. Adopting Poisson statistics, following a similar approach to Sect. \ref{sec:completness}, and taking the density of {distant} FXRTs as ${=}$6.5${\times}$10$^{-6}$~arcmin$^{-2}$ (i.e., considering the number of secure distant FXRTs found among all nonlocal \emph{Chandra} observations), we compute the probability of finding by chance a {distant} FXRT within the specific angular offset ($P_{\rm dist.FXRT}$; see Table~\ref{tab:prob} \emph{column 4}) of each {local} FXRT to its associated host galaxy (see Table~\ref{tab:prob} \emph{column 2}). Additionally, we calculated the Binomial probability [$P(k;N;P_{\rm dis.FXRT})$; see Table~\ref{tab:prob} \emph{column 6}], where the number of detections is taken to be the number of {local} FXRTs detected in a galaxy at least as bright as the specific associated {local} FXRT host at a distance less than or equal to the offset distance of each the specific {local} FXRT, and the number of trials is taken to be the number of \emph{Chandra} observations of local galaxies at least as bright as the specific associated {local} FXRT host (see Table~\ref{tab:prob} \emph{column 5}). Table~\ref{tab:prob} shows that only FXRT~1 has a significant probability (${\approx}$0.3) to be related with the {distant} sample; in the case of FXRT~11, it also shows a nonzero probability of being related with the {distant} sample (${\approx}$8.0$\times$10$^{-2}$), reinforcing the idea that this source is not likely associated with nearby galaxies. Thus, there remains some uncertainty as to which population FXRTs~1 and 11 belong. For the moment, we have tentatively assigned FXRT~1 to the {local} sample (due to the similarity of its associated host with others in that category) and FXRT~11 to the {distant} sample. However, we interpret them throughout leaving both possibilities open (see Sect. \ref{sec:flux}).

In summary, our results here reinforce the existence of two different populations of FXRTs and rules out a possible relation of FXRTs~2--6 with the {distant} sample.

\begin{figure*}
    \centering
    \includegraphics[scale=0.8]{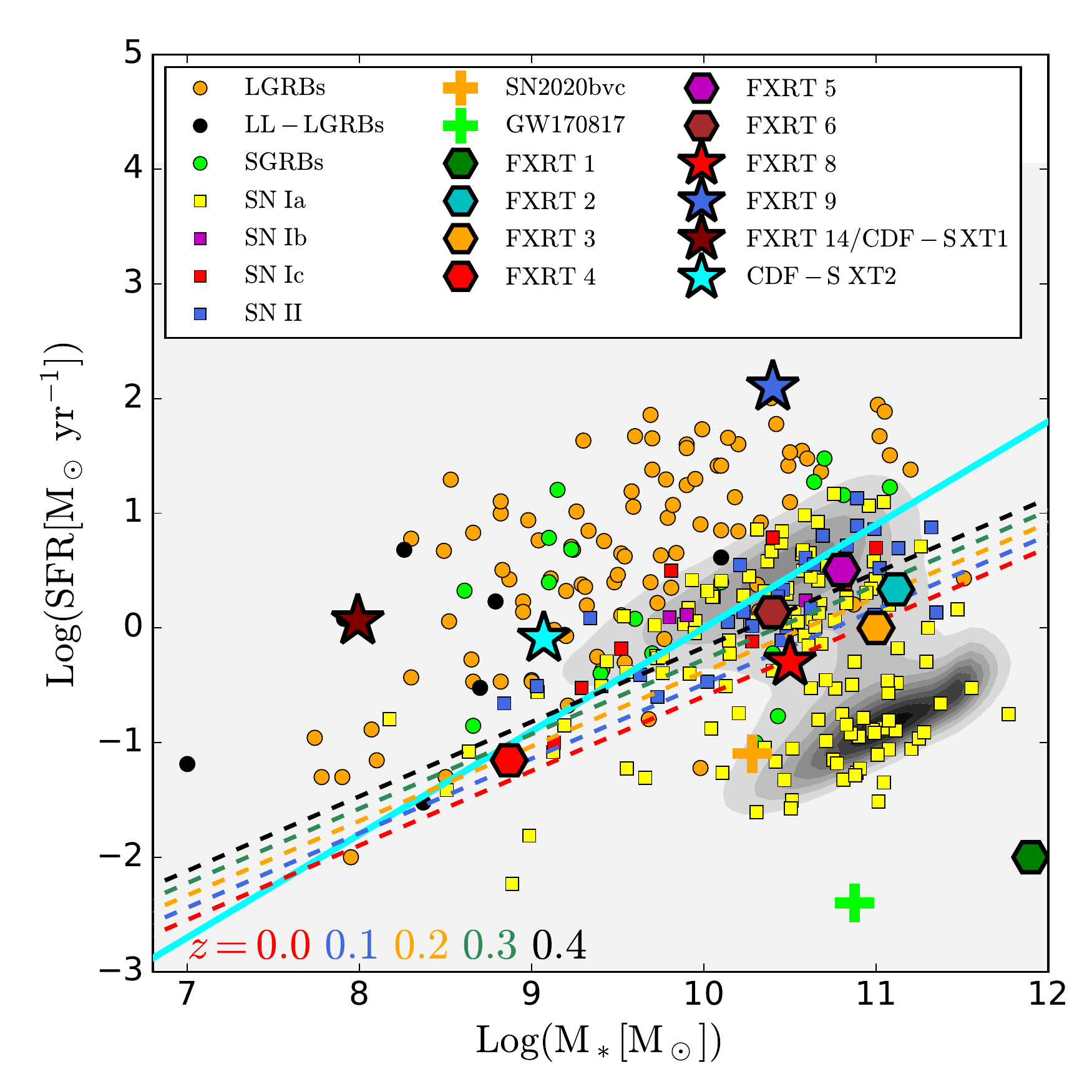}
    \vspace{-0.2cm}
    \caption{Comparison of the stellar mass ($M_*$) and SFR of host galaxies of different types of transients: nearby (\emph{colored hexagons}) and distant (\emph{colored stars}) FXRT candidates, LGRBs and SGRBs \citep{Li2016}, low-luminosity LGRBs \citep[LL-LGRBs; GRB\,980425, GRB\,020903, GRB\,030329, GRB\,031203, GRB\,050826, GRB\,060218, and GRB\,171205A;][]{Christensen2008,Michalowski2014,Levesque2014,Kruhler2017,Wiersema2007,Wang2018,Arabsalmani2019}, GW\,170817 \citep{Im2017}, SN\,2020bvc \citep{Chang2015,Izzo2020,Ho2020}, FXRT~14/CDF-S~XT1 \citep{Bauer2017} and CDF-S~XT2 \citep{Xue2019}, and supernovae events \citep[Types Ia, Ib, Ic, and II;][]{Tsvetkov1993,Galbany2014}. The \emph{solid cyan line} shows the best-fit local galaxy main sequence relation from \citet{Peng2010}. The \emph{gray background contours} are the galaxy distribution from the SDSS \citep[data taken from][]{Brinchmann2004}. The \emph{dashed colored lines} are the boundaries separating the star-forming and quiescent galaxies and its evolution with redshift \citep[at $z{=}$0.0, 0.1, 0.2, 0.3, and 0.4, from bottom to top;][]{Moustakas2013}.}
    \label{fig:Mass_SFR_plot}
\end{figure*}

%GRB\,980425, GRB\,020903, GRB\,030329, GRB\,031203, GRB\,050826, GRB\,060218, GRB\,171205A

 \begin{table*}
    \centering
    \scalebox{0.82}{
    \begin{tabular}{lllllllllllll}
    \hline\hline
    FXRT & ID & RA (deg) & DEC (deg) & Offset & $z$ or $d$(Mpc) & Log(Age/yr) & Log(M$_*/M_\odot$) & Log(SFR/($M_\odot$/yr)) & $A_V$ (mag) & References \\ \hline
    (1) & (2) & (3) & (4) & (5) & (6) & (7) & (8) & (9) & (10) & (11) \\ \hline
    \multicolumn{10}{c}{Parameters obtained from the literature} \\ \hline
    1 & XRT\,000519$\dagger$ & 186.54893 & 12.94622 & 12\farcm2 & 16.4 & -- & 11.89 & $-2.0$ & 0.081 & 1,15 \\
    2 & XRT\,010908 & 167.87904 & 55.67411 & 0\farcm4 & 9.3 & -- & 11.11 & 0.34 & 0.046 & 2,3,13,14 \\
    3 & XRT\,070530 & 201.36506 & -43.01911 & 5\farcm5 & 4.04 & -- & 11.0 & 0.0 & 0.315 & 4,5,13,16 \\
    4 & XRT\,071203 & 211.25671 & 53.66222 & 0\farcm4 & 6.9 & -- & 8.87 & $-1.15$ & 0.029 & 6,7,13,17 \\
    5 & XRT\,080331 & 170.06235 & 12.99154 & 1\farcm3 & 8.1 & -- & 10.80 & 0.51 & 0.091 & 8,9,13,14 \\ 
    6 & XRT\,130822 & 345.50403 & 15.96478 & 1\farcm2 & 29.3 & -- & 10.40 & 0.14 & 0.211 & 10,11,13 \\ 
    14 & XRT\,141001 & 53.16158 & -27.85938 & 0\farcs13 & 2.23$_{-1.84}^{+0.98}$ & -- & 7.99 & 0.06 & 0.021 & 12,18 \\ \hline
    \multicolumn{10}{c}{Parameters derived from photometric data using \texttt{BAGPIPES} \citep{Carnall2018}} \\ \hline
    8 & XRT\,041230 & 318.12653 & -63.49895 & 0\farcs7 & 0.61$_{-0.17}^{+0.13}$ & 9.6$_{-0.2}^{+0.2}$ & 10.5$_{-0.2}^{+0.2}$ & $-0.3_{-1.5}^{+1.3}$ & 0.9$_{-0.7}^{+0.8}$ & -- \\
    9 & XRT\,080819 & 175.00511 & -31.91749 & 0\farcs5 & 0.7$_{-0.10}^{+0.04}$ & 8.2$_{-0.1}^{+0.1}$ & 10.4$_{-0.1}^{+0.1}$ & 2.1$_{-0.1}^{+0.1}$ & 1.6$_{-0.1}^{+0.1}$ & -- \\
    \hline
    \end{tabular}
    }
    \caption{Parameters obtained from the literature and by our SED fitting to archival photometric data using the \texttt{BAGPIPES} package \citep{Carnall2018}.
    \emph{Column 1:} FXRT candidate number. 
    \emph{Column 2:} Candidate ID. 
    \emph{Column 3 and 4:} Right ascension and declination of the host galaxies. \emph{Column 5:} Angular offset between the transient and the host galaxy. 
    \emph{Column 6:} Host galaxy redshift or distance. 
    \emph{Columns 7, 8, 9:} Logarithmic values of the age of the stellar population, the stellar mass, and the   SFR from the host galaxies. 
    \emph{Column 10:} Dust attenuation.
    \emph{Column 11:} Literature references.\\
    References: (1)\citet{Rhode2007}, (2)\citet{Wiegert2015}, (3)\citet{Rhode2007}, (4)\citet{Espada2019}, (5)\citet{Rejkuba2011}, (6)\citet{Drozdovsky2000}, (7)\citet{Lanz2013}, (8)\citet{Buta2015}, (9)\citet{Beuther2017}, (10)\citet{Cappellari2011}, (11)\citet{Davis2014}, (12)\citet{Bauer2017}, (13)\citet{Helou1991}, (14)\citet{Sorce2014}, (15)\citet{Jonker2013}, (16)\citet{Crnojevic2016}, (17)\citet{Tully2013}, (18)\citet{Schlafly2011}.\\
    $\dagger$ Assuming the FXRT is associated with galaxy M86.}
    \label{tab:SED_para}
\end{table*}

\begin{figure}
    \centering
    \includegraphics[scale=0.8]{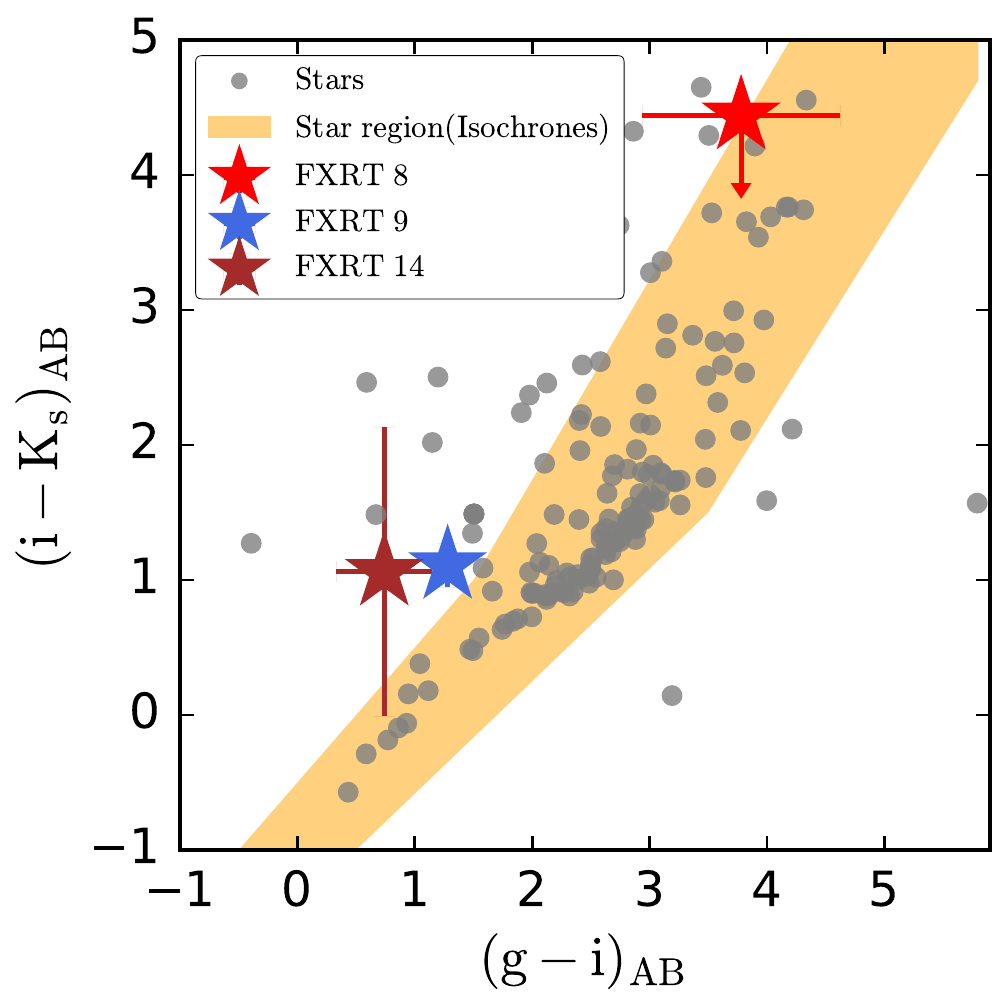}
    \vspace{-0.3cm}
    \caption{Color-color diagram of FXRTs~8, 9, and 14 and X-ray sources classified as stars according to {Criterion 2} (see Sect. \ref{sec:gaia}) with Pan-STARRS, DECam,  and 2MASS counterparts (\emph{gray filled circles}). The expected parameter space of stars with different ages ($\log(\rm{Age}){=}$7.0-10.3), metallicities (from $\rm{[Fe/H]}{=}$-3.0--0.5), and attenuations ($A_V{=}$0.0--5.0) taken from the MIST package (\citealp{Dotter2016},\citealp{Choi2016}) is overplotted as an \emph{orange} region. }
    \label{fig:color_color_2}
\end{figure}

\begin{figure*}
    \centering
    \includegraphics[scale=0.45]{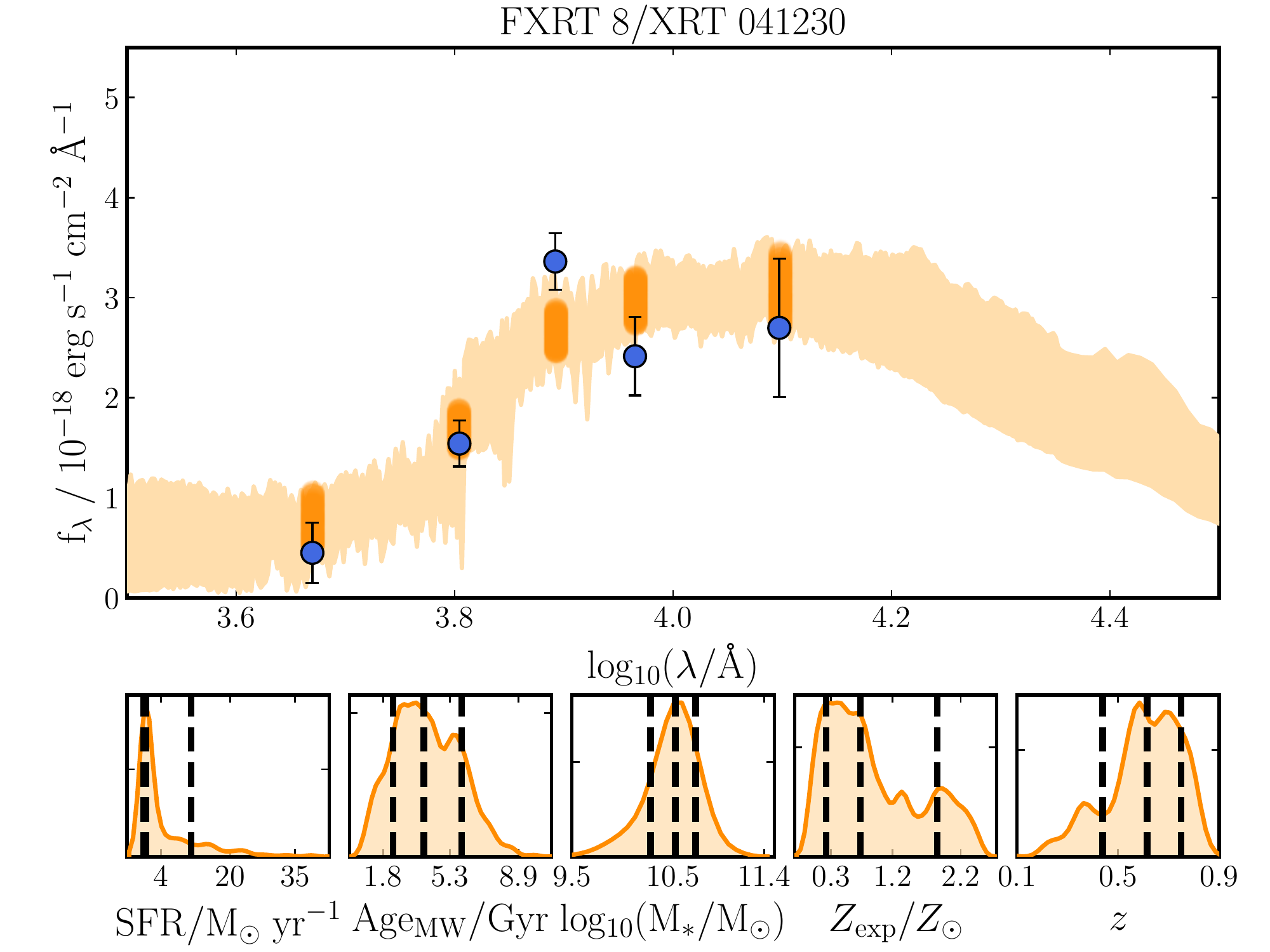}
    \includegraphics[scale=0.45]{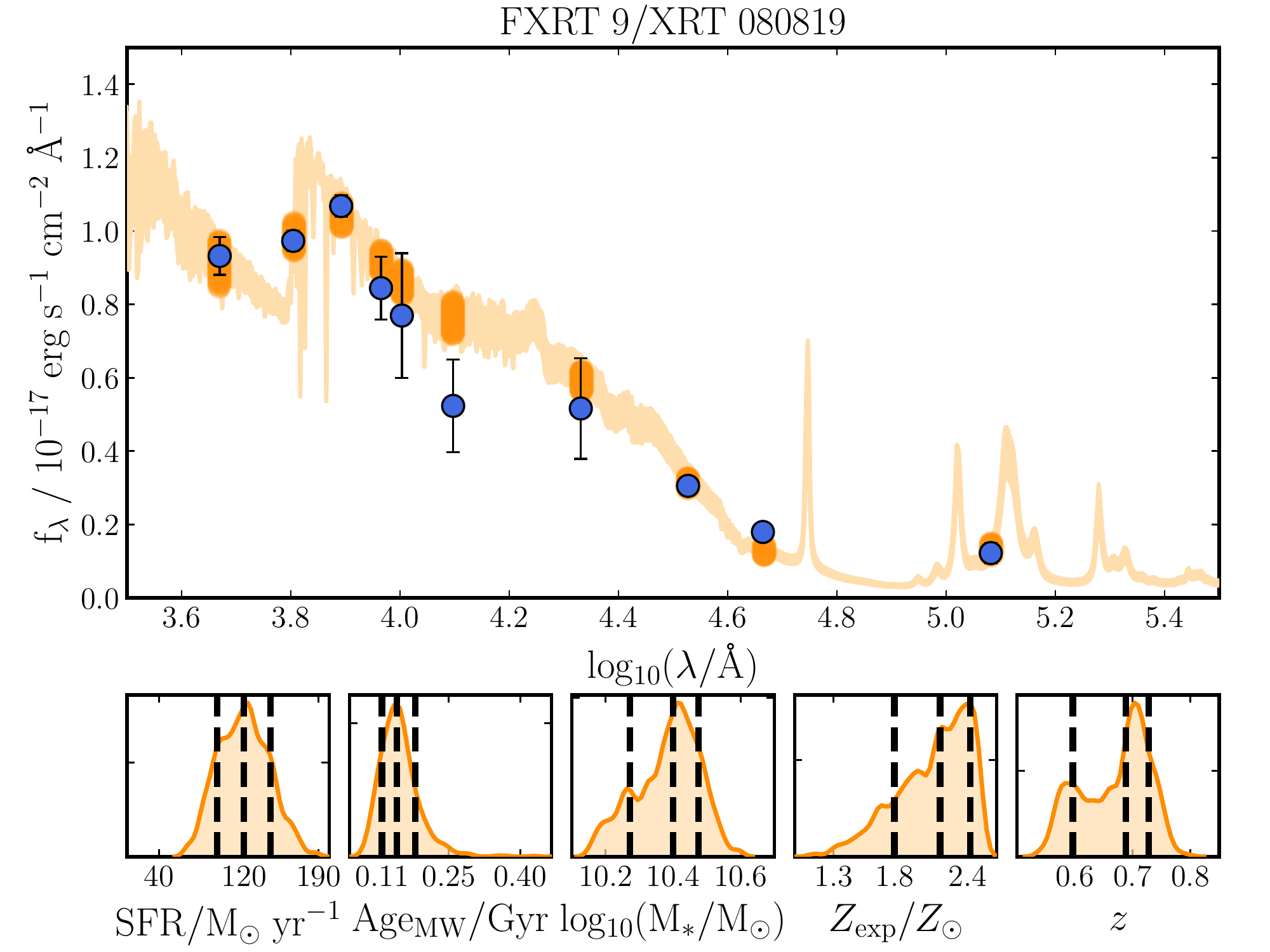}
    \vspace{-0.1cm}
    \caption{Best fitting SED models obtained using the \texttt{BAGPIPES} package \citep{Carnall2018} for FXRTs 8 (\emph{left panels}) and 9 (\emph{right panels}). \emph{Top panels:} 16th to 84th percentile range for the posterior spectrum and photometry (shaded \emph{orange}). The used photometric data and their uncertainties are given by the  \emph{blue markers}. \emph{Bottom panels:} Posterior distribution for the five fitted parameters (SFR, age, galaxy stellar mass, metallicity, and redshift). The 16th, 50th, and 84th percentile posterior values are indicated by the vertical \emph{dashed black lines}.}
    \label{fig:SED_models}
\end{figure*}

\section{Host galaxy features}\label{sec:counterpart_SED}

The host galaxy or host environment of an FXRT can provide important information on its nature. Five nearby FXRTs (2, 3, 4, 5 and 6) and three distant FXRTs (8, 9, and 14/CDF-S~XT1) have very probable associated optical/NIR host galaxy detections. It remains less certain whether FXRT 1 is associated with M86.

\subsection{Nearby extragalactic FXRT sample}\label{sec:counterpart_SED_nearby}

The nearby events (FXRTs~1, 2, 3, 4, 5, and 6) are located in well-studied local galaxies, although the association of FXRT~1 with M86 is less clear. We collect information from the literature in Table~\ref{tab:SED_para}, although we caution that this is a heterogeneous data set, and deriving consistent global parameters is beyond the scope of this work.

Figure~\ref{fig:Mass_SFR_plot} compares the host galaxy star-formation rate (SFR) versus~stellar mass (M$_*$) values of our sample to  hosts of other transients (LGRBs, low-luminosity LGRBs, SGRBs, CC-SNe, thermonuclear SNe, SN\,2020bvc, and GW\,170817/GRB\,170817A). It is clear that the different classes of transients fall in specific regions of the SFR-M$_*$ plane. For instance, thermonuclear type~Ia supernovae lie preferentially below the galaxy main sequence \citep[\emph{dashed cyan line};][]{Peng2010}, that is to say, they are related with older ({redder or recently quenched) stellar populations within galaxies}; meanwhile, CC-SNe \citep[type Ib, Ic, and II;][]{Tsvetkov1993,Galbany2014} fall closer the galaxy main sequence, highlighting their relation with ongoing star-formation in galaxies. LGRBs \citep{Li2016}, which are related to massive progenitor stars, lie above the galaxy main sequence (where so-called high-SFR {starburst} galaxies lie). In contrast, the location of SGRBs \citep{Li2016} shows a large spread in this figure (i.e., SGRBs occur in a mixed population of early-type and star-forming galaxies). 
GW\,170817/GRB\,170817A \citep{Im2017} is singled out among SGRBs, due to the unusually low SFR (${\approx}$0.001--0.01~M$_\odot$~yr$^{-1}$) of its host galaxy NGC~4993. We also single out the off-axis LGRB candidate SN\,2020bvc \citep[][]{Chang2015,Izzo2020,Ho2020}, the host galaxy UGC~9379 of which has a low SFR ($\approx$0.08~M$_\odot$~yr$^{-1}$) but a stellar mass M$_*{\approx}$1.9$\times$10$^{10}$~M$_\odot$ similar to the Milky Way and other large spirals \citep{Hjorth2012,Taggart2019,Ho2020}.

Among the nearby sample, the hosts of FXRTs~2, 3, 4, 5, and 6 fall just below (within a factor of two) of the local galaxy main sequence, implying they are probably related to active star formation processes or young stellar populations. In the cases of FXRTs~2, 3, 4, and 5, there are clear spatial associations with compact H{\sc ii} regions or young stellar clusters, strengthening the link to young, presumably massive stars. FXRT~1/XRT\,000519 is probably related with M86, which has a low SFR (${\approx}$0.01~M$_\odot$~yr$^{-1}$), as expected for elliptical galaxies.

\subsection{Distant extragalactic FXRT sample}\label{sec:counterpart_SED_distant}

The optical/NIR hosts of the distant events FXRTs~8 and 9 are classified as extended sources (galaxies) by {the VHS catalog} \citep{McMahon2013}, but their properties have not been analyzed previously. We used photometric data of their putative host galaxies to constrain the host properties through spectral energy distribution (SED) model fitting. 

We initially explored the spectral nature of FXRTs~8, 9, and 14 based on their $i-K_s$ versus $g-i$ colors in Fig.~\ref{fig:color_color_2}. FXRTs~8, 9, and 14 were compared to the counterparts of the X-ray sources classified as stars according to {Criterion 2} (\emph{gray points}; see Sect. \ref{sec:gaia}) and the expected parameter space for stars (\emph{orange region}) with different ages ($\log(\rm{Age/yr}){=}$7.0--10.3), metallicities (from $\rm{[Fe/H]}{=}-3.0$--0.5), and attenuations ($A_V{=}$0.0--5.0~mag) from theoretical stellar isochrones \citep[MIST;][]{Dotter2016,Choi2016}. The bulk of the stellar X-ray variables form a much tighter sequence than what is conceivably allowed by the full range of isochrones. The stellar X-ray sources that appear as outliers are identified as PNe, YSOs (e.g., eruptive variable stars, T Tauri stars), or emission-line stars. The FXRTs generally lie outside of or at the edge of the stellar region, away from the tight stellar locus, although the large error bars or limits in the NIR photometry preclude any definitive statements here. We conclude that the SED by itself is not a clear-cut discriminator and thus the spatially resolved nature of the counterparts remains vital to their confirmation.

Next, we employ the code \texttt{BAGPIPES} \citep[Bayesian Analysis of Galaxies for Physical Inference and Parameter EStimation;][]{Carnall2018}, which fits stellar-population models taking star-formation history and the transmission function of neutral/ionized ISM into account to broadband photometry and spectra using the \texttt{MultiNest} nested sampling algorithm \citep{Feroz2008,Feroz2009}, to derive constraints on the host-galaxy properties. \texttt{BAGPIPES} gives the posterior distributions for the host-galaxy redshift ($z$), age, extinction by dust ($A_V$), SFR, metallicity ($Z$), stellar mass ($M_*$), specific star formation rate. To account for dust attenuation in the SEDs, we use the parametrization developed by \citet{Calzetti2000}, where $A_V$ is a free parameter within the range 0.0 to 3.0~mag.

We assume an exponentially declining star formation history function parametrized by the star formation timescale (free parameter). The Table~\ref{tab:SED_para} provides the best-fit parameters obtained with \texttt{BAGPIPES} for the hosts of FXRTs~8 and 9, while Fig.~\ref{fig:SED_models} shows the 16th to 84th percentile range for the posterior spectrum and photometry. The posterior distribution for the fitted parameters is shown in the bottom panels. We have confirmed the obtained photometric redshifts of  0.61$_{-0.17}^{+0.13}$ and 0.7$_{-0.10}^{+0.04}$ for FXRTs~8 and 9 with our 2D spectra taken by X-SHOOTER (PIs: Quirola and Bauer, program ID: 105.20HY.001). A detailed analysis of the spectral data will be presented in future work.

FXRT~13 only has a single $i$-band DECam source associated with it. The non-detections in other bands may suggest that the $i$-band DECam image (${\approx}$6948--8645~\AA) includes a dominant flux contributions from a high equivalent width emission line. Considering the most important emission lines of galaxies (such as H$\alpha$, H$\beta$, [OIII]$\lambda\lambda$4959, 5007~\AA), the expected redshift range of FXRT~13 is $z{\approx}$0.2--1.1.

Finally, \citet{Bauer2017} associated FXRT~14 (CDF-S~XT1) with an extremely faint, small $z_{\rm photo}{\sim}2.23$ host galaxy with a relatively flat SED (see Tables~\ref{tab:photometric_data} and \ref{tab:SED_para}).

Among the FXRTs identified as distant candidates, FXRTs~8 and 14 are located above the galaxy main sequence, while FXRT~9 lies significantly below it. We also show the host of CDF-S~XT2, which also lies just above the galaxy main sequence. Thus, a sizable fraction of distant FXRTs appear to be associated with vigorous star formation; we should stress here, however, that the statistics are poor and the uncertainties from the SED model fits remain large (see Table~\ref{tab:SED_para}). 

\begin{figure*}
    \centering
    \includegraphics[scale=0.75]{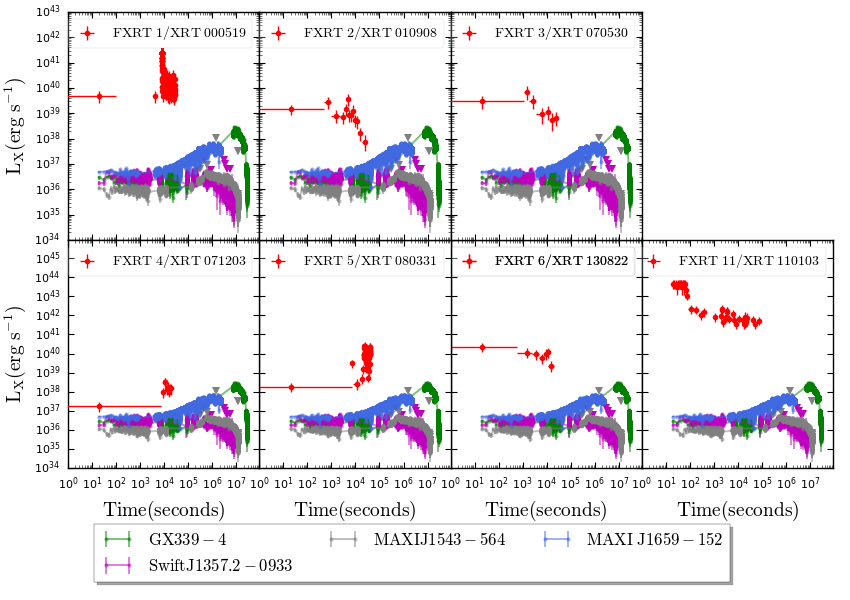}
    \caption{0.3--10 keV light curves of the five local CSC2 FXRTs, plus FXRTs~1 and 11, in luminosity units. The 0.3--10 keV light curves are obtained by multiplying the 0.3--7 keV light curves by the factor derived from extrapolating the best-fit PO model flux to the 0.5--7.0~keV spectrum to the 0.3--10 keV band and correcting it for the effects of Galactic plus intrinsic absorption. For comparison, we overplot flaring episodes for several individual well-known Galactic XRBs: GX339-4  \citep[9~kpc, \emph{green line};][]{Heida2017}, \emph{Swift}\,J1357.2-0933 \citep[8~kpc, \emph{magenta line};][]{Mata2015}, \emph{MAXI}\,J1543-564 \citep[5~kpc, \emph{gray line};][]{Stiele2012}, and \emph{MAXI}\,J1659-152 \citep[6~kpc, \emph{blue line};][]{Jonker2012c}. 
    The light curves of the comparison sources are taken from the 2SXPS catalog \citep{Evans2020b}.}
    \label{fig:lumin_comparison}
\end{figure*}

\begin{figure*}
    \centering
    \includegraphics[scale=0.75]{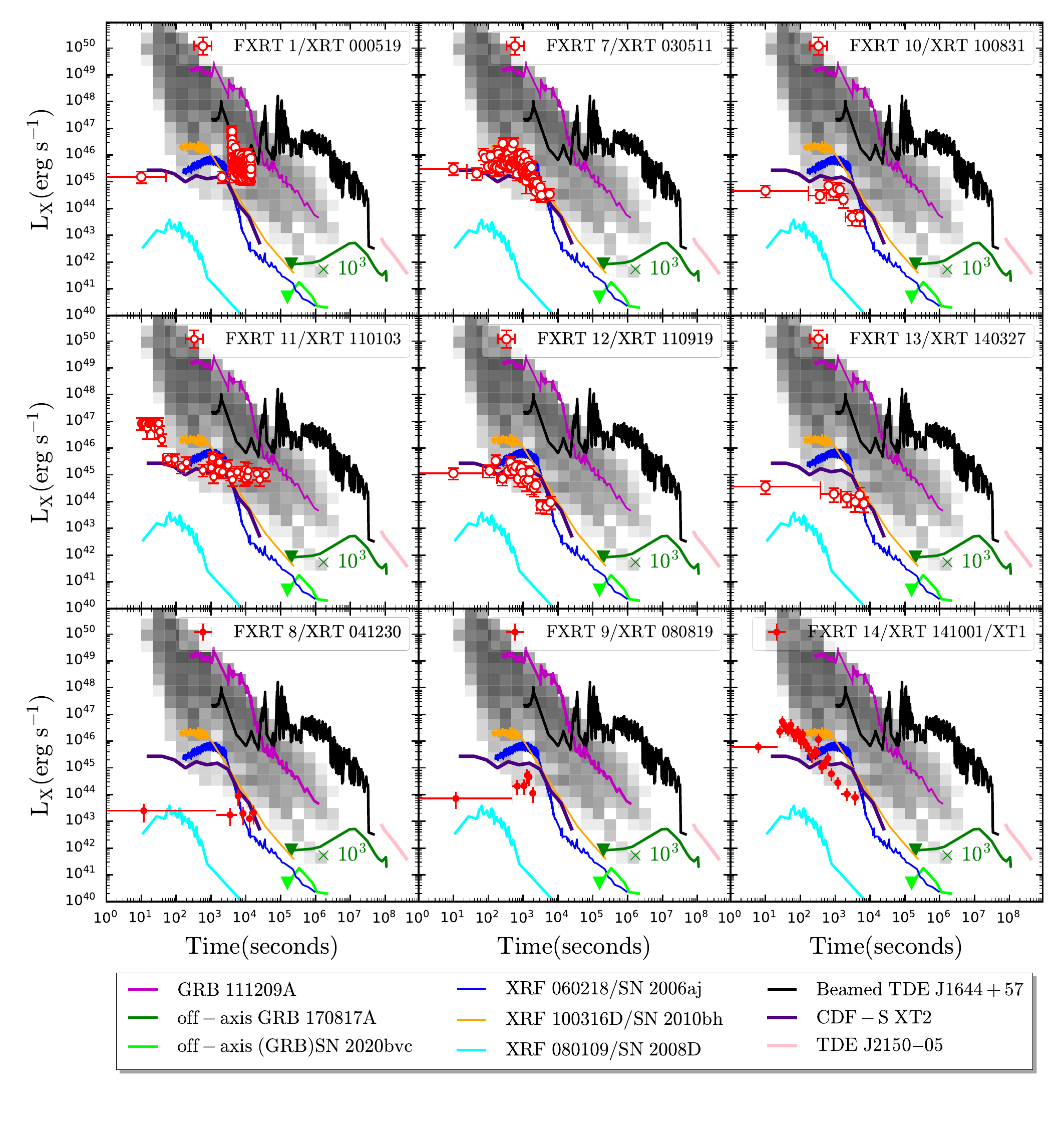}
    \vspace{-0.3 cm}
    \caption{0.3--10 keV light curves of the nine CSC2 FXRTs in luminosity units (as in Fig.~\ref{fig:lumin_comparison}, 0.3--10 keV light curves were converted from 0.5--7 keV ones). The \hbox{X-ray} afterglow light curves of 64 LGRBs plus 32 SGRBs \citep[taken from][]{Bernardini2012,Lu2015} are shown as a 2D histogram, as are the \hbox{X-ray} afterglows of GRB\,170817A \citep[off-axis SGRB, \emph{solid dark green line};][]{Nynka2018,DAvanzo2018,Troja2020,Troja2022}, SN~2020bvc \citep[the first off-axis LGRB candidate, \emph{solid light green line};][]{Izzo2020}, and the ultra-long duration GRB~111209A \citep[\emph{solid magenta line}, $z{=}0.677$;][]{Levan2014}. Additionally, several individual transients are overplotted: the low-luminosity supernova SBO XRF\,080109/SN\,2008D \citep[\emph{solid cyan lines}, 27~Mpc;), XRF\,060218/SN\,2006aj (\emph{solid blue lines}, 145~Mpc;), XRF\,100316D/SN\,2010bh (\emph{solid orange lines},263~Mpc;][]{Barniol2015,Starling2011,Modjaz2009,Evans2009,Soderberg2008,Evans2007,Campana2006}, the relativistically beamed TDE \emph{Swift}\,J1644+57 \citep[\emph{solid black lines}, $z{=}0.3543$;][]{Bloom2011,Levan2011}, the non-beamed TDE J2150-05 \citep[\emph{solid pink line}, $z{=}0.055$;][]{Lin2018}, and CDF-S~XT2 \citep[\emph{solid indigo lines};][]{Xue2019}. For FXRTs~1, 7, 10, 11, 12, and 13 (\emph{open symbols}), we assume $z{=}$1.0, we adopt $z_{\rm photo}{=}$2.23 for FXRT~14 from \citet{Bauer2017}, and for FXRTs~8 and 9 we consider the values from Table~\ref{tab:SED_para}.}
    \label{fig:flux_comparison}
\end{figure*}

\section{Possible interpretations}\label{sec:flux}

To understand the origin of our sample of FXRTs, we compare them with other well-known transients. We split our discussion here into {nearby} (Sect. \ref{sec:inter_nearby}) and {distant} (Sect. \ref{sec:inter_dist}) samples. The former have well-established distances, and therefore we can compare their light curves in luminosity units. As we do not know the redshift of several distant FXRTs, we compare their \hbox{X-ray} light curves in luminosity units assuming nominal distances. Given the uncertainty in the associations for FXRTs\,1 and 11, we discuss them under both the nearby and distant extragalactic scenarios. 

First, from the best-fit PL spectral model, we compute the X-ray peak flux (corrected for Galactic and intrinsic absorption; $F_{\rm peak}$)\footnote{Due to the lack of a standardized method to estimate the $F_{\rm peak}$, we consider the following. First, we find the shortest time interval during which 25\% of the counts are detected, and we compute a count rate during this shortest interval. Next, to convert the peak-count rates to fluxes, we multiply the flux from the time-averaged Spectral fits by the ratio between the peak and the time-averaged count rates (i.e., we assume no spectral evolution).}, the associated intrinsic X-ray peak luminosity ($L_{\rm X,peak}$), and the Eddington mass (defined as $M_{\rm Edd}{=}7.7{\times}10^{-39}L_{\rm X,peak}$ in solar mass units). We report these values in Table~\ref{tab:F_L_para} in the energy range 0.3--10.0~keV.

\subsection{Nearby extragalactic FXRT sample} \label{sec:inter_nearby}

The nearby FXRTs~ 2, 3, 4, 5, and 6 have peak isotropic \hbox{X-ray} luminosities in the range of $L_{\rm X,peak}{\approx}$10$^{38}$--10$^{40}$~erg~s$^{-1}$ (see Table~\ref{tab:F_L_para}). This appears inconsistent with origins as SBOs \citep[$L_{\rm X,peak}^{\rm SBOs}{\approx}$10$^{42}$--10$^{47}$~erg~s$^{-1}$;][]{Ensman1992,Soderberg2008,Modjaz2009,Waxman2017,Alp2020}, TDEs \citep[$L_{\rm X,peak}^{\rm TDEs}{\approx}$10$^{42}$--10$^{50}$~erg~s$^{-1}$;][considering a jetted emission]{Rees1988,MacLeod2014,Maguire2020,Saxton2021}, or on-axis GRBs \citep[$L_{\rm X,peak}^{\rm GRBs}{\approx}$10$^{47}$--10$^{51}$~erg~s$^{-1}$;][considering a jetted emission]{Berger2014,Bauer2017}. 

These lower luminosities fall into the realm of ULXs (extragalactic \hbox{X-ray} emitters located off-center of their host galaxy and with luminosities in excess of $L_X^{\rm ULX}{\approx}$10$^{39}$~erg~s$^{-1}$, if the emission is isotropic, well above the Eddington limit for neutron stars; \citealp{Bachetti2014,Kaaret2017}) and Galactic XRBs (\hbox{X-ray} emitters where a compact object accretes mass from a companion star with $L_X^{\rm XRB}{\lesssim}$10$^{39}$~erg~s$^{-1}$; \citealp{Remillard2006,van_den_Eijnden2018}). Most ULXs are semi-persistent \hbox{X-ray} emitters for years to decades \citep{Kaaret2017}, and in extreme cases can reach high luminosities such as NGC\, 5907\,ULX1 \citep[${\approx}$5$\times$10$^{40}$~erg~s$^{-1}$;][]{Walton2016}. The much shorter and stronger variability of our FXRTs compared to ULXs implies that they are caused by a different phenomenon. 

Another alternative could be XRBs. Figure~\ref{fig:lumin_comparison} shows the \hbox{X-ray} light curves of FXRTs~1, 2, 3, 4, 5, 6, and 11 one per panel, compared to several well-known XRB flaring episodes. XRBs in the Milky Way exhibit pronounced variability whereby the \hbox{X-ray} flux changes from quiescent to flare states on timescales of weeks to months \citep{Remillard2006}. Particularly, FXRTs~2 and 4 reach peak luminosities ($L_{\rm X,peak}{\approx}$ 10$^{38}$--10$^{39}$~erg~s$^{-1}$) similar to some XRBs' flares (e.g., the flare luminosity of GX339-4 of ${\approx}$4$\times$10$^{38}$~erg~s$^{-1}$), which suggests that these FXRTs could be related with the tip of longer flares. Nevertheless, they are not in agreement with the duration (in the order of weeks) and timescale evolution (following a slow PL decay $F_X{\propto}t^{-0.3}$) of XRBs flares. Thus, FXRTs~2, 3, 4 are unlikely to be related with XRBs. Meanwhile, FXRTs~1, 3, 5, 6, and 11 are not related with XRBs because of their high luminosity ($L_{\rm X,peak}{\gtrsim}$ 10$^{39}$~erg~s$^{-1}$). 

Next, we compare the FXRTs to SGRs and AXPs, which are both believed to be related to young, highly magnetic neutron stars \citep{Woods2006}. Soft gamma repeaters and AXPs are very faint in quiescence but can flare by factors of hundreds to thousands on timescales of tens of ms to seconds \citep{Gougucs1999,Aptekar2001} and on very rare occasions can generate giant flares by factors of $10^{5}$ over several-minute timescales \citep[e.g., SGR~1806-20, SGR~1900+14;][]{Hurley1999,Terasawa2005,Woods2006}. On such occasions they can reach \hbox{X-ray} luminosities as high as $10^{40}$--$10^{41}$ ergs~s$^{-1}$, although SGR~1900+14 experienced a giant flare with a peak luminosity of ${>}$10$^{44}$~erg~s$^{-1}$ \citep[at hard X-rays 40--700~keV;][]{Mazets1999,Feroci2001}, and generate bolometric outputs up to ${\sim}10^{46}$~erg in total \citep{Palmer2005,Terasawa2005,Strohmayer2005,Israel2005}. The weaker flares generally have a soft spectrum, while the giant flares are quite hard, and the flare durations follow a log-normal distribution \citep{Gougucs1999}. At the distances of our nearby FXRTs, we would presumably only see the most luminous portions of these rare giant bursts (a few seconds at most) and not be sensitive to the fainter bursts or quiescent emission. They should be quite spectrally hard \citep[e.g., SGR~1900+14 has a photon index range $\Gamma{\approx}$1.0--2.0;][]{Tamba2019} and given the relation to young, highly magnetic neutron stars, seen to be originating from young star clusters and H{\sc ii} regions \citep{Woods2006}. In this sense, FXRTs~2, 4, 5, and 6 share some similarities with the SGR and AXP phenomena. For instance, they seem related to star-formation galaxies (see Fig.~\ref{fig:Mass_SFR_plot}), although their spectra remain relatively soft and their light curve lengths last thousands of seconds. In the case of FXRT~5, we see multiple flares over ${\approx}$35\,ks. On the other hand, FXRTs~1, 3, and 11 are not associated with young star clusters, and thus seem far less likely to be explained by an SGR/AXP origin.

A final point of comparison is with the FXRTs discovered in NGC~4636 and NGC~5128 (Cen\,A) by \citet{Irwin2016} and NGC~4627 by \citet{Sivakoff2005}. All exhibit rapid ($\sim$50--100~s) flares with peak luminosities of ${\sim}$10$^{39}$--10$^{41}$~erg~s$^{-1}$, but remain detectable by \emph{Chandra} in quiescence. In two cases, multiple flares are observed across multiple observations, while two transients are spatially associated with globular clusters in their host galaxies (similar to FXRT~3, which intriguingly is also associated with NGC~5128). Overall, while the luminosities are comparable, the faster timescales, multiple outbursts, and quiescent detections are unlike the behavior seen among the nearby sample of FXRTs, although it could be the case that (some) FXRTs have quiescent fluxes well below the sensitivity of \emph{Chandra} and {\it XMM-Newton} and/or have not been observed frequently enough to see multiple outbursts (e.g., FXRT~6 has not been observed again by \emph{Chandra} or \emph{XMM-Newton}; see Fig.~\ref{fig:light_curves_1}).

In the case of FXRTs~1 and 11, their origin remains unclear. For FXRT~1, assuming the association with M86 
\citep[${\approx}$16.4~Mpc;][]{Jonker2013}, it is characterized by a peak luminosity of ${\approx}$6$\times$10$^{42}$~erg~s$^{-1}$ (see Table~\ref{tab:F_L_para}). According \citet{Jonker2013}, this X-ray flash (XRF) could have been caused by the disruption of a compact WD by a 4.9$\times$10$^4$~$M_\odot$ BH. Nevertheless, other scenarios such as a highly off-axis GRB \citep{Dado2019} cannot be discarded because of distance uncertainties. For FXRT~11, assuming the association with the galaxy cluster Abell~3581 \citep[${\approx}$94.9~Mpc;][]{Glennie2015,Johnstone2005}, its peak luminosity and Eddington mass are ${\approx}$2$\times$10$^{44}$~erg~s$^{-1}$ and ${\approx}$1.9${\times}$10$^{6}$~$M_\odot$ (see Table~\ref{tab:F_L_para}). \citet{Glennie2015} suggest that FXRT~11 could be consistent with the early \hbox{X-ray} emission typically seen in GRB light curves; however, its similarities with FXRT~1 also suggest that both events share the same origin.

Overall, the behavior of the nearby FXRTs appears to represent a genuinely new phase space of transient phenomena. The wide variety of observed properties strongly suggests that multiple physical origins may be at work.

\subsection{Distant extragalactic FXRT sample} \label{sec:inter_dist}

For the moment, we split the discussion of FXRTs into those with fairly secure hosts and reasonable distance estimates (see Sect. \ref{sec:dist_kn_FXRTs}) versus those with less certain or no clear hosts (see Sect. \ref{sec:dist_unkn_FXRTs}). Here we analyze the scenario where the FXRTs~1 and 11 are related to  distant extragalactic objects (see Sect. \ref{sec:dist_unkn_FXRTs}).

\subsubsection{FXRTs with known distances}\label{sec:dist_kn_FXRTs}

Using the photometric host redshifts calculated in Sect. \ref{sec:counterpart_SED}, FXRTs~8 and 9 reach peak \hbox{X-ray} luminosities of $L_{\rm X,peak}{\approx}$1.5$\times$10$^{44}$ and 1.3$\times$10$^{45}$~erg~s$^{-1}$, respectively (see Table~\ref{tab:F_L_para} and Fig.~\ref{fig:flux_comparison} for a light curve comparison). The FXRTs have an isotropic fluence of $E_X{\approx}$2.2${\times}$10$^{47}$ and 3.9${\times}$10$^{47}$~erg, respectively.

Such luminosities fall with the ranges predicted or detected for SBOs \citep[$L_{\rm X,peak}^{\rm SBOs}{\approx}\times$10$^{42}$--10$^{47}$~erg~s$^{-1}$;][]{Soderberg2008,Modjaz2009,Waxman2017,Alp2020}, although both FXRTs exhibit energy released that are at least one to two orders of magnitude higher than the energy predicted by SBO models \citep[e.g.,][]{Waxman2017} or detected from the enigmatic SBO XRT\,080109/SN\,2008D \citep[$E_X{\approx}2\times$10$^{46}$~erg;][]{Soderberg2008}. As such, we rule out an SBO interpretation for FXRTs~8 and 9.

Considering an on-axis GRB origin, we note that no gamma-ray signals detected near the time of discovery were associated with FXRTs~8 or 9, and neither exhibits a characteristic PL decay phase ($F_X^{\rm FXRTs8/9}{\propto}t^{-2.9/-2.8}$ associated with GRB afterglows $F_X^{\rm GRBs}{\propto}t^{-1.2}$; \citealp{Evans2009,Racusin2009}), although some GRBs show strong \hbox{X-ray} flaring in the tail of the \hbox{X-ray} afterglow distribution that could mimic the observed temporal behavior \citep{Barthelmy2005,Campana2006,Chincarini2010,Margutti2011}. Critically, Fig.~\ref{fig:flux_comparison} demonstrates that the \hbox{X-ray} light curves of both FXRTs are fainter than almost any known on-axis GRB \hbox{X-ray} afterglow over the same timescale, with (prompt) initial luminosities $>$3--4 dex below the luminosity ranges observed for GRBs ($L_{\rm X,peak}^{\rm GRBs}{\gtrsim}$10$^{47}$~erg~s$^{-1}$). Based on their best-fit \hbox{X-ray} spectral slopes of $\Gamma_{\rm FXRTs~8/9}{\approx}$2.7/3.0, they formally lie at the edge of the standard afterglow distribution \citep[$\Gamma_{\rm GRBs}{=}$1.5--3.0;][]{Berger2014,Wang2015,Bauer2017} overlapping at the 1$\sigma$ confidence level. In terms of their host-galaxy properties (see Fig~\ref{fig:Mass_SFR_plot} and Table~\ref{tab:SED_para}), FXRT~8's host has a low-SFR ($\approx$0.5~$M_\odot$~yr$^{-1}$) and old stellar population ($\gtrsim$1~Gyr). It is classified as a quiescent galaxy according to the criteria from \citet{Moustakas2013}, and is thus a potential host for an SGRB. Nevertheless, an association with low-luminosity LGRBs (LL-LGRBs) could be discarded due to the high stellar mass of its host galaxy.

FXRT~9's galaxy is a massive blue starburst galaxy ($\approx$120~$M_\odot$~yr$^{-1}$) with a young stellar population (${\approx}$0.15~Gyr). Hence, FXRT~9 might be related to an LGRB origin, although an association with SGRBs cannot be discarded. On the other hand, an association with LL-LGRBs could be ruled out because of the high stellar mass of its host galaxy relative to LL-LGRBs' hosts (higher than one order of magnitude; see Fig~\ref{fig:Mass_SFR_plot}). Unfortunately, the low angular resolution of the current archival images does not permit us to compute the offset from the host center.

Alternatively, these FXRTs could be related to ultra-long duration GRBs. Several ultra-long GRBs (longer than thousands of seconds) have been detected \citep{Thone2011,Campana2011,Gendre2013,Virgili2013,Stratta2013}. Their nature still unclear.
\citet{Gendre2013} and \citet{Levan2014} argued that ultra-long duration GRBs form another distinct group of GRBs; for example, \citet{Levan2014} argue that the long duration of this population of GRBs may be explained by engine driven explosions of stars of much larger radii than typical LGRB progenitors (which are thought to have compact Wolf-Rayet progenitor stars). Figure~\ref{fig:flux_comparison} shows a comparison of both FXRTs and the ultra-long GRB~111209A. At early times their luminosities are ${\approx}$6--7~dex lower than that of GRB~111209A. Nevertheless, we cannot discard an association with this population of GRBs because of the uncertainty in the zero point of our FXRTs, which when changed could match well with the temporal decay ($F_X{\propto}t^{-1.4/-5.3}$) and spectral trend ($\Gamma{\approx}2.4$ at $t{>}$40~ks) of this GRB.

The possibility of an off-axis orphan GRB origin still remains plausible, given the lack of an initial gamma-ray detection and lower luminosity. Here we compare to the light curves of XRF\,060218/SN\,2006aj \citep{Campana2006}, XRF\,100316D/SN\,2010bh \citep{Starling2011}, and SN\,2020bvc \citep{Izzo2020}, which have all been argued to be potential off-axis LGRBs, as well as GRB\,170817A \citep{Nynka2018, DAvanzo2018, Troja2020} and CDF-S XT2 \citep{Xue2019}, and thus possible off-axis SGRBs (see Fig.~\ref{fig:flux_comparison}). We note in particular that the plateau phases of FXRTs~8 and 9 are ${\approx}$ 1--3 dex lower than those of XRF\,060218, XRF\,100316D, and CDF-S~XT2, although the break and late-time light curves (to the extent that they can be quantified) appear to match reasonably well. By extension, SN\,2020bvc and GRB\,170817A appear to be even weaker, and join with the faint declining tails of the XRFs at very late times. We speculate that perhaps FXRTs~8 and 9 could be weaker or higher inclination versions of off-axis SGRB and LGRBs \citep[e.g.,][]{Granot2002}, respectively, somewhere intermediate between the XRFs and SN\,2020bvc/GRB\,170817A along the possible viewable parameter space of such events. Unfortunately, the poor count statistics (to constrain any spectral evolution) and the lack of additional EM counterparts do not permit us to analyze this picture in detail.

Finally, in the TDE scenario, if we interpret the peak luminosities as the Eddington luminosity, we derive masses of ${\gtrsim}$1.2$\times$10$^6$ and 1.0$\times$10$^7$~M$_\odot$ for FXRT~8 and FXRT~9, respectively. These masses fall in the supermassive black hole (SMBH) range \citep{Barack2019}, and assuming that a large fraction of the total stellar mass of the host galaxies as derived in Sect. \ref{sec:counterpart_SED_distant} is associated with a spheroid component, could be approximately consistent with the stellar velocity dispersion ($\sigma$) of a galaxy bulge and the mass of the SMBH ($M_{\rm BH}$) at its center \citep[$M_{\rm BH}-\sigma$ relation; e.g.,][]{Ferrarese2000}. These luminosities are in rough agreement with the recent sample of TDEs published by \citet{Sazonov2021}.

Alternatively, these FXRTs could be related with an IMBH--WD or IMBH--MS TDEs \citep[which could occur in dwarf galaxies and stellar systems such as globular clusters;][]{Jonker2012,Reines2013}, assuming the observed luminosities are super-Eddington or due to relativistic beaming. The FXRTs are offset from the nuclei of their associated optical and NIR sources by only 0\farcs5 and 0\farcs7 (or projected physical distances of 3 and 3.5~kpc), respectively, and hence remain consistent with both on-axis and off-axis scenarios within the positional uncertainties (see Fig.~\ref{fig:image_cutouts}).

\citet{Saxton2021} review the observed and theoretical \hbox{X-ray} properties of TDE candidates. Among confirmed SMBH--MS TDEs detected to date, several exhibit peak luminosities similar to those of FXRTs~8 and 9. However, the \hbox{X-ray} spectra of SMBH--MS TDEs are generally softer and none exhibit short-term \hbox{X-ray} variability comparable to what see from the FXRTs, but instead show much slower declines over timescales of months to years. For this reason, we disfavor such an explanation, but cannot completely rule out a possible detection bias here, given the limited sensitivity of current all-sky instruments. One intriguing possibility for generating higher luminosities, faster variability, and harder spectra is relativistic beaming from jetted TDEs such as \emph{Swift}~J1644+57 \citep{Bloom2011,Levan2011}. This could also significantly relax the mass and/or accretion rate limits quoted above. In the case of \emph{Swift}~J1644+57, shown in Fig.~\ref{fig:flux_comparison}, it has 
a peak luminosity of ${\approx}$10$^{48}$~erg~s$^{-1}$ and time-averaged photon index of $\Gamma{=}$1.6--1.8 \citep{Levan2011}, although the photon index increases and softens with decreasing flux \citep{Bloom2011}. Clearly FXRTs~8 and 9 remain $\sim$3 dex fainter, but otherwise have potentially consistent spectral and temporal properties. As neither has multiple \hbox{X-ray} observations, we cannot say anything about their long-term evolution. We can also compare the timing and spectral properties of the off-nuclear ultrasoft hyper-luminous 3XMM\,J215022.4-055108, an IMBH TDE candidate \citep[hereafter TDE\,J2150-05;][]{Lin2018,Lin2020}. TDE\,J2150-05 shows a peak luminosity of ${\approx}$1$\times$10$^{43}$~erg~s$^{-1}$, a light curve PL decay of $F_X{\propto}t^{-5/3}$ during ${\gtrsim}$14~yr (see Fig.\ref{fig:flux_comparison}), and ultrasoft \hbox{X-ray} spectra with $kT{\lesssim}$0.25~keV, which soften with time \citep{Lin2018,Lin2020}. This lies in stark contrast with FXRTs~8 and 9, which show a short and fast timescale variability, and somewhat hotter/harder \hbox{X-ray} spectra. In summary, FXRTs~8 and 9 do not conform to the "traditional" expectations of TDEs, in terms of slow temporal evolution or ultrasoft \hbox{X-ray} spectra, but relativistically beamed emission from an IMBH-TDE scenario cannot be discarded.

Unlike the other events, FXRT~14 has been constrained by multiwavelength counterparts \citep{Bauer2017}. The available data are consistent with expectations for off-axis SGRBs, although other possibilities might not be ruled out. For instance, \citet{Peng2019} argue for an IMBH--WD TDE, \citet{Sun2019} explain the \hbox{X-ray} emission considering a magnetar remnant after a BNS merger observer at an off-axis viewing angle, while \citet{Sarin2021} discuss an association with an off-axis afterglow of a BNS merger, without discarding that its \hbox{X-ray} properties could be related to compact object such as an asteroid hitting an isolated foreground neutron star \citep{Colgate1981,van_Buren1981,Campana2011}. It is important to mention that FXRT\,14/CDF-S XT1 and XT2 seem to fall in the same host's properties parameter space as the LL-LGRBs and SGRBs at lower stellar masses ($\lesssim$10$^9$~$M_\odot$; see Fig.~\ref{fig:Mass_SFR_plot}). This reinforces the likely association with SGRBs.

\subsubsection{FXRTs with unknown distances} \label{sec:dist_unkn_FXRTs}

FXRTs~7, 10, 12, and 13 do not have clear host associations as yet, and hence have wildly uncertain distances. Based on their typical optical and NIR upper limits (e.g., $m_r{\gtrsim}$23.3 and $m_z{\gtrsim}$22~AB~mag), and considering distances of other FXRT host galaxies such as FXRTs~8 and 9 ($z_{\rm photo/spec}{\sim}0.7$), FXRT~14 \citep[$m_R^{\rm FXRT~14}{=}$27.5~AB~mag and $z_{\rm photo}^{\rm FXRT~14}{=}$0.39--3.21;][]{Bauer2017}, and CDF-S~XT2 \citep[$m_{\rm F606W}^{\rm XT2}{=}$25.35~AB~mag and $z_{\rm spec}^{\rm XT2}{=}0.738$;][]{Xue2019}, we adopt a nominal redshift of $z{=}$1 for these sources. Figure~\ref{fig:flux_comparison} (\emph{open markers}) compares FXRTs~7, 10, 12 and 13 (at $z{=}$1.0) to several classes of transients.

We note that FXRTs~7, 10, and 12 have light curves that exhibit plateau phases of ${\approx}$1--3~ks, followed by PL decays ($F_X{\propto}t^{-2.4/-1.6}$) that are accompanied by possible softening of the spectra for FXRTs~7 and 12 (see Table~\ref{tab:fitting_para_Tbreak}). Spectral softening has been seen previously in SBOs (e.g., XRF\,080109/SN\,2008D), GRBs afterglows, TDEs \citep[e.g., ][]{MacLeod2014,Malyali2019}, and CDF-S~XT2 \citep{Xue2019}. FXRTs~7, 10 and 12 have photon indices (see Table~\ref{tab:spectral_para}) similar to the SBO XRF\,080109/SN\,2008D \citep[$\Gamma{\approx}2.3$;][]{Soderberg2008} and GRB afterglows \citep[$\Gamma{\approx}$1.5--3.0;][]{Berger2014,Wang2015} at a 1$\sigma$ confidence level. If these events lie at $z{\gtrsim}$0.5, we can discard the SBO scenario, however, due to their high \hbox{X-ray} luminosities ($L_{\rm X,peak}{\gtrsim}$10$^{44}$~erg~s$^{-1}$); an SBO association would only be expected at low redshift ($z{\lesssim}$0.5). The light curves (at $z{=}1.0$) also appear inconsistent with on-axis GRBs. Although they share similar luminosities and PL decays beyond ${\sim}10^{3}$~s, the early plateau phases of FXRTs are inconsistent with the typical PL or BPL decays of on-axis GRBs and afterglows. A subset of SGRBs exhibit plateau phases \citep{Rowlinson2010,Rowlinson2013}, although these generally have plateau luminosities ${\gtrsim}$10$^{46}$~erg~s$^{-1}$ (although if no redshift is known the mean SGRB redshift is assumed, $z{\sim}$0.72; energy band 0.3--10~keV), which are inconsistent with FXRTs~10 and 12 lying at $z{\lesssim}2.1$. An off-axis GRB afterglow scenario seems unlikely. To observe luminosities similar to SN\,2020bvc ($L_{\rm X,max}{\approx}$1.8${\times}$10$^{41}$~erg~s$^{-1}$) and GRB~170817 ($L_{\rm X,max}{\approx}$4${\times}$10$^{39}$~erg~s$^{-1}$), our sources must be at low redshifts, $z{\lesssim}$0.1, which could be discarded by the non-detection of hosts. Furthermore, Fig.~\ref{fig:flux_comparison} shows a comparison of these FXRTs with the ultra-long GRB~111209A. Assuming $z{=}1.0$, at early times their luminosities are orders of magnitude lower than GRB~111209A.

On the other hand, the luminosities and light curve shapes of FXRTs~7, 10, and 12 share remarkable similarities to \hbox{X-ray} flashes XRF\,060218/SN\,2006aj and XRF\,100316D/SN\,2010bh \citep[which may be related to shock breakout from choked GRB jets;][]{Campana2006,Bromberg2012,Nakar2012}, as well as CDF-S~XT2 \citep[which is consistent with being powered by a millisecond magnetar;][]{Xue2019,Sun2019}. The light curves of FXRTs~7, 10, and 12 follow the expected shape for IMBH--WD TDEs \citep[e.g., see][]{MacLeod2014,Malyali2019}. For instance, the photon index and flux PL decay of these FXRTs are similar to the IMBH TDE candidate TDE\,J2150-05 \citep[$\Gamma{\lesssim}$4.8 and $F_X{\propto}t^{-5/3}$;][]{Lin2018}. Assuming $z{=}$1, only FXRT~7 reaches a luminosity close to the beamed TDE \emph{Swift}~J1644+57 \citep[$L_{\rm X,peak}{\approx}$10$^{46}$--10$^{47}$~erg~s$^{-1}$; see Fig.~\ref{fig:flux_comparison};][]{Bloom2011,Levan2011}, but without flaring episodes.
Again, the poor count statistics (to constrain any spectral evolution) and the lack of host or additional EM counterparts do not permit us to analyze this picture in detail.

FXRT~13 exhibits a single PL light curve with a slow decay ($F_X{\propto}t^{-0.2}$). This seems to exclude a SBO nature for this FXRT. There is a faint optical source likely associated with this FXRT, only visible in $i$-band DECam images ($m_i{\approx}$24.7~AB~mag), which does not constrain its origin significantly.

Finally, assuming FXRTs 1 and 11 are actually background objects that randomly overlap with nearby sources, we find that their light curves remain unique. Given the uncertainties in their distances, we adopt nominal redshifts of $z{=}$1 as above (see Fig.~\ref{fig:flux_comparison}). Their \hbox{X-ray} luminosities of reach values $L_{\rm X,peak}{\approx}$10$^{47}$ and 5${\times}$10$^{47}$~erg~s$^{-1}$, respectively, ruling out an association with SBOs but falling in the range of XRFs \citep[e.g., XRF\,060218/SN\,2006aj and XRF\,100316D/SN\,2010bh;][]{Campana2006,Bromberg2012,Nakar2012} and beamed TDEs (e.g., TDE\,J1644+57; see Fig.~\ref{fig:flux_comparison}). The duration and shapes do not appear consistent with XRFs, but do resemble individual flares seen from TDE\,J1644+57.

Unfortunately, the unknown distances of these FXRTs do not permit better constraints on their origin. 

\begin{table}
    \centering
    \scalebox{0.75}{
    \begin{tabular}{lllll}
    \hline\hline
    FXRT & ID & $F_{\rm peak}$(erg~cm$^{-2}$~s$^{-1}$) & $L_{\rm X,peak}$(erg~s$^{-1}$) & $M_{\rm Edd}$($M_\odot$)  \\ \hline
    (1) & (2) & (3) & (4) & (5) \\ \hline
    \multicolumn{5}{c}{Nearby sample} \\ \hline
    1 & XRT\,000519$^{\dagger\dagger}$ & (1.9$\pm$0.1)${\times}$10$^{-10}$ & (6.1${\pm}$0.3)${\times}$10$^{42}$ & (4.9${\pm}$0.3)${\times}$10$^4$ \\
    2 & XRT\,010908$^{\dagger\dagger}$ & (1.7$\pm$0.5)${\times}$10$^{-13}$ & (1.8$\pm$0.5)${\times}$10$^{39}$ & 13.9$\pm$4.1 \\ 
    3 & XRT\,070530$^{\dagger\dagger}$ & (2.7$\pm$1.1)${\times}$10$^{-12}$ & (5.3$\pm$2.1)${\times}$10$^{39}$ & 41.8$\pm$17.0 \\ 
    4 & XRT\,071203$^{\dagger\dagger}$ & (6.4$\pm$2.5)${\times}$10$^{-14}$ & (3.6$\pm$1.4)${\times}$10$^{38}$ & 2.9$\pm$1.1 \\
    5 & XRT\,080331$^{\dagger\dagger}$ & (1.7$\pm$0.3)${\times}$10$^{-12}$ & (1.3$\pm$0.2)${\times}$10$^{40}$ & 105.9$\pm$18.7 \\
    6 & XRT\,130822$^{\dagger\dagger}$ & (2.3$\pm$0.9)${\times}$10$^{-13}$ & (2.4$\pm$0.9)${\times}$10$^{40}$ & 187.5$\pm$73.4 \\ \hline 
    \multicolumn{5}{c}{Distant sample} \\ \hline
    7 & XRT\,030511$^\dagger$ & (2.3$\pm$0.3)${\times}$10$^{-12}$ & (1.3$\pm$0.2)${\times}$10$^{46}$ & (9.9${\pm}$1.3)${\times}$10$^7$ \\
    8 & XRT\,041230$^{\dagger\dagger}$ & (6.9$\pm$3.4)${\times}$10$^{-14}$ & (1.1$\pm$0.5)${\times}$10$^{44}$ & (8.8$\pm$4.3)${\times}$10$^{5}$ \\
    9 & XRT\,080819$^{\dagger\dagger}$ & (6.5$\pm$2.9)${\times}$10$^{-13}$ & (1.5$\pm$0.7)${\times}$10$^{45}$ & (1.2$\pm$0.5)${\times}$10$^7$ \\
    10 & XRT\,100831$^\dagger$ & (8.9$\pm$3.4)${\times}$10$^{-13}$ & (4.8$\pm$1.8)${\times}$10$^{45}$ & (3.8${\pm}$1.5)${\times}$10$^7$ \\
    11 & XRT\,110103$^a$ & (2.2$\pm$0.2)${\times}$10$^{-10}$ & (2.4$\pm$0.2)${\times}$10$^{44}$ & (1.9$\pm$0.2)${\times}$10$^{6}$ \\
    12 & XRT\,110919$^\dagger$ & (5.6$\pm$1.3)${\times}$10$^{-13}$ & (3.0$\pm$0.7)${\times}$10$^{45}$ & (2.4${\pm}$0.6)${\times}$10$^7$ \\
    13 & XRT\,140327$^\dagger$ & (1.2$\pm$0.5)${\times}$10$^{-13}$ & (6.3$\pm$2.8)${\times}$10$^{44}$ & (4.9${\pm}$2.3)${\times}$10$^6$ \\
    14 & XRT\,141001$^b$ & (4.3$\pm$1.1)${\times}$10$^{-12}$ & (1.7$\pm$0.4)${\times}$10$^{47}$ & (1.3$\pm$0.3)${\times}$10$^9$ \\
    \hline
    \end{tabular}
    }
    \caption{Energetics of the FXRT sample (fluxes are corrected for Galactic and intrinsic absorption, and calculated over the energy range 0.3--10~keV). 
    \emph{Column 1:} FXRT candidate number.
    \emph{Column 2:} Candidate ID. 
    \emph{Column 3 and 4:} X-ray peak flux and isotropic luminosity in cgs units (corrected for Galactic and intrinsic absorption). 
    \emph{Column 5:} Eddington mass (defined as $M_{\rm Edd}{=}7.7{\times}10^{-39} L_{\rm X,peak}$) in solar mass units ($M_\odot$).\\
    $^a$ Assuming an association with Abell~3581 at 94.9~Mpc \citep{Glennie2015}.\\
    $^b$ Assuming a mean redshift of $z{=}2.23$ \citep{Bauer2017}.\\
    $^\dagger$ Assuming a redshift of $z{=}1$.\\
    $^{\dagger\dagger}$ The distance or redshift is taken from Table~\ref{tab:SED_para}.}
    \label{tab:F_L_para}
\end{table}

\section{Rates}\label{sec:rates}

We computed the event rates of FXRTs and compared them with those for other transients to explore possible associations and interpretations. We derived the event rates (deg$^{-2}$~yr$^{-1}$; Sect. \ref{sec:event_rate}), the volumetric rate for {nearby} and {distant} samples (yr$^{-1}$ Gpc$^{-3}$; Sect. \ref{sec:vol_rate}), the local density rate (Sect. \ref{sec:vol_local}), and the expected number of events for current and future \hbox{X-ray} missions (Sect. \ref{sec:event_future}).

\subsection{Event-rate estimation}\label{sec:event_rate}

We found 14 FXRTs \citep[including XRT\,000519, XRT\,110103 and CDF-S~XT1;][]{Jonker2013,Glennie2015,Bauer2017} within 160.96~Ms of CSC2 data. For a set of \emph{Chandra} observations, the number of transients can be written as
\begin{equation}
    \mathcal{N}=\sum_i \mathcal{R}_i\varepsilon_i\Omega_i t_i,
\end{equation}
where $\mathcal{R}_i$ is the event rate, $\Omega_i$ and $t_i$ are the FoV and exposure time, respectively, and $\varepsilon_i$ is an area correction factor, with the subscript $i$ denoting each \emph{Chandra} observation.

The area correction factor, $\varepsilon_i$, is important for the faintest FXRTs and captures the changes in sensitivity over the \emph{Chandra} detector. $\varepsilon_i$ is defined as the area within which we expect successful FXRT detections (S/N ratio$\gtrsim$3.0) normalized by the total detection area. To determine $\varepsilon_i$, we simulate 1,000 fake instances of each FXRT, randomly distributed in position (using \texttt{MARX} and \texttt{simulate\_psf} scripts taking into account the particular features per \emph{Chandra} observation) within \emph{Chandra}'s FoV for each individual observation. We compute the S/N ratio for fake FXRTs in the energy range of 0.3--10~keV.
Thus, $\varepsilon_i$ falls in the range $\varepsilon_i{\in}[0.0,1.0]$. For the brightest FXRTs, $\varepsilon_i{\approx}$1.0, meaning that they are detectable across the entire detector FoV, while for fainter FXRTs, $\varepsilon_i{\lesssim}$1.0, such that only a portion of the detector is sensitive to them.

We assume that $\mathcal{R}_i$ is constant (such that $\mathcal{R}_i{=}\mathcal{R}$), because the universe is isotropic on large scales and we are focusing on extragalactic sources \citep{Yang2019}. $\Omega_i$ depends on which chips of the detector are turned on; due to the degradation of the PSF at higher instrumental  off-axis angles, we consider only chips I0--I3 for ACIS-I and chips S1--S4 for ACIS-S, respectively. Therefore, the expected number of events depends on $\Omega_i$, $t_i$, and $\varepsilon_i$ per observation as
\begin{equation}
    \mathcal{N}=\mathcal{R}\sum_i\varepsilon_i\Omega_it_i,
\end{equation}
such that the event rate, $\mathcal{R}$, is
\begin{equation}
    \mathcal{R}=\frac{\mathcal{N}}{\sum_i\varepsilon_i\Omega_it_i}.
    \label{eq:final}
\end{equation}

We derive the rate of our sample considering two cases: 
$(i)$ five nearby events (seven if we include FXRT~1/XRT~000519 and FXRT~11/XRT~110103, which have unclear associations with M86 and the galaxy cluster Abell~3581, respectively; called \emph{Case I}), and $(ii)$ seven distant events (nine if we include FXRT~1/XRT~000519 and FXRT~11/XRT~110103; called \emph{Case II}). 
Because our algorithm does not have good efficiency in detecting objects in observations with exposure times ${<}$8~ks (in fact, we do not detect any candidates for such exposures), we do not consider such observations to derive the rates. Another consideration when estimating the event rates for both FXRT samples is to identify and isolate the fraction of observations that target nearby galaxies. While {distant} FXRTs can be detected in any \emph{Chandra} observation (i.e., in the background of nearby galaxy observations), {nearby} FXRTs can only be detected if nearby galaxies lie within the \emph{Chandra} FoV. Thus for \emph{Case II}, we consider just \emph{Chandra} observations that target non-nearby galaxies, while for \emph{Case I}, we only consider the fraction of \emph{Chandra} observations that target nearby galaxies at ${<}$100~Mpc (${\approx}$21\% of the total sample; see Sect. \ref{sec:population}).

Therefore, we estimate the event-rates (fully accounting for the ambiguity of FXRTs~1 and 11 in the errors) of {nearby} FXRTs to be $\mathcal{R}_{\rm Case~I}{=}$53.7$_{-15.1}^{+22.6}$~deg$^{-2}$~yr$^{-1}$; while for {distant} FXRTs it is $\mathcal{R}_{\rm Case~II}{=}$28.2$_{-6.9}^{+9.8}$~deg$^{-2}$~yr$^{-1}$ (for $F_{\rm X,peak}{\gtrsim}$1$\times$10$^{-13}$~erg~cm$^{-2}$~s$^{-1}$). The distant rate is consistent with the rate of $\mathcal{R}_{\rm Yang+19}{\approx}$59$_{-38}^{+77}$~deg$^{-2}$~yr$^{-1}$ at the Poisson 1$\sigma$ confidence level, as derived by \citet{Yang2019}, but is $\approx$0.9~dex higher than the rate of ${\approx}$3.4~deg$^{-2}$~yr$^{-1}$ derived by \citet{Glennie2015}. The latter discrepancy is not surprising, however, since \citet{Glennie2015} calculated the rate for a much higher peak flux of $F_{\rm X,peak}{\gtrsim}$10$^{-10}$~erg~cm$^{-2}$~s$^{-1}$.

It is essential to mention again that FXRTs previously discovered as CDF-S~XT2 \citep[XRT~150321;][]{Xue2019}, XRT~170831 \citep{Lin2019} and XRT~210423 \citep{Lin2021} are not part of this work because of the date cut-off of CSC2. As we showed in Sect. \ref{sec:completness}, the number of FXRTs that is removed from our sample by our selection criteria erroneously is probably less than 1. Therefore, the estimated event rates are robust results for FXRT candidates brighter than $\log(F_{\rm peak}){\gtrsim}-12.6$ for \emph{Chandra} observations with $T_{\rm exp}{>}$8~ks.

\begin{figure}
    \centering
    \includegraphics[scale=0.65]{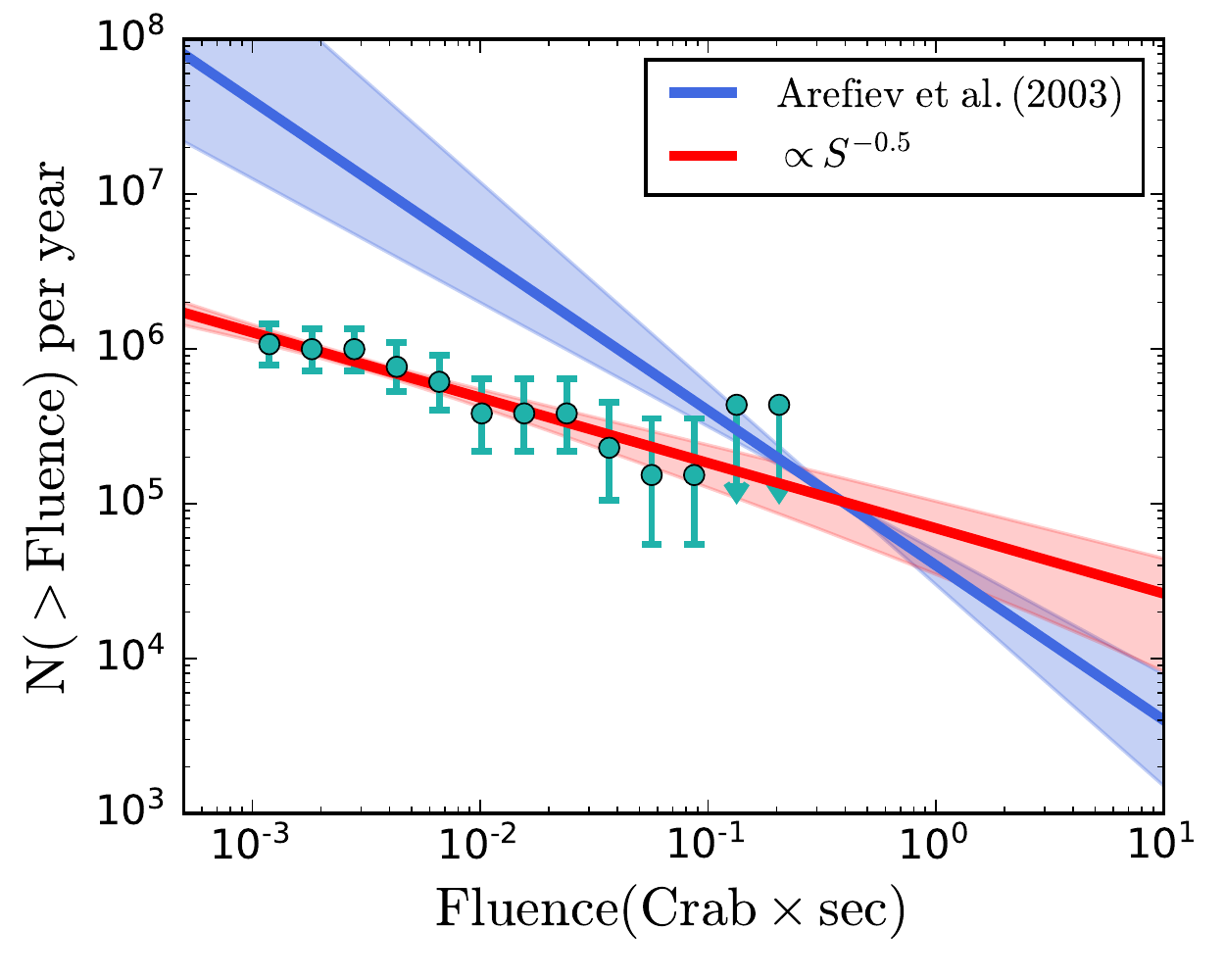}
    \vspace{-0.2cm}
    \caption{Observed cumulative log$\mathcal{N}$--log$S$ distribution of our sample of FXRTs as a function of fluence (in units of Crab$^\dagger$ $\times$ second). Also shown are two PL models, $N({>}S){\propto}S^{-\gamma}$, with slopes $\gamma{=}$0.5 (\emph{red line}) and 1.0 (\emph{blue line}). The $\gamma{=}0.5$ line denotes the best fit to the CSC2 sample. The $\gamma{=}1$ line represents the best fit and 1$\sigma$ error of \citet{Arefiev2003} based on bright FXRTs. The brightest sources in our sample appear to be consistent with this bright-end extrapolation, although our fainter sources fall $\sim$1 dex below, implying a break. For comparison with \citet{Arefiev2003}, we convert the fluence to 2--10~keV.\\
    $\dagger$ A Crab is a standard astrophotometric unit for measurement of the intensity of celestial X-ray sources.}
    \label{fig:fluence_plot}
\end{figure}

The event rate (event rate per dex of flux) behaves as a PL function as $\mathcal{R}{\propto}F_{\rm lim}^{-\gamma}$, where $\gamma$ is a positive value. In Fig.~\ref{fig:fluence_plot}, we plot the observed cumulative log$\mathcal{N}$--log$S$ distribution for our entire sample, which appears to follow $\gamma{\approx}0.5$ (\emph{red line}). We also plot the extrapolation of the best-fit slope, $\gamma{=}1.0$, based on the estimates of FXRTs at bright fluxes (${\gtrsim}10^{-10}$~erg~cm$^{-2}$~s$^{-1}$) from \citet{Arefiev2003}. We caution that \citet{Arefiev2003} do not specify an exact energy band and make no distinction between various potential Galactic and extragalactic classes, although it is noteworthy that the sky distribution at these bright fluxes is also isotropic. We see that the brightest sources in our CSC2 sample are consistent with this extrapolation, while the fainter sources fall well below it by ${\sim}$1~dex, implying a potential break around a fluence of $3\times10^{-8}$~erg~cm$^{-2}$ to our best-fit slope. 

For comparison, a spatially homogenous distribution of identical (standard candle) sources would yield a Euclidean slope of 1.5. Based on this, we adopt $\gamma{=}1.0$ when extrapolating to brighter fluxes, and $\gamma{=}0.5$ to fainter fluxes.

\subsection{Volumetric rate estimate} \label{sec:vol_rate}

In addition to the event rate on the sky (deg$^{-2}$), we compute the volumetric density rate $\rho(z)$, in units of yr$^{-1}$~Gpc$^{-3}$, to compare with other known transient classes (GRBs, SBOs, or TDEs).
Following \citet{Zhang_book_2018}, the number of FXRTs, $\mathcal{N}$, identified per unit (observing) time, $dt$, per unit redshift bin, $dz$, can be written as 
\begin{equation}
    \frac{d\mathcal{N}}{dtdz}=\frac{\rho(z)}{1+z}\frac{dV(z)}{dz},
\end{equation}
where $dV(z)/dz$ is the derivative of the volume with regards to $z$. Integrating the previous equation by $dt$ and $dz$, we can estimate the density rate at a particular redshift $z$ as
\begin{equation}
    \rho(z)=\frac{4\pi \mathcal{N}(1+z)}{\Omega TV_{\rm c,max}},
    \label{eq:015}
\end{equation}
where $V_{\rm c,max}$ is the maximum co-moving volume (at the maximum co-moving distance $D_{\rm c,max}$), while $\Omega$ and $T$ are the FoV and the exposure time used in this work (corrected by $\varepsilon_i$; see Sect. \ref{sec:event_rate}), respectively. 

For {Case~I} (between five and seven local FXRTs), the density rate at ${\lesssim}$100~Mpc is $\rho_{\rm Case~I}{=}$(5.9$_{-2.6}^{+2.5}$)${\times}$10$^{-2}$~yr$^{-1}$~Mpc$^{-3}$, at a 1$\sigma$ confidence level. Due to the small distance of these FXRTs, we can approximate this result at $z{\approx}$0, also called the {local density rate} (denoting as $\rho_0$), that is, $\rho_{\rm Case~I}{\approx}\rho_{0,\rm Case~I}$. This value is consistent with previously derived rates for ULXs, taking ULX M82 as an example \citep[1.75$\times$10$^{-2}$~yr$^{-1}$~Mpc$^{-3}$;][]{Kaaret2006,Swartz2011,Pradhan2020}.

\begin{figure*}
    \centering
    \includegraphics[scale=0.67]{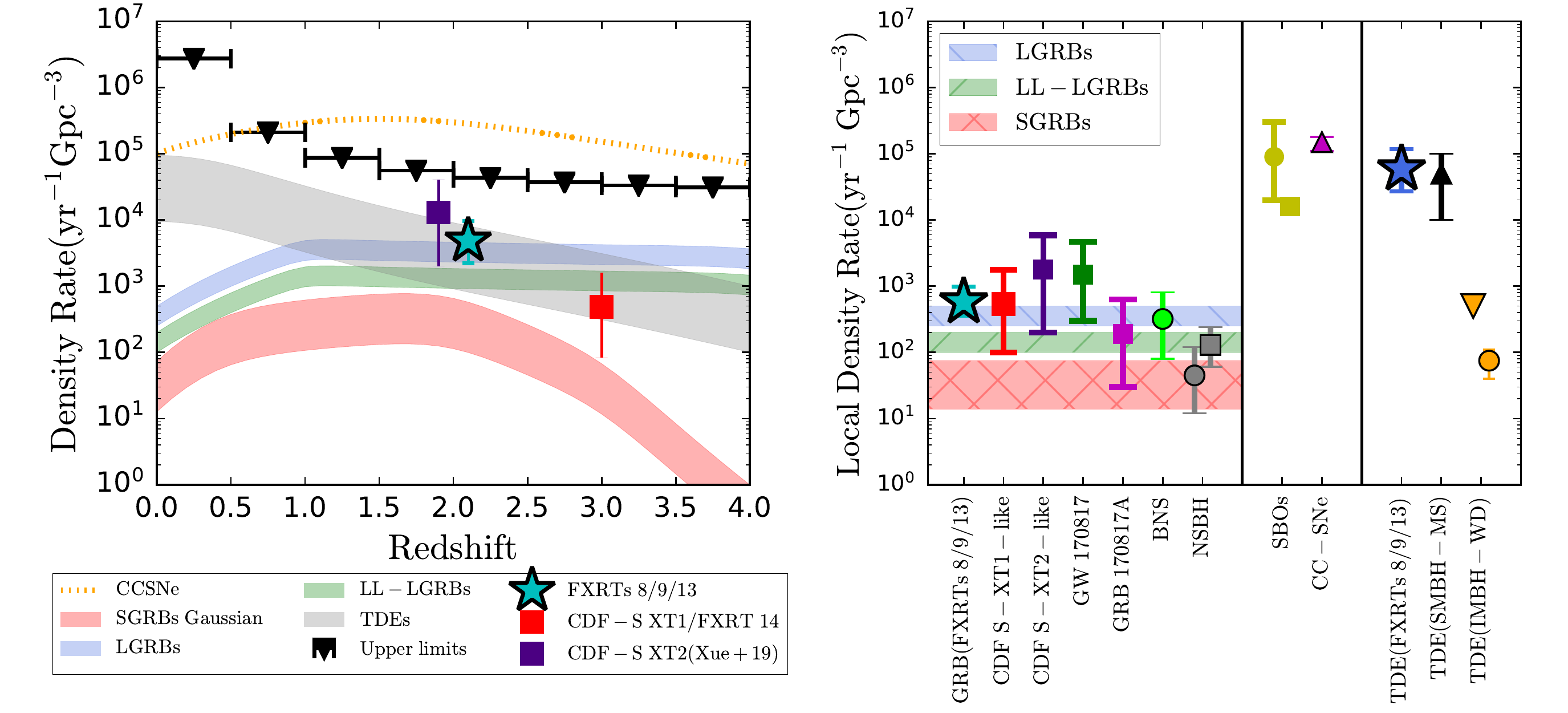}
    \vspace{-0.2cm}
    \caption{Density rate as a function of redshift for several known transient classes compared to our sample. \emph{Left panel:} Density rate as a function of redshift for FXRT~8/9/13 (\emph{cyan star}) and CDF-S~XT1/FXRT~14 (\emph{red square}), and upper limits (\emph{black triangle}) for FXRTs without measured redshifts. 
    We also show the density rate of CDF-S~XT2 \citep[\emph{purple square};][]{Xue2019}, CC-SNe \citep[\emph{dotted orange line};][]{Madau2014},
    the redshift-dependent intrinsic event rate densities of LGRBs (\emph{blue-filled region}; taken from \citealt{Sun2015} and \citealt{Wanderman2010}, normalized to the local universe value and corrected for jet-aperture as $\rho_{\rm 0,LGRBs}{\sim}$250--500~yr$^{-1}$~Gpc$^{-3}$), LL-LGRBs (\emph{green-filled region}; taken from \citealt{Zhang_book_2018}, normalized to the local universe value and corrected for jet-aperture as $\rho_{\rm 0,LGRBs}{\sim}$100--200~yr$^{-1}$~Gpc$^{-3}$), SGRBs considering a merger delay Gaussian model (\emph{red-filled region}; taken from \citealt{Sun2015} and \citealt{Wanderman2015}, normalized to the local universe value and corrected for jet-aperture as $\rho_{\rm 0,SGRBs}{\sim}$13--75~yr$^{-1}$~Gpc$^{-3}$), and TDEs (\emph{gray-filled region}, taken from \citealt{Sun2015}, normalized to the local universe as $\rho_{\rm 0,TDEs}{\sim}$10$^4$--10$^5$~yr$^{-1}$~Gpc$^{-3}$ at luminosities ${\sim}$10$^{42}$--10$^{44}$~erg~s$^{-1}$).
    \emph{Right panel:} Local density rate for FXRT~8/9/13 considering they are related to GRBs (\emph{cyan star}), TDEs (\emph{blue star}), or FXRT~14/CDF-S~XT1 (\emph{red square}) (see Sect. \ref{sec:vol_local}). As a comparison, we also plot the local event rate of CDF-S~XT2-like events \citep[\emph{purple square};][]{Xue2019}, GW\,170817 \citep[\emph{green square};][]{Abbott2017b}, and GRB\,170817A \citep[\emph{magenta square};][]{Zhang2018}, as well other kinds of transients, such as the new SBO candidate in \emph{XMM-Newton data} \citep[\emph{yellow square and circle};][]{Xu2008,Novara2020}, the TDE rate of SMBH--MS TDEs \citep[\emph{black triangle};][]{Sun2015} and the IMBH--WD TDE rate \citep[\emph{orange triangle and circle};][]{Malyali2019,Tanikawa2021}, the CC-SN rate (\emph{magenta triangle}), the merger rate of BNS systems \citep[\emph{light green circle};][]{Abbott2021a}, and the merger rate of neutron star and BH systems \citep[\emph{gray circles};][]{Abbott2021b}. The local event rate of LGRBs, LL-LGRBs, and SGRBs are plotted \citep[\emph{blue, green, and red horizontal shaded regions,} corrected for the jet aperture factor;][]{Zhang_book_2018}.}
    \label{fig:rate_plot}
\end{figure*}

For {Case~II} (distant FXRTs), redshift and cosmological effects become important. Currently, we only have photometric redshifts for FXRT~8 and FXRT~9 ($z_{\rm phot}{\approx}$0.7), and suspect that FXRT~13 must have a broadly similar redshift range ($z{\approx}$0.2--1.1 and $\overline{z}{\approx}$0.7; see Sect. \ref{sec:counterpart_SED}). Thus we only compute the cosmological rate for these FXRTs. Using Eq.~\ref{eq:015} and assuming that FXRTs~8, 9, and 13 occurred at $z{\approx}$0.7, the density rate of these three FXRTs is $\rho_{\rm FXRTs~8/9/13}{=}$(4.8$_{-2.6}^{+4.7}$)${\times}$10$^{3}$~yr$^{-1}$~Gpc$^{-3}$ at a maximum redshift of $z_{\rm max}{\approx}$2.1 (assuming a mean value of $F_{\rm X,peak}^{\rm lim}{\approx}$1$\times$10$^{-13}$~erg~cm$^{-2}$~s$^{-1}$ as the threshold limit detection and an isotropic luminosity $L_{\rm X,peak}^{\rm max}{\approx}$3$\times$10$^{45}$~erg~s$^{-1}$). In a similar way, we compute the density rate for CDF-S~XT1 at $z_{\rm max}{\approx}$3 of $\rho_{\rm CDF-S~XT1}{=}$(4.8$_{-4.0}^{+11.1}$)${\times}$10$^{2}$~yr$^{-1}$~Gpc$^{-3}$. In Fig.~\ref{fig:rate_plot}, \emph{left panel}, we compare the density rates for FXRTs~8/9/13 (\emph{cyan star}) and CDF-S~XT1 (\emph{red square}) to other transient classes. We note that these rate could increase by a factor of up to $\approx$2, considering that we do not have firm redshifts for ${\sim}$50\% of the FXRT candidates.

Considering first the density rate for FXRT~8/9/13, we find that $\rho_{\rm FXRT~8/9/13}$ is comparable to the rate of events like CDF-S~XT2, \citep[$\rho_{\rm XT2}(z_{\rm max}{=}1.9){=}$(1.3$_{-1.1}^{+2.8}$)${\times}$10$^{4}$~yr$^{-1}$~Gpc$^{-3}$; \emph{purple square}, Fig.~\ref{fig:rate_plot} \emph{left panel};][]{Xue2019}, at a similar redshift. Excluding FXRT~13 due to its uncertain redshift and computing the density rate only for FXRTs~8 and 9, the density rate drops by a factor of ${\sim}$1.6. $\rho_{\rm FXRT~8/9/13}$ remains a factor ${\gtrsim}$100 lower than the expected rate of CC-SNe \citep[\emph{dotted orange line};][]{Madau2014}, but is in good agreement with the density rate expected at $z{\sim}$2.0 for LGRBs [\emph{blue-filled region}; assuming $\rho_{\rm 0,LGRBs}{=}$250--500~yr$^{-1}$~Gpc$^{-3}$ from \citet{Wanderman2010} and \citet{Zhang_book_2018}, and the normalized rate redshift evolution from \citet{Sun2015}] and TDEs [\emph{gray-filled region}; assuming $\rho_{\rm 0,TDEs}{=}$10$^{4}$--10$^{5}$~yr$^{-1}$~Gpc$^{-3}$ from \citet{Sun2015} and luminosities $L{\approx}$10$^{42}$--10$^{44}$~erg~s$^{-1}$] reinforcing a possible association with these kinds of events. For LGRBs, we adopted a jet correction factor of ${\approx}$500 \citep{Frail2001,Zhang_book_2018}; however, other works argue for lower corrections of ${\approx}$50--100 \citep{Piran2004,Guetta2005}. The $\rho_{\rm FXRT~8/9/13}$ also overlaps with the expected density rate of LL-LGRBs [\emph{green-filled region}; with $L_{\rm min}{=}$5$\times$10$^{46}$~erg~s$^{-1}$, $\rho_{\rm 0,LL-LGRBs}{=}$100--200~yr$^{-1}$~Gpc$^{-3}$ and the normalized rate redshift evolution from \citet{Sun2015}], adopting a beaming correction of ${\approx}$1 since observations of LL-LGRBs do not show strong evidence of collimation, suggesting wider jet opening angles \citep{Virgili2009,Pescalli2015}.

On the other hand, $\rho_{\rm FXRT~8/9/13}$ is a factor of ${\approx}$1.5 higher than the estimated SGRB rate [\emph{red-filled regions}; assuming $\rho_{\rm 0,SGRBs}{=}$13--75~yr$^{-1}$~Gpc$^{-3}$ from \citet{Wanderman2015} and \citet{Zhang_book_2018}, a merger delay Gaussian model \citep{Virgili2011,Wanderman2015}, and the normalized rate redshift evolution from \citet{Sun2015}]. However, some SGRBs are collimated \citep{Burrows2006,DePasquale2010}, with typical jet aperture correction factors of ${\approx}$25 \citep{Fong2015}, although other authors claim a wider range of ${\approx}$70$\pm$40 \citep{Berger2014}. Given the large uncertainties, the rates of $\rho_{\rm FXRT~8/9/13}$ and SGRBs might remain compatible. 

We now consider the density rate for CDF-S~XT1/FXRT~14 ($\rho_{\rm CDF-S~XT1}{=}$(4.8$_{-4.0}^{+11.1}$)${\times}$10$^{2}$~yr$^{-1}$~Gpc$^{-3}$). It remains consistent with TDEs, falls on the high side of SGRBs at $z{\approx}$3, and is a factor of ${\approx}$2--5 lower than LGRBs, and falls on the low side of the LL-LGRBs. Keeping in mind the uncertainty in the jet aperture correction, it is impossible to discard the association with LGRBs. On the other hand, the rates are ${\approx}$2 order of magnitude lower than those of CC-SNe.

Finally, five potential distant FXRTs (FXRTs~1, 7, 10, 11, and 12) lack redshift constraints of any kind. To constrain their contribution to the density rate, we compute upper limits assuming that they all lie in a single redshift bin of $\Delta z{\approx}$0.5. Figure~\ref{fig:rate_plot}, \emph{left panel}, shows the resulting upper limits (\emph{black triangles}) on the rate of these FXRTs. These limits are consistent with the density rate computed for FXRTs~8, 9, and 13, CDF-S~XT1, and CDF-S~XT2 \citep{Xue2019}, but are inconsistent with CC-SNe beyond $z{\gtrsim}$0.5. Clearly, with firmer distance constraints on these objects, we will be able to pin down the density rates with higher precision.

\subsection{Local density rate}\label{sec:vol_local}

Additionally, we extrapolate the density rates of FXRTs~8, 9, and 13 and CDF-S~XT1 to the local universe (i.e., $z{\approx}$0) and compare them to other transients. The density rate of any transient evolves through redshift following \citet{Sun2015},
\begin{equation}
    \rho(z){=}\rho_0f(z),
\end{equation}
where $f(z)$ is a function that describes the density rate evolution (normalized to $z{=}$0) and $\rho_0$ is the density rate at $z{=}$0. Therefore, it is possible to determine the local density rate if $f(z)$ is known. 

We adopted $\rho_{\rm FXRT~8/9/13}{=}$(4.8$_{-2.6}^{+4.7}$)${\times}$10$^{3}$~yr$^{-1}$~Gpc$^{-3}$ at $z_{\rm max}{=}$2.1, and use the average $f(z)$ between LGRBs and SGRBs (considering the \emph{Gaussian merger delay model} because of the slight overlap with our result) taken from \citet{Sun2015}; there is not much difference between the relative evolutions of LGRBs and SGRBs. This yields a local density rate for FXRTs~8, 9, and 13 of $\rho_{\rm 0,FXRT~8/9/13}^{\rm GRBs}{=}$(5.8$_{-2.2}^{+4.0}$)${\times}$10$^{2}$~yr$^{-1}$~Gpc$^{-3}$. Meanwhile, for CDF-S~XT1, we find $\rho_{0,\rm CDF-S~XT1}^{\rm GRBs}{=}$(5.4$_{-4.4}^{+12.4}$)${\times}$10$^{2}$~yr$^{-1}$~Gpc$^{-3}$. The range of $\rho_{\rm FXRT~8/9/13}$ is also consistent with that of TDEs (see Fig.~\ref{fig:rate_plot}, \emph{left panel}), and based on the density rate evolution of TDEs from \citet{Sun2015}, this yields a local density rate for FXRTs~8/9/13 of $\rho_{\rm 0,FXRT~8/9/13}^{\rm TDEs}{=}$(5.9$_{-3.2}^{+5.8}$)${\times}$10$^{4}$~yr$^{-1}$~Gpc$^{-3}$. This value is ${\approx}$2 order of magnitude higher than the recent extended ROentgen Survey with an Imaging Telescope Array (eROSITA) average TDE volumetric rate reported by \citet{Sazonov2021} of ${\approx}$210~yr$^{-1}$~Gpc$^{-3}$ (at $z{=}$0.0--0.6).

The \emph{right panel} in Fig.~\ref{fig:rate_plot} shows the local density rate of FXRTs~8/9/13 assuming evolution as GRBs (\emph{cyan star}), TDEs (\emph{blue star}) and CDF-S~XT1-like events (\emph{red square}), as well as a comparison with CDF-S~XT2-like events \citep[\emph{purple square};][]{Xue2019} and other transients. The GRB local density rates of FXRTs~8, 9, and 13 ($\rho_{\rm 0,FXRT~8/9/13}^{\rm GRBs}$), CDF-S~XT1, and CDF-S~XT2-like events remain consistent with the values observed for most flavors of GRBs (given the large uncertainties), as well as GRB~170817A and GW~170817 \citep[\emph{green and magenta squares}, respectively;][]{Abbott2017b}, BNSs \citep[\emph{light green circle};][]{Abbott2021a}, and neutron star and BH mergers (NS--BH) \citep[\emph{gray circle};][]{Abbott2021b}. These rates are, however, 1--3 dex below those expected for SBOs \citep[\emph{yellow circle and square};][]{Madau2014,Novara2020}, CC-SNe \citep[\emph{magenta triangle};][]{Madau2014}, and SMBH-MS TDEs \citep{Sun2015}. The difference with SMBH-MS TDEs may simply be a consequence of the $f(z)$ assumption. 
Moreover, the TDE local density rate of FXRTs~8, 9, and 13 ($\rho_{\rm 0,FXRT~8/9/13}^{\rm TDEs}$) remains consistent with SMBH-MS TDE rates \citep{Sun2015} but not with IMBH-WD TDEs \citep{Tanikawa2021}, likely due to the different progenitor system.

\subsection{Expected events in current and future missions}\label{sec:event_future}

Taking the computed rates from Sect. \ref{sec:event_rate}, we examine the prospects for detecting FXRTs in other ongoing and future \hbox{X-ray} missions. The expected event rate of a {new mission} (called $\mathcal{R}_{\rm New}$) regarding our results using CSC2 is
\begin{equation}
    \mathcal{R}_{\rm New}=\left(\frac{F_{\rm New,lim}}{F_{\rm CSC2,lim}}\right)^{-\gamma}\mathcal{R}_{\rm CSC2},
    \label{eq:010}
\end{equation}
where $\mathcal{R}_{\rm New}$ and $F_{\rm New,lim}$ are the event rate and \hbox{X-ray} flux limit of the {new mission}, respectively. Then, the expected total number of events must be 
\begin{equation}
    \mathcal{N}_{\rm New}{=}\Omega_{\rm New} T_{\rm New}\mathcal{R}_{\rm New}=\Omega_{\rm New} T_{\rm New}\left(\frac{F_{\rm New,lim}}{F_{\rm CSC2,lim}}\right)^{-\gamma}\mathcal{R}_{\rm CSC2},
    \label{eq:011}
\end{equation}
where $\Omega_{\rm New}$ and $T_{\rm New}$ are the FoV and the operational time of a {new mission}, respectively. It is important to realize that Eq.~\ref{eq:011} takes into account the ratio between the {new mission} ($F_{\rm New,lim}$) and \emph{Chandra} (the limit imposed by our method $F_{\rm CSC2,lim}{=}$1.5$\times$10$^{-13}$~erg~cm$^{-2}$~s$^{-1}$) \hbox{X-ray} flux limits, respectively, which is a correction factor between both instruments. Given the low-count statistics, we quote estimates incorporating the Poisson 1$\sigma$ errors.

Current operating observatories such as \emph{XMM-Newton}, \emph{Swift}--XRT, and eROSITA have sufficient sensitivity and/or history in orbit to detect similar FXRTs to those found here. 

The European Photon Imaging Camera [EPIC; pn plus Metal Oxide Semi-conductor CCD arrays] on board the \emph{XMM-Newton} telescope have an instantaneous FoV${\approx}$0.25~deg$^2$, flux sensitivity of  ${\approx}$10$^{-14}$~erg~cm$^{-2}$~s$^{-1}$ in the energy range of 0.15--12~keV, and have an archive of roughly ${\approx}$476~Ms total exposure time during ${\sim}$20 years in orbit \citep[mean value between pn and MOS cameras;][]{Ehle2003}. Adopting a spectral slope of $\Gamma{=}1.7$, typical of FXRTs (e.g., CDF-S XT1), a correction factor to account for the contribution of background flares (assuming that 30--40\% of exposure time is affected by them) and a flux cutoff of $F_{\rm XMM,lim}{\sim}$10$^{-13}$~erg~cm$^{-2}$~s$^{-1}$ (to avoid effects from Poisson noise), we predict up to ${\approx}$68--135 {Case~I} and ${\approx}$37--68 {Case~II} FXRTs, respectively. 

Similarly, \emph{Swift}--XRT has a FoV${\approx}$0.15~deg$^2$, a flux sensitivity of ${\approx}$8${\times}$10$^{-14}$~erg~cm$^{-2}$~s$^{-1}$ in the energy band of 0.2--10~keV, and has accumulated ${\approx}$~315.4~Ms of archival data over ${\sim}$14 years operational time \citep{Hill2000,Burrows2003}. Adopting a flux limit of $F_{\rm XRT,lim}{\sim}$8${\times}$10$^{-13}$~erg~cm$^{-2}$~s$^{-1}$ (again, to avoid Poisson noise effects), the expected number of FXRTs are ${\approx}$27--55 {Case~I} and ${\approx}$15--27 {Case~II} events.

The above implies that there should be a substantial number of FXRTs hidden within the \emph{XMM-Newton} and \emph{Swift}--XRT archives and catalogs. The X-ray transient and variable sky (EXTraS) project \citep{DeLuca2016} and systematic searches such as \citet{Alp2020} have reported 136 and a dozen candidates to date, respectively, which presents a lower bound to the total numbers estimated above. Also, in the systematic search developed by the EPIC-pn \emph{XMM-Newton} Outburst Detector (EXOD) search project \citep{Pastor2020}, 2536 potential XRTs have been identified, but this large number is dominated by stellar flares, cataclysmic variables, type I X-ray bursts, supergiant FXRTs, SBOs, AGNs, and more.

Finally, the Spectrum-Roentgen-Gamma (\emph{SRG})--eROSITA mission, launched in July 2019, is scanning the entire sky in the \hbox{X-ray} band (0.2--10~keV) with a FoV${\approx}$0.833~deg$^2$ during \emph{SRG}--eROSITA's official 4-year survey phase. 
This should provide roughly equivalent coverage in sky area per time to the current \emph{XMM-Newton} archive. The \emph{SRG}--eROSITA all sky survey is expected to yield flux limits of ${\approx}$10$^{-14}$ and ${\approx}$10$^{-13}$~erg~cm$^{-2}$~s$^{-1}$ in the 0.5--2 and 2--10~keV energy bands, respectively. Avoiding Poisson noise effects as above, we adopt an \emph{SRG}--eROSITA 0.5--2~keV flux limit for FXRTs  of $F_{\rm eROSITA,lim}{\approx}$10$^{-13}$~erg~cm$^{-2}$~s$^{-1}$. Thus, during the 4-year survey, the expected number of FXRTs detected by \emph{SRG}--eROSITA (in the 0.5--2~keV band) should be ${\approx}$50--100 and 27--50~events for {Case~I }and {Case~II}, respectively.

Concerning future missions, {the Advanced Telescope for High ENergy Astrophysics (\emph{Athena})} has been selected by {European Space Agency} to characterize the hot and energetic universe, with an anticipated launch in the mid 2030s. It is projected to have an effective area of 0.25--2.0~m$^2$, energy range of 0.3--12~keV, and a nominal lifetime of five years, although consumables (such as fuel) have been rated for 10 years in the case of a mission extension \citep{Nandra2013,Barret2013}. The {Wide Field Imager (WFI)} is one of two detectors on board \emph{Athena}, with a spectral resolution of $\Delta E{<}170$~eV at 7~keV, spatial resolution of $\leq$10~arcsec PSF on-axis), and FoV of 0.44~deg$^2$ \citep{Rau2016}. To estimate the number of extragalactic FXRTs, we conservatively assume a flux threshold 10 times higher than the nominal 60~ks (longer than the expected duration of the FXRTs) flux limit due to Poisson fluctuations of $F_{\rm WFI,lim}{\approx}$10$^{-15}$~erg~cm$^{-2}$~s$^{-1}$ (where the point source detection limit is ${\approx}$10$^{-16}$~erg~cm$^{-2}$~s$^{-1}$ for the WFI deep fields). This flux limit is a factor of 100 deeper than the \emph{SRG}--eROSITA sky survey flux limit. Thus, during a ${\approx}$4~year mission, adopting $\gamma{=}$0.5 for the faint-end slope extrapolation the expected number of FXRTs detected by \emph{Athena} will ${\approx}$ 130--270 and 72--130 events for \emph{Case I} and \emph{Case II}, respectively. This sample size of bright and fainter events can be used to probe the multiwavelength properties with coordinated campaigns. Assuming that the WFI observations will be spread evenly during the mission and that those observations will also be performed during the \emph{Athena} ground contact, approximately one-sixth of the events (${\approx}$9 and 16) could have \emph{Athena} alerts with latencies ${<}$4~hours.

We also consider the \emph{Einstein Probe} (EP), which aims to monitor high-energy transient and variable phenomena in 0.5--4.0~keV band \citep{Yuan2015,Yuan2017}. The EP is scheduled for launch by the end of 2023, with a 3-year operational lifetime and 5-year goal \citep{Yuan2017}. EP will carry two scientific instruments, the Wide-field \hbox{X-ray} Telescope (WXT) with a large instantaneous FoV of 3600~deg$^{2}$ and a narrow-field Follow-up \hbox{X-ray} Telescope, as well as a fast alert downlink system \citep{Yuan2015}. To estimate the expected number of FXRTs, we consider just the WXT instrument, which has a threshold sensitivity of $F_{\rm WXT}{\approx}$5$\times$10$^{-11}$~erg~cm$^{-2}$~s$^{-1}$ at 1~ks, that is, ${\approx}$500 times higher than our flux limit and $\gamma{\approx}$1.0.

Thus, during the ${\approx}$3~year mission, the expected number of FXRTs detected by \emph{EP} should be ${\approx}$69--138 and 38--69 events for Case I and Case II, respectively.

\section{Conclusions and future work}\label{sec:conclusion}

In this work we search for extragalactic FXRTs hidden in CSC2. We have applied a modified version of the algorithm developed by \citet{Yang2019} to 214,701 \hbox{X-ray} sources identified in the CSC2 with $|b|{>}$10~deg (i.e., 5303 \emph{Chandra} observations, totaling $\approx$169.6~Ms and 592~deg$^2$). Considering additional criteria (analyzing further \hbox{X-ray} observations taken by \emph{Chandra}, \emph{XMM-Newton}, \emph{Swift}--XRT, \emph{Einstein,} and \emph{ROSAT}) and other astronomical catalogs (\emph{Gaia}, NED, SIMBAD, VHS, DES, Pan-STARRS, and others), we identify 14 FXRTs that remain consistent with an extragalactic origin.
We rediscover all (five) previously reported \emph{Chandra} events covered by CSC2: 
XRT~000519 \citep[previously identified by][]{Jonker2013},
XRT~110103 \citep[previously identified by][]{Glennie2015},
XRT~030511 and XRT~110919 \citep[previously identified by][]{Lin2019,Lin2022}, and XRT~141001/CDF-S~XT1 \citep[previously identified by][]{Bauer2017}.

Candidates have peak 0.5--7~keV fluxes between ${\approx}$1.0$\times$10$^{-13}$ and 2$\times$10$^{-10}$~erg~cm$^{-2}$~s$^{-1}$ and $T_{90}$ values from ${\approx}$4 to 40~ks. None of the FXRTs are detected in gamma rays near the time of the detection of the transient \hbox{X-ray} light. Based on multiwavelength constraints, we rule out a Galactic origin (e.g., as Galactic M or brown-dwarf stellar flares) in all but two cases (for these, 
%~10 and 12 
existing data cannot yet rule out extreme stellar \hbox{X-ray} flares). The origin of the extragalactic FXRT sample appears to be diverse: five events are robustly associated with local galaxies  (${\lesssim}$100~Mpc; called the {local sample}); seven are likely distant events (${\gtrsim}$100~Mpc; called the {distant sample}); and two events, XRT~000519 and XRT~110103, have nearby associations that remain somewhat ambiguous. 
Among the distant FXRTs, we identify hosts for four FXRTs, which span a wide range of magnitudes ($m_i{\approx}$20.6--27.0~AB~mag), while we can only place upper limits on five FXRTs. 

We have studied the spectral and timing properties of the FXRTs. 
The \hbox{X-ray} spectra can be well fitted by PLs with a median slope of $\Gamma{=}$2.5 and an overall range $\Gamma{\approx}$1.7--4.0.
Furthermore, we observe potential spectral softening for six FXRTs with time 
%(FXRTs~1, 2, 3, 7, 11, and 12) 
\citep[for XRT~000519 and XRT~110103, the softening is highly significant and occurs during the main flare;][]{Glennie2015}. In the case of timing properties, five FXRTs 
%(FXRTs~2, 3, 7, 10, and 12) 
show plateaus in their \hbox{X-ray} light curves, similar to CDF-S~XT2 \citep{Xue2019}, with durations of \hbox{${\sim}$2--10~ks} followed by PL decays with slopes ranging from ${\sim}$1.2 to 2.6. For three FXRTs we see, simultaneously with the plateau and decay, possible spectral softening (at 90\% confidence), similar to CDF-S~XT2 \citep{Xue2019}.
%FXRTs~2, 7, and 12.

The five local FXRTs 
%(FXRTs~2--6) 
have projected physical offsets between ${\approx}$0.7 and 9.4~kpc, with four being co-spatial with apparent star-forming regions or young star clusters. Adopting their host distances, these local events have peak isotropic \hbox{X-ray} luminosities of $L_{\rm X,peak}{\approx}$10$^{38}$--10$^{40}$~erg~s$^{-1}$, well below expectations for GRBs, TDEs, XRFs, and supernova SBOs. 
Such luminosities are comparable to those of ULXs and Galactic XRBs, although the durations and time variability properties of the local FXRTs are quite distinct. As such, we speculate that several may represent a new type of \hbox{X-ray} phenomenon related to massive stars.

Among the distant FXRT sample, two are associated with relatively bright optical and NIR extended sources, allowing us to derive galaxy properties using photometric archival data. The other two host associations are very faint extended sources; one is detected only in a single band, and hence lacks physical constraints, while the other is fortuitously observed by the \emph{HST} but has only weak constraints on its properties. Both bright hosts have similar redshifts ($z_{\rm phot}{\approx}$0.5--0.7) and stellar masses ($M_*{\approx}$3$\times$10$^{10}$~$M_\odot$), but starkly different SFRs (SFR${\approx}$0.5 vs. $\approx$125~$M_\odot$~yr$^{-1}$), and the faint \emph{HST} host has an uncertain redshift ($z_{\rm phot}{\approx}$0.4--3.2) and associated host properties \citep{Bauer2017}. Adopting $\bar{z}{=}$0.7 for all four events, the peak luminosities, energetics, and spectro-temporal properties robustly rule out an SBO origin but potentially remain consistent with origins as on-axis GRBs, and even off-axis GRBs in the tail of the \hbox{X-ray} afterglow, or TDEs involving an IMBH and a WD.

For the three FXRTs that lack optical and NIR host detections,
interpretations are broader. An association with SBOs remains possible at low redshifts ($z{\lesssim}$0.5), as long as potential hosts are low-mass, low-SFR dwarf galaxies. An on-axis GRB scenario remains possible 
for $z{\gtrsim}$1.0 and naturally explains the non-detection of faint host galaxies by existing optical and NIR facilities. An off-axis GRB afterglow scenario is also viable, except perhaps for very low redshifts ($z{\lesssim}$0.1), where the lack of any association with a host becomes problematic. Finally, a TDE scenario
remains possible across a broad redshift range, although the lack of a detectable host requires strong beaming, for instance, similar to \emph{Swift}~J1644+57.

Finally, we compute the event rates of local (\emph{Case~I}) and distant (\emph{Case~II}) FXRTs of $\mathcal{R}_{\rm Case~I}{=}$53.7$_{-15.1}^{+22.6}$  and $\mathcal{R}_{\rm Case~II}{=}$28.2$_{-6.9}^{+9.8}$~deg$^{-2}$~yr$^{-1}$, respectively.  Additionally, for three distant FXRTs (assuming  $\bar{z}{=}$0.7), we derive a volumetric rate (in units of yr$^{-1}$~Gpc$^{-3}$) of $\rho_{\rm FXRT~8/9/13}{=}$(4.8$_{-2.6}^{+4.7}$)${\times}$10$^3$~yr$^{-1}$~Gpc$^{-3}$ at $z_{\rm max}{=}$2.1. This value is in good agreement with the value derived by \citet{Xue2019} at a similar redshift ($z_{\rm max}{=}$1.9), as well as with other transient classes such as LGRBs, SGRBs, and TDEs. Nevertheless, this rate is ${\approx}$2 order of magnitude lower than that of CC-SNe.

Our investigation of 14 \emph{Chandra}-detected extragalactic FXRT candidates breaks new ground in terms of characterizing their diverse properties and nature, although the lack of firm distances and host properties for the distant subset clearly leaves much to speculation. The \emph{Chandra} sample provides the most accurate positions among existing \hbox{X-ray} missions, which is critical for pinpointing potential host galaxies and potential physical offsets. Given the low numbers of distant FXRTs (both found here and predicted in other archives) and the diverse range of host redshifts and properties, it will be critical to identify and follow up their associated host galaxies with dedicated spectroscopy and/or deep multiwavelength imaging in order to place extragalactic FXRTs in a proper physical and cosmological context. The contemporaneous multiwavelength nature of FXRTs remains completely unknown. Given the short duration of these events, progress here will crucially hinge upon the ability of current and future \hbox{X-ray} observatories to carry out efficient strategies for (onboard) detection and alert generation to trigger follow-up campaigns while the FXRTs are still active in \hbox{X-rays} and, presumably, at other wavelengths.  The launch of narrow- and wide-field observatories such as \emph{Athena} and \emph{EP} should provide a watershed moment for expanding samples. 

As future work, we plan to characterize this new sample of FXRTs using recent optical and NIR observations to catch their host galaxies and thus constraint their energetics. Also, we plan to extend our search to \emph{Chandra} data not considered in the CSC2 to identify new FXRTs and thus better understand their elusive nature.

\begin{acknowledgements}
We acknowledge support from: ANID grants Programa de Capital Humano Avanzado folio \#21180886 (J.Q--V), CATA-Basal AFB-170002 (J.Q--V, F.E.B.), 
FONDECYT Regular 1190818 (J.Q--V, F.E.B.), 1200495 (J.Q--V, F.E.B.)
and Millennium Science Initiative ICN12\_009 (J.Q--V, F.E.B.);
NSF grant AST-2106990 and \emph{Chandra} X-ray Center grant GO0-21080X (W.N.B.); the National Natural Science Foundation of China grant 11991053 (B.L.); support from NSFC grants 12025303 and 11890693 (Y.Q.X.); support from the George P.\ and Cynthia Woods Mitchell Institute for Fundamental Physics and Astronomy at Texas A\&M University, from the National Science Foundation through grants AST-1614668 and AST-2009442, and from the NASA/ESA/CSA James Webb Space Telescope through the Space Telescope Science Institute, which is operated by the Association of Universities for Research in Astronomy, Incorporated, under NASA contract NAS5-03127 (G.Y.). % Chandra
The scientific results reported in this article are based on observations made by the \emph{Chandra} \hbox{X-ray} Observatory. This research has made use of software provided by the \emph{Chandra} \hbox{X-ray} Center (CXC).
% NOIR Datalab
This research uses services or data provided by the Astro Data Lab at NSF's National Optical-Infrared Astronomy Research Laboratory. NOIRLab is operated by the Association of Universities for Research in Astronomy (AURA), Inc. under a cooperative agreement with the National Science Foundation.
\end{acknowledgements}

\bibliographystyle{aa}
\bibliography{AA_2022_43047}

\begin{thebibliography}{276}
\expandafter\ifx\csname natexlab\endcsname\relax\def\natexlab#1{#1}\fi

\bibitem[{Abbott {et~al.}(2017{\natexlab{a}})Abbott, Abbott, Abbott, Acernese,
  Ackley, Adams, Adams, Addesso, Adhikari, \& Adya}]{Abbott2017a}
Abbott, B.~P., Abbott, R., Abbott, T.~D., {et~al.} 2017{\natexlab{a}}, \apj,
  848, L12

\bibitem[{Abbott {et~al.}(2017{\natexlab{b}})Abbott, Abbott, Abbott, Acernese,
  Ackley, Adams, Adams, Addesso, Adhikari, Adya, Affeldt, Afrough, Agarwal, \&
  Agathos}]{Abbott2017b}
Abbott, B.~P., Abbott, R., Abbott, T.~D., {et~al.} 2017{\natexlab{b}}, \apj,
  848, L13

\bibitem[{{Abbott} {et~al.}(2021{\natexlab{a}}){Abbott}, {Abbott}, {Abraham},
  {Acernese}, {Ackley}, {Adams}, {Adams}, {Adhikari}, {Adya}, {Affeldt},
  {Agarwal}, {Agathos}, {Agatsuma}, {Aggarwal}, {Aguiar}, {Aiello}, {Ain},
  {Ajith}, {Akutsu}, {Aleman}, {Allen}, {Allocca}, {Altin}, {Amato}, {Anand},
  {Ananyeva}, {Anderson}, {Anderson}, {Ando}, {Angelova}, {Ansoldi}, {Antelis},
  {Antier}, {Appert}, {Arai}, {Arai}, {Arai}, {Araki}, {Araya}, {Araya},
  {Areeda}, {Ar{\`e}ne}, {Aritomi}, {Arnaud}, {Aronson}, {Arun}, {Asada},
  {Asali}, {Ashton}, {Aso}, {Aston}, {Astone}, {Aubin}, {Aufmuth}, {Aultoneal},
  {Austin}, {Babak}, {Badaracco}, {Bader}, {Bae}, {Bae}, {Baer}, {Bagnasco},
  {Bai}, {Baiotti}, {Baird}, {Bajpai}, {Ball}, {Ballardin}, {Ballmer}, {Bals},
  {Balsamo}, {Baltus}, {Banagiri}, {Bankar}, {Bankar}, {Barayoga}, {Barbieri},
  {Barish}, {Barker}, {Barneo}, {Barone}, {Barr}, {Barsotti}, {Barsuglia},
  {Barta}, {Bartlett}, {Barton}, {Bartos}, {Bassiri}, {Basti}, {Bawaj},
  {Bayley}, {Baylor}, {Bazzan}, {B{\'e}csy}, {Bedakihale}, {Bejger},
  {Belahcene}, {Benedetto}, {Beniwal}, {Benjamin}, {Benkel}, {Bennett},
  {Bentley}, {Benyaala}, {Bergamin}, {Berger}, {Bernuzzi}, {Berry},
  {Bersanetti}, {Bertolini}, {Betzwieser}, {Bhandare}, {Bhandari},
  {Bhattacharjee}, {Bhaumik}, {Bidler}, {Bilenko}, {Billingsley}, {Birney},
  {Birnholtz}, {Biscans}, {Bischi}, {Biscoveanu}, {Bisht}, {Biswas}, {Bitossi},
  {Bizouard}, {Blackburn}, {Blackman}, {Blair}, {Blair}, {Blair}, {Bobba},
  {Bode}, {Boer}, {Bogaert}, {Boldrini}, {Bondu}, {Bonilla}, {Bonnand},
  {Booker}, {Boom}, {Bork}, {Boschi}, {Bose}, {Bose}, {Bossilkov}, {Boudart},
  {Bouffanais}, {Bozzi}, {Bradaschia}, {Brady}, {Bramley}, {Branch},
  {Branchesi}, {Brau}, {Breschi}, {Briant}, {Briggs}, {Brillet}, {Brinkmann},
  {Brockill}, {Brooks}, {Brooks}, {Brown}, {Brunett}, {Bruno}, {Bruntz},
  {Bryant}, {Buikema}, {Bulik}, {Bulten}, {Buonanno}, {Buscicchio}, {Buskulic},
  {Byer}, {Cadonati}, {Caesar}, {Cagnoli}, {Cahillane}, {Cain}, {Calder{\'o}n
  Bustillo}, {Callaghan}, {Callister}, {Calloni}, {Camp}, {Canepa},
  {Cannavacciuolo}, {Cannon}, {Cao}, {Cao}, {Cao}, {Capocasa}, {Capote},
  {Carapella}, {Carbognani}, {Carlin}, {Carney}, {Carpinelli}, {Carullo},
  {Carver}, {Casanueva Diaz}, {Casentini}, {Castaldi}, {Caudill},
  {Cavagli{\`a}}, {Cavalier}, {Cavalieri}, {Cella}, {Cerd{\'a}-Dur{\'a}n},
  {Cesarini}, {Chaibi}, {Chakravarti}, {Champion}, {Chan}, {Chan}, {Chan},
  {Chan}, {Chandra}, {Chanial}, {Chao}, {Charlton}, {Chase},
  {Chassande-Mottin}, {Chatterjee}, {Chaturvedi}, {Chatziioannou}, {Chen},
  {Chen}, {Chen}, {Chen}, {Chen}, {Chen}, {Chen}, {Chen}, {Chen}, {Cheng},
  {Cheong}, {Cheung}, {Chia}, {Chiadini}, {Chiang}, {Chierici}, {Chincarini},
  {Chiofalo}, {Chiummo}, {Cho}, {Cho}, {Choate}, {Choudhary}, {Choudhary},
  {Christensen}, {Chu}, {Chu}, {Chu}, {Chua}, {Chung}, {Ciani}, {Ciecielag},
  {Cie{\'s}lar}, {Cifaldi}, {Ciobanu}, {Ciolfi}, {Cipriano}, {Cirone}, {Clara},
  {Clark}, {Clark}, {Clarke}, {Clearwater}, {Clesse}, {Cleva}, {Coccia},
  {Cohadon}, {Cohen}, {Cohen}, {Colleoni}, {Collette}, {Colpi}, {Compton},
  {Constancio}, {Conti}, {Cooper}, {Corban}, {Corbitt}, {Cordero-Carri{\'o}n},
  {Corezzi}, {Corley}, {Cornish}, {Corre}, {Corsi}, {Cortese}, {Costa},
  {Cotesta}, {Coughlin}, {Coughlin}, {Coulon}, {Countryman}, {Cousins},
  {Couvares}, {Covas}, {Coward}, {Cowart}, {Coyne}, {Coyne}, {Creighton},
  {Creighton}, {Criswell}, {Croquette}, {Crowder}, {Cudell}, {Cullen},
  {Cumming}, {Cummings}, {Cuoco}, {Cury{\l}o}, {Dal Canton}, {D{\'a}lya},
  {Dana}, {Daneshgaranbajastani}, {D'Angelo}, {Danilishin}, {D'Antonio},
  {Danzmann}, {Darsow-Fromm}, {Dasgupta}, {Datrier}, {Dattilo}, {Dave},
  {Davier}, {Davies}, {Davis}, {Daw}, {Dean}, {Debra}, {Deenadayalan},
  {Degallaix}, {de Laurentis}, {Del{\'e}glise}, {Del Favero}, {de Lillo}, {de
  Lillo}, {Del Pozzo}, {Demarchi}, {de Matteis}, {D'Emilio}, {Demos}, {Dent},
  {Depasse}, {de Pietri}, {De Rosa}, {de Rossi}, {Desalvo}, {de Simone},
  {Dhurandhar}, {D{\'\i}az}, {Diaz-Ortiz}, {Didio}, {Dietrich}, {di Fiore}, {di
  Fronzo}, {di Giorgio}, {di Giovanni}, {di Girolamo}, {di Lieto}, {Ding}, {di
  Pace}, {di Palma}, {di Renzo}, {Divakarla}, {Dmitriev}, {Doctor},
  {D'Onofrio}, {Donovan}, {Dooley}, {Doravari}, {Dorrington}, {Drago},
  {Driggers}, {Drori}, {Du}, {Ducoin}, {Dupej}, {Durante}, {D'Urso}, {Duverne},
  {Dwyer}, {Easter}, {Ebersold}, {Eddolls}, {Edelman}, {Edo}, {Edy}, {Effler},
  {Eguchi}, {Eichholz}, {Eikenberry}, {Eisenmann}, {Eisenstein}, {Ejlli},
  {Enomoto}, {Errico}, {Essick}, {Estell{\'e}s}, {Estevez}, {Etienne}, {Etzel},
  {Evans}, {Evans}, {Ewing}, {Fafone}, {Fair}, {Fairhurst}, {Fan}, {Farah},
  {Farinon}, {Farr}, {Farr}, {Farrow}, {Fauchon-Jones}, {Favata}, {Fays},
  {Fazio}, {Feicht}, {Fejer}, {Feng}, {Fenyvesi}, {Ferguson},
  {Fernandez-Galiana}, {Ferrante}, {Ferreira}, {Fidecaro}, {Figura}, {Fiori},
  {Fishbach}, {Fisher}, {Fittipaldi}, {Fiumara}, {Flaminio}, {Floden}, {Flynn},
  {Fong}, {Font}, {Fornal}, {Forsyth}, {Franke}, {Frasca}, {Frasconi},
  {Frederick}, {Frei}, {Freise}, {Frey}, {Fritschel}, {Frolov}, {Fronz{\'e}},
  {Fujii}, {Fujikawa}, {Fukunaga}, {Fukushima}, {Fulda}, {Fyffe}, {Gabbard},
  {Gadre}, {Gaebel}, {Gair}, {Gais}, {Galaudage}, {Gamba}, {Ganapathy},
  {Ganguly}, {Gao}, {Gaonkar}, {Garaventa}, {Garc{\'\i}a-N{\'u}{\~n}ez},
  {Garc{\'\i}a-Quir{\'o}s}, {Garufi}, {Gateley}, {Gaudio}, {Gayathri}, {Ge},
  {Gemme}, {Gennai}, {George}, {Gergely}, {Gewecke}, {Ghonge}, {Ghosh},
  {Ghosh}, {Ghosh}, {Ghosh}, {Ghosh}, {Giacomazzo}, {Giacoppo}, {Giaime},
  {Giardina}, {Gibson}, {Gier}, {Giesler}, {Giri}, {Gissi}, {Glanzer},
  {Gleckl}, {Godwin}, {Goetz}, {Goetz}, {Gohlke}, {Goncharov}, {Gonz{\'a}lez},
  {Gopakumar}, {Gosselin}, {Gouaty}, {Grace}, {Grado}, {Granata}, {Granata},
  {Grant}, {Gras}, {Grassia}, {Gray}, {Gray}, {Greco}, {Green}, {Green},
  {Gretarsson}, {Gretarsson}, {Griffith}, {Griffiths}, {Griggs}, {Grignani},
  {Grimaldi}, {Grimes}, {Grimm}, {Grote}, {Grunewald}, {Gruning}, {Guerrero},
  {Guidi}, {Guimaraes}, {Guix{\'e}}, {Gulati}, {Guo}, {Guo}, {Gupta}, {Gupta},
  {Gupta}, {Gustafson}, {Gustafson}, {Guzman}, {Ha}, {Haegel}, {Hagiwara},
  {Haino}, {Halim}, {Hall}, {Hamilton}, {Hammond}, {Han}, {Haney}, {Hanks},
  {Hanna}, {Hannam}, {Hannuksela}, {Hansen}, {Hansen}, {Hanson}, {Harder},
  {Hardwick}, {Haris}, {Harms}, {Harry}, {Harry}, {Hartwig}, {Hasegawa},
  {Haskell}, {Hasskew}, {Haster}, {Hattori}, {Haughian}, {Hayakawa}, {Hayama},
  {Hayes}, {Healy}, {Heidmann}, {Heintze}, {Heinze}, {Heinzel}, {Heitmann},
  {Hellman}, {Hello}, {Helmling-Cornell}, {Hemming}, {Hendry}, {Heng},
  {Hennes}, {Hennig}, {Hennig}, {Hernandez Vivanco}, {Heurs}, {Hild}, {Hill},
  {Himemoto}, {Hinderer}, {Hines}, {Hiranuma}, {Hirata}, {Hirose}, {Ho},
  {Hochheim}, {Hofman}, {Hohmann}, {Holgado}, {Holland}, {Hollows}, {Holmes},
  {Holt}, {Holz}, {Hong}, {Hopkins}, {Hough}, {Howell}, {Hoy}, {Hoyland},
  {Hreibi}, {Hsieh}, {Hsu}, {Huang}, {Huang}, {Huang}, {Huang}, {Huang},
  {Huang}, {H{\"u}bner}, {Huddart}, {Huerta}, {Hughey}, {Hui}, {Hui}, {Husa},
  {Huttner}, {Huxford}, {Huynh-Dinh}, {Ide}, {Idzkowski}, {Iess}, {Ikenoue},
  {Imam}, {Inayoshi}, {Inchauspe}, {Ingram}, {Inoue}, {Intini}, {Ioka}, {Isi},
  {Isleif}, {Ito}, {Itoh}, {Iyer}, {Izumi}, {Jaberianhamedan}, {Jacqmin},
  {Jadhav}, {Jadhav}, {James}, {Jan}, {Jani}, {Janssens}, {Janthalur},
  {Jaranowski}, {Jariwala}, {Jaume}, {Jenkins}, {Jeon}, {Jeunon}, {Jia},
  {Jiang}, {Jin}, {Johns}, {Jones}, {Jones}, {Jones}, {Jones}, {Jones},
  {Jonker}, {Ju}, {Jung}, {Jung}, {Junker}, {Kaihotsu}, {Kajita}, {Kakizaki},
  {Kalaghatgi}, {Kalogera}, {Kamai}, {Kamiizumi}, {Kanda}, {Kandhasamy},
  {Kang}, {Kanner}, {Kao}, {Kapadia}, {Kapasi}, {Karat}, {Karathanasis},
  {Karki}, {Kashyap}, {Kasprzack}, {Kastaun}, {Katsanevas}, {Katsavounidis},
  {Katzman}, {Kaur}, {Kawabe}, {Kawaguchi}, {Kawai}, {Kawasaki},
  {K{\'e}f{\'e}lian}, {Keitel}, {Key}, {Khadka}, {Khalili}, {Khan}, {Khan},
  {Khazanov}, {Khetan}, {Khursheed}, {Kijbunchoo}, {Kim}, {Kim}, {Kim}, {Kim},
  {Kim}, {Kim}, {Kimball}, {Kimura}, {King}, {Kinley-Hanlon}, {Kirchhoff},
  {Kissel}, {Kita}, {Kitazawa}, {Kleybolte}, {Klimenko}, {Knee}, {Knowles},
  {Knyazev}, {Koch}, {Koekoek}, {Kojima}, {Kokeyama}, {Koley}, {Kolitsidou},
  {Kolstein}, {Komori}, {Kondrashov}, {Kong}, {Kontos}, {Koper}, {Korobko},
  {Kotake}, {Kovalam}, {Kozak}, {Kozakai}, {Kozu}, {Kringel}, {Krishnendu},
  {Kr{\'o}lak}, {Kuehn}, {Kuei}, {Kumar}, {Kumar}, {Kumar}, {Kumar}, {Kume},
  {Kuns}, {Kuo}, {Kuo}, {Kuromiya}, {Kuroyanagi}, {Kusayanagi}, {Kwak},
  {Kwang}, {Laghi}, {Lalande}, {Lam}, {Lamberts}, {Landry}, {Landry}, {Lane},
  {Lang}, {Lange}, {Lantz}, {La Rosa}, {Lartaux-Vollard}, {Lasky}, {Laxen},
  {Lazzarini}, {Lazzaro}, {Leaci}, {Leavey}, {Lecoeuche}, {Lee}, {Lee}, {Lee},
  {Lee}, {Lee}, {Lee}, {Lehmann}, {Lema{\^\i}tre}, {Leon}, {Leonardi}, {Leroy},
  {Letendre}, {Levin}, {Leviton}, {Li}, {Li}, {Li}, {Li}, {Li}, {Li}, {Lin},
  {Lin}, {Lin}, {Lin}, {Lin}, {Linde}, {Linker}, {Linley}, {Littenberg}, {Liu},
  {Liu}, {Liu}, {Liu}, {Llorens-Monteagudo}, {Lo}, {Lockwood}, {Lollie},
  {London}, {Longo}, {Lopez}, {Lorenzini}, {Loriette}, {Lormand}, {Losurdo},
  {Lough}, {Lousto}, {Lovelace}, {L{\"u}ck}, {Lumaca}, {Lundgren}, {Luo},
  {Macas}, {Macinnis}, {MacLeod}, {MacMillan}, {Macquet}, {Maga{\~n}a
  Hernandez}, {Maga{\~n}a-Sandoval}, {Magazz{\`u}}, {Magee}, {Maggiore},
  {Majorana}, {Makarem}, {Maksimovic}, {Maliakal}, {Malik}, {Man}, {Mandic},
  {Mangano}, {Mango}, {Mansell}, {Manske}, {Mantovani}, {Mapelli},
  {Marchesoni}, {Marchio}, {Marion}, {Mark}, {M{\'a}rka}, {M{\'a}rka},
  {Markakis}, {Markosyan}, {Markowitz}, {Maros}, {Marquina}, {Marsat},
  {Martelli}, {Martin}, {Martin}, {Martinez}, {Martinez}, {Martinovic},
  {Martynov}, {Marx}, {Masalehdan}, {Mason}, {Massera}, {Masserot},
  {Massinger}, {Masso-Reid}, {Mastrogiovanni}, {Matas}, {Mateu-Lucena},
  {Matichard}, {Matiushechkina}, {Mavalvala}, {McCann}, {McCarthy},
  {McClelland}, {McClincy}, {McCormick}, {McCuller}, {McGhee}, {McGuire},
  {McIsaac}, {McIver}, {McManus}, {McRae}, {McWilliams}, {Meacher}, {Mehmet},
  {Mehta}, {Melatos}, {Melchor}, {Mendell}, {Menendez-Vazquez}, {Menoni},
  {Mercer}, {Mereni}, {Merfeld}, {Merilh}, {Merritt}, {Merzougui}, {Meshkov},
  {Messenger}, {Messick}, {Meyers}, {Meylahn}, {Mhaske}, {Miani}, {Miao},
  {Michaloliakos}, {Michel}, {Michimura}, {Middleton}, {Milano}, {Miller},
  {Millhouse}, {Mills}, {Milotti}, {Milovich-Goff}, {Minazzoli}, {Minenkov},
  {Mio}, {Mir}, {Mishkin}, {Mishra}, {Mishra}, {Mistry}, {Mitra}, {Mitrofanov},
  {Mitselmakher}, {Mittleman}, {Miyakawa}, {Miyamoto}, {Miyazaki}, {Miyo},
  {Miyoki}, {Mo}, {Mogushi}, {Mohapatra}, {Mohite}, {Molina}, {Molina-Ruiz},
  {Mondin}, {Montani}, {Moore}, {Moraru}, {Morawski}, {More}, {Moreno},
  {Moreno}, {Mori}, {Morisaki}, {Moriwaki}, {Mours}, {Mow-Lowry}, {Mozzon},
  {Muciaccia}, {Mukherjee}, {Mukherjee}, {Mukherjee}, {Mukherjee}, {Mukund},
  {Mullavey}, {Munch}, {Mu{\~n}iz}, {Murray}, {Musenich}, {Nadji}, {Nagano},
  {Nagano}, {Nagar}, {Nakamura}, {Nakano}, {Nakano}, {Nakashima}, {Nakayama},
  {Nardecchia}, {Narikawa}, {Naticchioni}, {Nayak}, {Nayak}, {Negishi}, {Neil},
  {Neilson}, {Nelemans}, {Nelson}, {Nery}, {Neunzert}, {Ng}, {Ng}, {Nguyen},
  {Nguyen}, {Nguyen}, {Nguyen Quynh}, {Ni}, {Nichols}, {Nishizawa}, {Nissanke},
  {Nocera}, {Noh}, {Norman}, {North}, {Nozaki}, {Nuttall}, {Oberling},
  {O'Brien}, {Obuchi}, {O'Dell}, {Ogaki}, {Oganesyan}, {Oh}, {Oh}, {Oh},
  {Ohashi}, {Ohishi}, {Ohkawa}, {Ohme}, {Ohta}, {Okada}, {Okutani}, {Okutomi},
  {Olivetto}, {Oohara}, {Ooi}, {Oram}, {O'Reilly}, {Ormiston}, {Ormsby},
  {Ortega}, {O'Shaughnessy}, {O'Shea}, {Oshino}, {Ossokine}, {Osthelder},
  {Otabe}, {Ottaway}, {Overmier}, {Pace}, {Pagano}, {Page}, {Pagliaroli},
  {Pai}, {Pai}, {Palamos}, {Palashov}, {Palomba}, {Pan}, {Panda}, {Pang},
  {Pang}, {Pankow}, {Pannarale}, {Pant}, {Paoletti}, {Paoli}, {Paolone},
  {Parisi}, {Park}, {Parker}, {Pascucci}, {Pasqualetti}, {Passaquieti},
  {Passuello}, {Patel}, {Patricelli}, {Payne}, {Pechsiri}, {Pedraza},
  {Pegoraro}, {Pele}, {Pe{\~n}a Arellano}, {Penn}, {Perego}, {Pereira},
  {Pereira}, {Perez}, {P{\'e}rigois}, {Perreca}, {Perri{\`e}s}, {Petermann},
  {Petterson}, {Pfeiffer}, {Pham}, {Phukon}, {Piccinni}, {Pichot},
  {Piendibene}, {Piergiovanni}, {Pierini}, {Pierro}, {Pillant}, {Pilo},
  {Pinard}, {Pinto}, {Piotrzkowski}, {Piotrzkowski}, {Pirello}, {Pitkin},
  {Placidi}, {Plastino}, {Pluchar}, {Poggiani}, {Polini}, {Pong}, {Ponrathnam},
  {Popolizio}, {Porter}, {Powell}, {Pracchia}, {Pradier}, {Prajapati},
  {Prasai}, {Prasanna}, {Pratten}, {Prestegard}, {Principe}, {Prodi},
  {Prokhorov}, {Prosposito}, {Prudenzi}, {Puecher}, {Punturo}, {Puosi},
  {Puppo}, {P{\"u}rrer}, {Qi}, {Quetschke}, {Quinonez}, {Quitzow-James},
  {Raab}, {Raaijmakers}, {Radkins}, {Radulesco}, {Raffai}, {Rail}, {Raja},
  {Rajan}, {Ramirez}, {Ramirez}, {Ramos-Buades}, {Rana}, {Rapagnani}, {Rapol},
  {Ratto}, {Ray}, {Raymond}, {Raza}, {Razzano}, {Read}, {Rees}, {Regimbau},
  {Rei}, {Reid}, {Reitze}, {Relton}, {Rettegno}, {Ricci}, {Richardson},
  {Richardson}, {Richardson}, {Ricker}, {Riemenschneider}, {Riles}, {Rizzo},
  {Robertson}, {Robie}, {Robinet}, {Rocchi}, {Rocha}, {Rodriguez},
  {Rodriguez-Soto}, {Rolland}, {Rollins}, {Roma}, {Romanelli}, {Romano},
  {Romel}, {Romero}, {Romero-Shaw}, {Romie}, {Rose}, {Rosi{\'n}ska},
  {Rosofsky}, {Ross}, {Rowan}, {Rowlinson}, {Roy}, {Roy}, {Rozza}, {Ruggi},
  {Ryan}, {Sachdev}, {Sadecki}, {Sadiq}, {Sago}, {Saito}, {Saito}, {Sakai},
  {Sakai}, {Sakellariadou}, {Sakuno}, {Salafia}, {Salconi}, {Saleem}, {Salemi},
  {Samajdar}, {Sanchez}, {Sanchez}, {Sanchez}, {Sanchis-Gual}, {Sanders},
  {Sanuy}, {Saravanan}, {Sarin}, {Sassolas}, {Satari}, {Sathyaprakash}, {Sato},
  {Sato}, {Sauter}, {Savage}, {Savant}, {Sawada}, {Sawant}, {Sawant}, {Sayah},
  {Schaetzl}, {Scheel}, {Scheuer}, {Schindler-Tyka}, {Schmidt}, {Schnabel},
  {Schneewind}, {Schofield}, {Sch{\"o}nbeck}, {Schulte}, {Schutz}, {Schwartz},
  {Scott}, {Scott}, {Seglar-Arroyo}, {Seidel}, {Sekiguchi}, {Sekiguchi},
  {Sellers}, {Sengupta}, {Sennett}, {Sentenac}, {Seo}, {Sequino}, {Sergeev},
  {Setyawati}, {Shaffer}, {Shahriar}, {Shams}, {Shao}, {Sharifi}, {Sharma},
  {Sharma}, {Shawhan}, {Shcheblanov}, {Shen}, {Shibagaki}, {Shikauchi},
  {Shimizu}, {Shimoda}, {Shimode}, {Shink}, {Shinkai}, {Shishido}, {Shoda},
  {Shoemaker}, {Shoemaker}, {Shukla}, {Shyamsundar}, {Sieniawska}, {Sigg},
  {Singer}, {Singh}, {Singh}, {Singha}, {Sintes}, {Sipala}, {Skliris},
  {Slagmolen}, {Slaven-Blair}, {Smetana}, {Smith}, {Smith}, {Somala}, {Somiya},
  {Son}, {Soni}, {Soni}, {Sorazu}, {Sordini}, {Sorrentino}, {Sorrentino},
  {Sotani}, {Soulard}, {Souradeep}, {Sowell}, {Spagnuolo}, {Spencer}, {Spera},
  {Srivastava}, {Srivastava}, {Staats}, {Stachie}, {Steer}, {Steinlechner},
  {Steinlechner}, {Stops}, {Stevenson}, {Stover}, {Strain}, {Strang},
  {Stratta}, {Strunk}, {Sturani}, {Stuver}, {S{\"u}dbeck}, {Sudhagar},
  {Sudhir}, {Sugimoto}, {Suh}, {Summerscales}, {Sun}, {Sun}, {Sunil}, {Sur},
  {Suresh}, {Sutton}, {Suzuki}, {Suzuki}, {Swinkels}, {Szczepa{\'n}czyk},
  {Szewczyk}, {Tacca}, {Tagoshi}, {Tait}, {Takahashi}, {Takahashi}, {Takamori},
  {Takano}, {Takeda}, {Takeda}, {Talbot}, {Tanaka}, {Tanaka}, {Tanaka},
  {Tanaka}, {Tanaka}, {Tanasijczuk}, {Tanioka}, {Tanner}, {Tao}, {Tapia},
  {Tapia San Martin}, {Tasson}, {Telada}, {Tenorio}, {Terkowski}, {Test},
  {Thirugnanasambandam}, {Thomas}, {Thomas}, {Thompson}, {Thondapu}, {Thorne},
  {Thrane}, {Tiwari}, {Tiwari}, {Tiwari}, {Toland}, {Tolley}, {Tomaru},
  {Tomigami}, {Tomura}, {Tonelli}, {Torres-Forn{\'e}}, {Torrie}, {Tosta E
  Melo}, {T{\"o}yr{\"a}}, {Trapananti}, {Travasso}, {Traylor}, {Tringali},
  {Tripathee}, {Troiano}, {Trovato}, {Trozzo}, {Trudeau}, {Tsai}, {Tsai},
  {Tsang}, {Tsang}, {Tsao}, {Tse}, {Tso}, {Tsubono}, {Tsuchida}, {Tsukada},
  {Tsuna}, {Tsutsui}, {Tsuzuki}, {Turconi}, {Tuyenbayev}, {Ubhi}, {Uchikata},
  {Uchiyama}, {Udall}, {Ueda}, {Uehara}, {Ueno}, {Ueshima}, {Ugolini},
  {Unnikrishnan}, {Uraguchi}, {Urban}, {Ushiba}, {Usman}, {Utina}, {Vahlbruch},
  {Vajente}, {Vajpeyi}, {Valdes}, {Valentini}, {Valsan}, {van Bakel}, {van
  Beuzekom}, {van den Brand}, {van den Broeck}, {Vander-Hyde}, {van der
  Schaaf}, {van Heijningen}, {Vanosky}, {van Putten}, {Vardaro}, {Vargas},
  {Varma}, {Vas{\'u}th}, {Vecchio}, {Vedovato}, {Veitch}, {Veitch},
  {Venkateswara}, {Venneberg}, {Venugopalan}, {Verkindt}, {Verma}, {Veske},
  {Vetrano}, {Vicer{\'e}}, {Viets}, {Villa-Ortega}, {Vinet}, {Vitale}, {Vo},
  {Vocca}, {von Reis}, {von Wrangel}, {Vorvick}, {Vyatchanin}, {Wade}, {Wade},
  {Wagner}, {Walet}, {Walker}, {Wallace}, {Wallace}, {Walsh}, {Wang}, {Wang},
  {Wang}, {Ward}, {Warner}, {Was}, {Washimi}, {Washington}, {Watchi}, {Weaver},
  {Wei}, {Weinert}, {Weinstein}, {Weiss}, {Weller}, {Wellmann}, {Wen},
  {We{\ss}els}, {Westhouse}, {Wette}, {Whelan}, {White}, {Whiting}, {Whittle},
  {Wilken}, {Williams}, {Williams}, {Williamson}, {Willis}, {Willke}, {Wilson},
  {Winkler}, {Wipf}, {Wlodarczyk}, {Woan}, {Woehler}, {Wofford}, {Wong}, {Wu},
  {Wu}, {Wu}, {Wu}, {Wysocki}, {Xiao}, {Xu}, {Yamada}, {Yamamoto}, {Yamamoto},
  {Yamamoto}, {Yamamoto}, {Yamashita}, {Yamazaki}, {Yang}, {Yang}, {Yang},
  {Yang}, {Yang}, {Yap}, {Yeeles}, {Yelikar}, {Ying}, {Yokogawa}, {Yokoyama},
  {Yokozawa}, {Yoon}, {Yoshioka}, {Yu}, {Yu}, {Yuzurihara}, {Zadro{\.z}ny},
  {Zanolin}, {Zappa}, {Zeidler}, {Zelenova}, {Zendri}, {Zevin}, {Zhan},
  {Zhang}, {Zhang}, {Zhang}, {Zhang}, {Zhang}, {Zhao}, {Zhao}, {Zhao}, {Zhao},
  {Zhou}, {Zhu}, {Zhu}, {Zimmerman}, {Zlochower}, {Zucker}, {Zweizig}, {Ligo
  Scientific Collaboration}, {VIRGO Collaboration}, \& {KAGRA
  Collaboration}}]{Abbott2021b}
{Abbott}, R., {Abbott}, T.~D., {Abraham}, S., {et~al.} 2021{\natexlab{a}},
  \apjl, 915, L5

\bibitem[{{Abbott} {et~al.}(2021{\natexlab{b}}){Abbott}, {Abbott}, {Abraham},
  {Acernese}, {Ackley}, {Adams}, {Adams}, {Adhikari}, {Adya}, {Affeldt},
  {Agathos}, {Agatsuma}, {Aggarwal}, {Aguiar}, {Aiello}, {Ain}, {Ajith},
  {Akcay}, {Allen}, {Allocca}, {Altin}, {Amato}, {Anand}, {Ananyeva},
  {Anderson}, {Anderson}, {Angelova}, {Ansoldi}, {Antelis}, {Antier}, {Appert},
  {Arai}, {Araya}, {Areeda}, {Ar{\`e}ne}, {Arnaud}, {Aronson}, {Arun}, {Asali},
  {Ascenzi}, {Ashton}, {Aston}, {Astone}, {Aubin}, {Aufmuth}, {AultONeal},
  {Austin}, {Avendano}, {Babak}, {Badaracco}, {Bader}, {Bae}, {Baer},
  {Bagnasco}, {Baird}, {Ball}, {Ballardin}, {Ballmer}, {Bals}, {Balsamo},
  {Baltus}, {Banagiri}, {Bankar}, {Bankar}, {Barayoga}, {Barbieri}, {Barish},
  {Barker}, {Barneo}, {Barnum}, {Barone}, {Barr}, {Barsotti}, {Barsuglia},
  {Barta}, {Bartlett}, {Bartos}, {Bassiri}, {Basti}, {Bawaj}, {Bayley},
  {Bazzan}, {Becher}, {B{\'e}csy}, {Bedakihale}, {Bejger}, {Belahcene},
  {Beniwal}, {Benjamin}, {Bennett}, {Bentley}, {Bergamin}, {Berger},
  {Bergmann}, {Bernuzzi}, {Berry}, {Bersanetti}, {Bertolini}, {Betzwieser},
  {Bhandare}, {Bhandari}, {Bhattacharjee}, {Bidler}, {Bilenko}, {Billingsley},
  {Birney}, {Birnholtz}, {Biscans}, {Bischi}, {Biscoveanu}, {Bisht}, {Bitossi},
  {Bizouard}, {Blackburn}, {Blackman}, {Blair}, {Blair}, {Blair}, {Blanch},
  {Bobba}, {Bode}, {Boer}, {Boetzel}, {Bogaert}, {Boldrini}, {Bondu},
  {Bonilla}, {Bonnand}, {Booker}, {Boom}, {Bork}, {Boschi}, {Bose},
  {Bossilkov}, {Boudart}, {Bouffanais}, {Bozzi}, {Bradaschia}, {Brady},
  {Bramley}, {Branchesi}, {Brau}, {Breschi}, {Briant}, {Briggs}, {Brighenti},
  {Brillet}, {Brinkmann}, {Brockill}, {Brooks}, {Brooks}, {Brown}, {Brunett},
  {Bruno}, {Bruntz}, {Buikema}, {Bulik}, {Bulten}, {Buonanno}, {Buscicchio},
  {Buskulic}, {Byer}, {Cabero}, {Cadonati}, {Caesar}, {Cagnoli}, {Cahillane},
  {Calder{\'o}n Bustillo}, {Callaghan}, {Callister}, {Calloni}, {Camp},
  {Canepa}, {Cannon}, {Cao}, {Cao}, {Carapella}, {Carbognani}, {Carney},
  {Carpinelli}, {Carullo}, {Carver}, {Casanueva Diaz}, {Casentini}, {Caudill},
  {Cavagli{\`a}}, {Cavalier}, {Cavalieri}, {Cella}, {Cerd{\'a}-Dur{\'a}n},
  {Cesarini}, {Chaibi}, {Chakravarti}, {Chan}, {Chan}, {Chandra}, {Chanial},
  {Chao}, {Charlton}, {Chase}, {Chassande-Mottin}, {Chatterjee},
  {Chattopadhyay}, {Chaturvedi}, {Chatziioannou}, {Chen}, {Chen}, {Chen},
  {Chen}, {Cheng}, {Cheong}, {Chia}, {Chiadini}, {Chierici}, {Chincarini},
  {Chiummo}, {Cho}, {Cho}, {Cho}, {Choate}, {Christensen}, {Chu}, {Chua},
  {Chung}, {Chung}, {Ciani}, {Ciecielag}, {Cie{\'s}lar}, {Cifaldi}, {Ciobanu},
  {Ciolfi}, {Cipriano}, {Cirone}, {Clara}, {Clark}, {Clark}, {Clarke},
  {Clearwater}, {Clesse}, {Cleva}, {Coccia}, {Cohadon}, {Cohen}, {Colleoni},
  {Collette}, {Collins}, {Colpi}, {Constancio}, {Conti}, {Cooper}, {Corban},
  {Corbitt}, {Cordero-Carri{\'o}n}, {Corezzi}, {Corley}, {Cornish}, {Corre},
  {Corsi}, {Cortese}, {Costa}, {Cotesta}, {Coughlin}, {Coughlin}, {Coulon},
  {Countryman}, {Cousins}, {Couvares}, {Covas}, {Coward}, {Cowart}, {Coyne},
  {Coyne}, {Creighton}, {Creighton}, {Croquette}, {Crowder}, {Cudell},
  {Cullen}, {Cumming}, {Cummings}, {Cunningham}, {Cuoco}, {Cury{\l}o},
  {Canton}, {D{\'a}lya}, {Dana}, {DaneshgaranBajastani}, {D'Angelo}, {Danila},
  {Danilishin}, {D'Antonio}, {Danzmann}, {Darsow-Fromm}, {Dasgupta}, {Datrier},
  {Dattilo}, {Dave}, {Davier}, {Davies}, {Davis}, {Daw}, {Dean}, {DeBra},
  {Deenadayalan}, {Degallaix}, {De Laurentis}, {Del{\'e}glise}, {Del Favero},
  {De Lillo}, {De Lillo}, {Del Pozzo}, {DeMarchi}, {De Matteis}, {D'Emilio},
  {Demos}, {Denker}, {Dent}, {Depasse}, {De Pietri}, {De Rosa}, {De Rossi},
  {DeSalvo}, {de Varona}, {Dhurandhar}, {D{\'\i}az}, {Diaz-Ortiz}, {Didio},
  {Dietrich}, {Di Fiore}, {DiFronzo}, {Di Giorgio}, {Di Giovanni}, {Di
  Giovanni}, {Di Girolamo}, {Di Lieto}, {Ding}, {Di Pace}, {Di Palma}, {Di
  Renzo}, {Divakarla}, {Dmitriev}, {Doctor}, {D'Onofrio}, {Donovan}, {Dooley},
  {Doravari}, {Dorrington}, {Downes}, {Drago}, {Driggers}, {Du}, {Ducoin},
  {Dupej}, {Durante}, {D'Urso}, {Duverne}, {Dwyer}, {Easter}, {Eddolls},
  {Edelman}, {Edo}, {Edy}, {Effler}, {Eichholz}, {Eikenberry}, {Eisenmann},
  {Eisenstein}, {Ejlli}, {Errico}, {Essick}, {Estell{\'e}s}, {Estevez},
  {Etienne}, {Etzel}, {Evans}, {Evans}, {Ewing}, {Fafone}, {Fair}, {Fairhurst},
  {Fan}, {Farah}, {Farinon}, {Farr}, {Farr}, {Fauchon-Jones}, {Favata}, {Fays},
  {Fazio}, {Feicht}, {Fejer}, {Feng}, {Fenyvesi}, {Ferguson},
  {Fernandez-Galiana}, {Ferrante}, {Ferreira}, {Fidecaro}, {Figura}, {Fiori},
  {Fiorucci}, {Fishbach}, {Fisher}, {Fishner}, {Fittipaldi}, {Fitz-Axen},
  {Fiumara}, {Flaminio}, {Floden}, {Flynn}, {Fong}, {Font}, {Forsyth},
  {Fournier}, {Frasca}, {Frasconi}, {Frei}, {Freise}, {Frey}, {Frey},
  {Fritschel}, {Frolov}, {Fronz{\'e}}, {Fulda}, {Fyffe}, {Gabbard}, {Gadre},
  {Gaebel}, {Gair}, {Gais}, {Galaudage}, {Gamba}, {Ganapathy}, {Ganguly},
  {Gaonkar}, {Garaventa}, {Garc{\'\i}a-Quir{\'o}s}, {Garufi}, {Gateley},
  {Gaudio}, {Gayathri}, {Gemme}, {Gennai}, {George}, {George}, {George},
  {Gergely}, {Ghonge}, {Ghosh}, {Ghosh}, {Ghosh}, {Giacomazzo}, {Giacoppo},
  {Giaime}, {Giardina}, {Gibson}, {Gier}, {Gill}, {Giri}, {Glanzer}, {Gleckl},
  {Godwin}, {Goetz}, {Goetz}, {Gohlke}, {Goncharov}, {Gonz{\'a}lez},
  {Gopakumar}, {Gossan}, {Gosselin}, {Gouaty}, {Grace}, {Grado}, {Granata},
  {Granata}, {Grant}, {Gras}, {Grassia}, {Gray}, {Gray}, {Greco}, {Green},
  {Green}, {Gretarsson}, {Griggs}, {Grignani}, {Grimaldi}, {Grimes}, {Grimm},
  {Grote}, {Grunewald}, {Gruning}, {Guerrero}, {Guidi}, {Guimaraes},
  {Guix{\'e}}, {Gulati}, {Guo}, {Gupta}, {Gupta}, {Gupta}, {Gustafson},
  {Gustafson}, {Guzman}, {Haegel}, {Halim}, {Hall}, {Hamilton}, {Hammond},
  {Haney}, {Hanke}, {Hanks}, {Hanna}, {Hannam}, {Hannuksela}, {Hannuksela},
  {Hansen}, {Hansen}, {Hanson}, {Harder}, {Hardwick}, {Haris}, {Harms},
  {Harry}, {Harry}, {Hartwig}, {Hasskew}, {Haster}, {Haughian}, {Hayes},
  {Healy}, {Heidmann}, {Heintze}, {Heinze}, {Heinzel}, {Heitmann}, {Hellman},
  {Hello}, {Helmling-Cornell}, {Hemming}, {Hendry}, {Heng}, {Hennes}, {Hennig},
  {Hennig}, {Hernandez Vivanco}, {Heurs}, {Hild}, {Hill}, {Hines}, {Hochheim},
  {Hofgard}, {Hofman}, {Hohmann}, {Holgado}, {Holland}, {Hollows}, {Holmes},
  {Holt}, {Holz}, {Hopkins}, {Horst}, {Hough}, {Howell}, {Hoy}, {Hoyland},
  {Huang}, {H{\"u}bner}, {Huddart}, {Huerta}, {Hughey}, {Hui}, {Husa},
  {Huttner}, {Hutzler}, {Huxford}, {Huynh-Dinh}, {Idzkowski}, {Iess},
  {Imperato}, {Inchauspe}, {Ingram}, {Intini}, {Isi}, {Iyer},
  {JaberianHamedan}, {Jacqmin}, {Jadhav}, {Jadhav}, {James}, {Jani},
  {Janssens}, {Janthalur}, {Jaranowski}, {Jariwala}, {Jaume}, {Jenkins},
  {Jeunon}, {Jiang}, {Johns}, {Johnson-McDaniel}, {Jones}, {Jones}, {Jones},
  {Jones}, {Jones}, {Jonker}, {Ju}, {Junker}, {Kalaghatgi}, {Kalogera},
  {Kamai}, {Kandhasamy}, {Kang}, {Kanner}, {Kapadia}, {Kapasi}, {Karathanasis},
  {Karki}, {Kashyap}, {Kasprzack}, {Kastaun}, {Katsanevas}, {Katsavounidis},
  {Katzman}, {Kawabe}, {K{\'e}f{\'e}lian}, {Keitel}, {Key}, {Khadka},
  {Khalili}, {Khan}, {Khan}, {Khazanov}, {Khetan}, {Khursheed}, {Kijbunchoo},
  {Kim}, {Kim}, {Kim}, {Kim}, {Kim}, {Kim}, {Kimball}, {King}, {Kinley-Hanlon},
  {Kirchhoff}, {Kissel}, {Kleybolte}, {Klimenko}, {Knowles}, {Knyazev}, {Koch},
  {Koehlenbeck}, {Koekoek}, {Koley}, {Kolstein}, {Komori}, {Kondrashov},
  {Kontos}, {Koper}, {Korobko}, {Korth}, {Kovalam}, {Kozak}, {Kr{\"a}mer},
  {Kringel}, {Krishnendu}, {Kr{\'o}lak}, {Kuehn}, {Kumar}, {Kumar}, {Kumar},
  {Kumar}, {Kuns}, {Kwang}, {Lackey}, {Laghi}, {Lalande}, {Lam}, {Lamberts},
  {Landry}, {Lane}, {Lang}, {Lange}, {Lantz}, {Lanza}, {La Rosa},
  {Lartaux-Vollard}, {Lasky}, {Laxen}, {Lazzarini}, {Lazzaro}, {Leaci},
  {Leavey}, {Lecoeuche}, {Lee}, {Lee}, {Lee}, {Lee}, {Lehmann}, {Leon},
  {Leroy}, {Letendre}, {Levin}, {Li}, {Li}, {Li}, {Li}, {Li}, {Linde},
  {Linker}, {Linley}, {Littenberg}, {Liu}, {Liu}, {Llorens-Monteagudo}, {Lo},
  {Lockwood}, {London}, {Longo}, {Lorenzini}, {Loriette}, {Lormand}, {Losurdo},
  {Lough}, {Lousto}, {Lovelace}, {L{\"u}ck}, {Lumaca}, {Lundgren}, {Ma},
  {Macas}, {MacInnis}, {Macleod}, {MacMillan}, {Macquet}, {Maga{\~n}a
  Hernandez}, {Maga{\~n}a-Sandoval}, {Magazz{\`u}}, {Magee}, {Majorana},
  {Maksimovic}, {Maliakal}, {Malik}, {Man}, {Mandic}, {Mangano}, {Mansell},
  {Manske}, {Mantovani}, {Mapelli}, {Marchesoni}, {Marion}, {M{\'a}rka},
  {M{\'a}rka}, {Markakis}, {Markosyan}, {Markowitz}, {Maros}, {Marquina},
  {Marsat}, {Martelli}, {Martin}, {Martin}, {Martinez}, {Martinez}, {Martynov},
  {Masalehdan}, {Mason}, {Massera}, {Masserot}, {Massinger}, {Masso-Reid},
  {Mastrogiovanni}, {Matas}, {Mateu-Lucena}, {Matichard}, {Matiushechkina},
  {Mavalvala}, {Maynard}, {McCann}, {McCarthy}, {McClelland}, {McCormick},
  {McCuller}, {McGuire}, {McIsaac}, {McIver}, {McManus}, {McRae}, {McWilliams},
  {Meacher}, {Meadors}, {Mehmet}, {Mehta}, {Melatos}, {Melchor}, {Mendell},
  {Menendez-Vazquez}, {Mercer}, {Mereni}, {Merfeld}, {Merilh}, {Merritt},
  {Merzougui}, {Meshkov}, {Messenger}, {Messick}, {Metzdorff}, {Meyers},
  {Meylahn}, {Mhaske}, {Miani}, {Miao}, {Michaloliakos}, {Michel}, {Middleton},
  {Milano}, {Miller}, {Millhouse}, {Mills}, {Milotti}, {Milovich-Goff},
  {Minazzoli}, {Minenkov}, {Mir}, {Mishkin}, {Mishra}, {Mistry}, {Mitra},
  {Mitrofanov}, {Mitselmakher}, {Mittleman}, {Mo}, {Mogushi}, {Mohapatra},
  {Mohite}, {Molina}, {Molina-Ruiz}, {Mondin}, {Montani}, {Moore}, {Moraru},
  {Morawski}, {Moreno}, {Morisaki}, {Mours}, {Mow-Lowry}, {Mozzon},
  {Muciaccia}, {Mukherjee}, {Mukherjee}, {Mukherjee}, {Mukherjee}, {Mukund},
  {Mullavey}, {Munch}, {Mu{\~n}iz}, {Murray}, {Nadji}, {Nagar}, {Nardecchia},
  {Naticchioni}, {Nayak}, {Neil}, {Neilson}, {Nelemans}, {Nelson}, {Nery},
  {Neunzert}, {Nitz}, {Ng}, {Ng}, {Nguyen}, {Nguyen}, {Nguyen}, {Nichols},
  {Nissanke}, {Nocera}, {Noh}, {North}, {Nothard}, {Nuttall}, {Oberling},
  {O'Brien}, {O'Dell}, {Oganesyan}, {Ogin}, {Oh}, {Oh}, {Ohme}, {Ohta},
  {Okada}, {Olivetto}, {Oppermann}, {Oram}, {O'Reilly}, {Ormiston}, {Ortega},
  {O'Shaughnessy}, {Ossokine}, {Osthelder}, {Ottaway}, {Overmier}, {Owen},
  {Pace}, {Pagano}, {Page}, {Pagliaroli}, {Pai}, {Pai}, {Palamos}, {Palashov},
  {Palomba}, {Pan}, {Panda}, {Pang}, {Pankow}, {Pannarale}, {Pant}, {Paoletti},
  {Paoli}, {Paolone}, {Parker}, {Pascucci}, {Pasqualetti}, {Passaquieti},
  {Passuello}, {Patel}, {Patricelli}, {Payne}, {Pechsiri}, {Pedraza},
  {Pegoraro}, {Pele}, {Penn}, {Perego}, {Perez}, {P{\'e}rigois}, {Perreca},
  {Perri{\`e}s}, {Petermann}, {Petterson}, {Pfeiffer}, {Pham}, {Phukon},
  {Piccinni}, {Pichot}, {Piendibene}, {Piergiovanni}, {Pierini}, {Pierro},
  {Pillant}, {Pilo}, {Pinard}, {Pinto}, {Piotrzkowski}, {Pirello}, {Pitkin},
  {Placidi}, {Plastino}, {Pluchar}, {Poggiani}, {Polini}, {Pong}, {Ponrathnam},
  {Popolizio}, {Porter}, {Poverman}, {Powell}, {Pracchia}, {Prajapati},
  {Prasai}, {Prasanna}, {Pratten}, {Prestegard}, {Principe}, {Prodi},
  {Prokhorov}, {Prosposito}, {Prudenzi}, {Puecher}, {Punturo}, {Puosi},
  {Puppo}, {P{\"u}rrer}, {Qi}, {Quetschke}, {Quinonez}, {Quitzow-James},
  {Raab}, {Raaijmakers}, {Radkins}, {Radulesco}, {Raffai}, {Rafferty}, {Rail},
  {Raja}, {Rajan}, {Rajbhandari}, {Rakhmanov}, {Ramirez}, {Ramirez},
  {Ramos-Buades}, {Rana}, {Rao}, {Rapagnani}, {Rapol}, {Ratto}, {Raymond},
  {Razzano}, {Read}, {Regimbau}, {Rei}, {Reid}, {Reitze}, {Rettegno}, {Ricci},
  {Richardson}, {Richardson}, {Richardson}, {Ricker}, {Riemenschneider},
  {Riles}, {Rizzo}, {Robertson}, {Robinet}, {Rocchi}, {Rocha}, {Rodriguez},
  {Rodriguez-Soto}, {Rolland}, {Rollins}, {Roma}, {Romanelli}, {Romano},
  {Romel}, {Romero}, {Romero-Shaw}, {Romie}, {Ronchini}, {Rose}, {Rose},
  {Rose}, {Rosell}, {Rosi{\'n}ska}, {Rosofsky}, {Ross}, {Rowan}, {Rowlinson},
  {Roy}, {Roy}, {Ruggi}, {Ryan}, {Sachdev}, {Sadecki}, {Sadiq},
  {Sakellariadou}, {Salafia}, {Salconi}, {Saleem}, {Samajdar}, {Sanchez},
  {Sanchez}, {Sanchez}, {Sanchis-Gual}, {Sanders}, {Sandles}, {Santiago},
  {Santos}, {Saravanan}, {Sarin}, {Sassolas}, {Sathyaprakash}, {Sauter},
  {Savage}, {Savant}, {Sawant}, {Sayah}, {Schaetzl}, {Schale}, {Scheel},
  {Scheuer}, {Schindler-Tyka}, {Schmidt}, {Schnabel}, {Schofield},
  {Sch{\"o}nbeck}, {Schreiber}, {Schulte}, {Schutz}, {Schwarm}, {Schwartz},
  {Scott}, {Scott}, {Seglar-Arroyo}, {Seidel}, {Sellers}, {Sengupta},
  {Sennett}, {Sentenac}, {Sequino}, {Sergeev}, {Setyawati}, {Shaffer},
  {Shahriar}, {Sharifi}, {Sharma}, {Sharma}, {Shawhan}, {Shen}, {Shikauchi},
  {Shink}, {Shoemaker}, {Shoemaker}, {Shukla}, {ShyamSundar}, {Sieniawska},
  {Sigg}, {Singer}, {Singh}, {Singh}, {Singha}, {Singhal}, {Sintes}, {Sipala},
  {Skliris}, {Slagmolen}, {Slaven-Blair}, {Smetana}, {Smith}, {Smith},
  {Somala}, {Son}, {Soni}, {Soni}, {Sorazu}, {Sordini}, {Sorrentino},
  {Sorrentino}, {Soulard}, {Souradeep}, {Sowell}, {Spencer}, {Spera},
  {Srivastava}, {Srivastava}, {Staats}, {Stachie}, {Steer}, {Steinhoff},
  {Steinke}, {Steinlechner}, {Steinlechner}, {Steinmeyer}, {Stevenson},
  {Stolle-McAllister}, {Stops}, {Stover}, {Strain}, {Stratta}, {Strunk},
  {Sturani}, {Stuver}, {S{\"u}dbeck}, {Sudhagar}, {Sudhir}, {Suh},
  {Summerscales}, {Sun}, {Sun}, {Sunil}, {Sur}, {Suresh}, {Sutton}, {Swinkels},
  {Szczepa{\'n}czyk}, {Tacca}, {Tait}, {Talbot}, {Tanasijczuk}, {Tanner},
  {Tao}, {Tapia}, {Tapia San Martin}, {Tasson}, {Taylor}, {Tenorio},
  {Terkowski}, {Thirugnanasambandam}, {Thomas}, {Thomas}, {Thomas}, {Thompson},
  {Thondapu}, {Thorne}, {Thrane}, {Tiwari}, {Tiwari}, {Tiwari}, {Toland},
  {Tolley}, {Tonelli}, {Tornasi}, {Torres-Forn{\'e}}, {Torrie}, {e Melo},
  {T{\"o}yr{\"a}}, {Tran}, {Trapananti}, {Travasso}, {Traylor}, {Tringali},
  {Tripathee}, {Trovato}, {Trudeau}, {Tsai}, {Tsang}, {Tse}, {Tso}, {Tsukada},
  {Tsuna}, {Tsutsui}, {Turconi}, {Ubhi}, {Udall}, {Ueno}, {Ugolini},
  {Unnikrishnan}, {Urban}, {Usman}, {Utina}, {Vahlbruch}, {Vajente}, {Vajpeyi},
  {Valdes}, {Valentini}, {Valsan}, {van Bakel}, {van Beuzekom}, {van den
  Brand}, {Van Den Broeck}, {Vander-Hyde}, {van der Schaaf}, {van Heijningen},
  {Vardaro}, {Vargas}, {Varma}, {Vass}, {Vas{\'u}th}, {Vecchio}, {Vedovato},
  {Veitch}, {Veitch}, {Venkateswara}, {Venneberg}, {Venugopalan}, {Verkindt},
  {Verma}, {Veske}, {Vetrano}, {Vicer{\'e}}, {Viets}, {Vijaykumar},
  {Villa-Ortega}, {Vinet}, {Vitale}, {Vo}, {Vocca}, {Vorvick}, {Vyatchanin},
  {Wade}, {Wade}, {Wade}, {Walet}, {Walker}, {Wallace}, {Wallace}, {Walsh},
  {Wang}, {Wang}, {Wang}, {Wang}, {Ward}, {Warner}, {Was}, {Washington},
  {Watchi}, {Weaver}, {Wei}, {Weinert}, {Weinstein}, {Weiss}, {Wellmann},
  {Wen}, {We{\ss}els}, {Westhouse}, {Wette}, {Whelan}, {White}, {White},
  {Whiting}, {Whittle}, {Wilken}, {Williams}, {Williams}, {Williamson},
  {Willis}, {Willke}, {Wilson}, {Wimmer}, {Winkler}, {Wipf}, {Woan}, {Woehler},
  {Wofford}, {Wong}, {Wrangel}, {Wright}, {Wu}, {Wysocki}, {Xiao}, {Yamamoto},
  {Yang}, {Yang}, {Yang}, {Yap}, {Yeeles}, {Yoon}, {Yu}, {Yu}, {Yuen},
  {Zadro{\.Z}ny}, {Zanolin}, {Zelenova}, {Zendri}, {Zevin}, {Zhang}, {Zhang},
  {Zhang}, {Zhang}, {Zhao}, {Zhao}, {Zheng}, {Zhou}, {Zhou}, {Zhu},
  {Zimmerman}, {Zlochower}, {Zucker}, {Zweizig}, {LIGO Scientific
  Collaboration}, \& {Virgo Collaboration}}]{Abbott2021a}
{Abbott}, R., {Abbott}, T.~D., {Abraham}, S., {et~al.} 2021{\natexlab{b}},
  Physical Review X, 11, 021053

\bibitem[{{Abbott} {et~al.}(2021{\natexlab{c}}){Abbott}, {Adam{\'o}w},
  {Aguena}, {Allam}, {Amon}, {Annis}, {Avila}, {Bacon}, {Banerji}, {Bechtol},
  {Becker}, {Bernstein}, {Bertin}, {Bhargava}, {Bridle}, {Brooks}, {Burke},
  {Carnero Rosell}, {Carrasco Kind}, {Carretero}, {Castander}, {Cawthon},
  {Chang}, {Choi}, {Conselice}, {Costanzi}, {Crocce}, {da Costa}, {Davis}, {De
  Vicente}, {DeRose}, {Desai}, {Diehl}, {Dietrich}, {Drlica-Wagner}, {Eckert},
  {Elvin-Poole}, {Everett}, {Evrard}, {Ferrero}, {Fert{\'e}}, {Flaugher},
  {Fosalba}, {Friedel}, {Frieman}, {Garc{\'\i}a-Bellido}, {Gaztanaga},
  {Gelman}, {Gerdes}, {Giannantonio}, {Gill}, {Gruen}, {Gruendl}, {Gschwend},
  {Gutierrez}, {Hartley}, {Hinton}, {Hollowood}, {Honscheid}, {Huterer},
  {James}, {Jeltema}, {Johnson}, {Kent}, {Kron}, {Kuehn}, {Kuropatkin},
  {Lahav}, {Li}, {Lidman}, {Lin}, {MacCrann}, {Maia}, {Manning}, {Maloney},
  {March}, {Marshall}, {Martini}, {Melchior}, {Menanteau}, {Miquel}, {Morgan},
  {Myles}, {Neilsen}, {Ogando}, {Palmese}, {Paz-Chinch{\'o}n}, {Petravick},
  {Pieres}, {Plazas}, {Pond}, {Rodriguez-Monroy}, {Romer}, {Roodman}, {Rykoff},
  {Sako}, {Sanchez}, {Santiago}, {Scarpine}, {Serrano}, {Sevilla-Noarbe},
  {Smith}, {Smith}, {Soares-Santos}, {Suchyta}, {Swanson}, {Tarle}, {Thomas},
  {To}, {Tremblay}, {Troxel}, {Tucker}, {Turner}, {Varga}, {Walker},
  {Wechsler}, {Weller}, {Wester}, {Wilkinson}, {Yanny}, {Zhang}, {Nikutta},
  {Fitzpatrick}, {Jacques}, {Scott}, {Olsen}, {Huang}, {Herrera}, {Juneau},
  {Nidever}, {Weaver}, {Adean}, {Correia}, {de Freitas}, {Freitas},
  {Singulani}, {Vila-Verde}, \& {Linea Science Server}}]{Abbott2021}
{Abbott}, T.~M.~C., {Adam{\'o}w}, M., {Aguena}, M., {et~al.}
  2021{\natexlab{c}}, \apjs, 255, 20

\bibitem[{{Ahumada} {et~al.}(2020){Ahumada}, {Prieto}, {Almeida}, {Anders},
  {Anderson}, {Andrews}, {Anguiano}, {Arcodia}, {Armengaud}, {Aubert}, {Avila},
  {Avila-Reese}, {Badenes}, {Balland}, {Barger}, {Barrera-Ballesteros}, {Basu},
  {Bautista}, {Beaton}, {Beers}, {Benavides}, {Bender}, {Bernardi}, {Bershady},
  {Beutler}, {Bidin}, {Bird}, {Bizyaev}, {Blanc}, {Blanton}, {Boquien},
  {Borissova}, {Bovy}, {Brandt}, {Brinkmann}, {Brownstein}, {Bundy}, {Bureau},
  {Burgasser}, {Burtin}, {Cano-D{\'\i}az}, {Capasso}, {Cappellari}, {Carrera},
  {Chabanier}, {Chaplin}, {Chapman}, {Cherinka}, {Chiappini}, {Doohyun Choi},
  {Chojnowski}, {Chung}, {Clerc}, {Coffey}, {Comerford}, {Comparat}, {da
  Costa}, {Cousinou}, {Covey}, {Crane}, {Cunha}, {Ilha}, {Dai}, {Damsted},
  {Darling}, {Davidson}, {Davies}, {Dawson}, {De}, {de la Macorra}, {De Lee},
  {Queiroz}, {Deconto Machado}, {de la Torre}, {Dell'Agli}, {du Mas des
  Bourboux}, {Diamond-Stanic}, {Dillon}, {Donor}, {Drory}, {Duckworth},
  {Dwelly}, {Ebelke}, {Eftekharzadeh}, {Davis Eigenbrot}, {Elsworth},
  {Eracleous}, {Erfanianfar}, {Escoffier}, {Fan}, {Farr},
  {Fern{\'a}ndez-Trincado}, {Feuillet}, {Finoguenov}, {Fofie},
  {Fraser-McKelvie}, {Frinchaboy}, {Fromenteau}, {Fu}, {Galbany}, {Garcia},
  {Garc{\'\i}a-Hern{\'a}ndez}, {Oehmichen}, {Ge}, {Maia}, {Geisler}, {Gelfand},
  {Goddy}, {Gonzalez-Perez}, {Grabowski}, {Green}, {Grier}, {Guo}, {Guy},
  {Harding}, {Hasselquist}, {Hawken}, {Hayes}, {Hearty}, {Hekker}, {Hogg},
  {Holtzman}, {Horta}, {Hou}, {Hsieh}, {Huber}, {Hunt}, {Chitham}, {Imig},
  {Jaber}, {Angel}, {Johnson}, {Jones}, {J{\"o}nsson}, {Jullo}, {Kim},
  {Kinemuchi}, {Kirkpatrick}, {Kite}, {Klaene}, {Kneib}, {Kollmeier}, {Kong},
  {Kounkel}, {Krishnarao}, {Lacerna}, {Lan}, {Lane}, {Law}, {Le Goff}, {Leung},
  {Lewis}, {Li}, {Lian}, {Lin}, {Long}, {Longa-Pe{\~n}a}, {Lundgren}, {Lyke},
  {Ted Mackereth}, {MacLeod}, {Majewski}, {Manchado}, {Maraston}, {Martini},
  {Masseron}, {Masters}, {Mathur}, {McDermid}, {Merloni}, {Merrifield},
  {M{\'e}sz{\'a}ros}, {Miglio}, {Minniti}, {Minsley}, {Miyaji}, {Mohammad},
  {Mosser}, {Mueller}, {Muna}, {Mu{\~n}oz-Guti{\'e}rrez}, {Myers}, {Nadathur},
  {Nair}, {Nandra}, {do Nascimento}, {Nevin}, {Newman}, {Nidever}, {Nitschelm},
  {Noterdaeme}, {O'Connell}, {Olmstead}, {Oravetz}, {Oravetz}, {Osorio},
  {Pace}, {Padilla}, {Palanque-Delabrouille}, {Palicio}, {Pan}, {Pan},
  {Parker}, {Paviot}, {Peirani}, {Ram{\'r}ez}, {Penny}, {Percival},
  {Perez-Fournon}, {P{\'e}rez-R{\`a}fols}, {Petitjean}, {Pieri},
  {Pinsonneault}, {Poovelil}, {Povick}, {Prakash}, {Price-Whelan}, {Raddick},
  {Raichoor}, {Ray}, {Rembold}, {Rezaie}, {Riffel}, {Riffel}, {Rix}, {Robin},
  {Roman-Lopes}, {Rom{\'a}n-Z{\'u}{\~n}iga}, {Rose}, {Ross}, {Rossi},
  {Rowlands}, {Rubin}, {Salvato}, {S{\'a}nchez}, {S{\'a}nchez-Menguiano},
  {S{\'a}nchez-Gallego}, {Sayres}, {Schaefer}, {Schiavon}, {Schimoia},
  {Schlafly}, {Schlegel}, {Schneider}, {Schultheis}, {Schwope}, {Seo},
  {Serenelli}, {Shafieloo}, {Shamsi}, {Shao}, {Shen}, {Shetrone}, {Shirley},
  {Aguirre}, {Simon}, {Skrutskie}, {Slosar}, {Smethurst}, {Sobeck}, {Sodi},
  {Souto}, {Stark}, {Stassun}, {Steinmetz}, {Stello}, {Stermer},
  {Storchi-Bergmann}, {Streblyanska}, {Stringfellow}, {Stutz}, {Su{\'a}rez},
  {Sun}, {Taghizadeh-Popp}, {Talbot}, {Tayar}, {Thakar}, {Theriault}, {Thomas},
  {Thomas}, {Tinker}, {Tojeiro}, {Toledo}, {Tremonti}, {Troup}, {Tuttle},
  {Unda-Sanzana}, {Valentini}, {Vargas-Gonz{\'a}lez}, {Vargas-Maga{\~n}a},
  {V{\'a}zquez-Mata}, {Vivek}, {Wake}, {Wang}, {Weaver}, {Weijmans}, {Wild},
  {Wilson}, {Wilson}, {Wolthuis}, {Wood-Vasey}, {Yan}, {Yang}, {Y{\`e}che},
  {Zamora}, {Zarrouk}, {Zasowski}, {Zhang}, {Zhao}, {Zhao}, {Zheng}, {Zheng},
  {Zhu}, \& {Zou}}]{Ahumada2019}
{Ahumada}, R., {Prieto}, C.~A., {Almeida}, A., {et~al.} 2020, \apjs, 249, 3

\bibitem[{{Ajello} {et~al.}(2019){Ajello}, {Arimoto}, {Axelsson}, {Baldini},
  {Barbiellini}, {Bastieri}, {Bellazzini}, {Bhat}, {Bissaldi}, {Blandford},
  {Bonino}, {Bonnell}, {Bottacini}, {Bregeon}, {Bruel}, {Buehler}, {Cameron},
  {Caputo}, {Caraveo}, {Cavazzuti}, {Chen}, {Cheung}, {Chiaro}, {Ciprini},
  {Costantin}, {Crnogorcevic}, {Cutini}, {Dainotti}, {D'Ammand o}, {de la Torre
  Luque}, {de Palma}, {Desai}, {Desiante}, {Di Lalla}, {Di Venere}, {Fana
  Dirirsa}, {Fegan}, {Franckowiak}, {Fukazawa}, {Funk}, {Fusco}, {Gargano},
  {Gasparrini}, {Giglietto}, {Giordano}, {Giroletti}, {Green}, {Grenier},
  {Grove}, {Guiriec}, {Hays}, {Hewitt}, {Horan}, {J{\'o}hannesson}, {Kocevski},
  {Kuss}, {Latronico}, {Li}, {Longo}, {Loparco}, {Lovellette}, {Lubrano},
  {Maldera}, {Manfreda}, {Mart{\'\i}-Devesa}, {Mazziotta}, {Mereu}, {Meyer},
  {Michelson}, {Mirabal}, {Mitthumsiri}, {Mizuno}, {Monzani}, {Moretti},
  {Morselli}, {Moskalenko}, {Negro}, {Nuss}, {Ohno}, {Omodei}, {Orienti},
  {Orlando}, {Palatiello}, {Paliya}, {Paneque}, {Persic}, {Pesce-Rollins},
  {Petrosian}, {Piron}, {Poolakkil}, {Poon}, {Porter}, {Principe}, {Racusin},
  {Rain{\`o}}, {Rando}, {Razzano}, {Razzaque}, {Reimer}, {Reimer}, {Reposeur},
  {Ryde}, {Serini}, {Sgr{\`o}}, {Siskind}, {Sonbas}, {Spandre}, {Spinelli},
  {Suson}, {Tajima}, {Takahashi}, {Tak}, {Thayer}, {Torres}, {Troja},
  {Valverde}, {Veres}, {Vianello}, {von Kienlin}, {Wood}, {Yassine}, {Zhu}, \&
  {Zimmer}}]{Ajello2019}
{Ajello}, M., {Arimoto}, M., {Axelsson}, M., {et~al.} 2019, \apj, 878, 52

\bibitem[{{Alp} \& {Larsson}(2020)}]{Alp2020}
{Alp}, D. \& {Larsson}, J. 2020, \apj, 896, 39

\bibitem[{Aptekar {et~al.}(2001)Aptekar, Frederiks, Golenetskii, Il\'inskii,
  Mazets, Pal\'shin, Butterworth, \& Cline}]{Aptekar2001}
Aptekar, R., Frederiks, D., Golenetskii, S., {et~al.} 2001, \apjs, 137, 227

\bibitem[{{Arabsalmani} {et~al.}(2019){Arabsalmani}, {Roychowdhury},
  {Starkenburg}, {Christensen}, {Le Floc'h}, {Kanekar}, {Bournaud}, {Zwaan},
  {Fynbo}, {M{\o}ller}, \& {Pian}}]{Arabsalmani2019}
{Arabsalmani}, M., {Roychowdhury}, S., {Starkenburg}, T.~K., {et~al.} 2019,
  \mnras, 485, 5411

\bibitem[{{Arefiev} {et~al.}(2003){Arefiev}, {Priedhorsky}, \&
  {Borozdin}}]{Arefiev2003}
{Arefiev}, V.~A., {Priedhorsky}, W.~C., \& {Borozdin}, K.~N. 2003, \apj, 586,
  1238

\bibitem[{{Arnaud}(1996)}]{Arnaud1996}
{Arnaud}, K.~A. 1996, Astronomical Society of the Pacific Conference Series,
  Vol. 101, {XSPEC: The First Ten Years}, ed. G.~H. {Jacoby} \& J.~{Barnes}, 17

\bibitem[{{Astropy Collaboration} {et~al.}(2018){Astropy Collaboration},
  {Price-Whelan}, {Sip{\H{o}}cz}, {G{\"u}nther}, {Lim}, {Crawford}, {Conseil},
  {Shupe}, {Craig}, {Dencheva}, {Ginsburg}, {Vand erPlas}, {Bradley},
  {P{\'e}rez-Su{\'a}rez}, {de Val-Borro}, {Aldcroft}, {Cruz}, {Robitaille},
  {Tollerud}, {Ardelean}, {Babej}, {Bach}, {Bachetti}, {Bakanov}, {Bamford},
  {Barentsen}, {Barmby}, {Baumbach}, {Berry}, {Biscani}, {Boquien}, {Bostroem},
  {Bouma}, {Brammer}, {Bray}, {Breytenbach}, {Buddelmeijer}, {Burke},
  {Calderone}, {Cano Rodr{\'\i}guez}, {Cara}, {Cardoso}, {Cheedella}, {Copin},
  {Corrales}, {Crichton}, {D'Avella}, {Deil}, {Depagne}, {Dietrich}, {Donath},
  {Droettboom}, {Earl}, {Erben}, {Fabbro}, {Ferreira}, {Finethy}, {Fox},
  {Garrison}, {Gibbons}, {Goldstein}, {Gommers}, {Greco}, {Greenfield},
  {Groener}, {Grollier}, {Hagen}, {Hirst}, {Homeier}, {Horton}, {Hosseinzadeh},
  {Hu}, {Hunkeler}, {Ivezi{\'c}}, {Jain}, {Jenness}, {Kanarek}, {Kendrew},
  {Kern}, {Kerzendorf}, {Khvalko}, {King}, {Kirkby}, {Kulkarni}, {Kumar},
  {Lee}, {Lenz}, {Littlefair}, {Ma}, {Macleod}, {Mastropietro}, {McCully},
  {Montagnac}, {Morris}, {Mueller}, {Mumford}, {Muna}, {Murphy}, {Nelson},
  {Nguyen}, {Ninan}, {N{\"o}the}, {Ogaz}, {Oh}, {Parejko}, {Parley}, {Pascual},
  {Patil}, {Patil}, {Plunkett}, {Prochaska}, {Rastogi}, {Reddy Janga},
  {Sabater}, {Sakurikar}, {Seifert}, {Sherbert}, {Sherwood-Taylor}, {Shih},
  {Sick}, {Silbiger}, {Singanamalla}, {Singer}, {Sladen}, {Sooley},
  {Sornarajah}, {Streicher}, {Teuben}, {Thomas}, {Tremblay}, {Turner},
  {Terr{\'o}n}, {van Kerkwijk}, {de la Vega}, {Watkins}, {Weaver}, {Whitmore},
  {Woillez}, {Zabalza}, \& {Astropy Contributors}}]{Astropy2018}
{Astropy Collaboration}, {Price-Whelan}, A.~M., {Sip{\H{o}}cz}, B.~M., {et~al.}
  2018, \aj, 156, 123

\bibitem[{{Astropy Collaboration} {et~al.}(2013){Astropy Collaboration},
  {Robitaille}, {Tollerud}, {Greenfield}, {Droettboom}, {Bray}, {Aldcroft},
  {Davis}, {Ginsburg}, {Price-Whelan}, {Kerzendorf}, {Conley}, {Crighton},
  {Barbary}, {Muna}, {Ferguson}, {Grollier}, {Parikh}, {Nair}, {Unther},
  {Deil}, {Woillez}, {Conseil}, {Kramer}, {Turner}, {Singer}, {Fox}, {Weaver},
  {Zabalza}, {Edwards}, {Azalee Bostroem}, {Burke}, {Casey}, {Crawford},
  {Dencheva}, {Ely}, {Jenness}, {Labrie}, {Lim}, {Pierfederici}, {Pontzen},
  {Ptak}, {Refsdal}, {Servillat}, \& {Streicher}}]{Astropy2013}
{Astropy Collaboration}, {Robitaille}, T.~P., {Tollerud}, E.~J., {et~al.} 2013,
  \aap, 558, A33

\bibitem[{{Avenhaus} {et~al.}(2012){Avenhaus}, {Schmid}, \&
  {Meyer}}]{Avenhaus2012}
{Avenhaus}, H., {Schmid}, H.~M., \& {Meyer}, M.~R. 2012, \aap, 548, A105

\bibitem[{{Bachetti} {et~al.}(2014){Bachetti}, {Harrison}, {Walton},
  {Grefenstette}, {Chakrabarty}, {F{\"u}rst}, {Barret}, {Beloborodov}, {Boggs},
  {Christensen}, {Craig}, {Fabian}, {Hailey}, {Hornschemeier}, {Kaspi},
  {Kulkarni}, {Maccarone}, {Miller}, {Rana}, {Stern}, {Tendulkar}, {Tomsick},
  {Webb}, \& {Zhang}}]{Bachetti2014}
{Bachetti}, M., {Harrison}, F.~A., {Walton}, D.~J., {et~al.} 2014, \nat, 514,
  202

\bibitem[{{Balberg} \& {Loeb}(2011)}]{Balberg2011}
{Balberg}, S. \& {Loeb}, A. 2011, \mnras, 414, 1715

\bibitem[{{Barack} {et~al.}(2019){Barack}, {Cardoso}, {Nissanke}, {Sotiriou},
  {Askar}, {Belczynski}, {Bertone}, {Bon}, {Blas}, {Brito}, {Bulik}, {Burrage},
  {Byrnes}, {Caprini}, {Chernyakova}, {Chru{\'s}ciel}, {Colpi}, {Ferrari},
  {Gaggero}, {Gair}, {Garc{\'\i}a-Bellido}, {Hassan}, {Heisenberg}, {Hendry},
  {Heng}, {Herdeiro}, {Hinderer}, {Horesh}, {Kavanagh}, {Kocsis}, {Kramer}, {Le
  Tiec}, {Mingarelli}, {Nardini}, {Nelemans}, {Palenzuela}, {Pani}, {Perego},
  {Porter}, {Rossi}, {Schmidt}, {Sesana}, {Sperhake}, {Stamerra}, {Stein},
  {Tamanini}, {Tauris}, {Urena-L{\'o}pez}, {Vincent}, {Volonteri}, {Wardell},
  {Wex}, {Yagi}, {Abdelsalhin}, {Aloy}, {Amaro-Seoane}, {Annulli},
  {Arca-Sedda}, {Bah}, {Barausse}, {Barakovic}, {Benkel}, {Bennett}, {Bernard},
  {Bernuzzi}, {Berry}, {Berti}, {Bezares}, {Juan Blanco-Pillado},
  {Bl{\'a}zquez-Salcedo}, {Bonetti}, {Bo{\v{s}}kovi{\'c}}, {Bosnjak},
  {Bricman}, {Br{\"u}gmann}, {Capelo}, {Carloni}, {Cerd{\'a}-Dur{\'a}n},
  {Charmousis}, {Chaty}, {Clerici}, {Coates}, {Colleoni}, {Collodel},
  {Comp{\`e}re}, {Cook}, {Cordero-Carri{\'o}n}, {Correia}, {de la
  Cruz-Dombriz}, {Czinner}, {Destounis}, {Dialektopoulos}, {Doneva}, {Dotti},
  {Drew}, {Eckner}, {Edholm}, {Emparan}, {Erdem}, {Ferreira}, {Ferreira},
  {Finch}, {Font}, {Franchini}, {Fransen}, {Gal'tsov}, {Ganguly}, {Gerosa},
  {Glampedakis}, {Gomboc}, {Goobar}, {Gualtieri}, {Guendelman}, {Haardt},
  {Harmark}, {Hejda}, {Hertog}, {Hopper}, {Husa}, {Ihanec}, {Ikeda}, {Jaodand},
  {Jetzer}, {Jimenez-Forteza}, {Kamionkowski}, {Kaplan}, {Kazantzidis},
  {Kimura}, {Kobayashi}, {Kokkotas}, {Krolik}, {Kunz}, {L{\"a}mmerzahl},
  {Lasky}, {Lemos}, {Levi Said}, {Liberati}, {Lopes}, {Luna}, {Ma}, {Maggio},
  {Mangiagli}, {Martinez Montero}, {Maselli}, {Mayer}, {Mazumdar}, {Messenger},
  {M{\'e}nard}, {Minamitsuji}, {Moore}, {Mota}, {Nampalliwar}, {Nerozzi},
  {Nichols}, {Nissimov}, {Obergaulinger}, {Obers}, {Oliveri}, {Pappas},
  {Pasic}, {Peiris}, {Petrushevska}, {Pollney}, {Pratten}, {Rakic}, {Racz},
  {Radia}, {Ramazano{\u{g}}lu}, {Ramos-Buades}, {Raposo}, {Rogatko},
  {Rosca-Mead}, {Rosinska}, {Rosswog}, {Ruiz-Morales}, {Sakellariadou},
  {Sanchis-Gual}, {Sharan Salafia}, {Samajdar}, {Sintes}, {Smole}, {Sopuerta},
  {Souza-Lima}, {Stalevski}, {Stergioulas}, {Stevens}, {Tamfal},
  {Torres-Forn{\'e}}, {Tsygankov}, {{\"U}nl{\"u}t{\"u}rk}, {Valiante}, {van de
  Meent}, {Velhinho}, {Verbin}, {Vercnocke}, {Vernieri}, {Vicente},
  {Vitagliano}, {Weltman}, {Whiting}, {Williamson}, {Witek}, {Wojnar}, {Yakut},
  {Yan}, {Yazadjiev}, {Zaharijas}, \& {Zilh{\~a}o}}]{Barack2019}
{Barack}, L., {Cardoso}, V., {Nissanke}, S., {et~al.} 2019, Classical and
  Quantum Gravity, 36, 143001

\bibitem[{{Barniol Duran} {et~al.}(2015){Barniol Duran}, {Nakar}, {Piran}, \&
  {Sari}}]{Barniol2015}
{Barniol Duran}, R., {Nakar}, E., {Piran}, T., \& {Sari}, R. 2015, \mnras, 448,
  417

\bibitem[{{Barret} {et~al.}(2013){Barret}, {Nandra}, {Barcons}, {Fabian}, {den
  Herder}, {Piro}, {Watson}, {Aird}, {Branduardi-Raymont}, {Cappi}, {Carrera},
  {Comastri}, {Costantini}, {Croston}, {Decourchelle}, {Done}, {Dovciak},
  {Ettori}, {Finoguenov}, {Georgakakis}, {Jonker}, {Kaastra}, {Matt}, {Motch},
  {O'Brien}, {Pareschi}, {Pointecouteau}, {Pratt}, {Rauw}, {Reiprich},
  {Sanders}, {Sciortino}, {Willingale}, \& {Wilms}}]{Barret2013}
{Barret}, D., {Nandra}, K., {Barcons}, X., {et~al.} 2013, in SF2A-2013:
  Proceedings of the Annual meeting of the French Society of Astronomy and
  Astrophysics, ed. L.~{Cambresy}, F.~{Martins}, E.~{Nuss}, \& A.~{Palacios},
  447--453

\bibitem[{{Barthelmy} {et~al.}(2005){Barthelmy}, {Chincarini}, {Burrows},
  {Gehrels}, {Covino}, {Moretti}, {Romano}, {O'Brien}, {Sarazin},
  {Kouveliotou}, {Goad}, {Vaughan}, {Tagliaferri}, {Zhang}, {Antonelli},
  {Campana}, {Cummings}, {D'Avanzo}, {Davies}, {Giommi}, {Grupe}, {Kaneko},
  {Kennea}, {King}, {Kobayashi}, {Melandri}, {Meszaros}, {Nousek}, {Patel},
  {Sakamoto}, \& {Wijers}}]{Barthelmy2005}
{Barthelmy}, S.~D., {Chincarini}, G., {Burrows}, D.~N., {et~al.} 2005, \nat,
  438, 994

\bibitem[{{Bauer} {et~al.}(2017){Bauer}, {Treister}, {Schawinski}, {Schulze},
  {Luo}, {Alexander}, {Brandt}, {Comastri}, {Forster}, {Gilli}, {Kann},
  {Maeda}, {Nomoto}, {Paolillo}, {Ranalli}, {Schneider}, {Shemmer}, {Tanaka},
  {Tolstov}, {Tominaga}, {Tozzi}, {Vignali}, {Wang}, {Xue}, \&
  {Yang}}]{Bauer2017}
{Bauer}, F.~E., {Treister}, E., {Schawinski}, K., {et~al.} 2017, \mnras, 467,
  4841

\bibitem[{{Berger}(2006)}]{Berger2006}
{Berger}, E. 2006, \apj, 648, 629

\bibitem[{{Berger}(2014)}]{Berger2014}
{Berger}, E. 2014, \araa, 52, 43

\bibitem[{{Bernardini} {et~al.}(2012){Bernardini}, {Margutti}, {Mao},
  {Zaninoni}, \& {Chincarini}}]{Bernardini2012}
{Bernardini}, M.~G., {Margutti}, R., {Mao}, J., {Zaninoni}, E., \&
  {Chincarini}, G. 2012, \aap, 539, A3

\bibitem[{{Beuther} {et~al.}(2017){Beuther}, {Meidt}, {Schinnerer}, {Paladino},
  \& {Leroy}}]{Beuther2017}
{Beuther}, H., {Meidt}, S., {Schinnerer}, E., {Paladino}, R., \& {Leroy}, A.
  2017, \aap, 597, A85

\bibitem[{{Bianchi} {et~al.}(2011){Bianchi}, {Herald}, {Efremova}, {Girardi},
  {Zabot}, {Marigo}, {Conti}, \& {Shiao}}]{Bianchi2011}
{Bianchi}, L., {Herald}, J., {Efremova}, B., {et~al.} 2011, \apss, 335, 161

\bibitem[{{Bloom} {et~al.}(2011){Bloom}, {Giannios}, {Metzger}, {Cenko},
  {Perley}, {Butler}, {Tanvir}, {Levan}, {O'Brien}, {Strubbe}, {De Colle},
  {Ramirez-Ruiz}, {Lee}, {Nayakshin}, {Quataert}, {King}, {Cucchiara},
  {Guillochon}, {Bower}, {Fruchter}, {Morgan}, \& {van der Horst}}]{Bloom2011}
{Bloom}, J.~S., {Giannios}, D., {Metzger}, B.~D., {et~al.} 2011, Science, 333,
  203

\bibitem[{{Bloom} {et~al.}(2002){Bloom}, {Kulkarni}, \&
  {Djorgovski}}]{Bloom2002}
{Bloom}, J.~S., {Kulkarni}, S.~R., \& {Djorgovski}, S.~G. 2002, \aj, 123, 1111

\bibitem[{{Brinchmann} {et~al.}(2004){Brinchmann}, {Charlot}, {White},
  {Tremonti}, {Kauffmann}, {Heckman}, \& {Brinkmann}}]{Brinchmann2004}
{Brinchmann}, J., {Charlot}, S., {White}, S.~D.~M., {et~al.} 2004, \mnras, 351,
  1151

\bibitem[{{Bromberg} {et~al.}(2012){Bromberg}, {Nakar}, {Piran}, \&
  {Sari}}]{Bromberg2012}
{Bromberg}, O., {Nakar}, E., {Piran}, T., \& {Sari}, R. 2012, \apj, 749, 110

\bibitem[{{Buchner} {et~al.}(2014){Buchner}, {Georgakakis}, {Nandra}, {Hsu},
  {Rangel}, {Brightman}, {Merloni}, {Salvato}, {Donley}, \&
  {Kocevski}}]{Buchner2014}
{Buchner}, J., {Georgakakis}, A., {Nandra}, K., {et~al.} 2014, A\&A, 564, A125

\bibitem[{{Burrows} {et~al.}(2006){Burrows}, {Grupe}, {Capalbi}, {Panaitescu},
  {Patel}, {Kouveliotou}, {Zhang}, {M{\'e}sz{\'a}ros}, {Chincarini}, {Gehrels},
  \& {Wijers}}]{Burrows2006}
{Burrows}, D.~N., {Grupe}, D., {Capalbi}, M., {et~al.} 2006, \apj, 653, 468

\bibitem[{Burrows {et~al.}(2003)Burrows, Hill, Nousek, Wells, Short, Turner,
  Citterio, Tagliaferri, \& Chincarini}]{Burrows2003}
Burrows, D.~N., Hill, J.~E., Nousek, J.~A., {et~al.} 2003, 662, 488

\bibitem[{{Buta} {et~al.}(2015){Buta}, {Sheth}, {Athanassoula}, {Bosma},
  {Knapen}, {Laurikainen}, {Salo}, {Elmegreen}, {Ho}, {Zaritsky}, {Courtois},
  {Hinz}, {Mu{\~n}oz-Mateos}, {Kim}, {Regan}, {Gadotti}, {Gil de Paz}, {Laine},
  {Men{\'e}ndez-Delmestre}, {Comer{\'o}n}, {Erroz Ferrer}, {Seibert},
  {Mizusawa}, {Holwerda}, \& {Madore}}]{Buta2015}
{Buta}, R.~J., {Sheth}, K., {Athanassoula}, E., {et~al.} 2015, \apjs, 217, 32

\bibitem[{{Calzetti} {et~al.}(2000){Calzetti}, {Armus}, {Bohlin}, {Kinney},
  {Koornneef}, \& {Storchi-Bergmann}}]{Calzetti2000}
{Calzetti}, D., {Armus}, L., {Bohlin}, R.~C., {et~al.} 2000, \apj, 533, 682

\bibitem[{{Campana} {et~al.}(2011){Campana}, {Lodato}, {D'Avanzo}, {Panagia},
  {Rossi}, {Della Valle}, {Tagliaferri}, {Antonelli}, {Covino}, {Ghirlanda},
  {Ghisellini}, {Melandri}, {Pian}, {Salvaterra}, {Cusumano}, {D'Elia},
  {Fugazza}, {Palazzi}, {Sbarufatti}, \& {Vergani}}]{Campana2011}
{Campana}, S., {Lodato}, G., {D'Avanzo}, P., {et~al.} 2011, \nat, 480, 69

\bibitem[{{Campana} {et~al.}(2006){Campana}, {Mangano}, {Blustin}, {Brown},
  {Burrows}, {Chincarini}, {Cummings}, {Cusumano}, {Della Valle}, {Malesani},
  {M{\'e}sz{\'a}ros}, {Nousek}, {Page}, {Sakamoto}, {Waxman}, {Zhang}, {Dai},
  {Gehrels}, {Immler}, {Marshall}, {Mason}, {Moretti}, {O'Brien}, {Osborne},
  {Page}, {Romano}, {Roming}, {Tagliaferri}, {Cominsky}, {Giommi}, {Godet},
  {Kennea}, {Krimm}, {Angelini}, {Barthelmy}, {Boyd}, {Palmer}, {Wells}, \&
  {White}}]{Campana2006}
{Campana}, S., {Mangano}, V., {Blustin}, A.~J., {et~al.} 2006, \nat, 442, 1008

\bibitem[{{Cappellari} {et~al.}(2011){Cappellari}, {Emsellem}, {Krajnovi{\'c}},
  {McDermid}, {Scott}, {Verdoes Kleijn}, {Young}, {Alatalo}, {Bacon}, {Blitz},
  {Bois}, {Bournaud}, {Bureau}, {Davies}, {Davis}, {de Zeeuw}, {Duc},
  {Khochfar}, {Kuntschner}, {Lablanche}, {Morganti}, {Naab}, {Oosterloo},
  {Sarzi}, {Serra}, \& {Weijmans}}]{Cappellari2011}
{Cappellari}, M., {Emsellem}, E., {Krajnovi{\'c}}, D., {et~al.} 2011, \mnras,
  413, 813

\bibitem[{{Carnall} {et~al.}(2018){Carnall}, {McLure}, {Dunlop}, \&
  {Dav{\'e}}}]{Carnall2018}
{Carnall}, A.~C., {McLure}, R.~J., {Dunlop}, J.~S., \& {Dav{\'e}}, R. 2018,
  \mnras, 480, 4379

\bibitem[{{Cash}(1979)}]{Cash1979}
{Cash}, W. 1979, \apj, 228, 939

\bibitem[{Chang {et~al.}(2015)Chang, van~der Wel, da~Cunha, \& Rix}]{Chang2015}
Chang, Y.-Y., van~der Wel, A., da~Cunha, E., \& Rix, H.-W. 2015, \apjs, 219, 8

\bibitem[{{Chincarini} {et~al.}(2010){Chincarini}, {Mao}, {Margutti},
  {Bernardini}, {Guidorzi}, {Pasotti}, {Giannios}, {Della Valle}, {Moretti},
  {Romano}, {D'Avanzo}, {Cusumano}, \& {Giommi}}]{Chincarini2010}
{Chincarini}, G., {Mao}, J., {Margutti}, R., {et~al.} 2010, \mnras, 406, 2113

\bibitem[{{Choi} {et~al.}(2016){Choi}, {Dotter}, {Conroy}, {Cantiello},
  {Paxton}, \& {Johnson}}]{Choi2016}
{Choi}, J., {Dotter}, A., {Conroy}, C., {et~al.} 2016, \apj, 823, 102

\bibitem[{{Christensen} {et~al.}(2008){Christensen}, {Vreeswijk}, {Sollerman},
  {Th{\"o}ne}, {Le Floc'h}, \& {Wiersema}}]{Christensen2008}
{Christensen}, L., {Vreeswijk}, P.~M., {Sollerman}, J., {et~al.} 2008, \aap,
  490, 45

\bibitem[{{Colbert} \& {Mushotzky}(1999)}]{Colbert1999}
{Colbert}, E. J.~M. \& {Mushotzky}, R.~F. 1999, \apj, 519, 89

\bibitem[{{Colgate} \& {Petschek}(1981)}]{Colgate1981}
{Colgate}, S.~A. \& {Petschek}, A.~G. 1981, \apj, 248, 771

\bibitem[{{Connaughton} {et~al.}(2015){Connaughton}, {Briggs}, {Goldstein},
  {Meegan}, {Paciesas}, {Preece}, {Wilson-Hodge}, {Gibby}, {Greiner}, {Gruber},
  {Jenke}, {Kippen}, {Pelassa}, {Xiong}, {Yu}, {Bhat}, {Burgess}, {Byrne},
  {Fitzpatrick}, {Foley}, {Giles}, {Guiriec}, {van der Horst}, {von Kienlin},
  {McBreen}, {McGlynn}, {Tierney}, \& {Zhang}}]{Connaughton2015}
{Connaughton}, V., {Briggs}, M.~S., {Goldstein}, A., {et~al.} 2015, \apjs, 216,
  32

\bibitem[{{Crnojevi{\'c}} {et~al.}(2016){Crnojevi{\'c}}, {Sand}, {Spekkens},
  {Caldwell}, {Guhathakurta}, {McLeod}, {Seth}, {Simon}, {Strader}, \&
  {Toloba}}]{Crnojevic2016}
{Crnojevi{\'c}}, D., {Sand}, D.~J., {Spekkens}, K., {et~al.} 2016, \apj, 823,
  19

\bibitem[{{Dado} \& {Dar}(2019)}]{Dado2019}
{Dado}, S. \& {Dar}, A. 2019, \apjl, 884, L44

\bibitem[{{Dai} {et~al.}(2018){Dai}, {McKinney}, {Roth}, {Ramirez-Ruiz}, \&
  {Miller}}]{Dai2018}
{Dai}, L., {McKinney}, J.~C., {Roth}, N., {Ramirez-Ruiz}, E., \& {Miller},
  M.~C. 2018, \apjl, 859, L20

\bibitem[{{D{\'a}lya} {et~al.}(2018){D{\'a}lya}, {Galg{\'o}czi}, {Dobos},
  {Frei}, {Heng}, {Macas}, {Messenger}, {Raffai}, \& {de Souza}}]{Dalya2018}
{D{\'a}lya}, G., {Galg{\'o}czi}, G., {Dobos}, L., {et~al.} 2018, \mnras, 479,
  2374

\bibitem[{{Dark Energy Survey Collaboration} {et~al.}(2016){Dark Energy Survey
  Collaboration}, {Abbott}, {Abdalla}, {Aleksi{\'c}}, {Allam}, {Amara},
  {Bacon}, {Balbinot}, {Banerji}, {Bechtol}, {Benoit-L{\'e}vy}, {Bernstein},
  {Bertin}, {Blazek}, {Bonnett}, {Bridle}, {Brooks}, {Brunner}, {Buckley-Geer},
  {Burke}, {Caminha}, {Capozzi}, {Carlsen}, {Carnero-Rosell}, {Carollo},
  {Carrasco-Kind}, {Carretero}, {Castander}, {Clerkin}, {Collett}, {Conselice},
  {Crocce}, {Cunha}, {D'Andrea}, {da Costa}, {Davis}, {Desai}, {Diehl},
  {Dietrich}, {Dodelson}, {Doel}, {Drlica-Wagner}, {Estrada}, {Etherington},
  {Evrard}, {Fabbri}, {Finley}, {Flaugher}, {Foley}, {Fosalba}, {Frieman},
  {Garc{\'\i}a-Bellido}, {Gaztanaga}, {Gerdes}, {Giannantonio}, {Goldstein},
  {Gruen}, {Gruendl}, {Guarnieri}, {Gutierrez}, {Hartley}, {Honscheid}, {Jain},
  {James}, {Jeltema}, {Jouvel}, {Kessler}, {King}, {Kirk}, {Kron}, {Kuehn},
  {Kuropatkin}, {Lahav}, {Li}, {Lima}, {Lin}, {Maia}, {Makler}, {Manera},
  {Maraston}, {Marshall}, {Martini}, {McMahon}, {Melchior}, {Merson}, {Miller},
  {Miquel}, {Mohr}, {Morice-Atkinson}, {Naidoo}, {Neilsen}, {Nichol}, {Nord},
  {Ogando}, {Ostrovski}, {Palmese}, {Papadopoulos}, {Peiris}, {Peoples},
  {Percival}, {Plazas}, {Reed}, {Refregier}, {Romer}, {Roodman}, {Ross},
  {Rozo}, {Rykoff}, {Sadeh}, {Sako}, {S{\'a}nchez}, {Sanchez}, {Santiago},
  {Scarpine}, {Schubnell}, {Sevilla-Noarbe}, {Sheldon}, {Smith}, {Smith},
  {Soares-Santos}, {Sobreira}, {Soumagnac}, {Suchyta}, {Sullivan}, {Swanson},
  {Tarle}, {Thaler}, {Thomas}, {Thomas}, {Tucker}, {Vieira}, {Vikram},
  {Walker}, {Wechsler}, {Weller}, {Wester}, {Whiteway}, {Wilcox}, {Yanny},
  {Zhang}, \& {Zuntz}}]{Abbott2016}
{Dark Energy Survey Collaboration}, {Abbott}, T., {Abdalla}, F.~B., {et~al.}
  2016, \mnras, 460, 1270

\bibitem[{{D'Avanzo} {et~al.}(2018){D'Avanzo}, {Campana}, {Salafia}, {Ghirland
  a}, {Ghisellini}, {Melandri}, {Bernardini}, {Branchesi}, {Chassande-Mottin},
  {Covino}, {D'Elia}, {Nava}, {Salvaterra}, {Tagliaferri}, \&
  {Vergani}}]{DAvanzo2018}
{D'Avanzo}, P., {Campana}, S., {Salafia}, O.~S., {et~al.} 2018, \aap, 613, L1

\bibitem[{{Davis} {et~al.}(2014){Davis}, {Young}, {Crocker}, {Bureau}, {Blitz},
  {Alatalo}, {Emsellem}, {Naab}, {Bayet}, {Bois}, {Bournaud}, {Cappellari},
  {Davies}, {de Zeeuw}, {Duc}, {Khochfar}, {Krajnovi{\'c}}, {Kuntschner},
  {McDermid}, {Morganti}, {Oosterloo}, {Sarzi}, {Scott}, {Serra}, \&
  {Weijmans}}]{Davis2014}
{Davis}, T.~A., {Young}, L.~M., {Crocker}, A.~F., {et~al.} 2014, \mnras, 444,
  3427

\bibitem[{{De Luca} {et~al.}(2021){De Luca}, {Salvaterra}, {Belfiore},
  {Carpano}, {D'Agostino}, {Haberl}, {Israel}, {Law-Green}, {Lisini},
  {Marelli}, {Novara}, {Read}, {Rodriguez-Castillo}, {Rosen}, {Salvetti},
  {Tiengo}, {Vianello}, {Watson}, {Delvaux}, {Dickens}, {Esposito}, {Greiner},
  {H{\"a}mmerle}, {Kreikenbohm}, {Kreykenbohm}, {Oertel}, {Pizzocaro}, {Pye},
  {Sandrelli}, {Stelzer}, {Wilms}, \& {Zagaria}}]{DeLuca2016}
{De Luca}, A., {Salvaterra}, R., {Belfiore}, A., {et~al.} 2021, \aap, 650, A167

\bibitem[{{De Luca} {et~al.}(2020){De Luca}, {Stelzer}, {Burgasser},
  {Pizzocaro}, {Ranalli}, {Raetz}, {Marelli}, {Novara}, {Vignali}, {Belfiore},
  {Esposito}, {Franzetti}, {Fumana}, {Gilli}, {Salvaterra}, \&
  {Tiengo}}]{DeLuca2020}
{De Luca}, A., {Stelzer}, B., {Burgasser}, A.~J., {et~al.} 2020, \aap, 634, L13

\bibitem[{{De Pasquale} {et~al.}(2010){De Pasquale}, {Schady}, {Kuin}, {Page},
  {Curran}, {Zane}, {Oates}, {Holland}, {Breeveld}, {Hoversten}, {Chincarini},
  {Grupe}, {Abdo}, {Ackermann}, {Ajello}, {Axelsson}, {Baldini}, {Ballet},
  {Barbiellini}, {Baring}, {Bastieri}, {Bechtol}, {Bellazzini}, {Berenji},
  {Bissaldi}, {Blandford}, {Bloom}, {Bonamente}, {Borgland}, {Bouvier},
  {Bregeon}, {Brez}, {Briggs}, {Brigida}, {Bruel}, {Burnett}, {Buson},
  {Caliandro}, {Cameron}, {Caraveo}, {Carrigan}, {Casandjian}, {Cecchi},
  {{\c{C}}elik}, {Chekhtman}, {Chiang}, {Ciprini}, {Claus}, {Cohen-Tanugi},
  {Connaughton}, {Conrad}, {Dermer}, {de Angelis}, {de Palma}, {Dingus},
  {Silva}, {Drell}, {Dubois}, {Dumora}, {Farnier}, {Favuzzi}, {Fegan},
  {Fishman}, {Focke}, {Frailis}, {Fukazawa}, {Funk}, {Fusco}, {Gargano},
  {Gasparrini}, {Gehrels}, {Germani}, {Giglietto}, {Giordano}, {Glanzman},
  {Godfrey}, {Granot}, {Greiner}, {Grenier}, {Grove}, {Guillemot}, {Guiriec},
  {Harding}, {Hayashida}, {Hays}, {Horan}, {Hughes}, {Jackson},
  {J{\'o}hannesson}, {Johnson}, {Johnson}, {Kamae}, {Katagiri}, {Kataoka},
  {Kawai}, {Kerr}, {Kippen}, {Kn{\"o}dlseder}, {Kocevski}, {Kuss}, {Lande},
  {Latronico}, {Lemoine-Goumard}, {Longo}, {Loparco}, {Lott}, {Lovellette},
  {Lubrano}, {Makeev}, {Mazziotta}, {McEnery}, {McGlynn}, {Meegan},
  {M{\'e}sz{\'a}ros}, {Meurer}, {Michelson}, {Mitthumsiri}, {Mizuno}, {Monte},
  {Monzani}, {Moretti}, {Morselli}, {Moskalenko}, {Murgia}, {Nolan}, {Norris},
  {Nuss}, {Ohno}, {Ohsugi}, {Omodei}, {Orlando}, {Ormes}, {Paciesas},
  {Paneque}, {Panetta}, {Parent}, {Pelassa}, {Pepe}, {Pesce-Rollins}, {Piron},
  {Porter}, {Preece}, {Rain{\`o}}, {Rando}, {Razzano}, {Reimer}, {Reimer},
  {Reposeur}, {Ritz}, {Rochester}, {Rodriguez}, {Roth}, {Ryde}, {Sadrozinski},
  {Sander}, {Saz Parkinson}, {Scargle}, {Schalk}, {Sgr{\`o}}, {Siskind},
  {Smith}, {Spandre}, {Spinelli}, {Stamatikos}, {Starck}, {Stecker},
  {Strickman}, {Suson}, {Tajima}, {Takahashi}, {Tanaka}, {Thayer}, {Thayer},
  {Thompson}, {Tibaldo}, {Toma}, {Torres}, {Tosti}, {Tramacere}, {Uchiyama},
  {Uehara}, {Usher}, {van der Horst}, {Vasileiou}, {Vilchez}, {Vitale}, {von
  Kienlin}, {Waite}, {Wang}, {Winer}, {Wood}, {Wu}, {Yamazaki}, {Ylinen}, \&
  {Ziegler}}]{DePasquale2010}
{De Pasquale}, M., {Schady}, P., {Kuin}, N.~P.~M., {et~al.} 2010, \apjl, 709,
  L146

\bibitem[{{Dotter}(2016)}]{Dotter2016}
{Dotter}, A. 2016, \apjs, 222, 8

\bibitem[{{Drozdovsky} \& {Karachentsev}(2000)}]{Drozdovsky2000}
{Drozdovsky}, I.~O. \& {Karachentsev}, I.~D. 2000, \aaps, 142, 425

\bibitem[{{Dye} {et~al.}(2018){Dye}, {Lawrence}, {Read}, {Fan}, {Kerr},
  {Varricatt}, {Furnell}, {Edge}, {Irwin}, {Hambly}, {Lucas}, {Almaini},
  {Chambers}, {Green}, {Hewett}, {Liu}, {McGreer}, {Best}, {Zhang}, {Sutorius},
  {Froebrich}, {Magnier}, {Hasinger}, {Lederer}, {Bold}, \& {Tedds}}]{Dye2018}
{Dye}, S., {Lawrence}, A., {Read}, M.~A., {et~al.} 2018, \mnras, 473, 5113

\bibitem[{{Eappachen} {et~al.}(2022){Eappachen}, {Jonker}, {Fraser}, {Torres},
  {Dhillon}, {Marsh}, {Littlefair}, {Quirola-Vasquez}, {Maguire}, {Mata
  Sanchez}, {Cannizzaro}, {Kostrzewa-Rutkowska}, {Wevers}, {Onori},
  {Inkenhaag}, \& {Brennan}}]{Eappachen2022}
{Eappachen}, D., {Jonker}, P.~G., {Fraser}, M., {et~al.} 2022, arXiv e-prints,
  arXiv:2204.10012

\bibitem[{Ehle {et~al.}(2003)Ehle, Breitfellner, Dahlem, Guainazzi, Rodriguez,
  Santos-Lleo, Schartel, \& Tomas}]{Ehle2003}
Ehle, M., Breitfellner, M., Dahlem, M., {et~al.} 2003, Issue, 2, 2003

\bibitem[{{Ensman} \& {Burrows}(1992)}]{Ensman1992}
{Ensman}, L. \& {Burrows}, A. 1992, \apj, 393, 742

\bibitem[{{Espada} {et~al.}(2019){Espada}, {Verley}, {Miura}, {Israel},
  {Henkel}, {Matsushita}, {Vila-Vilaro}, {Ott}, {Morokuma-Matsui}, {Peck},
  {Hirota}, {Aalto}, {Quillen}, {Hogerheijde}, {Neumayer}, {Vlahakis}, {Iono},
  \& {Kohno}}]{Espada2019}
{Espada}, D., {Verley}, S., {Miura}, R.~E., {et~al.} 2019, \apj, 887, 88

\bibitem[{{Evans} {et~al.}(2019){Evans}, {Allen}, {Anderson}, {Budynkiewicz},
  {Burke}, {Chen}, {Civano}, {D'Abrusco}, {Doe}, {Evans}, {Fabbiano}, {Gibbs},
  {Glotfelty}, {Graessle}, {Grier}, {Hain}, {Hall}, {Harbo}, {Houck}, {Lauer},
  {Laurino}, {Lee}, {Martinez-Galarza}, {McCollough}, {McDowell}, {Miller},
  {McLaughlin}, {Morgan}, {Mossman}, {Nguyen}, {Nichols}, {Nowak}, {Paxson},
  {Plummer}, {Primini}, {Rots}, {Siemiginowska}, {Sundheim}, {Tibbetts}, {Van
  Stone}, \& {Zografou}}]{Evans2019}
{Evans}, I.~N., {Allen}, C., {Anderson}, C.~S., {et~al.} 2019, in AAS/High
  Energy Astrophysics Division, Vol.~17, AAS/High Energy Astrophysics Division,
  114.01

\bibitem[{{Evans} {et~al.}(2010){Evans}, Primini, Glotfelty, Anderson,
  Bonaventura, Chen, Davis, Doe, Evans, Fabbiano, Galle, Gibbs, Grier, Hain,
  Hall, Harbo, He, Houck, Karovska, Kashyap, Lauer, McCollough, McDowell,
  Miller, Mitschang, Morgan, Mossman, Nichols, Nowak, Plummer, Refsdal, Rots,
  Siemiginowska, Sundheim, Tibbetts, Stone, Winkelman, \& Zografou}]{Evans2010}
{Evans}, I.~N., Primini, F.~A., Glotfelty, K.~J., {et~al.} 2010, \apjs, 189, 37

\bibitem[{{Evans} {et~al.}(2020{\natexlab{a}}){Evans}, {Primini}, {Miller},
  {Evans}, {Allen}, {Anderson}, {Becker}, {Budynkiewicz}, {Burke}, {Chen},
  {Civano}, {D'Abrusco}, {Doe}, {Fabbiano}, {Martinez Galarza}, {Gibbs},
  {Glotfelty}, {Graessle}, {Grier}, {Hain}, {Hall}, {Harbo}, {Houck}, {Lauer},
  {Laurino}, {Lee}, {McCollough}, {McDowell}, {McLaughlin}, {Morgan},
  {Mossman}, {Nguyen}, {Nichols}, {Nowak}, {Paxson}, {Perdikeas}, {Plummer},
  {Rots}, {Siemiginowska}, {Sundheim}, {Thong}, {Tibbetts}, {Van Stone},
  {Winkelman}, \& {Zografou}}]{Evans2020}
{Evans}, I.~N., {Primini}, F.~A., {Miller}, J.~B., {et~al.} 2020{\natexlab{a}},
  in American Astronomical Society Meeting Abstracts, Vol. 235, American
  Astronomical Society Meeting Abstracts \#235, 154.05

\bibitem[{{Evans} {et~al.}(2009){Evans}, {Beardmore}, {Page}, {Osborne},
  {O'Brien}, {Willingale}, {Starling}, {Burrows}, {Godet}, {Vetere}, {Racusin},
  {Goad}, {Wiersema}, {Angelini}, {Capalbi}, {Chincarini}, {Gehrels}, {Kennea},
  {Margutti}, {Morris}, {Mountford}, {Pagani}, {Perri}, {Romano}, \&
  {Tanvir}}]{Evans2009}
{Evans}, P.~A., {Beardmore}, A.~P., {Page}, K.~L., {et~al.} 2009, \mnras, 397,
  1177

\bibitem[{{Evans} {et~al.}(2007){Evans}, {Beardmore}, {Page}, {Tyler},
  {Osborne}, {Goad}, {O'Brien}, {Vetere}, {Racusin}, {Morris}, {Burrows},
  {Capalbi}, {Perri}, {Gehrels}, \& {Romano}}]{Evans2007}
{Evans}, P.~A., {Beardmore}, A.~P., {Page}, K.~L., {et~al.} 2007, \aap, 469,
  379

\bibitem[{{Evans} {et~al.}(2014){Evans}, {Osborne}, {Beardmore}, {Page},
  {Willingale}, {Mountford}, {Pagani}, {Burrows}, {Kennea}, {Perri},
  {Tagliaferri}, \& {Gehrels}}]{Evans2014}
{Evans}, P.~A., {Osborne}, J.~P., {Beardmore}, A.~P., {et~al.} 2014, \apjs,
  210, 8

\bibitem[{{Evans} {et~al.}(2020{\natexlab{b}}){Evans}, {Page}, {Osborne},
  {Beardmore}, {Willingale}, {Burrows}, {Kennea}, {Perri}, {Capalbi},
  {Tagliaferri}, \& {Cenko}}]{Evans2020b}
{Evans}, P.~A., {Page}, K.~L., {Osborne}, J.~P., {et~al.} 2020{\natexlab{b}},
  \apjs, 247, 54

\bibitem[{Fasano \& Franceschini(1987)}]{Fasano1987}
Fasano, G. \& Franceschini, A. 1987, \mnras, 225, 155

\bibitem[{{Feroci} {et~al.}(2001){Feroci}, {Hurley}, {Duncan}, \&
  {Thompson}}]{Feroci2001}
{Feroci}, M., {Hurley}, K., {Duncan}, R.~C., \& {Thompson}, C. 2001, \apj, 549,
  1021

\bibitem[{{Feroz} \& {Hobson}(2008)}]{Feroz2008}
{Feroz}, F. \& {Hobson}, M.~P. 2008, \mnras, 384, 449

\bibitem[{{Feroz} {et~al.}(2009){Feroz}, {Hobson}, \& {Bridges}}]{Feroz2009}
{Feroz}, F., {Hobson}, M.~P., \& {Bridges}, M. 2009, \mnras, 398, 1601

\bibitem[{{Ferrarese} \& {Merritt}(2000)}]{Ferrarese2000}
{Ferrarese}, L. \& {Merritt}, D. 2000, \apjl, 539, L9

\bibitem[{{Flewelling}(2018)}]{Flewelling2018}
{Flewelling}, H. 2018, in American Astronomical Society Meeting Abstracts, Vol.
  231, American Astronomical Society Meeting Abstracts \#231, 436.01

\bibitem[{{Flewelling} {et~al.}(2020){Flewelling}, {Magnier}, {Chambers},
  {Heasley}, {Holmberg}, {Huber}, {Sweeney}, {Waters}, {Calamida}, {Casertano},
  {Chen}, {Farrow}, {Hasinger}, {Henderson}, {Long}, {Metcalfe}, {Narayan},
  {Nieto-Santisteban}, {Norberg}, {Rest}, {Saglia}, {Szalay}, {Thakar},
  {Tonry}, {Valenti}, {Werner}, {White}, {Denneau}, {Draper}, {Hodapp},
  {Jedicke}, {Kaiser}, {Kudritzki}, {Price}, {Wainscoat}, {Builders},
  {Chastel}, {McLean}, {Postman}, \& {Shiao}}]{Flewelling2016}
{Flewelling}, H.~A., {Magnier}, E.~A., {Chambers}, K.~C., {et~al.} 2020, \apjs,
  251, 7

\bibitem[{{Fong} {et~al.}(2015){Fong}, {Berger}, {Margutti}, \&
  {Zauderer}}]{Fong2015}
{Fong}, W., {Berger}, E., {Margutti}, R., \& {Zauderer}, B.~A. 2015, \apj, 815,
  102

\bibitem[{{Frail} {et~al.}(2001){Frail}, {Kulkarni}, {Sari}, {Djorgovski},
  {Bloom}, {Galama}, {Reichart}, {Berger}, {Harrison}, {Price}, {Yost},
  {Diercks}, {Goodrich}, \& {Chaffee}}]{Frail2001}
{Frail}, D.~A., {Kulkarni}, S.~R., {Sari}, R., {et~al.} 2001, \apjl, 562, L55

\bibitem[{{Gaia Collaboration} {et~al.}(2018){Gaia Collaboration}, {Brown},
  {Vallenari}, {Prusti}, {de Bruijne}, {Babusiaux}, {Bailer-Jones}, {Biermann},
  {Evans}, {Eyer}, \& et~al.}]{Gaia2018}
{Gaia Collaboration}, {Brown}, A.~G.~A., {Vallenari}, A., {et~al.} 2018, \aap,
  616, A1

\bibitem[{{Gaia Collaboration} {et~al.}(2021){Gaia Collaboration}, {Brown},
  {Vallenari}, {Prusti}, {de Bruijne}, {Babusiaux}, {Biermann}, {Creevey},
  {Evans}, {Eyer}, {Hutton}, {Jansen}, {Jordi}, {Klioner}, {Lammers},
  {Lindegren}, {Luri}, {Mignard}, {Panem}, {Pourbaix}, {Randich}, {Sartoretti},
  {Soubiran}, {Walton}, {Arenou}, {Bailer-Jones}, {Bastian}, {Cropper},
  {Drimmel}, {Katz}, {Lattanzi}, {van Leeuwen}, {Bakker}, {Cacciari},
  {Casta{\~n}eda}, {De Angeli}, {Ducourant}, {Fabricius}, {Fouesneau},
  {Fr{\'e}mat}, {Guerra}, {Guerrier}, {Guiraud}, {Jean-Antoine Piccolo},
  {Masana}, {Messineo}, {Mowlavi}, {Nicolas}, {Nienartowicz}, {Pailler},
  {Panuzzo}, {Riclet}, {Roux}, {Seabroke}, {Sordo}, {Tanga}, {Th{\'e}venin},
  {Gracia-Abril}, {Portell}, {Teyssier}, {Altmann}, {Andrae}, {Bellas-Velidis},
  {Benson}, {Berthier}, {Blomme}, {Brugaletta}, {Burgess}, {Busso}, {Carry},
  {Cellino}, {Cheek}, {Clementini}, {Damerdji}, {Davidson}, {Delchambre},
  {Dell'Oro}, {Fern{\'a}ndez-Hern{\'a}ndez}, {Galluccio}, {Garc{\'\i}a-Lario},
  {Garcia-Reinaldos}, {Gonz{\'a}lez-N{\'u}{\~n}ez}, {Gosset}, {Haigron},
  {Halbwachs}, {Hambly}, {Harrison}, {Hatzidimitriou}, {Heiter},
  {Hern{\'a}ndez}, {Hestroffer}, {Hodgkin}, {Holl}, {Jan{\ss}en}, {Jevardat de
  Fombelle}, {Jordan}, {Krone-Martins}, {Lanzafame}, {L{\"o}ffler}, {Lorca},
  {Manteiga}, {Marchal}, {Marrese}, {Moitinho}, {Mora}, {Muinonen}, {Osborne},
  {Pancino}, {Pauwels}, {Petit}, {Recio-Blanco}, {Richards}, {Riello},
  {Rimoldini}, {Robin}, {Roegiers}, {Rybizki}, {Sarro}, {Siopis}, {Smith},
  {Sozzetti}, {Ulla}, {Utrilla}, {van Leeuwen}, {van Reeven}, {Abbas}, {Abreu
  Aramburu}, {Accart}, {Aerts}, {Aguado}, {Ajaj}, {Altavilla}, {{\'A}lvarez},
  {{\'A}lvarez Cid-Fuentes}, {Alves}, {Anderson}, {Anglada Varela}, {Antoja},
  {Audard}, {Baines}, {Baker}, {Balaguer-N{\'u}{\~n}ez}, {Balbinot}, {Balog},
  {Barache}, {Barbato}, {Barros}, {Barstow}, {Bartolom{\'e}}, {Bassilana},
  {Bauchet}, {Baudesson-Stella}, {Becciani}, {Bellazzini}, {Bernet}, {Bertone},
  {Bianchi}, {Blanco-Cuaresma}, {Boch}, {Bombrun}, {Bossini}, {Bouquillon},
  {Bragaglia}, {Bramante}, {Breedt}, {Bressan}, {Brouillet}, {Bucciarelli},
  {Burlacu}, {Busonero}, {Butkevich}, {Buzzi}, {Caffau}, {Cancelliere},
  {C{\'a}novas}, {Cantat-Gaudin}, {Carballo}, {Carlucci}, {Carnerero},
  {Carrasco}, {Casamiquela}, {Castellani}, {Castro-Ginard}, {Castro Sampol},
  {Chaoul}, {Charlot}, {Chemin}, {Chiavassa}, {Cioni}, {Comoretto}, {Cooper},
  {Cornez}, {Cowell}, {Crifo}, {Crosta}, {Crowley}, {Dafonte}, {Dapergolas},
  {David}, {David}, {de Laverny}, {De Luise}, {De March}, {De Ridder}, {de
  Souza}, {de Teodoro}, {de Torres}, {del Peloso}, {del Pozo}, {Delbo},
  {Delgado}, {Delgado}, {Delisle}, {Di Matteo}, {Diakite}, {Diener},
  {Distefano}, {Dolding}, {Eappachen}, {Edvardsson}, {Enke}, {Esquej}, {Fabre},
  {Fabrizio}, {Faigler}, {Fedorets}, {Fernique}, {Fienga}, {Figueras},
  {Fouron}, {Fragkoudi}, {Fraile}, {Franke}, {Gai}, {Garabato},
  {Garcia-Gutierrez}, {Garc{\'\i}a-Torres}, {Garofalo}, {Gavras}, {Gerlach},
  {Geyer}, {Giacobbe}, {Gilmore}, {Girona}, {Giuffrida}, {Gomel}, {Gomez},
  {Gonzalez-Santamaria}, {Gonz{\'a}lez-Vidal}, {Granvik},
  {Guti{\'e}rrez-S{\'a}nchez}, {Guy}, {Hauser}, {Haywood}, {Helmi}, {Hidalgo},
  {Hilger}, {H{\l}adczuk}, {Hobbs}, {Holland}, {Huckle}, {Jasniewicz},
  {Jonker}, {Juaristi Campillo}, {Julbe}, {Karbevska}, {Kervella}, {Khanna},
  {Kochoska}, {Kontizas}, {Kordopatis}, {Korn}, {Kostrzewa-Rutkowska},
  {Kruszy{\'n}ska}, {Lambert}, {Lanza}, {Lasne}, {Le Campion}, {Le Fustec},
  {Lebreton}, {Lebzelter}, {Leccia}, {Leclerc}, {Lecoeur-Taibi}, {Liao},
  {Licata}, {Lindstr{\o}m}, {Lister}, {Livanou}, {Lobel}, {Madrero Pardo},
  {Managau}, {Mann}, {Marchant}, {Marconi}, {Marcos Santos}, {Marinoni},
  {Marocco}, {Marshall}, {Martin Polo}, {Mart{\'\i}n-Fleitas}, {Masip},
  {Massari}, {Mastrobuono-Battisti}, {Mazeh}, {McMillan}, {Messina},
  {Michalik}, {Millar}, {Mints}, {Molina}, {Molinaro}, {Moln{\'a}r},
  {Montegriffo}, {Mor}, {Morbidelli}, {Morel}, {Morris}, {Mulone}, {Munoz},
  {Muraveva}, {Murphy}, {Musella}, {Noval}, {Ord{\'e}novic}, {Orr{\`u}},
  {Osinde}, {Pagani}, {Pagano}, {Palaversa}, {Palicio}, {Panahi}, {Pawlak},
  {Pe{\~n}alosa Esteller}, {Penttil{\"a}}, {Piersimoni}, {Pineau}, {Plachy},
  {Plum}, {Poggio}, {Poretti}, {Poujoulet}, {Pr{\v{s}}a}, {Pulone}, {Racero},
  {Ragaini}, {Rainer}, {Raiteri}, {Rambaux}, {Ramos}, {Ramos-Lerate}, {Re
  Fiorentin}, {Regibo}, {Reyl{\'e}}, {Ripepi}, {Riva}, {Rixon}, {Robichon},
  {Robin}, {Roelens}, {Rohrbasser}, {Romero-G{\'o}mez}, {Rowell}, {Royer},
  {Rybicki}, {Sadowski}, {Sagrist{\`a} Sell{\'e}s}, {Sahlmann}, {Salgado},
  {Salguero}, {Samaras}, {Sanchez Gimenez}, {Sanna}, {Santove{\~n}a},
  {Sarasso}, {Schultheis}, {Sciacca}, {Segol}, {Segovia}, {S{\'e}gransan},
  {Semeux}, {Shahaf}, {Siddiqui}, {Siebert}, {Siltala}, {Slezak}, {Smart},
  {Solano}, {Solitro}, {Souami}, {Souchay}, {Spagna}, {Spoto}, {Steele},
  {Steidelm{\"u}ller}, {Stephenson}, {S{\"u}veges}, {Szabados}, {Szegedi-Elek},
  {Taris}, {Tauran}, {Taylor}, {Teixeira}, {Thuillot}, {Tonello}, {Torra},
  {Torra}, {Turon}, {Unger}, {Vaillant}, {van Dillen}, {Vanel}, {Vecchiato},
  {Viala}, {Vicente}, {Voutsinas}, {Weiler}, {Wevers}, {Wyrzykowski}, {Yoldas},
  {Yvard}, {Zhao}, {Zorec}, {Zucker}, {Zurbach}, \& {Zwitter}}]{Gaia_DR3_2020}
{Gaia Collaboration}, {Brown}, A.~G.~A., {Vallenari}, A., {et~al.} 2021, \aap,
  649, A1

\bibitem[{{Galbany} {et~al.}(2014){Galbany}, {Stanishev}, {Mour{\~a}o},
  {Rodrigues}, {Flores}, {Garc{\'\i}a-Benito}, {Mast}, {Mendoza},
  {S{\'a}nchez}, {Badenes}, {Barrera-Ballesteros}, {Bland-Hawthorn},
  {Falc{\'o}n-Barroso}, {Garc{\'\i}a-Lorenzo}, {Gomes}, {Gonz{\'a}lez Delgado},
  {Kehrig}, {Lyubenova}, {L{\'o}pez-S{\'a}nchez}, {de Lorenzo-C{\'a}ceres},
  {Marino}, {Meidt}, {Moll{\'a}}, {Papaderos}, {P{\'e}rez-Torres},
  {Rosales-Ortega}, \& {van de Ven}}]{Galbany2014}
{Galbany}, L., {Stanishev}, V., {Mour{\~a}o}, A.~M., {et~al.} 2014, \aap, 572,
  A38

\bibitem[{{Garc{\'\i}a-Alvarez} {et~al.}(2008){Garc{\'\i}a-Alvarez}, {Drake},
  {Kashyap}, {Lin}, \& {Ball}}]{Garcia2008}
{Garc{\'\i}a-Alvarez}, D., {Drake}, J.~J., {Kashyap}, V.~L., {Lin}, L., \&
  {Ball}, B. 2008, \apj, 679, 1509

\bibitem[{{Gendre} {et~al.}(2013){Gendre}, {Stratta}, {Atteia}, {Basa},
  {Bo{\"e}r}, {Coward}, {Cutini}, {D'Elia}, {Howell}, {Klotz}, \&
  {Piro}}]{Gendre2013}
{Gendre}, B., {Stratta}, G., {Atteia}, J.~L., {et~al.} 2013, \apj, 766, 30

\bibitem[{{Glennie} {et~al.}(2015){Glennie}, {Jonker}, {Fender}, {Nagayama}, \&
  {Pretorius}}]{Glennie2015}
{Glennie}, A., {Jonker}, P.~G., {Fender}, R.~P., {Nagayama}, T., \&
  {Pretorius}, M.~L. 2015, \mnras, 450, 3765

\bibitem[{G{\"o}{\u{g}}{\"u}{\c{s}} {et~al.}(1999)G{\"o}{\u{g}}{\"u}{\c{s}},
  Woods, Kouveliotou, van Paradijs, Briggs, Duncan, \& Thompson}]{Gougucs1999}
G{\"o}{\u{g}}{\"u}{\c{s}}, E., Woods, P.~M., Kouveliotou, C., {et~al.} 1999,
  \apjl, 526, L93

\bibitem[{{Gonz{\'a}lez-Fern{\'a}ndez}
  {et~al.}(2018){Gonz{\'a}lez-Fern{\'a}ndez}, {Hodgkin}, {Irwin},
  {Gonz{\'a}lez-Solares}, {Koposov}, {Lewis}, {Emerson}, {Hewett},
  {Yolda{\c{s}}}, \& {Riello}}]{Gonzalez2018}
{Gonz{\'a}lez-Fern{\'a}ndez}, C., {Hodgkin}, S.~T., {Irwin}, M.~J., {et~al.}
  2018, \mnras, 474, 5459

\bibitem[{{Granot} {et~al.}(2018{\natexlab{a}}){Granot}, {De Colle}, \&
  {Ramirez-Ruiz}}]{Granot2018b}
{Granot}, J., {De Colle}, F., \& {Ramirez-Ruiz}, E. 2018{\natexlab{a}}, \mnras,
  481, 2711

\bibitem[{{Granot} {et~al.}(2018{\natexlab{b}}){Granot}, {Gill}, {Guetta}, \&
  {De Colle}}]{Granot2018a}
{Granot}, J., {Gill}, R., {Guetta}, D., \& {De Colle}, F. 2018{\natexlab{b}},
  \mnras, 481, 1597

\bibitem[{{Granot} {et~al.}(2002){Granot}, Panaitescu, Kumar, \&
  Woosley}]{Granot2002}
{Granot}, J., Panaitescu, A., Kumar, P., \& Woosley, S.~E. 2002, \apjl, 570,
  L61

\bibitem[{{G{\"u}del} \& {Naz{\'e}}(2009)}]{Gudel2009}
{G{\"u}del}, M. \& {Naz{\'e}}, Y. 2009, \aapr, 17, 309

\bibitem[{Guetta {et~al.}(2005)Guetta, Piran, \& Waxman}]{Guetta2005}
Guetta, D., Piran, T., \& Waxman, E. 2005, \apj, 619, 412

\bibitem[{{Guo} {et~al.}(2013){Guo}, {Ferguson}, {Giavalisco}, {Barro},
  {Willner}, {Ashby}, {Dahlen}, {Donley}, {Faber}, {Fontana}, {Galametz},
  {Grazian}, {Huang}, {Kocevski}, {Koekemoer}, {Koo}, {McGrath}, {Peth},
  {Salvato}, {Wuyts}, {Castellano}, {Cooray}, {Dickinson}, {Dunlop}, {Fazio},
  {Gardner}, {Gawiser}, {Grogin}, {Hathi}, {Hsu}, {Lee}, {Lucas}, {Mobasher},
  {Nandra}, {Newman}, \& {van der Wel}}]{Guo2013}
{Guo}, Y., {Ferguson}, H.~C., {Giavalisco}, M., {et~al.} 2013, \apjs, 207, 24

\bibitem[{{Hajela} {et~al.}(2022){Hajela}, {Margutti}, {Bright}, {Alexander},
  {Metzger}, {Nedora}, {Kathirgamaraju}, {Margalit}, {Radice}, {Guidorzi},
  {Berger}, {MacFadyen}, {Giannios}, {Chornock}, {Heywood}, {Sironi},
  {Gottlieb}, {Coppejans}, {Laskar}, {Cendes}, {Duran}, {Eftekhari}, {Fong},
  {McDowell}, {Nicholl}, {Xie}, {Zrake}, {Bernuzzi}, {Broekgaarden},
  {Kilpatrick}, {Terreran}, {Villar}, {Blanchard}, {Gomez}, {Hosseinzadeh},
  {Matthews}, \& {Rastinejad}}]{Hajela2021}
{Hajela}, A., {Margutti}, R., {Bright}, J.~S., {et~al.} 2022, \apjl, 927, L17

\bibitem[{{Hawley} {et~al.}(2002){Hawley}, {Covey}, {Knapp}, {Golimowski},
  {Fan}, {Anderson}, {Gunn}, {Harris}, {Ivezi{\'c}}, {Long}, {Lupton},
  {McGehee}, {Narayanan}, {Peng}, {Schlegel}, {Schneider}, {Spahn}, {Strauss},
  {Szkody}, {Tsvetanov}, {Walkowicz}, {Brinkmann}, {Harvanek}, {Hennessy},
  {Kleinman}, {Krzesinski}, {Long}, {Neilsen}, {Newman}, {Nitta}, {Snedden}, \&
  {York}}]{Hawley2002}
{Hawley}, S.~L., {Covey}, K.~R., {Knapp}, G.~R., {et~al.} 2002, \aj, 123, 3409

\bibitem[{Heida {et~al.}(2017)Heida, Jonker, Torres, \& Chiavassa}]{Heida2017}
Heida, M., Jonker, P.~G., Torres, M. A.~P., \& Chiavassa, A. 2017, \apj, 846,
  132

\bibitem[{{Helou} {et~al.}(1991){Helou}, {Madore}, {Schmitz}, {Bicay}, {Wu}, \&
  {Bennett}}]{Helou1991}
{Helou}, G., {Madore}, B.~F., {Schmitz}, M., {et~al.} 1991, Astrophysics and
  Space Science Library, Vol. 171, {The NASA/IPAC extragalactic database.}, ed.
  M.~A. {Albrecht} \& D.~{Egret}, 89--106

\bibitem[{{Hewett} {et~al.}(2006){Hewett}, {Warren}, {Leggett}, \&
  {Hodgkin}}]{Hewett2006}
{Hewett}, P.~C., {Warren}, S.~J., {Leggett}, S.~K., \& {Hodgkin}, S.~T. 2006,
  \mnras, 367, 454

\bibitem[{Hickox \& Markevitch(2006)}]{Hickox2006}
Hickox, R.~C. \& Markevitch, M. 2006, \apj, 645, 95

\bibitem[{Hill {et~al.}(2000)Hill, Zugger, Shoemaker, Witherite, Koch, Chou,
  Case, \& Burrows}]{Hill2000}
Hill, J.~E., Zugger, M.~E., Shoemaker, J., {et~al.} 2000, in X-Ray and
  Gamma-Ray Instrumentation for Astronomy XI, Vol. 4140, International Society
  for Optics and Photonics, 87--98

\bibitem[{{Hjorth} \& {Bloom}(2012)}]{Hjorth2012}
{Hjorth}, J. \& {Bloom}, J.~S. 2012, {The Gamma-Ray Burst - Supernova
  Connection}, 169--190

\bibitem[{{Ho} {et~al.}(2020){Ho}, {Kulkarni}, {Perley}, {Cenko}, {Corsi},
  {Schulze}, {Lunnan}, {Sollerman}, {Gal-Yam}, {Anand}, {Barbarino}, {Bellm},
  {Bruch}, {Burns}, {De}, {Dekany}, {Delacroix}, {Duev}, {Frederiks},
  {Fremling}, {Goldstein}, {Golkhou}, {Graham}, {Hale}, {Kasliwal}, {Kupfer},
  {Laher}, {Martikainen}, {Masci}, {Neill}, {Ridnaia}, {Rusholme}, {Savchenko},
  {Shupe}, {Soumagnac}, {Strotjohann}, {Svinkin}, {Taggart}, {Tartaglia},
  {Yan}, \& {Zolkower}}]{Ho2020}
{Ho}, A. Y.~Q., {Kulkarni}, S.~R., {Perley}, D.~A., {et~al.} 2020, \apj, 902,
  86

\bibitem[{{Hurley} {et~al.}(2011){Hurley}, {Atteia}, {Barraud},
  {P{\'e}langeon}, {Bo{\"e}r}, {Vanderspek}, {Ricker}, {Mazets}, {Golenetskii},
  {Frederiks}, {Pal'shin}, {Aptekar}, {Smith}, {Wigger}, {Hajdas}, {Rau}, {von
  Kienlin}, {Mitrofanov}, {Golovin}, {Kozyrev}, {Litvak}, {Sanin}, {Boynton},
  {Fellows}, {Harshman}, {Barthelmy}, {Cline}, {Cummings}, {Gehrels}, {Krimm},
  {Yamaoka}, {Fukazawa}, {Hanabata}, {Ohno}, {Takahashi}, {Tashiro}, {Terada},
  {Murakami}, {Makishima}, {Guidorzi}, {Frontera}, {Montanari}, {Rossi},
  {Trombka}, {McClanahan}, {Starr}, {Goldsten}, \& {Gold}}]{Hurley2011}
{Hurley}, K., {Atteia}, J.~L., {Barraud}, C., {et~al.} 2011, \apjs, 197, 34

\bibitem[{Hurley {et~al.}(1999)Hurley, Cline, Mazets, Barthelmy, Butterworth,
  Marshall, Palmer, Aptekar, Golenetskii, Il'Inskii, {et~al.}}]{Hurley1999}
Hurley, K., Cline, T., Mazets, E., {et~al.} 1999, Nature, 397, 41

\bibitem[{{Ide} {et~al.}(2020){Ide}, {Hayashida}, {Noda}, {Kurubi}, {Yoneyama},
  \& {Matsumoto}}]{Ide2020}
{Ide}, S., {Hayashida}, K., {Noda}, H., {et~al.} 2020, \pasj, 72, 40

\bibitem[{{Im} {et~al.}(2017){Im}, {Yoon}, {Lee}, {Lee}, {Kim}, {Lee}, {Kim},
  {Troja}, {Choi}, {Lim}, {Ko}, \& {Shim}}]{Im2017}
{Im}, M., {Yoon}, Y., {Lee}, S.-K.~J., {et~al.} 2017, \apjl, 849, L16

\bibitem[{Irwin {et~al.}(2016)Irwin, Maksym, Sivakoff, Romanowsky, Lin,
  Speegle, Prado, Mildebrath, Strader, Liu, {et~al.}}]{Irwin2016}
Irwin, J.~A., Maksym, W.~P., Sivakoff, G.~R., {et~al.} 2016, Nature, 538, 356

\bibitem[{{Ishak}(2017)}]{Ishak2014}
{Ishak}, B. 2017, Contemporary Physics, 58, 99

\bibitem[{{Israel} {et~al.}(2005){Israel}, {Belloni}, {Stella}, {Rephaeli},
  {Gruber}, {Casella}, {Dall'Osso}, {Rea}, {Persic}, \&
  {Rothschild}}]{Israel2005}
{Israel}, G.~L., {Belloni}, T., {Stella}, L., {et~al.} 2005, \apjl, 628, L53

\bibitem[{Ivezi{\'c} {et~al.}(2014)Ivezi{\'c}, Connolly, VanderPlas, \&
  Gray}]{Ivezic2014}
Ivezi{\'c}, {\v{Z}}., Connolly, A.~J., VanderPlas, J.~T., \& Gray, A. 2014,
  Statistics, data mining, and machine learning in astronomy (Princeton
  University Press)

\bibitem[{{Izzo} {et~al.}(2020){Izzo}, {Auchettl}, {Hjorth}, {De Colle},
  {Gall}, {Angus}, {Raimundo}, \& {Ramirez-Ruiz}}]{Izzo2020}
{Izzo}, L., {Auchettl}, K., {Hjorth}, J., {et~al.} 2020, \aap, 639, L11

\bibitem[{Johnstone {et~al.}(2005)Johnstone, Fabian, Morris, \&
  Taylor}]{Johnstone2005}
Johnstone, R.~M., Fabian, A.~C., Morris, R.~G., \& Taylor, G.~B. 2005, \mnras,
  356, 237

\bibitem[{{Jonker} {et~al.}(2013){Jonker}, {Glennie}, {Heida}, {Maccarone},
  {Hodgkin}, {Nelemans}, {Miller-Jones}, {Torres}, \& {Fender}}]{Jonker2013}
{Jonker}, P.~G., {Glennie}, A., {Heida}, M., {et~al.} 2013, \apj, 779, 14

\bibitem[{{Jonker} {et~al.}(2012{\natexlab{a}}){Jonker}, {Heida}, {Torres},
  {Miller-Jones}, {Fabian}, {Ratti}, {Miniutti}, {Walton}, \&
  {Roberts}}]{Jonker2012}
{Jonker}, P.~G., {Heida}, M., {Torres}, M.~A.~P., {et~al.} 2012{\natexlab{a}},
  \apj, 758, 28

\bibitem[{{Jonker} {et~al.}(2012{\natexlab{b}}){Jonker}, {Miller-Jones},
  {Homan}, {Tomsick}, {Fender}, {Kaaret}, {Markoff}, \& {Gallo}}]{Jonker2012c}
{Jonker}, P.~G., {Miller-Jones}, J.~C.~A., {Homan}, J., {et~al.}
  2012{\natexlab{b}}, \mnras, 423, 3308

\bibitem[{{Kaaret} {et~al.}(2017){Kaaret}, {Feng}, \& {Roberts}}]{Kaaret2017}
{Kaaret}, P., {Feng}, H., \& {Roberts}, T.~P. 2017, \araa, 55, 303

\bibitem[{{Kaaret} {et~al.}(2006){Kaaret}, {Simet}, \& {Lang}}]{Kaaret2006}
{Kaaret}, P., {Simet}, M.~G., \& {Lang}, C.~C. 2006, \apj, 646, 174

\bibitem[{{Kaastra}(2017)}]{Kaastra2017a}
{Kaastra}, J.~S. 2017, \aap, 605, A51

\bibitem[{{Kalberla} {et~al.}(2005){Kalberla}, {Burton}, {Hartmann}, {Arnal},
  {Bajaja}, {Morras}, \& {P{\"o}ppel}}]{Kalberla2005}
{Kalberla}, P.~M.~W., {Burton}, W.~B., {Hartmann}, D., {et~al.} 2005, \aap,
  440, 775

\bibitem[{{Kalberla} \& {Haud}(2015)}]{Kalberla2015}
{Kalberla}, P.~M.~W. \& {Haud}, U. 2015, \aap, 578, A78

\bibitem[{Kolmogorov(1933)}]{Kolmogorov1933}
Kolmogorov, A. 1933, Inst. Ital. Attuari, Giorn., 4, 83

\bibitem[{{Kraft} {et~al.}(1991){Kraft}, {Burrows}, \& {Nousek}}]{Kraft1991}
{Kraft}, R.~P., {Burrows}, D.~N., \& {Nousek}, J.~A. 1991, \apj, 374, 344

\bibitem[{Krishnamoorthy \& Thomson(2004)}]{Krishnamoorthy2004}
Krishnamoorthy, K. \& Thomson, J. 2004, Journal of Statistical Planning and
  Inference, 119, 23

\bibitem[{{Kr{\"u}hler} {et~al.}(2017){Kr{\"u}hler}, {Kuncarayakti}, {Schady},
  {Anderson}, {Galbany}, \& {Gensior}}]{Kruhler2017}
{Kr{\"u}hler}, T., {Kuncarayakti}, H., {Schady}, P., {et~al.} 2017, \aap, 602,
  A85

\bibitem[{{Lamb} {et~al.}(2021){Lamb}, {Fern{\'a}ndez}, {Hayes}, {Kong}, {Lin},
  {Tanvir}, {Hendry}, {Heng}, {Saha}, \& {Veitch}}]{Lamb2021}
{Lamb}, G.~P., {Fern{\'a}ndez}, J.~J., {Hayes}, F., {et~al.} 2021, Universe, 7,
  329

\bibitem[{{Lanz} {et~al.}(2013){Lanz}, {Zezas}, {Brassington}, {Smith},
  {Ashby}, {da Cunha}, {Fazio}, {Hayward}, {Hernquist}, \&
  {Jonsson}}]{Lanz2013}
{Lanz}, L., {Zezas}, A., {Brassington}, N., {et~al.} 2013, \apj, 768, 90

\bibitem[{{Levan} {et~al.}(2011){Levan}, {Tanvir}, {Cenko}, {Perley},
  {Wiersema}, {Bloom}, {Fruchter}, {de Ugarte Postigo}, {O'Brien}, {Butler},
  {van der Horst}, {Leloudas}, {Morgan}, {Misra}, {Bower}, {Farihi},
  {Tunnicliffe}, {Modjaz}, {Silverman}, {Hjorth}, {Th{\"o}ne}, {Cucchiara},
  {Cer{\'o}n}, {Castro-Tirado}, {Arnold}, {Bremer}, {Brodie}, {Carroll},
  {Cooper}, {Curran}, {Cutri}, {Ehle}, {Forbes}, {Fynbo}, {Gorosabel},
  {Graham}, {Hoffman}, {Guziy}, {Jakobsson}, {Kamble}, {Kerr}, {Kasliwal},
  {Kouveliotou}, {Kocevski}, {Law}, {Nugent}, {Ofek}, {Poznanski}, {Quimby},
  {Rol}, {Romanowsky}, {S{\'a}nchez-Ram{\'\i}rez}, {Schulze}, {Singh}, {van
  Spaandonk}, {Starling}, {Strom}, {Tello}, {Vaduvescu}, {Wheatley}, {Wijers},
  {Winters}, \& {Xu}}]{Levan2011}
{Levan}, A.~J., {Tanvir}, N.~R., {Cenko}, S.~B., {et~al.} 2011, Science, 333,
  199

\bibitem[{{Levan} {et~al.}(2014){Levan}, {Tanvir}, {Starling}, {Wiersema},
  {Page}, {Perley}, {Schulze}, {Wynn}, {Chornock}, {Hjorth}, {Cenko},
  {Fruchter}, {O'Brien}, {Brown}, {Tunnicliffe}, {Malesani}, {Jakobsson},
  {Watson}, {Berger}, {Bersier}, {Cobb}, {Covino}, {Cucchiara}, {de Ugarte
  Postigo}, {Fox}, {Gal-Yam}, {Goldoni}, {Gorosabel}, {Kaper}, {Kr{\"u}hler},
  {Karjalainen}, {Osborne}, {Pian}, {S{\'a}nchez-Ram{\'\i}rez}, {Schmidt},
  {Skillen}, {Tagliaferri}, {Th{\"o}ne}, {Vaduvescu}, {Wijers}, \&
  {Zauderer}}]{Levan2014}
{Levan}, A.~J., {Tanvir}, N.~R., {Starling}, R.~L.~C., {et~al.} 2014, \apj,
  781, 13

\bibitem[{{Levesque}(2014)}]{Levesque2014}
{Levesque}, E.~M. 2014, \pasp, 126, 1

\bibitem[{{Li} {et~al.}(2018){Li}, {Wu}, {Lei}, {Dai}, {Liang}, \&
  {Ryde}}]{Li2018}
{Li}, L., {Wu}, X.-F., {Lei}, W.-H., {et~al.} 2018, \apjs, 236, 26

\bibitem[{{Li} {et~al.}(2016){Li}, {Zhang}, \& {L{\"u}}}]{Li2016}
{Li}, Y., {Zhang}, B., \& {L{\"u}}, H.-J. 2016, \apjs, 227, 7

\bibitem[{{Lin} {et~al.}(2019){Lin}, {Irwin}, \& {Berger}}]{Lin2019}
{Lin}, D., {Irwin}, J., \& {Berger}, E. 2019, The Astronomer's Telegram, 13171,
  1

\bibitem[{{Lin} {et~al.}(2021){Lin}, {Irwin}, \& {Berger}}]{Lin2021}
{Lin}, D., {Irwin}, J.~A., \& {Berger}, E. 2021, The Astronomer's Telegram,
  14599, 1

\bibitem[{{Lin} {et~al.}(2022){Lin}, {Irwin}, {Berger}, \& {Nguyen}}]{Lin2022}
{Lin}, D., {Irwin}, J.~A., {Berger}, E., \& {Nguyen}, R. 2022, \apj, 927, 211

\bibitem[{{Lin} {et~al.}(2018){Lin}, {Strader}, {Carrasco}, {Page},
  {Romanowsky}, {Homan}, {Irwin}, {Remillard}, {Godet}, {Webb}, {Baumgardt},
  {Wijnands}, {Barret}, {Duc}, {Brodie}, \& {Gwyn}}]{Lin2018}
{Lin}, D., {Strader}, J., {Carrasco}, E.~R., {et~al.} 2018, Nature Astronomy,
  2, 656

\bibitem[{{Lin} {et~al.}(2020){Lin}, {Strader}, {Romanowsky}, {Irwin}, {Godet},
  {Barret}, {Webb}, {Homan}, \& {Remillard}}]{Lin2020}
{Lin}, D., {Strader}, J., {Romanowsky}, A.~J., {et~al.} 2020, \apjl, 892, L25

\bibitem[{{Lin} {et~al.}(2012){Lin}, {Webb}, \& {Barret}}]{Lin2012}
{Lin}, D., {Webb}, N.~A., \& {Barret}, D. 2012, \apj, 756, 27

\bibitem[{{Lindegren} {et~al.}(2018){Lindegren}, {Hern{\'a}ndez}, {Bombrun},
  {Klioner}, {Bastian}, {Ramos-Lerate}, {de Torres}, {Steidelm{\"u}ller},
  {Stephenson}, {Hobbs}, {Lammers}, {Biermann}, {Geyer}, {Hilger}, {Michalik},
  {Stampa}, {McMillan}, {Casta{\~n}eda}, {Clotet}, {Comoretto}, {Davidson},
  {Fabricius}, {Gracia}, {Hambly}, {Hutton}, {Mora}, {Portell}, {van Leeuwen},
  {Abbas}, {Abreu}, {Altmann}, {Andrei}, {Anglada}, {Balaguer-N{\'u}{\~n}ez},
  {Barache}, {Becciani}, {Bertone}, {Bianchi}, {Bouquillon}, {Bourda},
  {Br{\"u}semeister}, {Bucciarelli}, {Busonero}, {Buzzi}, {Cancelliere},
  {Carlucci}, {Charlot}, {Cheek}, {Crosta}, {Crowley}, {de Bruijne}, {de
  Felice}, {Drimmel}, {Esquej}, {Fienga}, {Fraile}, {Gai}, {Garralda},
  {Gonz{\'a}lez-Vidal}, {Guerra}, {Hauser}, {Hofmann}, {Holl}, {Jordan},
  {Lattanzi}, {Lenhardt}, {Liao}, {Licata}, {Lister}, {L{\"o}ffler},
  {Marchant}, {Martin-Fleitas}, {Messineo}, {Mignard}, {Morbidelli}, {Poggio},
  {Riva}, {Rowell}, {Salguero}, {Sarasso}, {Sciacca}, {Siddiqui}, {Smart},
  {Spagna}, {Steele}, {Taris}, {Torra}, {van Elteren}, {van Reeven}, \&
  {Vecchiato}}]{Lindegren2018}
{Lindegren}, L., {Hern{\'a}ndez}, J., {Bombrun}, A., {et~al.} 2018, \aap, 616,
  A2

\bibitem[{{Liu}(2011)}]{Liu2011}
{Liu}, J. 2011, \apjs, 192, 10

\bibitem[{{Lopes de Oliveira} {et~al.}(2007){Lopes de Oliveira}, {Motch},
  {Smith}, {Negueruela}, \& {Torrej\'on}}]{Lopes2007}
{Lopes de Oliveira}, R., {Motch}, C., {Smith}, M.~A., {Negueruela}, I., \&
  {Torrej\'on}, J.~M. 2007, A\&A, 474, 983

\bibitem[{{L{\"u}} {et~al.}(2019){L{\"u}}, {Yuan}, {Lan}, {Zhang}, {Zou}, \&
  {Liang}}]{Lu2019}
{L{\"u}}, H.-J., {Yuan}, Y., {Lan}, L., {et~al.} 2019, arXiv e-prints,
  arXiv:1904.06664

\bibitem[{{L{\"u}} {et~al.}(2015){L{\"u}}, {Zhang}, {Lei}, {Li}, \&
  {Lasky}}]{Lu2015}
{L{\"u}}, H.-J., {Zhang}, B., {Lei}, W.-H., {Li}, Y., \& {Lasky}, P.~D. 2015,
  \apj, 805, 89

\bibitem[{{Luo} {et~al.}(2014){Luo}, {Brandt}, \& {Bauer}}]{Luo2014}
{Luo}, B., {Brandt}, N., \& {Bauer}, F. 2014, The Astronomer's Telegram, 6541,
  1

\bibitem[{{Luo} {et~al.}(2017){Luo}, {Brandt}, {Xue}, {Lehmer}, {Alexander},
  {Bauer}, {Vito}, {Yang}, {Basu-Zych}, {Comastri}, {Gilli}, {Gu},
  {Hornschemeier}, {Koekemoer}, {Liu}, {Mainieri}, {Paolillo}, {Ranalli},
  {Rosati}, {Schneider}, {Shemmer}, {Smail}, {Sun}, {Tozzi}, {Vignali}, \&
  {Wang}}]{Luo2017}
{Luo}, B., {Brandt}, W.~N., {Xue}, Y.~Q., {et~al.} 2017, \apjs, 228, 2

\bibitem[{{Lyons} {et~al.}(2010){Lyons}, {O'Brien}, {Zhang}, {Willingale},
  {Troja}, \& {Starling}}]{Lyons2010}
{Lyons}, N., {O'Brien}, P.~T., {Zhang}, B., {et~al.} 2010, \mnras, 402, 705

\bibitem[{MacLeod {et~al.}(2014)MacLeod, Goldstein, Ramirez-Ruiz, Guillochon,
  \& Samsing}]{MacLeod2014}
MacLeod, M., Goldstein, J., Ramirez-Ruiz, E., Guillochon, J., \& Samsing, J.
  2014, \apj, 794, 9

\bibitem[{{Madau} \& {Dickinson}(2014)}]{Madau2014}
{Madau}, P. \& {Dickinson}, M. 2014, \araa, 52, 415

\bibitem[{{Maguire} {et~al.}(2020){Maguire}, {Eracleous}, {Jonker}, {MacLeod},
  \& {Rosswog}}]{Maguire2020}
{Maguire}, K., {Eracleous}, M., {Jonker}, P.~G., {MacLeod}, M., \& {Rosswog},
  S. 2020, \ssr, 216, 39

\bibitem[{Malyali {et~al.}(2019)Malyali, Rau, \& Nandra}]{Malyali2019}
Malyali, A., Rau, A., \& Nandra, K. 2019, \mnras, 489, 5413

\bibitem[{{Margutti} {et~al.}(2011){Margutti}, {Bernardini}, {Barniol Duran},
  {Guidorzi}, {Shen}, \& {Chincarini}}]{Margutti2011}
{Margutti}, R., {Bernardini}, G., {Barniol Duran}, R., {et~al.} 2011, \mnras,
  410, 1064

\bibitem[{{Margutti} \& {Chornock}(2021)}]{Margutti2021}
{Margutti}, R. \& {Chornock}, R. 2021, \araa, 59

\bibitem[{{Marocco} {et~al.}(2021){Marocco}, {Eisenhardt}, {Fowler},
  {Kirkpatrick}, {Meisner}, {Schlafly}, {Stanford}, {Garcia}, {Caselden},
  {Cushing}, {Cutri}, {Faherty}, {Gelino}, {Gonzalez}, {Jarrett}, {Koontz},
  {Mainzer}, {Marchese}, {Mobasher}, {Schlegel}, {Stern}, {Teplitz}, \&
  {Wright}}]{Marocco2021}
{Marocco}, F., {Eisenhardt}, P. R.~M., {Fowler}, J.~W., {et~al.} 2021, \apjs,
  253, 8

\bibitem[{Massey~Jr(1951)}]{Massey1951}
Massey~Jr, F.~J. 1951, Journal of the American statistical Association, 46, 68

\bibitem[{Mata~S\'anchez {et~al.}(2015)Mata~S\'anchez, Mu$\tilde{n}$oz-Darias,
  Casares, Corral-Santana, \& Shahbaz}]{Mata2015}
Mata~S\'anchez, D., Mu$\tilde{n}$oz-Darias, T., Casares, J., Corral-Santana,
  J.~M., \& Shahbaz, T. 2015, \mnras, 454, 2199

\bibitem[{{Matzner} \& {McKee}(1999)}]{Matzner1999}
{Matzner}, C.~D. \& {McKee}, C.~F. 1999, \apj, 510, 379

\bibitem[{{Mazets} {et~al.}(1999){Mazets}, {Cline}, {Aptekar'}, {Butterworth},
  {Frederiks}, {Golenetskii}, {Il'Inskii}, \& {Pal'Shin}}]{Mazets1999}
{Mazets}, E.~P., {Cline}, T.~L., {Aptekar'}, R.~L., {et~al.} 1999, Astronomy
  Letters, 25, 635

\bibitem[{{Mazzali} {et~al.}(2008){Mazzali}, {Valenti}, {Della Valle},
  {Chincarini}, {Sauer}, {Benetti}, {Pian}, {Piran}, {D'Elia}, {Elias-Rosa},
  {Margutti}, {Pasotti}, {Antonelli}, {Bufano}, {Campana}, {Cappellaro},
  {Covino}, {D'Avanzo}, {Fiore}, {Fugazza}, {Gilmozzi}, {Hunter}, {Maguire},
  {Maiorano}, {Marziani}, {Masetti}, {Mirabel}, {Navasardyan}, {Nomoto},
  {Palazzi}, {Pastorello}, {Panagia}, {Pellizza}, {Sari}, {Smartt},
  {Tagliaferri}, {Tanaka}, {Taubenberger}, {Tominaga}, {Trundle}, \&
  {Turatto}}]{Mazzali2008}
{Mazzali}, P.~A., {Valenti}, S., {Della Valle}, M., {et~al.} 2008, Science,
  321, 1185

\bibitem[{{McMahon} {et~al.}(2013){McMahon}, {Banerji}, {Gonzalez}, {Koposov},
  {Bejar}, {Lodieu}, {Rebolo}, \& {VHS Collaboration}}]{McMahon2013}
{McMahon}, R.~G., {Banerji}, M., {Gonzalez}, E., {et~al.} 2013, The Messenger,
  154, 35

\bibitem[{{Micha{\l}owski} {et~al.}(2014){Micha{\l}owski}, {Hunt}, {Palazzi},
  {Savaglio}, {Gentile}, {Rasmussen}, {Baes}, {Basa}, {Bianchi}, {Berta},
  {Burlon}, {Castro Cer{\'o}n}, {Covino}, {Cuby}, {D'Elia}, {Ferrero},
  {G{\"o}tz}, {Jhorth}, {Koprowski}, {Le Borgne}, {Le Floc'h}, {Malesani},
  {Murphy}, {Pian}, {Piranomonte}, {Rossi}, {Sollerman}, {Tanvir}, {de Ugarte
  Postigo}, {Watson}, {van der Werf}, {Vergani}, \& {Xu}}]{Michalowski2014}
{Micha{\l}owski}, M.~J., {Hunt}, L.~K., {Palazzi}, E., {et~al.} 2014, \aap,
  562, A70

\bibitem[{{Mineo} {et~al.}(2012){Mineo}, {Gilfanov}, \& {Sunyaev}}]{Mineo2012}
{Mineo}, S., {Gilfanov}, M., \& {Sunyaev}, R. 2012, \mnras, 419, 2095

\bibitem[{{Miniutti} {et~al.}(2019){Miniutti}, {Saxton}, {Giustini},
  {Alexander}, {Fender}, {Heywood}, {Monageng}, {Coriat}, {Tzioumis}, {Read},
  {Knigge}, {Gandhi}, {Pretorius}, \& {Ag{\'\i}s-Gonz{\'a}lez}}]{Miniutti2019}
{Miniutti}, G., {Saxton}, R.~D., {Giustini}, M., {et~al.} 2019, \nat, 573, 381

\bibitem[{{Mitra-Kraev} {et~al.}(2005){Mitra-Kraev}, {Harra}, {G{\"u}del},
  {Audard}, {Branduardi-Raymont}, {Kay}, {Mewe}, {Raassen}, \& {van
  Driel-Gesztelyi}}]{Mitra-Kraev2005}
{Mitra-Kraev}, U., {Harra}, L.~K., {G{\"u}del}, M., {et~al.} 2005, \aap, 431,
  679

\bibitem[{{Modjaz} {et~al.}(2009){Modjaz}, {Li}, {Butler}, {Chornock},
  {Perley}, {Blondin}, {Bloom}, {Filippenko}, {Kirshner}, {Kocevski},
  {Poznanski}, {Hicken}, {Foley}, {Stringfellow}, {Berlind}, {Barrado y
  Navascues}, {Blake}, {Bouy}, {Brown}, {Challis}, {Chen}, {de Vries},
  {Dufour}, {Falco}, {Friedman}, {Ganeshalingam}, {Garnavich}, {Holden},
  {Illingworth}, {Lee}, {Liebert}, {Marion}, {Olivier}, {Prochaska},
  {Silverman}, {Smith}, {Starr}, {Steele}, {Stockton}, {Williams}, \&
  {Wood-Vasey}}]{Modjaz2009}
{Modjaz}, M., {Li}, W., {Butler}, N., {et~al.} 2009, \apj, 702, 226

\bibitem[{{Moustakas} {et~al.}(2013){Moustakas}, {Coil}, {Aird}, {Blanton},
  {Cool}, {Eisenstein}, {Mendez}, {Wong}, {Zhu}, \& {Arnouts}}]{Moustakas2013}
{Moustakas}, J., {Coil}, A.~L., {Aird}, J., {et~al.} 2013, \apj, 767, 50

\bibitem[{{Nakar}(2020)}]{Nakar2020}
{Nakar}, E. 2020, \physrep, 886, 1

\bibitem[{{Nakar} \& {Sari}(2010)}]{Nakar2010}
{Nakar}, E. \& {Sari}, R. 2010, \apj, 725, 904

\bibitem[{{Nakar} \& {Sari}(2012)}]{Nakar2012}
{Nakar}, E. \& {Sari}, R. 2012, \apj, 747, 88

\bibitem[{{Nandra} {et~al.}(2013){Nandra}, {Barret}, {Barcons}, {Fabian}, {den
  Herder}, {Piro}, {Watson}, {Adami}, {Aird}, {Afonso}, {Alexander},
  {Argiroffi}, {Amati}, {Arnaud}, {Atteia}, {Audard}, {Badenes}, {Ballet},
  {Ballo}, {Bamba}, {Bhardwaj}, {Stefano Battistelli}, {Becker}, {De Becker},
  {Behar}, {Bianchi}, {Biffi}, {B{\^\i}rzan}, {Bocchino}, {Bogdanov}, {Boirin},
  {Boller}, {Borgani}, {Borm}, {Bouch{\'e}}, {Bourdin}, {Bower}, {Braito},
  {Branchini}, {Branduardi-Raymont}, {Bregman}, {Brenneman}, {Brightman},
  {Br{\"u}ggen}, {Buchner}, {Bulbul}, {Brusa}, {Bursa}, {Caccianiga},
  {Cackett}, {Campana}, {Cappelluti}, {Cappi}, {Carrera}, {Ceballos},
  {Christensen}, {Chu}, {Churazov}, {Clerc}, {Corbel}, {Corral}, {Comastri},
  {Costantini}, {Croston}, {Dadina}, {D'Ai}, {Decourchelle}, {Della Ceca},
  {Dennerl}, {Dolag}, {Done}, {Dovciak}, {Drake}, {Eckert}, {Edge}, {Ettori},
  {Ezoe}, {Feigelson}, {Fender}, {Feruglio}, {Finoguenov}, {Fiore}, {Galeazzi},
  {Gallagher}, {Gandhi}, {Gaspari}, {Gastaldello}, {Georgakakis},
  {Georgantopoulos}, {Gilfanov}, {Gitti}, {Gladstone}, {Goosmann}, {Gosset},
  {Grosso}, {Guedel}, {Guerrero}, {Haberl}, {Hardcastle}, {Heinz}, {Alonso
  Herrero}, {Herv{\'e}}, {Holmstrom}, {Iwasawa}, {Jonker}, {Kaastra}, {Kara},
  {Karas}, {Kastner}, {King}, {Kosenko}, {Koutroumpa}, {Kraft}, {Kreykenbohm},
  {Lallement}, {Lanzuisi}, {Lee}, {Lemoine-Goumard}, {Lobban}, {Lodato},
  {Lovisari}, {Lotti}, {McCharthy}, {McNamara}, {Maggio}, {Maiolino}, {De
  Marco}, {de Martino}, {Mateos}, {Matt}, {Maughan}, {Mazzotta}, {Mendez},
  {Merloni}, {Micela}, {Miceli}, {Mignani}, {Miller}, {Miniutti}, {Molendi},
  {Montez}, {Moretti}, {Motch}, {Naz{\'e}}, {Nevalainen}, {Nicastro}, {Nulsen},
  {Ohashi}, {O'Brien}, {Osborne}, {Oskinova}, {Pacaud}, {Paerels}, {Page},
  {Papadakis}, {Pareschi}, {Petre}, {Petrucci}, {Piconcelli}, {Pillitteri},
  {Pinto}, {de Plaa}, {Pointecouteau}, {Ponman}, {Ponti}, {Porquet}, {Pounds},
  {Pratt}, {Predehl}, {Proga}, {Psaltis}, {Rafferty}, {Ramos-Ceja}, {Ranalli},
  {Rasia}, {Rau}, {Rauw}, {Rea}, {Read}, {Reeves}, {Reiprich}, {Renaud},
  {Reynolds}, {Risaliti}, {Rodriguez}, {Rodriguez Hidalgo}, {Roncarelli},
  {Rosario}, {Rossetti}, {Rozanska}, {Rovilos}, {Salvaterra}, {Salvato}, {Di
  Salvo}, {Sanders}, {Sanz-Forcada}, {Schawinski}, {Schaye}, {Schwope},
  {Sciortino}, {Severgnini}, {Shankar}, {Sijacki}, {Sim}, {Schmid}, {Smith},
  {Steiner}, {Stelzer}, {Stewart}, {Strohmayer}, {Str{\"u}der}, {Sun}, {Takei},
  {Tatischeff}, {Tiengo}, {Tombesi}, {Trinchieri}, {Tsuru}, {Ud-Doula},
  {Ursino}, {Valencic}, {Vanzella}, {Vaughan}, {Vignali}, {Vink}, {Vito},
  {Volonteri}, {Wang}, {Webb}, {Willingale}, {Wilms}, {Wise}, {Worrall},
  {Young}, {Zampieri}, {In't Zand}, {Zane}, {Zezas}, {Zhang}, \&
  {Zhuravleva}}]{Nandra2013}
{Nandra}, K., {Barret}, D., {Barcons}, X., {et~al.} 2013, arXiv e-prints,
  arXiv:1306.2307

\bibitem[{{Narayana Bhat} {et~al.}(2016){Narayana Bhat}, {Meegan}, {von
  Kienlin}, {Paciesas}, {Briggs}, {Burgess}, {Burns}, {Chaplin}, {Cleveland},
  {Collazzi}, {Connaughton}, {Diekmann}, {Fitzpatrick}, {Gibby}, {Giles},
  {Goldstein}, {Greiner}, {Jenke}, {Kippen}, {Kouveliotou}, {Mailyan},
  {McBreen}, {Pelassa}, {Preece}, {Roberts}, {Sparke}, {Stanbro}, {Veres},
  {Wilson-Hodge}, {Xiong}, {Younes}, {Yu}, \& {Zhang}}]{Narayana2016}
{Narayana Bhat}, P., {Meegan}, C.~A., {von Kienlin}, A., {et~al.} 2016, \apjs,
  223, 28

\bibitem[{{Nidever} {et~al.}(2021){Nidever}, {Dey}, {Fasbender}, {Juneau},
  {Meisner}, {Wishart}, {Scott}, {Matt}, {Nikutta}, \& {Pucha}}]{Nidever2020}
{Nidever}, D.~L., {Dey}, A., {Fasbender}, K., {et~al.} 2021, \aj, 161, 192

\bibitem[{{Novara} {et~al.}(2020){Novara}, {Esposito}, {Tiengo}, {Vianello},
  {Salvaterra}, {Belfiore}, {De Luca}, {D'Avanzo}, {Greiner}, {Scodeggio},
  {Rosen}, {Delvaux}, {Pian}, {Campana}, {Lisini}, {Mereghetti}, \&
  {Israel}}]{Novara2020}
{Novara}, G., {Esposito}, P., {Tiengo}, A., {et~al.} 2020, \apj, 898, 37

\bibitem[{{Nynka} {et~al.}(2018){Nynka}, {Ruan}, {Haggard}, \&
  {Evans}}]{Nynka2018}
{Nynka}, M., {Ruan}, J.~J., {Haggard}, D., \& {Evans}, P.~A. 2018, \apjl, 862,
  L19

\bibitem[{{Ochsenbein} {et~al.}(2000){Ochsenbein}, {Bauer}, \&
  {Marcout}}]{Ochsenbein2000}
{Ochsenbein}, F., {Bauer}, P., \& {Marcout}, J. 2000, \aaps, 143, 23

\bibitem[{{Pallavicini} {et~al.}(1990){Pallavicini}, {Tagliaferri}, \&
  {Stella}}]{Pallavicini1990}
{Pallavicini}, R., {Tagliaferri}, G., \& {Stella}, L. 1990, \aap, 228, 403

\bibitem[{Palmer {et~al.}(2005)Palmer, Barthelmy, Gehrels, Kippen, Cayton,
  Kouveliotou, Eichler, Wijers, Woods, Granot, {et~al.}}]{Palmer2005}
Palmer, D.~M., Barthelmy, S., Gehrels, N., {et~al.} 2005, Nature, 434, 1107

\bibitem[{{Pandey} \& {Singh}(2008)}]{Pandey2008}
{Pandey}, J.~C. \& {Singh}, K.~P. 2008, \mnras, 387, 1627

\bibitem[{{Park} {et~al.}(2006){Park}, {Kashyap}, {Siemiginowska}, {van Dyk},
  {Zezas}, {Heinke}, \& {Wargelin}}]{Park2006}
{Park}, T., {Kashyap}, V.~L., {Siemiginowska}, A., {et~al.} 2006, \apj, 652,
  610

\bibitem[{{Pastor-Marazuela} {et~al.}(2020){Pastor-Marazuela}, {Webb},
  {Wojtowicz}, \& {van Leeuwen}}]{Pastor2020}
{Pastor-Marazuela}, I., {Webb}, N.~A., {Wojtowicz}, D.~T., \& {van Leeuwen}, J.
  2020, \aap, 640, A124

\bibitem[{Peacock(1983)}]{Peacock1983}
Peacock, J.~A. 1983, \mnras, 202, 615

\bibitem[{{Peng} {et~al.}(2010){Peng}, {Lilly}, {Kova{\v{c}}}, {Bolzonella},
  {Pozzetti}, {Renzini}, {Zamorani}, {Ilbert}, {Knobel}, {Iovino}, {Maier},
  {Cucciati}, {Tasca}, {Carollo}, {Silverman}, {Kampczyk}, {de Ravel},
  {Sanders}, {Scoville}, {Contini}, {Mainieri}, {Scodeggio}, {Kneib}, {Le
  F{\`e}vre}, {Bardelli}, {Bongiorno}, {Caputi}, {Coppa}, {de la Torre},
  {Franzetti}, {Garilli}, {Lamareille}, {Le Borgne}, {Le Brun}, {Mignoli},
  {Perez Montero}, {Pello}, {Ricciardelli}, {Tanaka}, {Tresse}, {Vergani},
  {Welikala}, {Zucca}, {Oesch}, {Abbas}, {Barnes}, {Bordoloi}, {Bottini},
  {Cappi}, {Cassata}, {Cimatti}, {Fumana}, {Hasinger}, {Koekemoer},
  {Leauthaud}, {Maccagni}, {Marinoni}, {McCracken}, {Memeo}, {Meneux}, {Nair},
  {Porciani}, {Presotto}, \& {Scaramella}}]{Peng2010}
{Peng}, Y.-j., {Lilly}, S.~J., {Kova{\v{c}}}, K., {et~al.} 2010, \apj, 721, 193

\bibitem[{{Peng} {et~al.}(2019){Peng}, {Yang}, {Shen}, {Wang}, {Zou}, \&
  {Zhang}}]{Peng2019}
{Peng}, Z.-K., {Yang}, Y.-S., {Shen}, R.-F., {et~al.} 2019, \apjl, 884, L34

\bibitem[{Peretz \& Behar(2018)}]{Peretz2018}
Peretz, U. \& Behar, E. 2018, \mnras, 481, 3563

\bibitem[{Pescalli {et~al.}(2015)Pescalli, Ghirlanda, Salafia, Ghisellini,
  Nappo, \& Salvaterra}]{Pescalli2015}
Pescalli, A., Ghirlanda, G., Salafia, O.~S., {et~al.} 2015, \mnras, 447, 1911

\bibitem[{{Phillips} {et~al.}(2020){Phillips}, {Tremblin}, {Baraffe},
  {Chabrier}, {Allard}, {Spiegelman}, {Goyal}, {Drummond}, \&
  {H{\'e}brard}}]{Phillips2020}
{Phillips}, M.~W., {Tremblin}, P., {Baraffe}, I., {et~al.} 2020, \aap, 637, A38

\bibitem[{{Piran}(2004)}]{Piran2004}
{Piran}, T. 2004, Reviews of Modern Physics, 76, 1143

\bibitem[{{Pradhan} {et~al.}(2020){Pradhan}, {Falcone}, {Kennea}, \&
  {Burrows}}]{Pradhan2020}
{Pradhan}, P., {Falcone}, A.~D., {Kennea}, J.~A., \& {Burrows}, D.~N. 2020,
  Journal of Astronomical Telescopes, Instruments, and Systems, 6, 038002

\bibitem[{{Pye} {et~al.}(2015){Pye}, {Rosen}, {Fyfe}, \&
  {Schr{\"o}der}}]{Pye2015}
{Pye}, J.~P., {Rosen}, S., {Fyfe}, D., \& {Schr{\"o}der}, A.~C. 2015, \aap,
  581, A28

\bibitem[{{Racusin} {et~al.}(2009){Racusin}, {Liang}, {Burrows}, {Falcone},
  {Sakamoto}, {Zhang}, {Zhang}, {Evans}, \& {Osborne}}]{Racusin2009}
{Racusin}, J.~L., {Liang}, E.~W., {Burrows}, D.~N., {et~al.} 2009, \apj, 698,
  43

\bibitem[{{Rau} {et~al.}(2005){Rau}, {Kienlin}, {Hurley}, \&
  {Lichti}}]{Rau2005}
{Rau}, A., {Kienlin}, A.~V., {Hurley}, K., \& {Lichti}, G.~G. 2005, \aap, 438,
  1175

\bibitem[{{Rau} {et~al.}(2016){Rau}, {Nandra}, {Aird}, {Comastri}, {Dauser},
  {Merloni}, {Pratt}, {Reiprich}, {Fabian}, {Georgakakis}, {G{\"u}del},
  {R{\'o}{\.Z}a{\'n}ska}, {Sanders}, {Sasaki}, {Vaughan}, {Wilms}, \&
  {Meidinger}}]{Rau2016}
{Rau}, A., {Nandra}, K., {Aird}, J., {et~al.} 2016, in Society of Photo-Optical
  Instrumentation Engineers (SPIE) Conference Series, Vol. 9905, Space
  Telescopes and Instrumentation 2016: Ultraviolet to Gamma Ray, ed. J.-W.~A.
  {den Herder}, T.~{Takahashi}, \& M.~{Bautz}, 99052B

\bibitem[{{Rees}(1988)}]{Rees1988}
{Rees}, M.~J. 1988, \nat, 333, 523

\bibitem[{{Reines} {et~al.}(2013){Reines}, {Greene}, \& {Geha}}]{Reines2013}
{Reines}, A.~E., {Greene}, J.~E., \& {Geha}, M. 2013, \apj, 775, 116

\bibitem[{{Rejkuba} {et~al.}(2011){Rejkuba}, {Harris}, {Greggio}, \&
  {Harris}}]{Rejkuba2011}
{Rejkuba}, M., {Harris}, W.~E., {Greggio}, L., \& {Harris}, G.~L.~H. 2011,
  \aap, 526, A123

\bibitem[{{Remillard} \& {McClintock}(2006)}]{Remillard2006}
{Remillard}, R.~A. \& {McClintock}, J.~E. 2006, \araa, 44, 49

\bibitem[{{Rhode} {et~al.}(2007){Rhode}, {Zepf}, {Kundu}, \&
  {Larner}}]{Rhode2007}
{Rhode}, K.~L., {Zepf}, S.~E., {Kundu}, A., \& {Larner}, A.~N. 2007, \aj, 134,
  1403

\bibitem[{{Robrade} {et~al.}(2010){Robrade}, {Poppenhaeger}, \&
  {Schmitt}}]{Robrade2010}
{Robrade}, J., {Poppenhaeger}, K., \& {Schmitt}, J.~H.~M.~M. 2010, \aap, 513,
  A12

\bibitem[{{Rosen} {et~al.}(2016){Rosen}, {Webb}, {Watson}, {Ballet}, {Barret},
  {Braito}, {Carrera}, {Ceballos}, {Coriat}, {Della Ceca}, {Denkinson},
  {Esquej}, {Farrell}, {Freyberg}, {Gris{\'e}}, {Guillout}, {Heil},
  {Koliopanos}, {Law-Green}, {Lamer}, {Lin}, {Martino}, {Michel}, {Motch},
  {Nebot Gomez-Moran}, {Page}, {Page}, {Page}, {Pakull}, {Pye}, {Read},
  {Rodriguez}, {Sakano}, {Saxton}, {Schwope}, {Scott}, {Sturm}, {Traulsen},
  {Yershov}, \& {Zolotukhin}}]{Rosen2016}
{Rosen}, S.~R., {Webb}, N.~A., {Watson}, M.~G., {et~al.} 2016, \aap, 590, A1

\bibitem[{{Rots} \& {Budav{\'a}ri}(2011)}]{Rots2010}
{Rots}, A.~H. \& {Budav{\'a}ri}, T. 2011, \apjs, 192, 8

\bibitem[{{Rowlinson} {et~al.}(2013){Rowlinson}, {O'Brien}, {Metzger},
  {Tanvir}, \& {Levan}}]{Rowlinson2013}
{Rowlinson}, A., {O'Brien}, P.~T., {Metzger}, B.~D., {Tanvir}, N.~R., \&
  {Levan}, A.~J. 2013, \mnras, 430, 1061

\bibitem[{Rowlinson {et~al.}(2010)Rowlinson, O'Brien, Tanvir, Zhang, Evans,
  Lyons, Levan, Willingale, Page, Onal, Burrows, Beardmore, Ukwatta, Berger,
  Hjorth, Fruchter, Tunnicliffe, Fox, \& Cucchiara}]{Rowlinson2010}
Rowlinson, A., O'Brien, P.~T., Tanvir, N.~R., {et~al.} 2010, \mnras, 409, 531

\bibitem[{{Sakamoto} {et~al.}(2008){Sakamoto}, {Barthelmy}, {Barbier},
  {Cummings}, {Fenimore}, {Gehrels}, {Hullinger}, {Krimm}, {Markwardt},
  {Palmer}, {Parsons}, {Sato}, {Stamatikos}, {Tueller}, {Ukwatta}, \&
  {Zhang}}]{Sakamoto2008}
{Sakamoto}, T., {Barthelmy}, S.~D., {Barbier}, L., {et~al.} 2008, \apjs, 175,
  179

\bibitem[{{Sapir} {et~al.}(2013){Sapir}, {Katz}, \& {Waxman}}]{Sapir2013}
{Sapir}, N., {Katz}, B., \& {Waxman}, E. 2013, \apj, 774, 79

\bibitem[{{Sarin} {et~al.}(2021){Sarin}, {Ashton}, {Lasky}, {Ackley}, {Mong},
  \& {Galloway}}]{Sarin2021}
{Sarin}, N., {Ashton}, G., {Lasky}, P.~D., {et~al.} 2021, arXiv e-prints,
  arXiv:2105.10108

\bibitem[{{Saxton} {et~al.}(2021){Saxton}, {Komossa}, {Auchettl}, \&
  {Jonker}}]{Saxton2021}
{Saxton}, R., {Komossa}, S., {Auchettl}, K., \& {Jonker}, P.~G. 2021, \ssr,
  217, 18

\bibitem[{{Saxton} {et~al.}(2008){Saxton}, {Read}, {Esquej}, {Freyberg},
  {Altieri}, \& {Bermejo}}]{Saxton2008}
{Saxton}, R.~D., {Read}, A.~M., {Esquej}, P., {et~al.} 2008, \aap, 480, 611

\bibitem[{{Sazonov} {et~al.}(2021){Sazonov}, {Gilfanov}, {Medvedev}, {Yao},
  {Khorunzhev}, {Semena}, {Sunyaev}, {Burenin}, {Lyapin}, {Meshcheryakov},
  {Uskov}, {Zaznobin}, {Postnov}, {Dodin}, {Belinski}, {Cherepashchuk},
  {Eselevich}, {Dodonov}, {Grokhovskaya}, {Kotov}, {Bikmaev}, {Zhuchkov},
  {Gumerov}, {van Velzen}, \& {Kulkarni}}]{Sazonov2021}
{Sazonov}, S., {Gilfanov}, M., {Medvedev}, P., {et~al.} 2021, \mnras, 508, 3820

\bibitem[{{Sazonov} \& {Khabibullin}(2017)}]{Sazonov2017}
{Sazonov}, S. \& {Khabibullin}, I. 2017, \mnras, 466, 1019

\bibitem[{{Schlafly} \& {Finkbeiner}(2011)}]{Schlafly2011}
{Schlafly}, E.~F. \& {Finkbeiner}, D.~P. 2011, \apj, 737, 103

\bibitem[{{Schlafly} {et~al.}(2019){Schlafly}, {Meisner}, \&
  {Green}}]{Schlafly2019}
{Schlafly}, E.~F., {Meisner}, A.~M., \& {Green}, G.~M. 2019, \apjs, 240, 30

\bibitem[{{Schmitt} \& {Liefke}(2004)}]{Schmitt2004}
{Schmitt}, J.~H.~M.~M. \& {Liefke}, C. 2004, \aap, 417, 651

\bibitem[{{Sivakoff} {et~al.}(2005){Sivakoff}, {Sarazin}, \&
  {Jord{\'a}n}}]{Sivakoff2005}
{Sivakoff}, G.~R., {Sarazin}, C.~L., \& {Jord{\'a}n}, A. 2005, \apjl, 624, L17

\bibitem[{{Skrutskie} {et~al.}(2006){Skrutskie}, {Cutri}, {Stiening},
  {Weinberg}, {Schneider}, {Carpenter}, {Beichman}, {Capps}, {Chester},
  {Elias}, {Huchra}, {Liebert}, {Lonsdale}, {Monet}, {Price}, {Seitzer},
  {Jarrett}, {Kirkpatrick}, {Gizis}, {Howard}, {Evans}, {Fowler}, {Fullmer},
  {Hurt}, {Light}, {Kopan}, {Marsh}, {McCallon}, {Tam}, {Van Dyk}, \&
  {Wheelock}}]{Skrutskie2006}
{Skrutskie}, M.~F., {Cutri}, R.~M., {Stiening}, R., {et~al.} 2006, \aj, 131,
  1163

\bibitem[{{Soderberg} {et~al.}(2008){Soderberg}, {Berger}, {Page}, {Schady},
  {Parrent}, {Pooley}, {Wang}, {Ofek}, {Cucchiara}, {Rau}, {Waxman}, {Simon},
  {Bock}, {Milne}, {Page}, {Barentine}, {Barthelmy}, {Beardmore}, {Bietenholz},
  {Brown}, {Burrows}, {Burrows}, {Bryngelson}, {Cenko}, {Chand ra}, {Cummings},
  {Fox}, {Gal-Yam}, {Gehrels}, {Immler}, {Kasliwal}, {Kong}, {Krimm},
  {Kulkarni}, {Maccarone}, {M{\'e}sz{\'a}ros}, {Nakar}, {O'Brien}, {Overzier},
  {de Pasquale}, {Racusin}, {Rea}, \& {York}}]{Soderberg2008}
{Soderberg}, A.~M., {Berger}, E., {Page}, K.~L., {et~al.} 2008, \nat, 454, 246

\bibitem[{{Sorce} {et~al.}(2014){Sorce}, {Tully}, {Courtois}, {Jarrett},
  {Neill}, \& {Shaya}}]{Sorce2014}
{Sorce}, J.~G., {Tully}, R.~B., {Courtois}, H.~M., {et~al.} 2014, \mnras, 444,
  527

\bibitem[{{Starling} {et~al.}(2011){Starling}, {Wiersema}, {Levan}, {Sakamoto},
  {Bersier}, {Goldoni}, {Oates}, {Rowlinson}, {Campana}, {Sollerman}, {Tanvir},
  {Malesani}, {Fynbo}, {Covino}, {D'Avanzo}, {O'Brien}, {Page}, {Osborne},
  {Vergani}, {Barthelmy}, {Burrows}, {Cano}, {Curran}, {de Pasquale}, {D'Elia},
  {Evans}, {Flores}, {Fruchter}, {Garnavich}, {Gehrels}, {Gorosabel}, {Hjorth},
  {Holland}, {van der Horst}, {Hurkett}, {Jakobsson}, {Kamble}, {Kouveliotou},
  {Kuin}, {Kaper}, {Mazzali}, {Nugent}, {Pian}, {Stamatikos}, {Th{\"o}ne}, \&
  {Woosley}}]{Starling2011}
{Starling}, R.~L.~C., {Wiersema}, K., {Levan}, A.~J., {et~al.} 2011, \mnras,
  411, 2792

\bibitem[{Stiele {et~al.}(2012)Stiele, Muñoz-Darias, Motta, \&
  Belloni}]{Stiele2012}
Stiele, H., Muñoz-Darias, T., Motta, S., \& Belloni, T.~M. 2012, \mnras, 422,
  679

\bibitem[{{Stratta} {et~al.}(2013){Stratta}, {Gendre}, {Atteia}, {Bo{\"e}r},
  {Coward}, {De Pasquale}, {Howell}, {Klotz}, {Oates}, \& {Piro}}]{Stratta2013}
{Stratta}, G., {Gendre}, B., {Atteia}, J.~L., {et~al.} 2013, \apj, 779, 66

\bibitem[{{Strohmayer} \& {Watts}(2005)}]{Strohmayer2005}
{Strohmayer}, T.~E. \& {Watts}, A.~L. 2005, \apjl, 632, L111

\bibitem[{{Sun} {et~al.}(2019){Sun}, {Li}, {Zhang}, {Zhang}, {Bauer}, {Xue}, \&
  {Yuan}}]{Sun2019}
{Sun}, H., {Li}, Y., {Zhang}, B.-B., {et~al.} 2019, \apj, 886, 129

\bibitem[{{Sun} {et~al.}(2015){Sun}, {Zhang}, \& {Li}}]{Sun2015}
{Sun}, H., {Zhang}, B., \& {Li}, Z. 2015, \apj, 812, 33

\bibitem[{{Swartz} {et~al.}(2011){Swartz}, {Soria}, {Tennant}, \&
  {Yukita}}]{Swartz2011}
{Swartz}, D.~A., {Soria}, R., {Tennant}, A.~F., \& {Yukita}, M. 2011, \apj,
  741, 49

\bibitem[{{Taggart} \& {Perley}(2021)}]{Taggart2019}
{Taggart}, K. \& {Perley}, D.~A. 2021, \mnras, 503, 3931

\bibitem[{{Tamba} {et~al.}(2019){Tamba}, {Bamba}, {Odaka}, \&
  {Enoto}}]{Tamba2019}
{Tamba}, T., {Bamba}, A., {Odaka}, H., \& {Enoto}, T. 2019, \pasj, 71, 90

\bibitem[{Tananbaum {et~al.}(2014)Tananbaum, Weisskopf, Tucker, Wilkes, \&
  Edmonds}]{Tananbaum2014}
Tananbaum, H., Weisskopf, M., Tucker, W., Wilkes, B., \& Edmonds, P. 2014,
  Reports on Progress in Physics, 77, 066902

\bibitem[{{Tanikawa} {et~al.}(2021){Tanikawa}, {Giersz}, \& {Arca
  Sedda}}]{Tanikawa2021}
{Tanikawa}, A., {Giersz}, M., \& {Arca Sedda}, M. 2021, arXiv e-prints,
  arXiv:2103.14185

\bibitem[{{Teplitz} {et~al.}(2010){Teplitz}, {Capak}, {Brooke}, {Shenoy},
  {Brinkworth}, {Desai}, {Khan}, \& {Laher}}]{Teplitz2010}
{Teplitz}, H.~I., {Capak}, P., {Brooke}, T., {et~al.} 2010, Astronomical
  Society of the Pacific Conference Series, Vol. 434, {The Spitzer Source
  List}, ed. Y.~{Mizumoto}, K.~I. {Morita}, \& M.~{Ohishi}, 437

\bibitem[{Terasawa {et~al.}(2005)Terasawa, Tanaka, Takei, Kawai, Yoshida,
  Nomoto, Yoshikawa, Saito, Kasaba, Takashima, {et~al.}}]{Terasawa2005}
Terasawa, T., Tanaka, Y.~T., Takei, Y., {et~al.} 2005, Nature, 434, 1110

\bibitem[{{Th{\"o}ne} {et~al.}(2011){Th{\"o}ne}, {de Ugarte Postigo}, {Fryer},
  {Page}, {Gorosabel}, {Aloy}, {Perley}, {Kouveliotou}, {Janka}, {Mimica},
  {Racusin}, {Krimm}, {Cummings}, {Oates}, {Holland}, {Siegel}, {de Pasquale},
  {Sonbas}, {Im}, {Park}, {Kann}, {Guziy}, {Hern{\'a}ndez-Garc{\'\i}a},
  {Llorente}, {Bundy}, {Choi}, {Jeong}, {Korhonen}, {Kub{\`a}nek}, {Lim},
  {Moskvitin}, {Mu{\~n}oz-Darias}, {Pak}, \& {Parrish}}]{Thone2011}
{Th{\"o}ne}, C.~C., {de Ugarte Postigo}, A., {Fryer}, C.~L., {et~al.} 2011,
  \nat, 480, 72

\bibitem[{{Tinney} {et~al.}(2014){Tinney}, {Faherty}, {Kirkpatrick}, {Cushing},
  {Morley}, \& {Wright}}]{Tinney2014}
{Tinney}, C.~G., {Faherty}, J.~K., {Kirkpatrick}, J.~D., {et~al.} 2014, \apj,
  796, 39

\bibitem[{{Traulsen} {et~al.}(2019){Traulsen}, {Schwope}, {Lamer}, {Ballet},
  {Carrera}, {Coriat}, {Freyberg}, {Michel}, {Motch}, {Rosen}, {Webb},
  {Ceballos}, {Koliopanos}, {Kurpas}, {Page}, \& {Watson}}]{Traulsen2019}
{Traulsen}, I., {Schwope}, A.~D., {Lamer}, G., {et~al.} 2019, \aap, 624, A77

\bibitem[{{Troja} {et~al.}(2019){Troja}, {Castro-Tirado}, {Becerra
  Gonz{\'a}lez}, {Hu}, {Ryan}, {Cenko}, {Ricci}, {Novara},
  {S{\'a}nchez-R{\'a}mirez}, {Acosta-Pulido}, {Ackley}, {Caballero
  Garc{\'\i}a}, {Eikenberry}, {Guziy}, {Jeong}, {Lien}, {M{\'a}rquez},
  {Pandey}, {Park}, {Sakamoto}, {Tello}, {Sokolov}, {Sokolov}, {Tiengo},
  {Valeev}, {Zhang}, \& {Veilleux}}]{Troja2019}
{Troja}, E., {Castro-Tirado}, A.~J., {Becerra Gonz{\'a}lez}, J., {et~al.} 2019,
  \mnras, 489, 2104

\bibitem[{{Troja} {et~al.}(2007){Troja}, {Cusumano}, {O'Brien}, {Zhang},
  {Sbarufatti}, {Mangano}, {Willingale}, {Chincarini}, {Osborne}, {Marshall},
  {Burrows}, {Campana}, {Gehrels}, {Guidorzi}, {Krimm}, {La Parola}, {Liang},
  {Mineo}, {Moretti}, {Page}, {Romano}, {Tagliaferri}, {Zhang}, {Page}, \&
  {Schady}}]{Troja2007}
{Troja}, E., {Cusumano}, G., {O'Brien}, P.~T., {et~al.} 2007, \apj, 665, 599

\bibitem[{{Troja} {et~al.}(2022){Troja}, {O'Connor}, {Ryan}, {Piro}, {Ricci},
  {Zhang}, {Piran}, {Bruni}, {Cenko}, \& {van Eerten}}]{Troja2022}
{Troja}, E., {O'Connor}, B., {Ryan}, G., {et~al.} 2022, \mnras, 510, 1902

\bibitem[{{Troja} {et~al.}(2020){Troja}, {van Eerten}, {Zhang}, {Ryan}, {Piro},
  {Ricci}, {O'Connor}, {Wieringa}, {Cenko}, \& {Sakamoto}}]{Troja2020}
{Troja}, E., {van Eerten}, H., {Zhang}, B., {et~al.} 2020, \mnras, 498, 5643

\bibitem[{{Tsvetkov} \& {Bartunov}(1993)}]{Tsvetkov1993}
{Tsvetkov}, D.~Y. \& {Bartunov}, O.~S. 1993, Bulletin d'Information du Centre
  de Donnees Stellaires, 42, 17

\bibitem[{{Tully} {et~al.}(2013){Tully}, {Courtois}, {Dolphin}, {Fisher},
  {H{\'e}raudeau}, {Jacobs}, {Karachentsev}, {Makarov}, {Makarova},
  {Mitronova}, {Rizzi}, {Shaya}, {Sorce}, \& {Wu}}]{Tully2013}
{Tully}, R.~B., {Courtois}, H.~M., {Dolphin}, A.~E., {et~al.} 2013, \aj, 146,
  86

\bibitem[{{van Buren}(1981)}]{van_Buren1981}
{van Buren}, D. 1981, \apj, 249, 297

\bibitem[{{van den Eijnden} {et~al.}(2018){van den Eijnden}, {Degenaar},
  {Russell}, {Wijnands}, {Miller-Jones}, {Sivakoff}, \& {Hern{\'a}ndez
  Santisteban}}]{van_den_Eijnden2018}
{van den Eijnden}, J., {Degenaar}, N., {Russell}, T.~D., {et~al.} 2018, \nat,
  562, 233

\bibitem[{Virgili {et~al.}(2009)Virgili, Liang, \& Zhang}]{Virgili2009}
Virgili, F.~J., Liang, E.-W., \& Zhang, B. 2009, \mnras, 392, 91

\bibitem[{{Virgili} {et~al.}(2013){Virgili}, {Mundell}, {Pal'shin}, {Guidorzi},
  {Margutti}, {Melandri}, {Harrison}, {Kobayashi}, {Chornock}, {Henden},
  {Updike}, {Cenko}, {Tanvir}, {Steele}, {Cucchiara}, {Gomboc}, {Levan},
  {Cano}, {Mottram}, {Clay}, {Bersier}, {Kopa{\v{c}}}, {Japelj}, {Filippenko},
  {Li}, {Svinkin}, {Golenetskii}, {Hartmann}, {Milne}, {Williams}, {O'Brien},
  {Fox}, \& {Berger}}]{Virgili2013}
{Virgili}, F.~J., {Mundell}, C.~G., {Pal'shin}, V., {et~al.} 2013, \apj, 778,
  54

\bibitem[{{Virgili} {et~al.}(2011){Virgili}, {Zhang}, {O'Brien}, \&
  {Troja}}]{Virgili2011}
{Virgili}, F.~J., {Zhang}, B., {O'Brien}, P., \& {Troja}, E. 2011, \apj, 727,
  109

\bibitem[{{Vito} {et~al.}(2016){Vito}, {Gilli}, {Vignali}, {Brandt},
  {Comastri}, {Yang}, {Lehmer}, {Luo}, {Basu-Zych}, {Bauer}, {Cappelluti},
  {Koekemoer}, {Mainieri}, {Paolillo}, {Ranalli}, {Shemmer}, {Trump}, {Wang},
  \& {Xue}}]{Vito2016}
{Vito}, F., {Gilli}, R., {Vignali}, C., {et~al.} 2016, \mnras, 463, 348

\bibitem[{{von Kienlin} {et~al.}(2014){von Kienlin}, {Meegan}, {Paciesas},
  {Bhat}, {Bissaldi}, {Briggs}, {Burgess}, {Byrne}, {Chaplin}, {Cleveland},
  {Connaughton}, {Collazzi}, {Fitzpatrick}, {Foley}, {Gibby}, {Giles},
  {Goldstein}, {Greiner}, {Gruber}, {Guiriec}, {van der Horst}, {Kouveliotou},
  {Layden}, {McBreen}, {McGlynn}, {Pelassa}, {Preece}, {Rau}, {Tierney},
  {Wilson-Hodge}, {Xiong}, {Younes}, \& {Yu}}]{von_Kienlin2014}
{von Kienlin}, A., {Meegan}, C.~A., {Paciesas}, W.~S., {et~al.} 2014, \apjs,
  211, 13

\bibitem[{{Walton} {et~al.}(2016){Walton}, {F{\"u}rst}, {Bachetti}, {Barret},
  {Brightman}, {Fabian}, {Gehrels}, {Harrison}, {Heida}, {Middleton}, {Rana},
  {Roberts}, {Stern}, {Tao}, \& {Webb}}]{Walton2016}
{Walton}, D.~J., {F{\"u}rst}, F., {Bachetti}, M., {et~al.} 2016, \apjl, 827,
  L13

\bibitem[{{Wanderman} \& {Piran}(2010)}]{Wanderman2010}
{Wanderman}, D. \& {Piran}, T. 2010, \mnras, 406, 1944

\bibitem[{Wanderman \& Piran(2015)}]{Wanderman2015}
Wanderman, D. \& Piran, T. 2015, \mnras, 448, 3026

\bibitem[{{Wang} {et~al.}(2018){Wang}, {Zhu}, {Xu}, {Xin}, {Deng}, {Qiu},
  {Qiu}, {Wang}, {Zhang}, \& {Wei}}]{Wang2018}
{Wang}, J., {Zhu}, Z.~P., {Xu}, D., {et~al.} 2018, \apj, 867, 147

\bibitem[{{Wang} {et~al.}(2016){Wang}, {Liu}, {Qiu}, {Bai}, {Yang}, {Guo}, \&
  {Zhang}}]{Wang2016}
{Wang}, S., {Liu}, J., {Qiu}, Y., {et~al.} 2016, \apjs, 224, 40

\bibitem[{{Wang} {et~al.}(2015){Wang}, {Zhang}, {Liang}, {Gao}, {Li}, {Deng},
  {Qin}, {Tang}, {Kann}, {Ryde}, \& {Kumar}}]{Wang2015}
{Wang}, X.-G., {Zhang}, B., {Liang}, E.-W., {et~al.} 2015, \apjs, 219, 9

\bibitem[{{Warren} {et~al.}(2007){Warren}, {Cross}, {Dye}, {Hambly}, {Almaini},
  {Edge}, {Hewett}, {Hodgkin}, {Irwin}, {Jameson}, {Lawrence}, {Lucas},
  {Mortlock}, {Adamson}, {Bryant}, {Collins}, {Davis}, {Emerson}, {Evans},
  {Gonzales-Solares}, {Hirst}, {Kerr}, {Lewis}, {Mann}, {Rawlings}, {Read},
  {Riello}, {Sutorius}, \& {Varricatt}}]{Warren2007}
{Warren}, S.~J., {Cross}, N.~J.~G., {Dye}, S., {et~al.} 2007, arXiv e-prints,
  astro

\bibitem[{{Waxman} \& {Katz}(2017)}]{Waxman2017}
{Waxman}, E. \& {Katz}, B. 2017, {Shock Breakout Theory}, ed. A.~W. {Alsabti}
  \& P.~{Murdin}, 967

\bibitem[{{Webb} {et~al.}(2020){Webb}, {Coriat}, {Traulsen}, {Ballet}, {Motch},
  {Carrera}, {Koliopanos}, {Authier}, {de la Calle}, {Ceballos}, {Colomo},
  {Chuard}, {Freyberg}, {Garcia}, {Kolehmainen}, {Lamer}, {Lin}, {Maggi},
  {Michel}, {Page}, {Page}, {Perea-Calderon}, {Pineau}, {Rodriguez}, {Rosen},
  {Santos Lleo}, {Saxton}, {Schwope}, {Tom{\'a}s}, {Watson}, \&
  {Zakardjian}}]{Webb2020}
{Webb}, N.~A., {Coriat}, M., {Traulsen}, I., {et~al.} 2020, \aap, 641, A136

\bibitem[{{Welsh} {et~al.}(2007){Welsh}, {Wheatley}, {Seibert}, {Browne},
  {West}, {Siegmund}, {Barlow}, {Forster}, {Friedman}, {Martin}, {Morrissey},
  {Small}, {Wyder}, {Schiminovich}, {Neff}, \& {Rich}}]{Welsh2007}
{Welsh}, B.~Y., {Wheatley}, J.~M., {Seibert}, M., {et~al.} 2007, \apjs, 173,
  673

\bibitem[{{Wenger} {et~al.}(2000){Wenger}, {Ochsenbein}, {Egret}, {Dubois},
  {Bonnarel}, {Borde}, {Genova}, {Jasniewicz}, {Lalo{\"e}}, {Lesteven}, \&
  {Monier}}]{Wenger2000}
{Wenger}, M., {Ochsenbein}, F., {Egret}, D., {et~al.} 2000, \aaps, 143, 9

\bibitem[{{Whitmore} {et~al.}(2016){Whitmore}, {Allam}, {Budav{\'a}ri},
  {Casertano}, {Downes}, {Donaldson}, {Fall}, {Lubow}, {Quick}, {Strolger},
  {Wallace}, \& {White}}]{Whitmore2016}
{Whitmore}, B.~C., {Allam}, S.~S., {Budav{\'a}ri}, T., {et~al.} 2016, \aj, 151,
  134

\bibitem[{{Wiegert} {et~al.}(2015){Wiegert}, {Irwin}, {Miskolczi}, {Schmidt},
  {Mora}, {Damas-Segovia}, {Stein}, {English}, {Rand}, {Santistevan},
  {Walterbos}, {Krause}, {Beck}, {Dettmar}, {Kepley}, {Wezgowiec}, {Wang},
  {Heald}, {Li}, {MacGregor}, {Johnson}, {Strong}, {DeSouza}, \&
  {Porter}}]{Wiegert2015}
{Wiegert}, T., {Irwin}, J., {Miskolczi}, A., {et~al.} 2015, \aj, 150, 81

\bibitem[{{Wiersema} {et~al.}(2007){Wiersema}, {Savaglio}, {Vreeswijk},
  {Ellison}, {Ledoux}, {Yoon}, {M{\o}ller}, {Sollerman}, {Fynbo}, {Pian},
  {Starling}, \& {Wijers}}]{Wiersema2007}
{Wiersema}, K., {Savaglio}, S., {Vreeswijk}, P.~M., {et~al.} 2007, \aap, 464,
  529

\bibitem[{Wilkes \& Tucker(2019)}]{Wilkes2019}
Wilkes, B. \& Tucker, W., eds. 2019, The Chandra X-ray Observatory, 2514-3433
  (IOP Publishing)

\bibitem[{{Woods} \& {Thompson}(2006)}]{Woods2006}
{Woods}, P.~M. \& {Thompson}, C. 2006, {Soft gamma repeaters and anomalous
  X-ray pulsars: magnetar candidates}, Vol.~39, 547--586

\bibitem[{{Wright} {et~al.}(2010){Wright}, {Eisenhardt}, {Mainzer}, {Ressler},
  {Cutri}, {Jarrett}, {Kirkpatrick}, {Padgett}, {McMillan}, {Skrutskie},
  {Stanford}, {Cohen}, {Walker}, {Mather}, {Leisawitz}, {Gautier}, {McLean},
  {Benford}, {Lonsdale}, {Blain}, {Mendez}, {Irace}, {Duval}, {Liu}, {Royer},
  {Heinrichsen}, {Howard}, {Shannon}, {Kendall}, {Walsh}, {Larsen}, {Cardon},
  {Schick}, {Schwalm}, {Abid}, {Fabinsky}, {Naes}, \& {Tsai}}]{Wright2010}
{Wright}, E.~L., {Eisenhardt}, P. R.~M., {Mainzer}, A.~K., {et~al.} 2010, \apj,
  140, 1868

\bibitem[{{Xiao} {et~al.}(2019){Xiao}, {Zhang}, \& {Dai}}]{Xiao2019}
{Xiao}, D., {Zhang}, B.-B., \& {Dai}, Z.-G. 2019, \apjl, 879, L7

\bibitem[{{Xu} {et~al.}(2008){Xu}, {Watson}, {Fynbo}, {Fan}, {Zou}, \&
  {Hjorth}}]{Xu2008}
{Xu}, D., {Watson}, D., {Fynbo}, J., {et~al.} 2008, in 37th COSPAR Scientific
  Assembly, Vol.~37, 3512

\bibitem[{{Xue} {et~al.}(2016){Xue}, {Luo}, {Brandt}, {Alexander}, {Bauer},
  {Lehmer}, \& {Yang}}]{Xue2016}
{Xue}, Y.~Q., {Luo}, B., {Brandt}, W.~N., {et~al.} 2016, \apjs, 224, 15

\bibitem[{{Xue} {et~al.}(2019){Xue}, {Zheng}, {Li}, {Brandt}, {Zhang}, {Luo},
  {Zhang}, {Bauer}, {Sun}, {Lehmer}, {Wu}, {Yang}, {Kong}, {Li}, {Sun}, {Wang},
  \& {Vito}}]{Xue2019}
{Xue}, Y.~Q., {Zheng}, X.~C., {Li}, Y., {et~al.} 2019, \nat, 568, 198

\bibitem[{{Yang} {et~al.}(2016){Yang}, {Brandt}, {Luo}, {Xue}, {Bauer}, {Sun},
  {Kim}, {Schulze}, {Zheng}, {Paolillo}, {Shemmer}, {Liu}, {Schneider},
  {Vignali}, {Vito}, \& {Wang}}]{Yang2016}
{Yang}, G., {Brandt}, W.~N., {Luo}, B., {et~al.} 2016, \apj, 831, 145

\bibitem[{{Yang} {et~al.}(2019){Yang}, {Brandt}, {Zhu}, {Bauer}, {Luo}, {Xue},
  \& {Zheng}}]{Yang2019}
{Yang}, G., {Brandt}, W.~N., {Zhu}, S.~F., {et~al.} 2019, \mnras, 1535

\bibitem[{{Yi} {et~al.}(2014){Yi}, {Dai}, {Wu}, \& {Wang}}]{Yi2014}
{Yi}, S.~X., {Dai}, Z.~G., {Wu}, X.~F., \& {Wang}, F.~Y. 2014, arXiv e-prints,
  arXiv:1401.1601

\bibitem[{{Yi} {et~al.}(2016){Yi}, {Xi}, {Yu}, {Wang}, {Mu}, {L{\"u}}, \&
  {Liang}}]{Yi2016}
{Yi}, S.-X., {Xi}, S.-Q., {Yu}, H., {et~al.} 2016, \apjs, 224, 20

\bibitem[{Yuan {et~al.}(2015)Yuan, Zhang, Feng, Zhang, Ling, Zhao, Deng, Qiu,
  Osborne, O'Brien, Willingale, \& Lapington}]{Yuan2015}
Yuan, W., Zhang, C., Feng, H., {et~al.} 2015, PoS, SWIFT 10, 006

\bibitem[{{Yuan} {et~al.}(2017){Yuan}, {Zhang}, {Ling}, {Zhao}, {Chen}, {Lu},
  \& {Zhang}}]{Yuan2017}
{Yuan}, W., {Zhang}, C., {Ling}, Z., {et~al.} 2017, in The X-ray Universe 2017,
  ed. J.-U. {Ness} \& S.~{Migliari}, 240

\bibitem[{{Zhang}(2013)}]{Zhang2013}
{Zhang}, B. 2013, \apjl, 763, L22

\bibitem[{{Zhang}(2018)}]{Zhang_book_2018}
{Zhang}, B. 2018, {The Physics of Gamma-Ray Bursts} (Cambridge University
  Press)

\bibitem[{{Zhang} {et~al.}(2018){Zhang}, {Zhang}, {Sun}, {Lei}, {Gao}, {Li},
  {Shao}, {Zhao}, {Hu}, {L{\"u}}, {Wu}, {Fan}, {Wang}, {Castro-Tirado},
  {Zhang}, {Yu}, {Cao}, \& {Liang}}]{Zhang2018}
{Zhang}, B.~B., {Zhang}, B., {Sun}, H., {et~al.} 2018, Nature Communications,
  9, 447

\bibitem[{{Zheng} {et~al.}(2017){Zheng}, {Xue}, {Brandt}, {Li}, {Paolillo},
  {Yang}, {Zhu}, {Luo}, {Sun}, {Hughes}, {Bauer}, {Vito}, {Wang}, {Liu},
  {Vignali}, \& {Shu}}]{Zheng2017}
{Zheng}, X.~C., {Xue}, Y.~Q., {Brandt}, W.~N., {et~al.} 2017, \apj, 849, 127

\end{thebibliography}

%-------------------------------------------------------------
%                 A figure as large as the width of the column
%-------------------------------------------------------------

\begin{appendix} %First appendix

\section{Spatial location and duration of X-ray events}

\begin{figure*}
    \centering
    \includegraphics[scale=0.6]{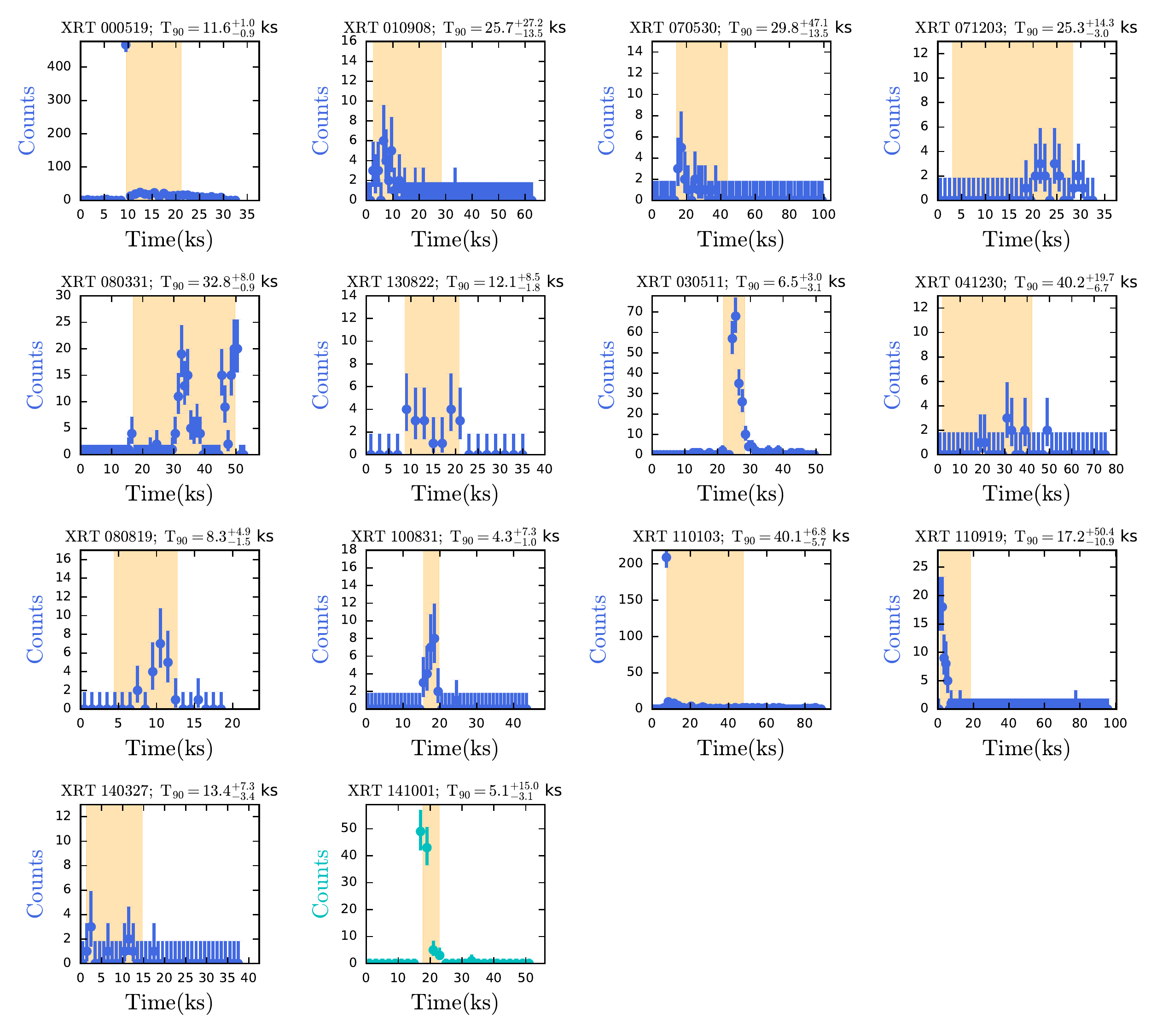}
    \vspace{-0.5cm}
    \caption{
    Light curves for each FXRT candidate in units of counts and the region covering the $T_{90}$ (which measures the time over which the event emits from 5\% to 95\% of its total measured counts; \emph{orange region}) The light curves have a bin width of 1~ks.}
    \label{fig:duration_t90}
\end{figure*}

To estimate the duration of the final sample of FXRTs, we computed the $T_{90}$ duration parameter. $T_{90}$ measures the time over which the event emits from 5\% to 95\% of its total measured counts (in the 0.5--7.0~keV band in our case). Figure~\ref{fig:duration_t90} shows the $T_{90}$ duration (\emph{orange region}) for each event, as well as their light curves (with a bin time of 1~ks) in unit of counts.

Furthermore, Fig.~\ref{fig:lissajaus} confirms that the final sample of FXRT candidates are real celestial sources in the sky rather than detector artifacts. Due to \emph{Chandra}'s Lissajous dither pattern, executed during observation, the X-ray photons of the FXRTs are distributed over dozens to hundreds of individual pixels on the detector. The \emph{first column} of the figure shows the light curves, color-coded by the phase in the light curve evolution. The \emph{second column} shows the spatial location in $x$ and $y$ chip detector coordinates, also color-coded by time, tracing out a sinusoidal-like evolution in $x$ and $y$ coordinates over time. The \emph{third and fourth columns} show the $x$ and $y$ position changes (in \emph{blue} and \emph{purple}, respectively, over time, with the light curve superimposed in dark gray.

\begin{figure*}
    \centering
    \includegraphics[width=17cm,height=22cm]{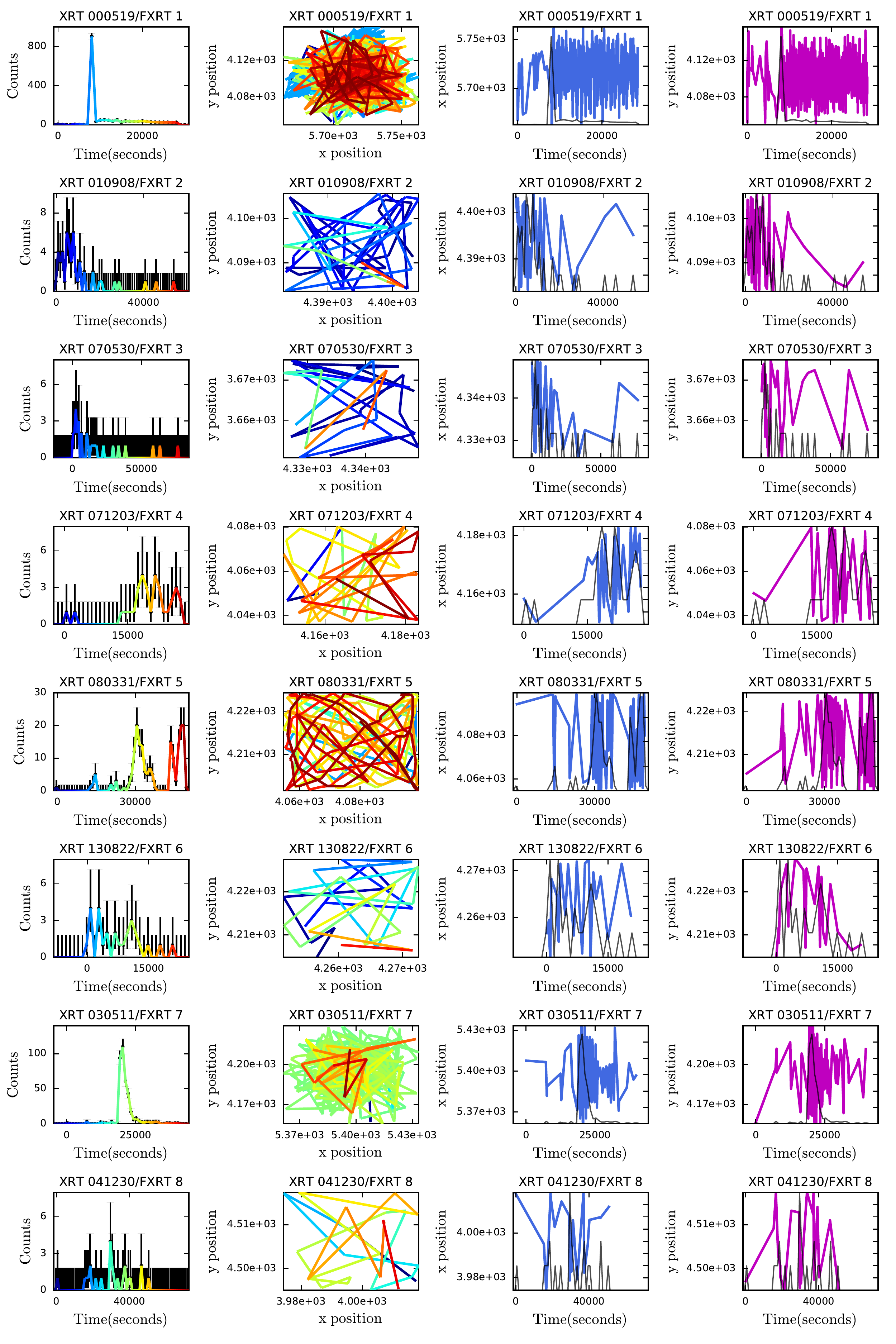}
    \vspace{-0.3cm}
    \caption{Lissajous dither pattern in detector coordinates. \emph{First column:} FXRT 0.5--7.0~keV light curves in count units, color-coded as a function of time. 
    \emph{Second column:} \emph{Chandra} 0.5--7.0~keV images in detector coordinates, with the same color-coding as a function of time, demonstrating the temporal movement of the source on the detector in response to the Lissajous dither pattern. A flaring pixel would appear as a point on these plots. \emph{Third and fourth columns:} x (\emph{blue}) and y (\emph{purple}) detector coordinates, respectively, of the detected X-ray photons from the FXRTs as a function of time, with the candidate light curves superimposed as solid dark gray lines.}
    \label{fig:lissajaus}
\end{figure*}

\begin{figure*}
    \centering
    \ContinuedFloat
    \includegraphics[width=17cm,height=16.5cm]{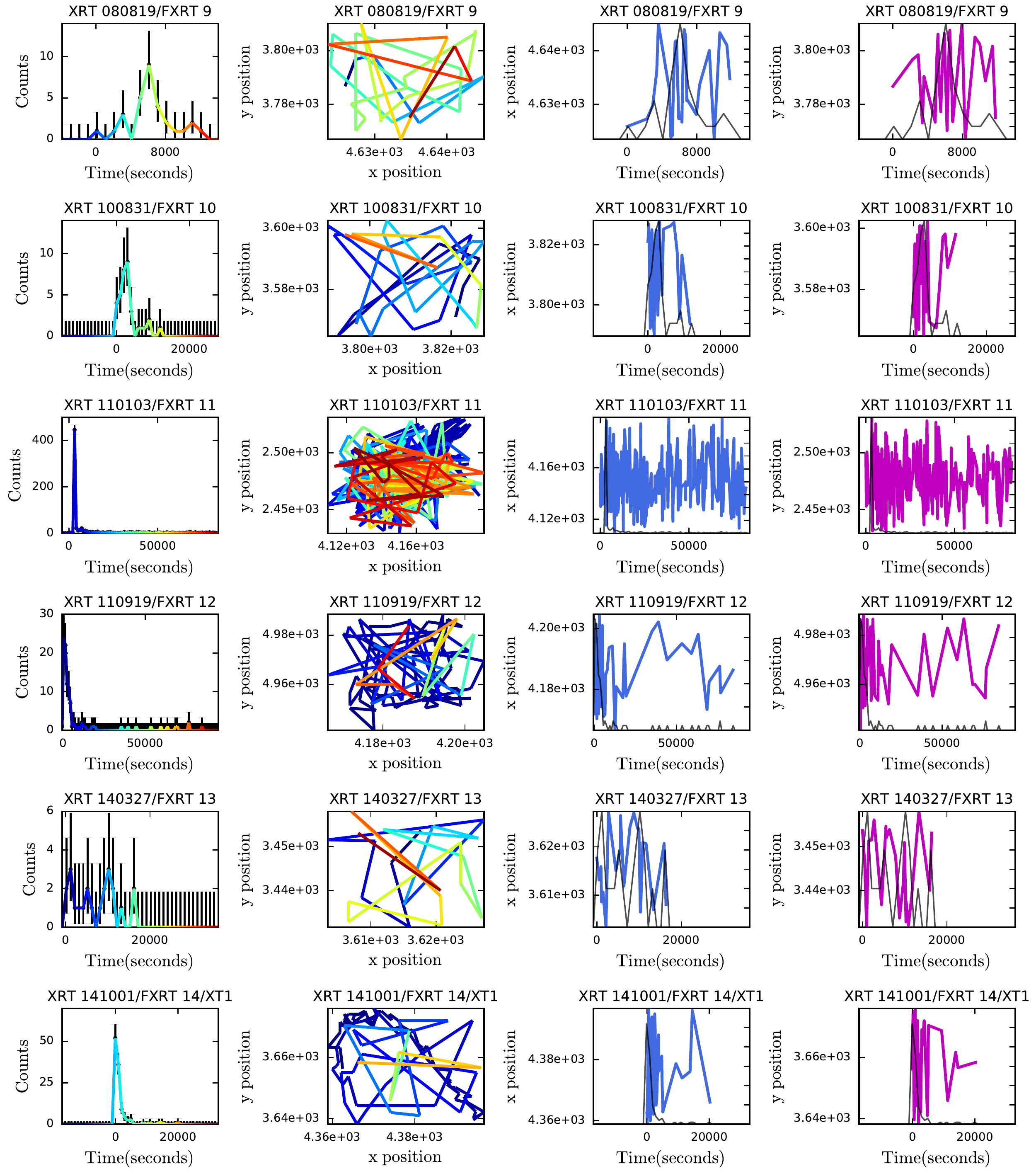}
    \caption[]{(continued)}
\end{figure*}

\section{Color-magnitude diagram of stellar matches}

\begin{figure*}
    \centering
    \includegraphics[width=17cm,height=7.0cm]{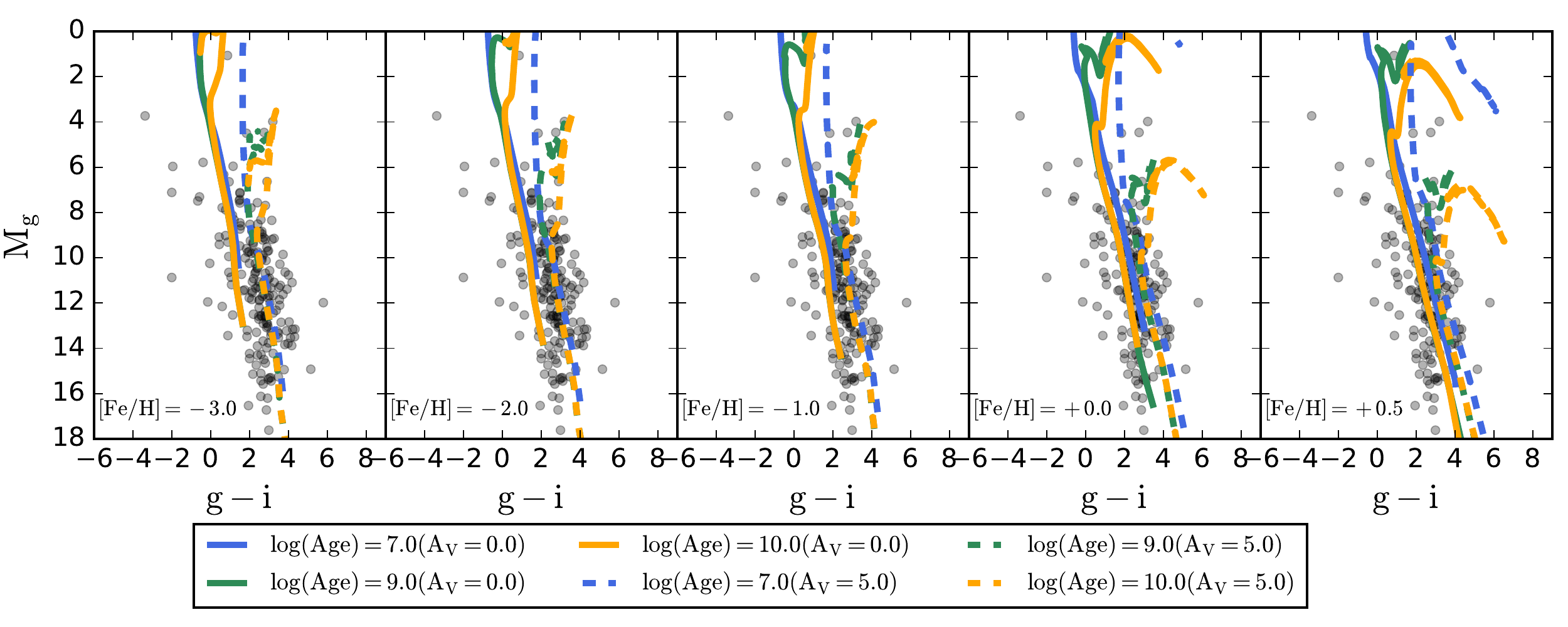}
    \vspace{-0.3cm}
    \caption{Color-magnitude diagrams, considering only Pan-STARRS and DECam counterparts (\emph{gray background points}) of X-ray sources classified as stars according to {Criterion 2} (see Sect. \ref{sec:gaia}). As a comparison, we overplot isochrones with different ages (from $\log(\rm{Age}){=}7.0-10.0$) taken from the MIST package \citep{Dotter2016,Choi2016}. Each panel represents different metallicities (from $\mathrm{[Fe/H]}{=}-3.0$ to $+0.5$), while \emph{solid} and \emph{dashed lines} are isochrones with attenuations of $A_V{=}$0.0 and 5.0, respectively.}
    \label{fig:color_color}
\end{figure*}

To further demonstrate the stellar-like nature of the star candidates (beyond identification by \emph{Gaia}), we show an example $M_{\rm g}$ versus $g-i$ color-magnitude diagram (see Fig.~\ref{fig:color_color}) considering all Pan-STARRS and DECam counterparts of X-ray sources classified as stars according to  {Criterion 2} (see Sect. \ref{sec:gaia}). Isochrones with different ages (from $\log(\rm{Age}){=}7.0-10.0$) taken from the MIST package \citep{Dotter2016,Choi2016} are overplotted, with each panel representing different metallicities (from $\mathrm{[Fe/H]}{=}-3.0$ to $+0.5$). \emph{Solid} and \emph{dashed lines} denote isochrones with attenuations of $A_V{=}$0.0 and 5.0, respectively. The vast majority of the stars fall on these tracks. According to SIMBAD, the outliers are identified as PNe, YSOs, or emission-line stars. We additionally stress that the Pan-STARRS and DECam colors are not necessarily taken in a purely simultaneous manner; in the case of Pan-STARRS, they are averaged over the duration of the survey, while for DECam they come from only a few disjoint epochs.

\end{appendix}

\end{document}